%% file: thesis.tex
\documentclass[10pt,b5paper,twoside,english]{myphdthesis}
\pdfoutput=1

\usepackage{ThALV} 
\usepackage[utf8]{inputenc} 

\usepackage{palatino} 
\usepackage{babel} 

\usepackage{lettrine}
\usepackage{epigraph} 
\usepackage{verbatim}

\makeatletter
\renewcommand{\@dotsep}{10000} 
\makeatother
\usepackage[nottoc]{tocbibind} 

\usepackage{deluxetable}  

\usepackage[figuresright]{rotating} 
\usepackage{subfigure}
\graphicspath{{./figure/}} 
\usepackage[small,bf]{caption} 
\usepackage{epsfig}
\usepackage[all]{xy}

\usepackage{amsmath,amsfonts,amssymb,amsthm}  
\usepackage{latexsym}
\usepackage{indentfirst}
\usepackage{slashed}
\usepackage{textcomp}
\usepackage{graphicx}
\usepackage{lmodern}
\usepackage{mathrsfs}
\usepackage{amssymb}
\usepackage{color}

\def\utw{\smash{\rlap{\lower5pt\hbox{$\sim$}}}}
\def\udtw{\smash{\rlap{\lower6pt\hbox{$\approx$}}}}

\def\beq{\begin{equation}}
\def\eeq{\end{equation}}
\def\beqa{\begin{eqnarray}}
\def\eeqa{\end{eqnarray}}
\def\a{{\alpha}}

\def\g{{\gamma}}

\def\d{{\delta}}

\def\m{{\mu}}
\def\n{{\nu}}
\def\r{{\rho}}
\def\s{{\sigma}}

\def\bfone{\relax{\rm 1\kern-.35em 1}}

\newcommand{\be}{\begin{equation}}
\newcommand{\ee}{\end{equation}}
\newcommand{\ben}{\begin{displaymath}}
\newcommand{\een}{\end{displaymath}}
\newcommand{\bea}{\begin{eqnarray}}
\newcommand{\eea}{\end{eqnarray}}

\newcommand{\bean}{\begin{eqnarray*}}
\newcommand{\eean}{\end{eqnarray*}}

\newcommand{\ma}[1]{\mbox{$\mathcal{#1}$}}
\newcommand{\mas}[1]{\mbox{$\mathscr{#1}$}}
\newcommand{\masf}[1]{\mbox{$\mathsf{#1}$}}
\newcommand{\maf}[1]{\mbox{$\mathfrak{#1}$}}
\newcommand{\mrm}[1]{\mbox{$\mathrm{#1}$}}
\newcommand{\D}{{\rm d}}

\newlength{\bibitemsep}
\newlength{\bibparskip}
\let\oldthebibliography\thebibliography
\renewcommand\thebibliography[1]{%
  \oldthebibliography{#1}%
  \setlength{\parskip}{\bibitemsep}%
  \setlength{\itemsep}{\bibparskip}%
}

\usepackage{fancyhdr}
\renewcommand{\chaptermark}[1]{\markboth{#1}{}}
\renewcommand{\sectionmark}[1]{\markright{\thesection\ #1}}

\renewcommand{\headrulewidth}{0.2pt}
\renewcommand{\footrulewidth}{0pt}

\fancypagestyle{plain}{
\fancyhf{} 
\fancyfoot[C]{\bfseries \thepage} 
\renewcommand{\headrulewidth}{0pt}
\renewcommand{\footrulewidth}{0pt}}

\begin{document}

\frontmatter
\setcounter{page}{1}
\maketitle

\tableofcontents 
\thispagestyle{plain}


\renewcommand{\chaptermark}[1]{\markboth{#1}{}}
\renewcommand{\sectionmark}[1]{\markright{\thesection\ #1}}
\fancyhead[RO]{\bf \thepage}
\fancyhead[LE]{\bf \thepage}
\fancyhead[LO]{{\it Introduction}}
\fancyhead[RE]{\it Introduction}
\renewcommand{\headrulewidth}{0.2pt}
\renewcommand{\footrulewidth}{0pt}
\addcontentsline{toc}{chapter}{Introduction}
\input{Intro} \thispagestyle{fancy}


\mainmatter
\renewcommand{\chaptermark}[1]{\markboth{#1}{}}
\renewcommand{\sectionmark}[1]{\markright{\thesection\ #1}}
\fancyhead[RO]{\bf \thepage}
\fancyhead[LE]{\bf \thepage}
\fancyhead[LO]{{\it \leftmark}}
\fancyhead[RE]{\it \rightmark}

\renewcommand{\headrulewidth}{0.2pt}
\renewcommand{\footrulewidth}{0pt}

\renewcommand{\theequation}{\arabic{chapter}.\arabic{equation}}
\input{chap1}
\input{chap2}
\input{chap3}

\input{chap4}

\input{chap5}

\input{chap6}


\addcontentsline{toc}{part}{\large Appendices} \thispagestyle{empty}
\input{App}\thispagestyle{empty}

\appendix
\makeatletter
\@addtoreset{equation}{section}  
\makeatother
\renewcommand{\theequation}{\thesection.\arabic{equation}}

\renewcommand{\theequation}{A.\arabic{equation}}
\input{App1}
\renewcommand{\theequation}{\thesection.\arabic{equation}}
\input{App2}



\backmatter

\input{pubs}
\addcontentsline{toc}{chapter}{List of Publications} 
\pagestyle{fancy}
\fancyhf{}
\renewcommand{\chaptermark}[1]{\markboth{#1}{}}
\renewcommand{\sectionmark}[1]{\markright{\thesection\ #1}}
\fancyhead[LE]{\bf \thepage}
\fancyhead[RO]{\bf \thepage}
\fancyhead[LO]{{\it List of Publications}}
\fancyhead[RE]{\it  List of Publications}
\renewcommand{\headrulewidth}{0.2pt}
\renewcommand{\footrulewidth}{0pt}

\listoftables 
\renewcommand{\chaptermark}[1]{\markboth{#1}{}}
\renewcommand{\sectionmark}[1]{\markright{\thesection\ #1}}
\fancyhead[RE,RO]{\bf \thepage}
\fancyhead[LO]{{\it List of Tables}}
\fancyhead[LE]{\it  List of Tables}
\renewcommand{\headrulewidth}{0.2pt}
\renewcommand{\footrulewidth}{0pt}

\listoffigures 
\renewcommand{\chaptermark}[1]{\markboth{#1}{}}
\renewcommand{\sectionmark}[1]{\markright{\thesection\ #1}}
\fancyhead[LE]{\bf \thepage}
\fancyhead[RO]{\bf \thepage}
\fancyhead[LO]{{\it List of Figures}}
\fancyhead[RE]{\it  List of Figures}
\renewcommand{\headrulewidth}{0.2pt}
\renewcommand{\footrulewidth}{0pt}

\pagestyle{fancy}
\fancyhf{}
\renewcommand{\chaptermark}[1]{\markboth{#1}{}}
\renewcommand{\sectionmark}[1]{\markright{\thesection\ #1}}
\fancyhead[LE]{\bf \thepage}
\fancyhead[RO]{\bf \thepage}
\fancyhead[LO]{{\it Bibliography}}
\fancyhead[RE]{\it Bibliography}
\renewcommand{\headrulewidth}{0.2pt}
\renewcommand{\footrulewidth}{0pt}

\bibliographystyle{utphys} 
\bibliography{thesis}




\end{document}

%% file: Intro.tex
\chapter*{Introduction}
\thispagestyle{plain}

The two main purposes of this thesis are to present the results on supersymmetric objects in gauged supergravities published during the three years of doctoral studies and organize them in the current research on theoretical physics of fundamental interactions. The general perspective will be that one defined by the system of theories, ideas and techniques going under the name of string theory, while the concrete framework considered will be supergravity theories and AdS/CFT correspondence. These provide in fact the most suitable environments decribing the low-energy regime of those systems whose microscopic origin is driven by gravitational interaction.

The Introduction of this thesis is composed by four parts. In the first we discuss the necessity of a theoretical understanding of quantum gravity in relation to the main experimental discoveries in fundamental physics of the last twenty years. String theory is introduced as the most complete setup fitting with this scenario. In the second part we take in consideration supersymmetry as the necessary condition to formulate quantum theories of gravity. Discussing the implications of some recent conjectures, we try to motivate the study of supersymmetric objects in the low-energy regime and, at the same time, to explain the limits deriving by the lack of a well-understood mechanism of supersymmetry breaking. In the third part we focus on supersymmetric objects in gauged supergravities and we relate them to the AdS/CFT correspondence. Finally, in the fourth part the outline of the thesis is presented.

\section*{String Theory and Experimental Results}

The formulation of a unified description of fundamental interactions has always been the most relevant and intriguing challenge in physics.
If we focus on the last twenty years, there have been at least three experimental discoveries with crucial implications in the research path towards a grand unification theory.
The first consists in the combined measurements of the cosmological constant $\Lambda$ coming from the observation of the redshift of the light rays emitted by supernovae \cite{Riess:1998cb,Perlmutter:1998np}, the Cosmic Microwave Background (CMB) \cite{Jaffe:2000tx} and the Baryonic Acoustic Oscillations (BAO) \cite{Tegmark:2003ud}. These measurements imply that the observed expansion of our universe is accelerated and driven by a positive cosmological constant.
The second is the detection at CERN of the Higgs boson \cite{Aad:2012tfa}. With this result the origin of masses of fundamental particles has been proven to come from a spontaneous symmetry breaking mechanism and the Standard Model has been confirmed as the theory unifying electromagnetism and nuclear forces.
The third is the detection of gravitational waves \cite{Abbott:2016blz}. In addition to the confirmation of the prediction of General Relativity on the existence of gravitational waves, this measurement opens a new path in the experimental research on systems that are strongly coupled to gravity. In other words, the study of gravitational waves' spectrum could produce in the future new decisive insights on the microscopic origin underlying macroscopic gravitational systems.

From a theoretical point of view, the understanding of the deep implications of these three experimental discoveries should involve a radical change of our actual point of view on fundamental interactions in relation to the nature of spacetime.
Alternatively it is manifest that, in order to formulate an interpretation, at the same time realistic and fundamental, of the measurements discussed above, a microscopic description of gravity and its consequent unification to the other three fundamental forces is needed. 

Such a description has to reproduce the predictions of General Relativity in the low-energy limit and has to be $\mrm{UV}$-complete in the high-energy limit. Moreover the other three fundamental forces should be included in this framework in order to realize a unified picture. Finally the origin of thermodynamic observables of macroscopic gravitational objects should be explained in terms of quantum states associated to the microscopic configurations of gravity. A theory realizing these three properties would be called {\itshape theory of quantum gravity}.

The consistent setup which seeks to reproduce these three properties is given by {\itshape string theory}. In this context the non-renormalizability of the classical gravity theory is resolved in the high-energy limit in terms of contributions coming from the physics of extra dimensions. In other words the more the energy scale grows up, the more extra spacetime dimensions arise resolving singularities that typically characterize classical gravity systems. In this way the $\mrm{UV}$ origin of macroscopic objects (and of their singularities) is explained in terms of extended fundamental objects, like strings and branes, characterized by non-local interactions and appearing when the spacetime is decompactified as non-perturbative contributions to the gravitational field.  

The typical UV regime describing these fundamental objects is defined by the Planck length $l_P$ or by the Planck energy $E_P$,
\begin{equation}
l_P\simeq 1,61 \times 10^{-35} \text{m}\, , \qquad \qquad    E_P\simeq 1,22 \times 10^{19} \text{Gev}\,.
\end{equation}
In this regime one is left with five possible ten-dimensional theories of strings that, in turn, are described in their strong-coupling limit by an eleven-dimensional non-lagrangian theory called {\itshape M-theory} \cite{Witten:1995ex}. 

From the point of view of string theory, all fundamental particles (included those of the Standard Model) have an origin in terms of strings' oscillation modes and, at the Planck scale, one has a completely divergence-free description including all possible quantum effects appearing in the low-energy limit (the IR regime\footnote{The definitions of UV and IR regimes could be misleading. In this thesis the UV regime, or high-energy limit, will be that one determined by the Planck energy $E_P$. The low-energy limit or IR regime will be understood in the most general way as possible as that scale in which quantum effects involving the gravitational field are negligible.}).

This formulation of quantum gravity manifests some intrinsic issues and it is far from being complete, but it needs to be said that all these issues open to some fundamental questions on the deep nature of the gravitational interaction, on the macroscopic objects that can be observed experimentally and on the range of validity of the theory itself.

\section*{The Role of Supersymmetry}

One of the main open problems arising in string theory concerns the dynamical {\itshape supersymmetry breaking} in relation to the stability of gravitational objects.
In this framework the formulation of quantum theories of gravity requires supersymmetry (SUSY) in the UV regime in order to have stable extended objects. Since at our scales we tipically observe non-supersymmetric configurations, a mechanism describing the breaking of supersymmetry when going at low-energies must be understood. There have been many attempts in this direction and, between these, of great relevance are those deriving de Sitter (dS) backgrounds that turn out to constitute realistic models of our universe (see for example \cite{Kachru:2003aw}). Unfortunately in all known examples of dS backgrounds produced by string configurations instabilities appear\footnote{Also the case of non-supersymmetric Anti de Sitter (AdS) backrounds in string theory is similarly problematic. A relevant example describing the dynamical SUSY breaking producing metastable AdS vacua can be found in \cite{Argurio:2006ny}.}.

A complementary approach to this problem consists in the investigation of the conditions that have to be respected by a low-energy theory in order to obtain a UV-completion in string theory. The set of low-energy models with an embedding in string theory and thus with a consistent UV completion is usually called {\itshape string landscape} while the set of models that look to be consistent at certain scales, but have not a UV completion when coupled to gravity is called {\itshape swampland}. There are many criteria\footnote{For a review of swampland criteria see \cite{Brennan:2017rbf}.} proposed to classify the effective theories in these two sets and, between these, that one based on the so-called {\itshape weak gravity conjecture} (WGC) could be the most intriguing for its implications in relation to the role of supersymmetry in the transition to the IR regime \cite{ArkaniHamed:2006dz}.



The WGC conjecture is based on the intuitive fact that gravity must be the weakest force and it places in the swampland all the effective models that don't respect this property. Consider for example a four-dimensional effective theory coupled to gravity and including the electromagnetic interaction, then the WGC states that it must exist a charged state such that
\begin{equation}
 \frac{m}{m_P}\leq q\,,
 \label{WGC}
\end{equation}
where $m$ and $q$ are the mass and the charge\footnote{A motivation for the WGC is related to another swampland criterium excluding from the landscape the effective gravity theories with global symmetries. In our example the ungauging limit is realized if the $\mrm{U(1)}$ gauge charge goes to zero, the inequality \eqref{WGC} prevents from this imposing a cutoff on the charge.} of the state and $m_P$ is the Planck mass.
The apparently unnatural equality in \eqref{WGC} is clearly satisfied by BPS states. These states saturate the BPS bound that, in turn, can be derived as a unitarity condition for supersymmetry\footnote{The anti-commutator of two supercharges in a supersymmetric theory is well-defined only for states respecting the BPS bound.}. Moreover they enjoy a crucial property: they are the only physical objects sharing macroscopic and microscopic properties. In fact they appear in string theory as microscopic supersymmetric states and their observables (as charges) are protected by corrections in the transition from the UV to the IR allowing, among other things, their counting \cite{Strominger:1996sh}.
On the converse, BPS states constitute a limiting case of the inequality \eqref{WGC} and then are included in the intersection between \eqref{WGC} and the correspondent reversed inequality defining the BPS bound.

Following this trajectory, a recent sharpened version of the WGC proposed in \cite{Ooguri:2016pdq} states that \eqref{WGC} is saturated if and only if the states are BPS and the underlying effective theory is supersymmetric. This sharpened version of the conjecture implies that any non-supersymmetric state would unavoidably manifest instabilities in the UV regime\footnote{Even if the assumption of the WGC is intuitive, the implications are powerful and restrictive. For example it follows that all non-supersymmetric AdS vacua produced in string theory would be unstable \cite{Ooguri:2016pdq} and this should be proven case by case.} of string theory. 

In this scenario also dS backgrounds are conjectured to be in the swampland and this shows that a theoretical understanding of the cosmological measurements cited above is still lacking. 
In order to formulate a cosmology based on a solid fundamental interpretation of gravity, a deeper understanding of the nature of $\Lambda$ at different energy scales is thus required. This would explain, for example, why the main contributions determining the dynamics of the universe seem paradoxically not to be given by the ordinary energy sources contained in the universe itself. 

Since at our scales physics appears to be non-supersymmetric, it is reasonable to suppose that the SUSY breaking could be somehow related to thermodynamic effects. These are absent in the UV and determinant in the IR where the classical gravity solutions describing realistic phenomena are tipically driven by a non-trivial thermodynamics and, at the same time, characterized by the absence of a well-defined embedding in string theory. 
More concretely, an explanation of the deep origin of the supersymmetry breaking mechanism in relation to thermodynamic excitations appearing when going to low energies could be crucial for a complete understanding of the microscopic origin of ``physical'' black objects. An example of these could be some non-extremal black hole solutions in General Relativity and their merging that constitute a good classical description of the sources of gravitational waves detected in \cite{Abbott:2016blz}.

\section*{Supergravities, BPS Objects and AdS/CFT Correspondence}

For the arguments proposed above the only gravitational objects with a clear UV interpretation are the supersymmetric ones and this thesis is devoted to the study of their low-energy regime. 
In this case the low-energy limit of string theories and $\mrm{M}$-theory reproduces stable macroscopic objects that are well-described as supersymmetric solutions of {\itshape supergravity theories}. These theories are determined by the massless spectrum of strings and, in general, they can be defined as classical and interacting gravity theories with local supersymmetry \cite{Freedman:1976xh}.

Since our scale of energy is characterized by a four-dimensional physics, it follows that the ten- or eleven- dimensional fields coming from the massless spectrum of strings will be defined on a background with some directions wrapping compact manifolds. Thus, in this picture, our four-dimensional spacetime has to be interpreted as the result of a compactification of a higher-dimensional background. This implies that, among the many things, by reducing higher-dimensional supergravity theories, a rich plethora of different lower-dimensional supergravities will be produced for a given dimension and number of preserved supersymmetries.

Clearly also supersymmetric objects, described in the low-energy regime by classical solutions in higher-dimensional supergravities, will have a lower-dimensional realization as solutions in a supergravity theory and typically they will describe macroscopic objects with a strong coupling to gravity, like black holes and domain walls.

Not all supergravity theories have the right properties needed for a realistic lower dimensional description since the compactification procedure usually produces many free massless scalar fields without a well-defined vacuum expectation value. This is another swampland criterium and, for this reason, in this thesis only {\itshape gauged supergravities} will be considered. These are defined by the gauging of some of their global symmetries and this procedure implies, as a main effect, the production of a suitable scalar potential stabilizing the scalars to its critical values. The field configurations corresponding to the critical values of the scalar potential define a vacuum with a possibly non-zero negative $\Lambda$. Joining the arguments discussed above on the WGC with the assumption on the absence of free parameters, it follows that the minimal energy configurations considered in this thesis will be given by supersymmetric AdS backgrounds.

It follows that the SUSY solutions in gauged supergravity interpolating between AdS vacua will be the main objects of our interest. In these cases the uplift to higher dimensions is consistently described by classical solutions that are associated to the IR regime of non-perturbative states of string theory, like branes and solitonic objects, and include string AdS vacua in their near-horizon (or in some particular value of their coordinates).
These solitonic states manifest a remarkable low-energy behavior consisting in the decoupling between the degrees of freedom coming respectively from closed and open strings that, in turn, include these solitons in their spectra as non-perturbative excitations. This decoupling produces two dual and equivalent descriptions of supersymmetric AdS vacua: one as a classical supergravity solution associated to the closed strings' degrees of freedom and the other in terms of a superconformal quantum field theory (SCFT) describing the open strings' degrees of freedom. This is the celebrated {\itshape AdS/CFT correspondence} \cite{Maldacena:1997re,Witten:1998qj} and constitutes the most relevant development in string theory of the last twenty years since it allows to extrapolate many important properties of non-pertubative effects of quantum gravity just by studying their low-energy regime.

The existence of the AdS/CFT correspondence does not give us the access to the physics of the entire strings' spectrum, but for sure gives a concrete perspective on the leading effects of quantum gravity and on the non-perturbative contributions to the gravitational field. For this reason the correspondence constitutes the conceptual starting point of this thesis since, thanks to this duality, from a classical solution in supergravity describing a supersymmetric object we can capture many relevant informations on the underlying quantum system.

\section*{Outline of the Thesis}

This thesis could be divided in two parts. The first one aims to provide a synthetic overview on string theory and supergravities as theories describing the low-energy limit of strings. In the second part three particular examples of lower-dimensional gauged supergravities are considered and studied in relation to some of their supersymmetric solutions. 
Even if the second part is more technical, it has been made the constant effort to link the lower-dimensonal world with the higher-dimensional one, and then to relate the features of higher-dimensional physics, introduced in the first part, to the lower-dimensional results presented in the second.

Chapter \ref{strings} is entirely dedicated on string theory. The ten-dimensional superstring theories are introduced and their quantization is discussed. The web of dualities linking them is discussed especially in relation to the non-perturbative states like branes and other solitonic objects appearing in the spectrum of superstrings. Furthermore $\mrm{M}$-theory is introduced as the $D=11$ theory describing the $\mrm{UV}$-completion of ten-dimensional superstring theories.

Chapter \ref{effectivesupergravity} is focused on the low-energy limit of string theory. Higher-dimensional supergravities are introduced as the theories defined by the massless spectrum of superstrings and $\mrm{M}$-theory. Moreover a brief review of compactification techniques is presented and the plethora of different lower-dimensional supergravities obtained is classified. Then the concept of supersymmetric solution in supergravity as the low-energy description of ``stringy" objects is presented especially in relation to its Killing spinor analysis and to some crucial concepts like extremality. In this context vacua and supergravity solutions describing branes and their intersections are discussed starting by some important examples. Finally the $\mrm{AdS/CFT}$ correspondence is formulated in generality and then applied to two relevant cases of $\mrm{AdS}_4/\mrm{CFT}_3$ and  $\mrm{AdS}_5/\mrm{CFT}_4$.

In chapter \ref{gaugedsugras} matter-coupled $\ma N=2$ gauged supergravities in $d=4, 5$ are presented and studied in relation to the scalar geometries parametrized by the scalar fields of the matter multiplets. The ``stringy" origin of these two theories is also discussed firstly by considering the compactifications on Calabi-Yau threefolds and then considering the $\mrm{M}$-theory $\ma N=2$ truncations on Sasaki-Einstein manifolds.

Chapter \ref{flows} deals with extremal black hole and black string solutions respectively in $d=4, 5$ matter-coupled $\ma N=2$ gauged supergravities. The properties of staticity and spherical/hyperbolic symmetry are required in the analysis through suitable Ansatz{\"e} and, for these two objects, the general expressions of the first-order equations are derived by using the Hamilton-Jacobi approach. Moreover in the $d=4$ case a symplectic covariant formulation of the first-order equations is presented. Furthermore the near-horizon properties of these solutions are analyzed. The attractor mechanism is formulated for $d=4$ black holes in presence of hypermultiplets while the near-horizon geometry of a generic BPS black string is solved and the general expression of its central charge is presented. Finally Freudenthal duality as a non-linear symmetry of the Bekenstein-Hawking entropy of the $d=4$ black holes is formulated in the case of $\ma N=2$ gauged supergravity.

In chapter \ref{blackholes}, two BPS black hole solutions in $d=4$, $\ma N=2$ gauged supergravity are derived by solving the first-order equations and their properties are studied. The first is an asymptotically $\mrm{AdS}_4$ black hole in a realization of the theory describing a non-homogeneous deformation of the $STU$ model. This object is dyonic and  described by a Fayet-Iliopoulos gauging. The second solution in an asymptotically hyperscaling-violating black hole derived in a model defined by the prepotential $F=-i X^0 X^1$ and by the coupling to the universal hypermultiplet with an abelian gauging of the type $\mathbb{R}\times\mrm{U(1)}$.

Finally chapter \ref{7Dobjects} considers minimal $\ma N =1$ gauged supergravity in $d=7$ and includes a class of supersymmetric solutions in this theory characterized by a running 3-form gauge potential. For these solutions the interpretation in $\mrm{M}$-theory is explicitly discussed in relation to a particular bound state of $\mrm{M}2$ and $\mrm{M}5$ branes preserving half of the supersymmetries. Some of these flows are asymptotically described by an $\mrm{AdS}_7$ geometry and their seven-dimensional backgrounds are defined by an $\mrm{AdS}_3$ slicing. By taking in consideration one particular solution within this class, its uplift in massive $\mrm{IIA}$ supergravity is presented and the holographic interpretation in terms of a defect $\ma N=(4, 0)$ $\mrm{SCFT}_2$ within the $\ma N=(1, 0)$ $\mrm{SCFT}_6$ is produced in relation to the $\mrm{AdS}_3$ slicing within the $\mrm{AdS}_7$ background described asymptotically by the flow.

%% file: chap1.tex

\chapter{Superstrings and M-theory}
\label{strings}
\thispagestyle{plain}


The origin of string theory comes from the late sixties as a tool in the study of strong interactions. The basic idea was to formulate the quantum description of a one-dimensional extended object where
the specific particles corresponded to the quantized oscillation modes. 
With the develop of $\mathrm{QCD}$, the approach to the strong force based on the strings fell out of favor, but at the same time, because of the presence of spin 2 particles in the strings' spectrum, string theory
turned out to be the most suitable enviroment giving a unified picture of gravity with the other fundamental forces. 

In this chapter\footnote{This chapter is partially based on \cite{Green:2012oqa, Ortin:2015hya, Becker:2007zj, Dibitetto:2012xd}.} we will introduce some generalities about string theory as the main candidate for a microscopic description of gravity. The bosonic string and its supersymmetric generalizations will be introduced.
We will then briefly discuss their quantum description in the {\itshape background approach}. From this point of view the dynamics of a string propagating on a flat background is affected by the {\itshape backreaction} of the background on the string itself through a non-zero curvature and couplings to the fields describing the oscillation modes of the string.
This in turn ``feels" the modifications of the background and couples to the fields producing these modifications.

In the second part of the chapter, branes and other non-perturbative objects will be presented in relation to the web of dualities linking them. Finally we will explain how to embed the various superstring theories in
$\mathrm{M}$-theory that will be introduced as the main candidate for a {\itshape theory of everything}. 

\section{Classical Strings}

In this section we will first introduce the main properties of the bosonic string including the action, the equations of motion and their symmetries; then we will consider the supersymmetric extensions of the bosonic action and, also in these cases, we will present the action of the superstring, the equations of motion and their solutions.

\subsection{The Bosonic String}
\label{bosonicstring}

Let's firstly consider the classical bosonic string and its action. Given a $D$-dimensional background parametrized by the coordinates $\{X^{\mu}(\tau, \sigma) \}$ with $\mu=0,1\cdots (D-1)$, the propagation of the string is described by
a bidimensional {\itshape worldsheet} embedded in the background and parametrized by two coordinates $(\tau,\sigma)$. 
The action has the well-known form of a non-linear sigma model. Given the metric on the
worldsheet $h_{\alpha\beta}$\footnote{We define
$h=|\det (h_{\alpha\beta})|$.} with $\alpha, \beta=0,1$, the action has the general form
\begin{equation}
 S=-\frac{T}{2}\int \D \tau \, \D \sigma \, \sqrt{h}\,h^{\alpha\beta}\,g_{\mu\nu}(X)\,\partial_{\alpha}\,X^{\mu}\,\partial_{\beta}\,X^{\nu}\,,
 \label{polyakov}
\end{equation}
where $g_{\mu\nu}$ is the metric\footnote{In this thesis we will use the mostly-plus convention on the metrics, i.e. $\eta_{\mu\nu}=\mathrm{diag(-1,+1\cdots+1)}$.} of the background and $T$ the tension of the string measured as a mass on unit of volume.
The worldsheet, identified by its metric $h_{\alpha\beta}$ and by the embedding coordinates $X^{\mu}$, is invariant under the following symmetries:
\begin{itemize}
 \item The Lorentz-Poincar{\'e} group.
\item The diffeomorphisms of the background.
\item  The Weyl rescalings.
\end{itemize}
Thanks to the last two symmetries one can always parametrize the worldsheet in the {\itshape conformal frame}, i.e. $h_{\alpha\beta}=\eta_{\alpha \beta}=\mathrm{diag}(-1,1)$. In this particular gauge the equation of motion describing the 
propagation of the string is given by the two-dimensional wave equation\footnote{We use the notation $\Box_2=-\partial^2_{\tau}+\partial_{\sigma}^2$.},
\begin{equation}
 \Box_2\,X^\mu=0\,.
 \label{eomstring}
\end{equation}
The stress-energy tensor associated to the dynamical variables $X^{\mu}$ given by 
\begin{equation}
 T_{\alpha\beta}=-\frac{2}{T}\,\frac{1}{\sqrt{h}}\,\frac{\delta \, S}{\delta \, h^{\alpha\beta}}\,,
 \label{stresstensorstring}
\end{equation}
is conserved thanks to \eqref{eomstring} and it is associated to the equation of motion for $h_{\alpha\beta}$ given by $T_{\alpha\beta}=0$.

In order to solve \eqref{eomstring} one must impose a set of suitable boundary conditions on the endpoints of the string. These can be of three types:
\begin{itemize}
 \item {\itshape Neumann $\mathrm{(N)}$ conditions for open strings}: $\partial_{\sigma}X^{\mu}|_{\sigma=0,\pi}=0$.
 \item {\itshape Dirichlet $\mathrm{(D)}$ conditions for open strings}: $X^{\mu}|_{\sigma=0}=X_0^{\mu}$ and $X^{\mu}|_{\sigma=\pi}=X_{\pi}^{\mu}$ with $X_0^{\mu}$, $X_{\pi}^{\mu}$ constants.
  \item {\itshape Periodic conditions for closed strings}: $X^{\mu}(\tau,\sigma)=X^{\mu}(\tau,\sigma+\pi)$\,.
\end{itemize}
The general solution of \eqref{eomstring} has the form
\begin{equation}
 X^{\mu}(\tau, \sigma)=X^{\mu}_R(\tau-\sigma)+X^{\mu}_L(\tau+\sigma)\,,
 \label{generalbosonicsolution}
\end{equation}
where $X^{\mu}_R$ and $X^{\mu}_L$ are respectively called {\itshape right}- and {\itshape left movers}.

Imposing the periodic boundary conditions for closed strings, one can expand in Fourier modes the right- and left-movers,
\begin{equation}
 \begin{split}
  &X^{\mu}_R(\tau, \sigma)=\frac{1}{2}\,x^\mu+\alpha^\prime\,p^\mu\,(\tau-\sigma)+i\,\sqrt{\frac{\alpha^\prime}{2}}\,\sum_{n\neq 0}\,\frac{1}{n}\alpha^\mu_n\,\mathrm{e}^{-2\,i\,n\,(\tau-\sigma)}\,, \\
   & X^{\mu}_L(\tau, \sigma)=\frac{1}{2}\,x^\mu+\alpha^\prime\,p^\mu\,(\tau+\sigma)+i\,\sqrt{\frac{\alpha^\prime}{2}}\,\sum_{n\neq 0}\,\frac{1}{n}\tilde{\alpha}^\mu_n\,\mathrm{e}^{-2\,i\,n\,(\tau+\sigma)}\,,
   \label{modesexpansion}
 \end{split}
\end{equation}
where
\begin{equation}
 \alpha^\prime=\frac{1}{2\pi\,T}=\frac{l_s^2}{2}
 \label{reggeslope}
 \end{equation}
 is the {\itshape Regge slope parameter} and $l_s$ the lenght of the string. The Fourier-modes $\alpha^\mu_n$ and $\tilde{\alpha}^\mu_n$ respect a set of reality conditions, 
$\alpha^\mu_n=(\alpha^\mu_{-n})^*$ and $\tilde{\alpha}^\mu_n=(\tilde{\alpha}^\mu_{-n})^*$, while $x^\mu$ and $p^\mu$ are real and describe respectively the position and the momentum of the string's centre of mass.

We conclude the discussion about the classical bosonic string considering the algebra of the Poisson brackets associated to the Fourier-modes\footnote{We set $\alpha_0^\mu=\tilde{\alpha}_0^\mu=\sqrt{\frac{\alpha^\prime}{2}}\,p^\mu$.},
\begin{equation}
 \begin{split}
 &[\alpha_m^\mu,\alpha_n^\nu ]_{\scriptsize\mrm{PB}}=i\,m\,\eta^{\mu\nu}\,\delta_{m+n,0}\,,\\
 &[\tilde{\alpha}_m^\mu,\tilde{\alpha}_n^\nu ]_{\scriptsize\mrm{PB}}=-i\,m\,\eta^{\mu\nu}\,\delta_{m+n,0}\,,\\
 &[\alpha_m^\mu,\tilde{\alpha}_n^\nu ]_{\scriptsize\mrm{PB}}=0\,.
 \end{split}
\end{equation}
All the physical observables describing the string can be expressed in terms of its Fourier-modes. In particular, we recall the Fourier-modes of \eqref{stresstensorstring}, $L_m$ and $\tilde{L}_m$, satisfying 
the {\itshape Virasoro algebra}
\begin{equation}
 [L_m,L_n ]_{\scriptsize\mrm{PB}}=i\,(m-n)\,L_{m+n}\,,
 \label{virasoro}
\end{equation}
where the same brackets hold for $\tilde{L}_m$ and $[L_m,\tilde{L}_n]=0$. The appearance of this algebra is due to the fact that the conformal gauge does not fix completely the reparametrization symmetry of the action \eqref{polyakov}. The symmetries
generated by \eqref{virasoro} are Weyl rescalings and they are infinite since the worldsheet is bidimensional.

As regards the open strings, the general solution obtained by imposing Neumann or Dirichelet conditions is given by an expansion in terms of Fourier-modes whose expression is 
similar to $\eqref{modesexpansion}$. The main difference is the presence of one 
single set of oscillators $\alpha_n^\mu$ and this fact is due to the open string boundary conditions that force the right- and the left-movers to combine into standing waves.

\subsection{The Superstring}
\label{superstring}

Let's extend the description of the bosonic string by considering fermionic degrees of freedom of the worldsheet and a supersymmetric formulation of the string. In order to introduce the action of the superstring two main approaches are possibile:
\begin{itemize}
 \item {\itshape The Ramond-Neveu-Schwarz (RNS) formalism}: Supersymmetry is introduced at the level of the worldsheet through the fermionic coordinates $\psi^{\mu}$.
 \item {\itshape The Green-Schwarz (GS) formalism}: Based on the superspace formulation of the worldsheet.
\end{itemize}
We will consider only the first approach where $\{\psi^\mu\}$ are a set of $D$ Majorana spinors belonging to the vector representation of $\mrm{SO}(D-1,1)$. The off-shell action of the superstring propagating in a flat background is given by
\begin{equation}
\begin{split}
  S=&-\frac{1}{2\,\pi\,\alpha^\prime}\int \D \tau \, \D \sigma \,\biggl( \partial_{\alpha}\,X^{\mu}\,\partial^{\alpha}\,X_{\mu}-i\,\bar{\psi}^\mu\,\rho^{\alpha}\,\partial_\alpha \,\psi_\mu \\
 &+2\,\bar{\chi}_\alpha\,\rho^\beta\rho^\alpha\,\psi^\mu\,\partial_\beta X_\mu+\frac12\,(\bar{\psi}^\mu\,\psi_\mu)\,(\bar{\chi}_\alpha\,\rho^\beta\rho^\alpha\,\chi_\beta) \biggr)\,,
  \end{split}
  \label{polyakovSUSY}
\end{equation}
where $\alpha^\prime$ is the Regge slope parameter introduced in \eqref{reggeslope} and the matrices $\rho^\alpha$ define a bidimensional Clifford algebra associated to the Lorentz invariance on the worldsheet, i.e. $\{\rho^\alpha,\rho^\beta \}=2\,\eta^{\alpha\beta}$. The fermionic field $\chi_\alpha$ is auxiliary and
it is needed to compensate the degrees of freedom of $h_{\alpha\beta}$. 

The action \eqref{polyakovSUSY} is supersymmetric, in fact it can be shown that if one introduces an arbitrary Majorana spinor $\varepsilon$ on the worldsheet, the $\mathrm{SUSY}$ transformations
\begin{equation}
 \begin{split}
  &\delta\,X^{\mu}=\bar{\varepsilon}\,\psi^{\mu}\,,\\
  &\delta \, \psi^{\mu}=-i\,\rho^\alpha\,\varepsilon\,\left( \partial_\alpha X^\mu-\bar{\psi}^\mu \chi_\alpha   \right)\,,\\
  &\delta \,h_{\alpha\beta}=-2i\,\bar{\varepsilon}\,\rho_\alpha\,\chi_\beta\,,\\
  &\delta \,\chi_\alpha=\partial_\alpha\varepsilon\,,
 \end{split}
 \label{susystring}
\end{equation}
preserve \eqref{polyakovSUSY}.

As in the bosonic case, one can choose the conformal gauge and set $h_{\alpha\beta}=\eta_{\alpha\beta}$. In an analogous way one can set $\chi_\alpha=0$ using \eqref{susystring} with a further symmetry of \eqref{polyakovSUSY} given by 
$\delta_\eta \chi_\alpha=i\,\rho_\alpha\,\eta$ with $\eta$ Majorana spinor.
The equations of motion for $X^{\mu}$ and $\psi^\mu$ in this gauge are given by
\begin{equation}
 \Box_2\,X^\mu=0\,,\qquad \qquad \rho^\alpha\partial_\alpha\,\psi^\mu=0\,
 \label{eomstringSUSY}
\end{equation}
and imply the vanishing of the stress-energy tensor \eqref{stresstensorstring} and of the supercurrent,
\begin{equation}
 J_\alpha=\frac12\,\rho^\beta\rho_\alpha\,\psi^{\mu}\partial_\beta\,X^\mu=0\,.
 \label{supercurrent}
\end{equation}
The bosonic solutions of \eqref{eomstringSUSY} are the same of \eqref{generalbosonicsolution} while as concern to the fermionic equation, the general solution is given by
\begin{equation}
\psi^{\mu}(\tau, \sigma)=\psi^{\mu}_R(\tau+\sigma)+\psi^{\mu}_L(\tau-\sigma)\,.
\end{equation}
Two kind of boundary conditions can be imposed:
\begin{itemize}
 \item {\itshape Ramond conditions $\mathrm{(R)}$ }: $\quad\psi^{\mu}_R(\tau,\pi)=+\psi^{\mu}_L(\tau,\pi)$\,.
  \item {\itshape Neveu-Schwarz conditions $\mathrm{(NS)}$ }: $\quad\psi^{\mu}_R(\tau,\pi)=-\,\psi^{\mu}_L(\tau,\pi)$\,.
\end{itemize}
Considering the case of closed superstrings and imposing $\mathrm{R}$ conditions, one obtains
\begin{equation}
\begin{split}
& \psi_L=\frac{1}{\sqrt{2}}\,\sum_{n \in \mathbb{Z}}d^\mu_n\, \mathrm{e}^{-i\,n\,(\tau-\sigma)}\,,\\
& \psi_R=\frac{1}{\sqrt{2}}\,\sum_{n \in \mathbb{Z}}\tilde{d}^\mu_n\, \mathrm{e}^{-i\,n\,(\tau+\sigma)}\,.
 \end{split}
\end{equation}
Requiring $\mathrm{NS}$ conditions one has
\begin{equation}
\begin{split}
& \psi_L=\frac{1}{\sqrt{2}}\,\sum_{r \in \mathbb{Z}+\frac12}\,b^\mu_r\, \mathrm e^{-i\,r\,(\tau-\sigma)}\,,\\
& \psi_R=\frac{1}{\sqrt{2}}\,\sum_{r \in \mathbb{Z}+\frac12}\,\tilde{b}^\mu_r\, \mathrm e^{-i\,r\,(\tau+\sigma)}\,,
\label{modeexpansionfermion}
 \end{split}
\end{equation}
where in both cases we imposed conventionally $\psi_R(\tau,0)=+\psi_L(\tau,0)$ and the Fourier-modes obey a set of reality conditions as the bosonic oscillators given by $d^\mu_n=(d^\mu_{-n})^*$, $\tilde{d}^\mu_n=(\tilde{d}^\mu_{-n})^*$, $b^\mu_r=(b^\mu_{-r})^*$ and $\tilde{b}^\mu_r=(\tilde{b}^\mu_{-r})^*$.

One can impose the conditions $\mrm R$ and $\mrm{NS}$ separately on the right- and on left-movers, obtaining four different pairings corresponding to four distinct closed string sectors. The sectors $\mrm{NS}$-$\mrm{NS}$ and
$\mrm R$-$\mrm R$ are purely bosonic, while $\mrm{NS}$-$\mrm{R}$ and $\mrm{R}$-$\mrm{NS}$ are fermionic.

The oscillators $d^\mu_n$, $\tilde{d}^\mu_n$ and $b^\mu_r$, $\tilde{b}^\mu_r$ respect a set of ``anticommuting Poisson brackets" given by
\begin{equation}
\{d_m^\mu,d_n^\nu \}_{\scriptsize\mrm{PB}}=i\,\eta^{\mu\nu}\,\delta_{m+n,0}\, \quad \text{and} \quad \{b_r^\mu,b_s^\nu \}_{\scriptsize\mrm{PB}}=i\,\eta^{\mu\nu}\,\delta_{r+s,0}\,,
\end{equation}
where the same brackets hold also for the tilded oscillators. Moreover the following anticommuting relations hold, $\{d_m^\mu,\tilde{d}_n^\nu \}=0$ and $\{b_r^\mu,\tilde{b}_s^\nu \}=0$.
In this context one can construct two inequivalent supersymmetric extensions of the Virasoro algebra \eqref{virasoro}, called {\itshape super-Virasoro algebras}, since the inclusion of the Fourier-modes of the supercurrent \eqref{supercurrent} is dependent on the particular sector,
$\mrm R$ or $\mrm{NS}$, considered.

\section{The Quantization of Strings}
\label{quantum}

In this section we will discuss the quantization of strings and superstrings. In particular we will present the quantization procedure of the bosonic string and we will see how the requirement of absence of ghosts in the spectrum implies
an {\itshape higer-dimensional description}. We will see why supersymmetry is crucial to avoid the presence of tachyonic states and we will classify all the superstring theories.

The quantization procedure in string theory follows the usual idea: the Fourier-modes describing the propagation of the (super)string are promoted to creators and annihilators on an Hilbert space with a well-defined vacuum $|0,p^\mu\rangle$ and the Poisson brackets are
promoted to the (anti)commutators between operators,
\begin{equation}
 [\cdots]_{\scriptsize\mrm{PB}}\longrightarrow i\,[\cdots] \quad \text{and} \quad \{\cdots\}_{\scriptsize\mrm{PB}}\longrightarrow i\,\{\cdots\}\,.
\end{equation}
The action of the creators on the vacuum is described by quantum fields and this is the key property making string theory a theory of unification: the quantum oscillation modes of the string are described by fundamental particles.

Historically three different approaches have been followed in quantizing the strings: the old covariant method, the $\mathrm{BRST}$ method and the light-cone quantization. In the first one the procedure of quantization is similar to the usual
Gubta-Bleuler quantization of electrodynamics, the second one is based on the inclusion of Faddeev-Popov ghosts and the last one starts by breaking the Lorentz covariance.

\subsection{Tachyons and Ghosts}
\label{tachyons}

Let's briefly discuss about the spectrum of bosonic string \eqref{polyakov}. 
With the promotion of the dynamical variables to operators the Virasoro algebra (or super-Virasoro in the case of superstring) is deformed by its central extension. In the case of the bosonic string one has the quantum Virasoro algebra given by
\begin{equation}
  [L_m,L_n ]=\,(m-n)\,L_{m+n}+\frac{D}{12}\,m\,(m^2-1)\,\delta_{m+n,0}\,.
  \label{wick}
\end{equation}
Considering the closed string and acting with the creators $\alpha_m^\mu$ and $\tilde{\alpha}_m^\mu$ on the vacuum one produces the complete spectrum. It is immediate to verify that there are two kinds of states that are unphysical:
\begin{itemize}
 \item {\itshape Ghosts}: States with negative Hilbert norm.
  \item {\itshape Tachyons}: States with imaginary mass.
\end{itemize}
The presence of these states in the spectrum makes the string unstable, thus one should wonder if certain particular conditions eliminating these states from the spectrum exist.
By studying the action of the quantum Virasoro algebra on the string states, one finds out that the spectrum does not contain ghosts if the dimension of the background is given by the critical value $D=26$. 

It follows that the higher dimensional formulation of string theory is necessary 
for the consistence of the quantum theory of gravity.
In fact if one considers the first excited states of the closed string one finds out the presence of a symmetric tensor field, the graviton, describing the metric of the background. This fact is crucial since, as we will see, only this first level of the spectrum is relevant in the low-energy limit and the presence of the graviton guarantees the right number of degrees of freedom to define an effective gravity theory.
In particular the first level of the spectrum is composed by 576 degrees of freedom that can be organized as follows
\begin{equation}
|\Omega^{ij}\rangle= \alpha^i_{-1}\,\tilde{\alpha}^j_{-1}\,|0,p^\mu\rangle\,,
\label{1levelbosonic} 
\end{equation}
where the indices $i,j$ are related to the transverse components (in the sense of the light-cone gauge). The object \eqref{1levelbosonic} lives in a massless representation of $\mathrm{SO}(24)$ which is defined by a scalar field $\Phi$, the dilaton, by
a symmetric tensor field $g_{\mu\nu}$, the graviton and by a 2-form $B_{\mu\nu}$, the Kalb-Ramond field or $B$-field.

The vacuum of the bosonic closed string is tachyonic. It is described by a scalar field with negative mass $M^2=-\frac{4}{\alpha^\prime}$ and its presence makes the bosonic string unstable. The same happens for the ground state
of the open string.
In this case we point out the presence of massless vectors at the first excited level described by states of the form
\begin{equation}
 |V^{i}\rangle= \alpha^i_{-1}\,|0,p^\mu\rangle\,.
\end{equation}
and this fact will be important when we will introduce the non-perturbative states included in the strings' spectrum.

\subsection{Classification of Superstring Theories}
\label{classification}

In order to construct a consistent quantum theory without ghosts and tachyonic states one has to take in consideration the superstring \eqref{polyakovSUSY}. Promoting the oscillators $d_n^\mu$, $b_r^\mu$, $\tilde{d}_n^\mu$ and $\tilde{b}_r^\mu$
to fermionic operators, it is possibile to construct a ``tower of states" describing the spectrum of the superstring. As we already mentioned at the end of section \ref{superstring}, the spectrum is composed by the four sectors $\mrm{NS}$-$\mrm{NS}$, 
$\mrm R$-$\mrm R$, $\mrm{NS}$-$\mrm{R}$ and $\mrm{R}$-$\mrm{NS}$, where the last two contain only fermionic fields and the firsts are purely bosonic.
As regards the ghosts, in analogy to the bosonic case, the study of the action of the super-Virasoro algebra on the superstring states forces
us to set the dimension of the background to the critical value $D=10$.

As in the bosonic case the vacua are tachyonic and their presence breaks supersymmetry since a fermion with the same mass as the tachyon is not present in the spectrum. 
From the analysis of the various superstring's states it follows that massless $s=3/2$ fields\footnote{These are stetes with spin $3/2$ under massless irreps of the ten-dimensional super-Poincar{\'e} algebra.} called {\itshape gravitinos} are included in the spectrum.
As we will see these are the gauge fields for local supersymmetry and thus unbroken supersymmetry is crucial for a consistent interacting theory.

The main peculiarity of superstring theory respect to the purely bosonic theory consists in the possibility to ``project out" the tachyonic degrees of freedom and the remakable fact is that this projection, called {\itshape GSO projection} \cite{GLIOZZI1977253}, leads to a completely supersymmetric spectrum in $D=10$.
The GSO projection is thus essential for the consistency of the theory: it eliminates the tachyons from the spectrum and leaves an equal number of bosons and fermions at each mass level.
After the projection the ground state associated to $\mathrm{R}$ boundary conditions is a massless spinor while the vacuum corresponding to $\mathrm{NS}$ boundary conditions is a massless vector.

Let's consider the four possibile sectors of the closed superstring spectrum. In order to match with
the number of bosonic degrees of freedom the massless spinor must be in an irreducible representation of the ten-dimensional super-Poincar{\'e} group, thus one can have two different theories depending on the chirality of the fermionic ground state. These are called {\itshape type IIA} and {\itshape type IIB Superstring Theories}. These theories are maximally supersymmetric in ten dimensions and this means that they preserve 32 real supercharges, i.e. they are $\ma N=2$ theories.
Each of the four sectors contains 64 degrees of freedom that are organized in $\mathrm{SO}(8)$ irreps as follows:

\begin{itemize}

 \item $\mrm{NS}$-$\mrm{NS}$ sector: It is the same for type $\mathrm{IIA}$ and type $\mathrm{IIB}$, and contains the dilaton $\Phi$ (1 state), the Kalb-Ramond field $B_{\mu\nu}$ (28 states) and the graviton $g_{\mu\nu}$ (35 states).

 \item $\mrm{NS}$-$\mrm{R}$ and $\mrm{R}$-$\mrm{NS}$ sectors: Each of these sectors contains a gravitino $\Psi_\mu$ (56 states) and a spin 1/2 fermion, the dilatino $\lambda$ (8 states). In type $\mathrm{IIB}$ the gravitinos have the same chirality
 whilst in $\mathrm{IIA}$ opposite chirality.
 
 \item $\mrm R$-$\mrm R$ sector: This sector is purely bosonic and organized in p-form:
 
\begin{itemize}

\item Type $\mathrm{IIA}$: 1-form $C^{(1)}_\mu$ (8 states), 3-form $C^{(3)}_{\mu\nu\rho}$ (56 states).
\item Type $\mathrm{IIB}$: 0-form $C^{(0)}$ (1 state), 2-form $C^{(2)}_{\mu\nu}$  (28 states), 4-form $C^{(4)}_{\mu\nu\rho\sigma}$ (35 states) with a self-dual field strength.
 
\end{itemize}

\end{itemize}

Also string theories with $\ma N=1$ supersymmetry (16 real supercharges) can be constructed. In particular one has the following theories:

\begin{itemize}

 \item {\itshape Type I Superstring Theory}: This theory can be obtained by modding out type $\mathrm{II}$ theories with respect to the parity simmetry ($\mathbb{Z}_2$) of the worldsheet coordinates.
 Both closed and open strings are present.
 
 \item {\itshape $SO(32)$ Heterotic Superstring Theory}: Obtained by combining bosonic left-movers with fermionic right-movers. The theory has an internal gauge symmetry given by $\mathrm{SO}(32)$ arising from the 
 reduction from $D=26$ to $D=10$.
 
 \item {\itshape $E_8\times E_8$ Heterotic Superstring Theory}: Obtained as the $\mathrm{SO}(32)$ Heterotic Superstring Theory, but in this case the gauge group arising is  $\mathrm{E}_8\times \mathrm{E}_8$.
 
\end{itemize}

These five theories are all the possible superstring theories that can be constructed in $D=10$ and, as we will see, they are related each others by an intricated web of {\itshape dualities}.

\section{Branes, dualities and M-theory}

In the nineties the so-called ``Second Superstring Revolution" took place with two main discoveries:

\begin{itemize}

 \item {\itshape Branes and dualities}: The string's spectrum contains solitonic states described by supersymmetric extended objects called {\itshape branes} \cite{Polchinski:1995mt}. The discovery of these objects came from an other key step in understading string theory that is the 
 discovery of a web of {\itshape dualities}\footnote{More information on dualities relating superstring theories can be found in \cite{Giveon:1994fu,Sen:1998kr}.} relating the five superstring theories presented in section \ref{classification}.
 
 \item {{\itshape M-theory}}: Type $\mathrm{IIA}$ and $\mathrm{E}_8\times \mathrm{E}_8$ Heterotic exhibit an eleven-dimensional description in their strong-coupling regime. Thanks to the dualities it follows that also the other three theories can be embedded in 
 a fundamental theory in $D=11$ \cite{Witten:1995ex}.
 This eleven-dimensional description is given by a {\itshape non-lagrangian theory} called $\mathrm{M}$-theory and provides a unified picture of all the superstring theories.
 
\end{itemize}

Let's introduce in more detail what these dualities in string theory are and why they suggest the existence of branes and other solitonic states. Thus, at the end of the section, we will discuss how to embed superstring theories in $\mrm M$-theory.

\subsection{T, S, and U-duality}
\label{dualities}

One of the main peculiarity of string theory is the presence of two coupling constants describing the interactions.
The first parameter is the Regge slope parameter $\alpha^\prime$ defined in \eqref{reggeslope}. The expansion with respect to $\alpha^\prime$ describes the ``stringy" deviation respect to the point-particle limit. It can be viewed as a quantum mechanical expansion in the $\mathrm{QFT}$
 defined by the worldsheet theory of the embedding coordinates \eqref{modesexpansion} and \eqref{modeexpansionfermion}, or an higher-derivative expansion respect to the fields describing the backreaction of the background.
 
The second is the {\itshape string coupling constant} $g_s$. It is determined by the expectation value of the dilaton field and describes the quantum corrections to the field theories defined by the field content of the spectrum. From the worldsheet point of view
 the expansion with respect to $g_s$ gives the contributions associated to the string interactions in terms of string loops.
 
It is well known that in physics a duality is generically a symmetry connecting theories in different regimes of their parameters and such a simmetry can be always explained in terms of field transformations.
The remarkable fact discovered in the 90s is the presence of a web of dualities relating different regimes of $\alpha^\prime$ 
and $g_s$ of all the superstring theories in ten dimensions \cite{Sen:1998kr}. 

The first duality that we are going to present is called {\itshape T-duality} and it is a perturbative duality with respect to $g_s$. For example let us take in consideration the closed string in type $\mathrm{IIA}$ on the background $\mathbb{R}^{1,8}\times S^1$, with a radius $R_A$ of the circle. $\mathrm{T}$-duality exchanges the momentum of the string and the winding modes indicating the number of times the string winds around the circle. 
The theory obtained acting with $\mrm{T}$-duality is type $\mathrm{IIB}$ on the background $\mathbb{R}^{1,8}\times S^1$ with the circle of radius
\begin{equation}
 R_B=\frac{\alpha^\prime}{R_A}\,.
\end{equation}
The duality acts on the embedding coordinates of the string by changing the sign of the righ-movers along the compact direction of the circle and it is intrinsically ``stringy" in the sense that it always involves configurations characterized by large $\alpha^\prime$ values. In the case of a single compact direction the group
describing this action is given by $\mathrm{O}(1,1;\mathbb{Z})$, while when more coordinates are compact and define a torus $T^n$, $\mathrm{T}$-duality acts as  $\mathrm{O}(n,n;\mathbb{Z})$.

The other fundamental duality linking different theories is called {\itshape S-duality}. This simmetry is non-perturbative in $g_s$ in the sense that it relates a theory described by $g_s$ to a theory with $1/g_s$. The intuitive way to 
understand this symmetry is thinking about the electromagnetic duality in classical electromagnetism or the strong-weak duality  in Yang-Mills theories. Since $\mathrm{S}$-duality is non-perturbative in $g_s$, its general proof is difficult 
and usually one has to perform some particular tests, like the comparison of the levels of the spectra of the dual theories or the matching between the tensions of the respective fundamental strings and solitonic objects. Then, even thow S-duality is non-perturbative in $g_s$, it can be defined order by order in $\alpha^\prime$ and thus it is achievable even just on the fields composing the ground state of superstring and this is the main difference with respect to T-duality.

An example is given by the relation between type $\mathrm{I}$ theory and $\mathrm{SO}(32)$ 
Heterotic theory. These two theories can be mapped into one another by changing the sign of the dilaton field and leaving all other bosonic field unchanged\footnote{The metric must be recasted in the Einstein frame by a Weyl
transformation.}. Since $g_s$ is given by the expectation value of the dilaton, the sign flipping of the dilaton is equivalent to the transformation $g_s \rightarrow 1/g_s$.

 We conclude the brief analysis on $\mathrm{S}$-duality by mentioning that $\mathrm{IIB}$ string theory has an interesting behavior under this symmetry, in fact $\mathrm{S}$-duality maps the $\mathrm{IIB}$ theory in itself and 
 can be extended to a group of discrete non-perturbative symmetries given by $\mathrm{SL}(2,\mathbb{Z})$.
  
Finally for type $\mathrm{II}$ theories, one can introduce an other symmetry called {\itshape U-duality} given by the combination of perturbative and non-perturbative dualities. This duality is crucial when compactifying string theories 
since it defines the largest group of internal symmetries of the lower dimensional theories.

\subsection{Branes, Open Strings and BPS States}
\label{branes1}

In the previous section we introduced the web of dualities relating superstring theories in $D=10$. The key point is that, thanks to the existence of dualities, it is possibile to show that the spectrum of strings is not only determined by the oscillation modes, but it also contains some solitonic states that are intimately non-perturbative.

The main example between these states are branes and the most intuitive way to make manifest their presence in the spectrum of strings is by using $\mathrm{T}$-duality. Let's consider the bosonic open string with 
Neumann boundary conditions at their extrema. If we put the open string with $\mathrm{N}$ conditions on a circle, the winding number is a meaningless concept since the open strings are topologically contractible in a point and, thus, the string is described in the compact direction
only by its momentum.
The effect of $\mathrm{T}$-duality is to map the open string with Neumann conditions on a circle of radius $R_A$ into the open string on a circle with radius 
$ R_B=\frac{\alpha^\prime}{R_A}$ with Dirichelet boundary conditions. The dual string will be characterized in the compact direction exactly by the opposite properties: it will be described by absence 
of momentum and non-vanishing winding number that now
is a well-defined quantity for the dual string since the endpoints are fixed by the $\mathrm{D}$ conditions.
The hyperplane where the dual string ends is a physical object since it is associated to the position of the extrema of the open string and, thus, to its dynamics and its spectrum. 

Also strings can be viewed as solitonic objects and, following this analogy, one can define the embedding of branes into a background in terms of a {\itshape worldvolume}. In this case we have two types of coordinates: the worldvolume and the transverse coordinates. 
  \begin{figure}[htbp]
\centering
\vspace{0.1cm}
\includegraphics[height=4.5cm, width=6cm]{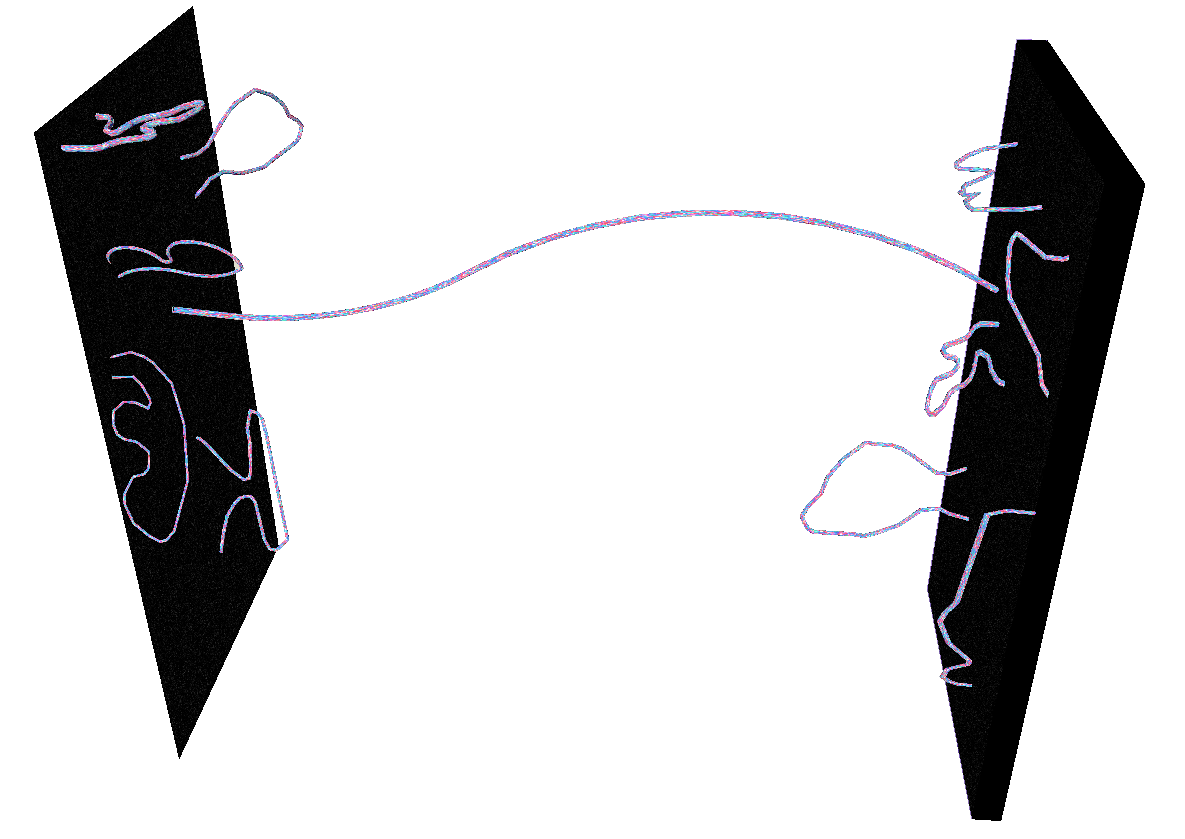}
\vspace{0.1cm}
\caption[Branes and strings. Image from en.wikipedia.org/wiki/Brane.]{Intuitive representation of open strings with the extrema ending on branes in presence of closed strings. Image from en.wikipedia.org/wiki/Brane.}
\label{brane}
\end{figure}
In other words the brane breaks the Lorentz invariance of the background in the following way,
\begin{equation}
 \mathrm{SO}(1,D-1)\longrightarrow \mathrm{SO}(1,p)\times \mathrm{SO}(D-p-1)\,,
 \label{breakinglorentz}
\end{equation}
where $p$ is the spatial dimension of the worldvolume and this implies that the quantum fields composing the spectrum of the open string are defined only on the worldvolume of the brane where the string is ending on. Furthermore the brane couples to the background whose backreaction is described in terms of non-trivial interactions between the fields on the worldvolume and the fields describing the background. 

As we said in section \ref{tachyons}, the spectrum of bosonic open string is tachyonic and this implies that branes in a purely bosonic string theory are unstable objects. Let's now consider type $\mathrm{II}$ superstrings and suppose that 
the background is defined in terms of fields belonging to the spectrum of closed strings. Since the spacetime background is usually a bosonic configuration of gravity with minimum energy, it will be described only by the massless fields of the closed string sectors $\mrm{NS}$-$\mrm{NS}$ and $\mrm R$-$\mrm R$. This will react to the presence of the brane and the interactions will include, in the most general scenario, fields coming from the entire spectra of open and closed superstrings. 

If the energy of the brane is low with respect to the energy associated to the open strings' motion (the brane is heavy), then only the massless spectrum of closed strings will describe the modifications of the background. The brane, in turn, will play the role of source for the gauge potentials included in the bosonic sectors of closed strings. In particular we call
 $\mathrm{D}p$-branes (or $p$-branes) those branes that couple electrically under the $\mrm R$-$\mrm R$ fields
$C^{(p+1)}$ or magnetically under $C^{(7-p)}$. In the case of $\mrm{NS}$-$\mrm{NS}$ fields, the Kalb-Ramond field is coupled electrically to the fundamental\footnote{It is the the usual string, we will call it ``fundamental" to make a distinction with other 1-dimensional objects.} string 
$\mathrm{NS1}$ and magnetically to a particular 5-brane called $\mathrm{NS5}$.
Regarding $\mathrm{D}p$-branes, they are classified in terms of the spatial dimensions $p$ of their worldvolume. In type $\mathrm{IIA}$ $p$ is even since the $\mrm R$-$\mrm R$ sector is composed by $p$-forms with $p$ odd,
viceversa in $\mathrm{IIB}$
 only branes with odd-dimensional worldvolume exist. In particular one has\footnote{Some of these branes are peculiar. The existence of the $\mathrm{D}8$ requires a non-dynamical 10-form field 
 strength and we will analyze this particular case later in section \ref{highersugras}. The $\mathrm{D}9$ is completely spacetime-filling, while the 0-form in $\mathrm{IIB}$ should coupled electrically to
a $(-1)$-brane. This object is istantonic since it is completely localized in time.

Moreover due to the $\mathrm{SL}(2,\mathbb{Z})$ symmetry, particular states obtained as sources of mixed fields exist in $\mathrm{IIB}$ theory. For example we mention the 
$(p, q)$5-branes describing the superposition between a $\mathrm{D}5$ and a $\mathrm{NS5}$, and the $(p,q)$-strings associated to the superposition of a $\mathrm{D}1$ with the fundamental string.}: 

\begin{itemize}
 \item {\itshape The $p$-branes in type IIA}: \qquad    $\mathrm{D}0$\,, \quad $\mathrm{D}2$\,,\quad $\mathrm{D}4$\,,\quad $\mathrm{D}6$\,,\quad $\mathrm{D}8$ .
  \item {\itshape The $p$-branes in type IIB}: \qquad    \,$\mathrm{D}1$\,, \quad $\mathrm{D}3$\,,\quad $\mathrm{D}5$\,, \,\, $\mathrm{D}7$\,, \, $\mathrm{D}9$ .
\end{itemize}
We conclude this section with a brief discussion about supersymmetry as condition for stability of branes in type II superstrings. As we mentioned above when branes are included, in order to avoid instabilities, one has to require the absence of tachyons in the spectrum of open strings ending on them.
The branes satisfying this property are called {\itshape BPS\footnote{Bogomol'nyi-Prasad-Sommerfield.} branes}. These are fully stable and preserve half of the total supersymmetry of the background ($\mathrm{BPS}/2$), since their presence breaks the Lorentz-invariance as in \eqref{breakinglorentz}.
Explicitly, given two supercharges $ Q_1$ and $ Q_2$ of the string theory considered\footnote{These are Majorana-Weyl spinors with opposite chirality
in $\mathrm{IIA}$ and with the same chirality in $\mathrm{IIB}$.} and a $\mathrm{D}p$-brane along the direction $0\cdots p$, the supercharge preserved by the brane is given by
\begin{equation}
 Q= Q_1+ \gamma^{01\cdots p}\, Q_2\,,
\end{equation}
where $\gamma^{01\cdots p}$ belongs to the ten-dimensional Clifford algebra.

The reason why we called them $\mathrm{BPS}$ is that they saturate the {\itshape BPS bound} on their masses. Given the mass of a brane at rest $M$ and the central charge $\ma Z$ realizing the central extension of type $\mathrm{II}$
supersymmetry algebras, it can be shown that the inequality
\begin{equation}
 M \geq |\ma Z|
 \label{bpsbound}
\end{equation}
must be valid to gain a consistent quantization of the supercharges. The BPS bound \eqref{bpsbound} characterizes all massive supersymmetric configurations since it guarantees the possibility to quantize consistently\footnote{The anti-commutators between the supercharges that are associated to the mass of the state could be negative if (\ref{bpsbound}) is violated.} the $\mrm{SUSY}$ algebra (implying that the central extensions $\ma Z$ must be included in order to have massive 
supersymmetric states). Then \eqref{bpsbound} is a crucial condition that explains the stability of a supersymmetric state giving a characterization of it as the state with the minimal mass allowed for a given value of the central charge. For these arguments supersymmetric states are called also {\itshape BPS states}, they are organized in short supermultiplets of the super-Poincar{\'e} algebra and they are fully stable.

\subsection{Effective Action and Worldvolume Theory}
\label{branes}

In addition to branes many other solitonic objects are present in the non-perturbative sector of type II superstrings. In particular we mention the {\itshape orientifold planes} ($\mathrm{O}p$-planes) that are extended objects with negative 
tension. These states are non-dynamical and they usually identify a fixed locus in the background inducing a discrete symmetry on the theory. Other objects are the {\itshape KK monopoles} that are charged 
under mixed symmetry fields. Moreover also {\itshape bound states} of branes are possible and these are crucial to understand the microscopic origin of lower-dimensional objectes. 

Generally the systems considered in string theory explaining the origin of macroscopic configurations, like black holes, are intricated bound states of branes and other solitonic objects whose worldvolume 
is {\itshape wrapped} on some particular manifolds. Also the wrapping of the 
worldvolume can be explained in terms of the interactions between different branes composing a bound state. The leading effect of these interactions
is described by a non-zero curvature of the worldvolume and, often, this curvature induces a spontaneous compactification on the wrapped directions. This holds also for the transverse directions that can be compact and thus defining a lower-dimensional description of a given system.

All these states can be obtained and related each other by making use of dualities. In fact since $\mathrm T$- and $\mathrm{S}$-duality relate different regimes of $\alpha^\prime$ and $g_s$, the brane's tension transforms
under dualities and this allows to relates different solitonic states and to describe their physics in terms of the physics of dual objects.

In order to describe concretely these non-perturbative objects, it is necessary to formulate an action capturing their dynamics. This is possible only in some cases and in a particular energy regime.

Let's consider for simplicity the case of a D$p$-brane on a closed string background and come back to the quantum fields living on their worldvolume mentioned in section \ref{branes1}. As we said, these fields are related to the modes of open strings with the extrema ending on the brane. If we consider the limit in which the energy of the brane is low with respect to that of the open strings, the dynamics of the brane is completely determined by the open strings' massless modes. Hence, there is a $(p+1)$-dimensional theory of massless fields that can be used to construct an effective action of the brane. 

The bosonic part of this action for a D$p$-brane on a background defined by the $\mrm{NS}$-$\mrm{NS}$ fields $(g, B, \Phi)$ is given by
\begin{equation}
\begin{split}
 S_{\mathrm{D}p}=&T_{\mathrm{D}p}\,\int \D ^{p+1}\sigma\,e^{-\Phi}\,\sqrt{-\det\left(g+B+2\pi\,\alpha^\prime\, F+ (2\pi\,\alpha^\prime)^2\,(\partial\,\Phi^i)^2 \right)}\\
 &+\mu_p\,\int \,e^{B+2\pi\,\alpha^\prime F}\wedge\,\sum_k C^{(k)} \,,
 \label{dbi}
 \end{split}
\end{equation}
where the coordinates of the worldvolume are denoted by $\sigma=(X^0,\cdots,X^{p+1})$. The fields appearing in the square root are all expressed in terms of their pull-back on the worldvolume, 
e.g. $g_{\alpha\beta}=g_{\mu\nu}\partial_\alpha X^{\mu}\partial_\beta X^{\nu}$. In particular, coupled to the metric of the background $g$, the $B$-field and the dilaton $\Phi$, there are higer-order contributions in $\alpha^\prime$ given by 
the field strength $F$ associated to the (abelian\footnote{It is also possible to obtain a more general formulation of \eqref{dbi} with non-abelian vector fields.}) vectors belonging to the massless sector of open strings, and by the scalars $\Phi^i$ with $i=(p+2),\cdots,D$ describing the transverse coordinates of the brane.
The first term is called {\itshape Dirac-Born-Infeld action} 
($\mathrm{DBI}$) and it describes the interactions between the brane and the $\mrm{NS}$-$\mrm{NS}$ sector of closed string. 

The second term is called {\itshape Wess-Zumino action} ($\mathrm{WZ}$) and generalizes to extended objects the electromagnetic coupling between
the gauge potential and point-particles. The coupling $\mu_p$ denotes the charge of the $\mathrm{D}p$-brane under the $\mrm R$-$\mrm R$ form $C^{(p+1)}$. Clearly the sum in the  $\mathrm{WZ}$ term must be performed respectively
over all the odd and even values respectively for type $\mathrm{IIA}$ and $\mathrm{IIB}$. 
Furthermore from \eqref{dbi} one can deduce the tension of the brane in the minimal coupling case, i.e. $B=C^{(p)}=0$,
\begin{equation}
 T_{\mathrm{D}p}=\frac{1}{g_s\,(2\pi)^p\,(\alpha^\prime)^{(p+1)/2}}\,,
 \label{tensionbrane}
\end{equation}
thus, when $g_s\rightarrow \infty$ or $\alpha^\prime \rightarrow \infty$ the $\mathrm{D}p$-branes are light. From \eqref{tensionbrane} it is manifest that branes are non-perturbative excitations of string theory. In fact in the limit $g_s\rightarrow 0$, i.e. when the energies are low, they become heavy and completely rigid objects. We will come back on the low-energy regime of string theory in section \ref{decoupling} for a more detailed discussion.

\subsection{M-Theory and UV-Completion}
\label{mtheory}

In section \ref{dualities} we discussed about dualities connecting different superstring theories. In this section we introduce $\mrm{M}$-theory as the eleven-dimensional
theory describing the strong-coupling limit of type $\mathrm{IIA}$ string theory \cite{Witten:1995ex}. Moreover, thanks to the web of dualities, we will show that $\mrm{M}$-theory contains all superstring theories in ten dimensions giving rise to a unified 
picture of all theories of quantum gravity.

It is well established in quantum field theory that, when one considers the strong-coupling regime of a given theory, often divegences arise. In a renormalizable theory one expects that these divergences disappear for two reason. The first  
consists in the perturbative regularization of the theory that can be realized by making use of tecniques like dimensional regularization. The second is the emergence of non-perturbative effects that cure some divergences describing new physical
effects arising only in the strong-coupling regime and that are invisible to perturbation theory. These new effects contribute with new degrees of freedom and, tipically, they manifest the emergence of a fundamental theory in which the divergences 
disappear. Such a fundamental theory constitutes the so-called {\itshape UV-completion} and often unifies the description of phenomena that, at low energies, were featuring different theories. 

As we said, the main example of non-perturbative effects in string theory is given by branes. In the case of type $\mathrm{IIA}$ string theory the strong regime is described by the limit $g_s\rightarrow + \infty$ and, for simplicity, 
we can start by considering $n$ $\mrm{D}0$-branes as non-perturbative\footnote{It can be shown by making use of \eqref{tensionbrane} that their tension diverges if $g_s\rightarrow 0$.} excitations of type $\mathrm{IIA}$. It follows that a tower of states of mass
\begin{equation}
 M_{D0_n}=\frac{n}{l_s\,g_s}\,,
 \label{kkstates}
\end{equation}
exists in the spectrum of open strings associated to the $\mrm{D}0$s.

Now we can consider an $S^1$-fibration over the ten-dimensional background defined in type $\mathrm{IIA}$ string theory. The key point consists in the interpretation of
the states \eqref{kkstates} as Kaluza-Klein\footnote{We will discuss in more detail Kaluza-Klein reductions in section \ref{geometry}.} excitations associated to the spontaneous compactification of an eleven-dimensional theory on the $S^1$ whose radius is given by
\begin{equation}
 R_{11}=l_s\,g_s\,.
 \label{Mtheoryradius}
\end{equation}
Since \eqref{Mtheoryradius} is proportional to the string coupling constant, the perturbative regime corresponds to the limit $R_{11}\rightarrow 0$ and it is described by type $\mathrm{IIA}$ string theory. Viceversa the strong-coupling regime is described
by the decompactification of the circular eleventh dimension. The eleven-dimensional theory obtained in this limit is called M-theory.

One can include in the M-theory picture all the perturbative objects of type $\mathrm{IIA}$ at the same footing of the $\mrm D0$-branes. There exist two types of $\mrm{BPS}$ objects in eleven dimensions describing the embedding of non-perturbative objects of type $\mathrm{IIA}$ string theory:
\begin{itemize}

 \item {\itshape The M2 brane}: It describes the embedding of the fundamental string $\mrm{NS1}$ and of the $\mrm{D}2$. In particular the wrapping of the $\mrm{M}2$ along the eleventh compact direction is given by the $\mrm{NS1}$, while the 
 wrapping along a transverse coordinate is associated to the $\mrm{D}2$ in type $\mathrm{IIA}$.
 
  \item {\itshape The M5 brane}: It describes the strong-coupling limit of a $\mrm{D}4$ wrapping the eleventh compact direction or an $\mrm{NS5}$ wrapping a transverse coordinate.
 
\end{itemize}
As regards\footnote{The case of the $\mrm{D}8$ will be treated separately, since its interpretation in $\mrm M$-theory is a puzzle. See section \ref{highersugras}.} the $\mrm{D}6$, it is the magnetic dual of the $\mrm{D}0$ and it has a ``pure  geometric" description in $\mrm M$-theory in  terms of a $\mrm{KK}$-monopole.

Thus the claim is the existence of theory of quantum gravity in $D=11$ called $\mrm{M}$-theory whose degrees of freedom are associated to the $\mrm{M}2$ and the $\mrm{M}5$ branes. This theory constitutes the $\mathrm{UV}$-completion of 
 type $\mathrm{IIA}$ string theory and it has not free coupling constants since all the parametes describing type $\mathrm{IIA}$ have been resolved in geometrical quantities like the radius $R_{11}$. 
 By construction $\mrm{M}$-theory has to be completely divergence-free since all the non-perturbative effects\footnote{At least excluding the $\mrm{D}8$ brane.} have been embedded in the $\mrm{M}2$ and $\mrm{M}5$ branes or in pure geometry.
 
   \begin{figure}[htbp]
\centering
\vspace{0.1cm}
\includegraphics[height=4.5cm, width=12cm]{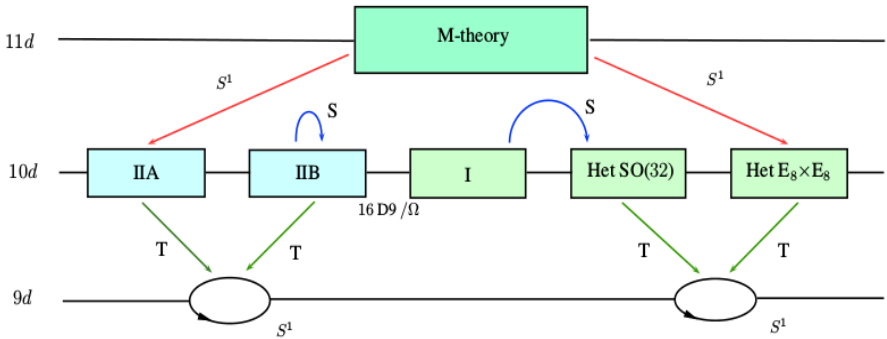}
\vspace{0.1cm}
\caption[Dualities and M-theory. Image from physics.uu.se/research/theoretical-physics/research-areas/string-cosmology-en/.]{The web of dualities linking superstring theories in $D=10$ and their embedding in $\mrm M$-theory.  Image from physics.uu.se/research/theoretical-physics/.}
\label{dualitiestab}
\end{figure}

 There is another direct way to make manifest the existence of $\mrm{M}$-theory that is through the strongly-coupled regime of the $\mathrm{E}_8\times \mathrm{E}_8$ heterotic string theory \cite{Horava:1995qa}. If one considers the $S^1$-fibration 
 over the ten-dimensional background defined in type $\mathrm{IIA}$ string theory and mods out respect to the $\mathbb{Z}_2$ symmetry reversing the sign of the compact coordinate $x^{11}$, 
 one obtains an orbifold fibration on the interval $S^1/\mathbb{Z}_2$ that breaks supersymmetry of the background to $\mathcal{N}=1$ and preserves exactly the spectrum of $\mathrm{E}_8\times \mathrm{E}_8$ heterotic string theory.
 The heterotic coupling constant is given by $g_s=R_{11}/l_s$ and the extrema of the segment $S^1/\mathbb{Z}_2$ are fixed point given by $x^{11}=0$ and $x^{11}=\pi\,R_{11}$. Also in this case $\mrm M$-theory emerges as the strong-coupling limit
 given by $g_s\rightarrow \infty$.
 
 The two boundaries determined by the extrema of the orbifold are called
 {\itshape end of the world 9-branes} and each of them carries on their worldvolume an $\mathrm{E}_8$ gauge supermultiplet and this guarantees the absence of anomalies at the boundaries.
 
 By using dualities there are many ways to relate $\mrm{M}$-theory to the other string theories and
an interesting test of the eleven-dimensional completion can be performed through the matching of the tensions of different $\mrm{BPS}$ states with the tensions of $\mrm M2$ and $\mrm M5$ branes. The main examples are given by the following procedures:
 
 \begin{itemize}
 
  \item Compactifying $\mrm{M}$-theory on a $T^2$, one obtains type $\mrm{IIB}$ on a cirle and this holds since $\mrm{IIB}$ on a circle is $\mrm T$-dual to type $\mrm{IIA}$ on circle.
  
  \item From $\mrm{SO}(32)$ heterotic string theory going to $\mathrm{E}_8\times \mathrm{E}_8$ heterotic string theory through an $\mrm S$-duality transformation,
  
  \item $\mrm S$-duality between $\mrm{SO}(32)$ heterotic string theory and type $\mrm I$ superstring.
  
 \end{itemize}
We conclude this section on $\mrm{M}$-theory with a brief discussion on the non-lagrangianity. The embedding of superstrings in eleven dimensions should implies the existence of supersymmetric sigma-models describing the $\mrm M2$ and $\mrm M5$ 
branes, but one verifies that such sigma-models cannot be quantized consistently. The lack of a lagrangian description is the main peculiarity distinguishing the quantum theory of gravity to the other quantum field theories describing 
fundamental interactions. This fact is manifest also in superstring theories in $D=10$ where it is impossible to construct a consistent quantum string action for backgrounds including $\mrm R$-$\mrm R$ fields. Anyway it is clear that one of the 
gratest challenge towards the unification of gravity to the other forces is finding out a new way in treating non-lagrangian theories.

%% file: chap2.tex

\chapter{Supergravity and Effective Description of Strings}
\label{effectivesupergravity}
\thispagestyle{plain}


The key elements in understanding the microscopic behaviour of gravity are the information coming from non-perturbative effects. 
As we briefly discussed in chapter \ref{strings}, $\mrm{M}$-theory emerges as the $\mrm{UV}$-completion of ten-dimensional string theories and this can be directly seen by studying solitonic states and their eleven-dimensional embedding.

The main issue consists in the lack of a general way in treating non-perturbative effects taking into account the contributions coming from the whole strings' spectrum and the interactions between open and closed superstrings. This puzzle
is probably due the formulation adopted in constructing string theories which is intimately dependent on the background that, in turn, describes the ground state of some particular configurations of superstrings.

At this stage one could conclude that quantum gravity is inaccessible since the higer-order contributions and non-perturbative states are not completely under control. Fortunately in the low-energy limit 
open and closed strings decouple and the background is described in terms of classical massless fields belonging to the ground states of closed strings and defining a {\itshape theory of supergravity}. 
In this context solitonic objects are represented 
by classical solutions in supergravity theories. Regarding the fields originating from the open strings' spectrum, they define a dual quantum description of the solitonic state in terms of a supersymmetric $\mathrm{QFT}$.

Thus the low-energy regime captures many crucial information about non-perturbative states and these can be studied in supergravity theories which are tractable and relatively ``simple".

In this chapter we will discuss about the low-energy limit of strings and their effective description in terms of supergravities. In particular we will firstly introduce the eleven- and ten-dimensional supergravities and secondly 
we will sketch how to produce many lower-dimensional\footnote{In this thesis we will consider lower-dimensional those theories with $d<D=10,11$. } supergravities by dimensional reductions and flux compactification. Then we will introduce the concept of solution in supergravity and we will use it to discuss the $\mrm{AdS/CFT}$ correspondence
as the best framework giving the effective description of non-perturbative states in quantum gravity.

\section{Supergravities as Effective Theories}

In this section we are going to discuss about the effective description of strings' in terms of {\itshape higer-dimensional supergravities}. We will define and discuss the low-energy limit in string theory and, consequently, we will introduce higer-dimensional supergravities in relation to the superstring theories presented in chapter \ref{strings}.

\subsection{Low-Energy Regime and Decoupling of Branes}
\label{decoupling}

We already introduced superstring theories and $\mrm M$-theory as consistent theories of quantum gravity. In particular in section \ref{dualities} we saw that for ten-dimensional string theories there exist two different 
couplings, the string coupling constant $g_s$ and the Regge slope parameter $\alpha^\prime$. The parameter $g_s$ gives rise to the quantum corrections to each mass level of the spectrum. The coupling $\alpha^\prime$  determines the corrections to the string sigma-model and it is related at the same time to the masses of fields in the excited levels of the closed string sector and to the open strings' contribution when non-perturbative effects are included\footnote{We point out that the contributions of the open strings to
the $\mrm{DBI}$ written in \eqref{dbi} are of higer order in $\alpha^\prime$.}.

In the general and non-trivial picture given by branes (or bound states of branes) interacting with a background associated to the $\mrm{NS}$-$\mrm{NS}$ and $\mrm{R}$-$\mrm{R}$ sectors of closed strings, a perturbative description at the level of fields in the spectrum fails since dualities, that relate different energy scales, automatically include non-perturbative effects.

The simplest strategy adopted to describe non-perturbative contributions in string theory is to consider the low-energy limit given by
\begin{equation}
\alpha^\prime\rightarrow 0 \qquad \text{and} \qquad g_s\rightarrow 0 \,.
\end{equation}
As we mentioned at the end of section \ref{branes}, this limit implies that the tensions of the solitonic states\footnote{ In the case of the $\mrm{D}p$-branes the tension has been written in (\ref{tensionbrane}). It is immediate to see that in the low-energy limit $ T_{\mathrm{D}p}$ becomes large.} becomes large. From this it follows that the dynamics of the brane decouples from the background or, in other words, the interactions between open strings and closed strings describing the background vanish.
This regime defines the low-energy limit of strings: all the masses of the excited states of the spectrum diverge and thus the fields of the excited levels decouple. 

In particular when $\alpha^\prime\rightarrow 0$ the dynamics of closed superstring is completely determined by its ground state that 
in turn gives rise, at the zero-order of $g_s$,  to an effective description in terms of a massless and supersymmetric classical theory of gravity, the supergravity theory.

A theory of supergravity can be generically defined as an interacting supersymmetric field theory including gravity in its field content and such that the $\mrm{SUSY}$ is gauged. As we already mention in section \ref{classification}, 
the gauge field corresponding to local supersymmetry is the gravitino which is a $s=3/2$ field\footnote{There are various way to show that the presence of the graviton is automatically included when 
supersymmetry is promoted to a local symmetry. For example, given minimal global SUSY, the anti-commutator between two supercharges is related to the generators of translations of the background. When supersymmetry is gauged, from the anti-commutators of supercharges we obtain the diffeomorphisms of the spacetime and, then, the graviton.}.
Such a theory determines as its particular solution a dynamical spacetime background in terms of a metric tensor and other massless
fields. Hence, from the supergravity point of view, the brane is associated to a configuration of classical fields describing how the background is modified interacting with the brane itself. Since a supergravity theory is a classical theory including gravity it will be non-renormalizable. This fact should not surprise since in this context a classical theory of gravity constitutes an effective description of a fundamental theory.
Moving away from the decoupling limit, perturbative and non-perturbative contributions corrects the divergences and, in the high-energy limit, they give rise to the $\mrm{UV}$-completion of the effective theory.

In the case of D$p$-branes the decoupling between closed and open strings can be understood\footnote{This is not completely clear for other solitonic states like the NS and M branes. Perhaps this could be the cause of the exotic nature of their worldvolume theories.} by considering the effective brane action presented in (\ref{dbi}).
Since the contribution of fields coming from the open strings' spectrum to (\ref{dbi}) is of higher-order in $\alpha^\prime$,  the decoupling of the open strings' modes defines a quantum field theory living on the worldvolume whose interactions with the fields of the background vanish. This leads to an alternative description of the non-perturbative state in terms of a $\mathrm{QFT}$ living on the worldvolume of the brane. This quantum field theory does not contain the graviton, is globally supersymmetric and generally strongly-coupled since its coupling constant is related to $g_s$ that goes to zero in the regime considered.
Furthermore it is a gauge theory and this can be seen directly by studying (\ref{dbi}) in the low-energy regime. If $\alpha^\prime$ is small, it is possibile to expand (\ref{dbi}) and obtain, among the various terms, the following contribution
 \begin{equation}
  -\frac{1}{g_{YM}^2} \int \D^{p+1}\, \sigma F_{\alpha \beta}F^{\alpha \beta}\,.
  \label{gymgs}
 \end{equation}
The coupling $g_{YM}^2$ is proportional to the dimensionless open string's coupling that, in turn, is related to $g_s$ coming from the $e^{-\Phi}$ factor multiplying all the contributions to the DBI.

Hence, the worldvolume theories for D$p$-branes are generally gauge theories with global SUSY. All the properties of these theories can be related to features of the correspondent solitonic state. In particular the R-symmetry group is directly determined by the symmetries of the transverse space.

\begin{table}[htbp]
\label{worldvolumetheories}
\begin{center}
\begin{tabular}{c|c|c}
D$p$-brane         &   $\mrm{SO}(9-p)$   &  R-symmetry        \\
\hline
D7          &  $ \mrm{SO(2)}$ & $\mrm{U}(1)_R$   \\
\hline
D6          &  $ \mrm{SO(3)}$ & $\mrm{SU}(2)_R$  \\
\hline
D5          &  $ \mrm{SO(4)}$ & $(\mrm{SU}(2) \times \mrm{SU}(2))_R $    \\
\hline
D4          &  $ \mrm{SO(5)}$ & $\mrm{USp}(4)_R$    \\
\hline
D3          &  $ \mrm{SO(6)}$ & $\mrm{SU}(4)_R$ 
\end{tabular}

\caption[R-symmetry groups of maximal worldvolume theories.]{\it The R-symmetry groups for maximal supersymmetric theories of D$p$-branes with $p=3,\cdots , 7$. In the cases of D8 and D9 there is no R-symmetry group. The case of the D2 is more complicated because of the duality between vectors and scalars in $d=3$.}
\end{center}
\end{table}
In particular for D$p$-branes with $p=3,\cdots, 7$, the worldvolume theories are $\mrm{SYM}_{p+1}$ theories with maximal supersymmetry. The scalars $\Phi^i$, associated to the transverse coordinates of the brane in \eqref{dbi}, play an important role in this picture since they transform in the fundamental representation of $\mrm{SO}(D-p-1)$ describing the rotational symmetry along the transverse directions. This group in turn defines the R-symmetry of the correspondent worldvolume theory.

In order to produce less supersymmetric worldvolume theories, instead of D$p$-branes of the same type, one has to consider bound states of different branes and solitonic states. In this case the determination of the worldvolume theory is for sure more complicated and has to be considered case by case.

\subsection{Higher-Dimensional Supergravities}
\label{highersugras}

 In this section\footnote{This section is based on \cite{Ortin:2015hya, Becker:2007zj}} we consider the simplest examples of supergravity theories that are those obtained directly by applying the decoupling limit on superstring theories in $D=10$ and on $\mrm{M}$-theory. All these theories are massless since
 they are defined by the fields belonging to the ground state of superstring theories.
 
 Let's start with the low-energy regime of $\mrm{M}$-theory given by {\itshape eleven-dimensional} {\itshape supergravity} \cite{Cremmer:1978km}. This theory contains one
 eleven-dimensional Majorana\footnote{In $D=11$ the irreducible spinors respect the Majorana condition. It has not physical sense to consider extended supersymmetry in $D=11$ and this can 
 be also understood by considering the four-dimensional toroidal reduction of eleven-dimensional supergravity. In fact a reduction of a $D=11$ extended supergravity would produce a $D=4$ supergravity with $\ma N >8$, thus described by
 $s=5/2$ fields. The same argument could be also applied to $D>12$ supergravities. See \cite{Nahm:1977tg}.} spinor with 
 32 independent components, the gravitino gauging supersymmetry, $\Psi_M$ (128 states). The bosonic field content must carry the same amount of degrees of freedom and it is given by the gravitational field $g_{MN}$ (44 states)
 and a three-form\footnote{The indices $M,N,P\cdots=0\cdots10$ are defined by the eleven-dimensional (curved) coordinates' basis. The indices $A,B,C\cdots$ will be used for the flat basis.} $A^{(3)}_{MNP}$ (84 states). The bosonic part of the action is
 \begin{equation}
  S_{11}=\frac{1}{2\kappa_{11}^2}\left [\,\int \D^{11}x\,\sqrt{-g}\,\left (R-\frac12 |F_4|^2 \right)- \frac16 \int A^{(3)}\wedge F_4 \wedge F_4 \right]\,,
 \label{11action}
 \end{equation}
where $\kappa_{11}^2=8\pi G_{11}$ is related to the Newton constant $G_{11}$ that is defined by the Plank scale $l_P$ through the relation $16\pi\,G_{11}=(2\pi)^8\, l_P^9$. The quantity $F_4$ is the field strength of the three form\footnote{We use the notation
 $|A_p|^2=\frac{1}{p !}A_{p\,M_1\cdots M_p}A_p^{\,M_1\cdots M_p}$, with $A_p$ $p$-form.}. Introducing an eleven-dimensional Majorana spinor the action (\ref{11action}) is invariant under the following $\mrm{SUSY}$ transformations,
\begin{equation}
 \begin{split}
  &\delta\, E_M^A=\bar{\epsilon} \, \Gamma^A \,\Psi_M\,,\\
  &\delta \,A^{(3)}_{MNP}=-3\,\bar{\epsilon} \,\Gamma_{[MN}\,\Psi_{P]}\,,\\
  &\delta \, \Psi_M= \nabla_M \epsilon+\frac{1}{12}\,\left(\frac{1}{4!}\,\Gamma_M \Gamma^{NPQR}\,F_{NPQR}-\frac{1}{2}\,\Gamma^{NPQ}\,F_{MNPQ}\right )\,,
  \label{susyvariation11d}
 \end{split}
\end{equation}
where $\nabla_M \epsilon=\partial_M \epsilon+\frac14 \omega_{MAB}\Gamma^{AB}\epsilon$ and $\omega_{MAB}$ is the spin-connection defined in terms of the eleven-dimensional vielbein $E_M^A$ and $\{\Gamma^A\}$ constitute a realization of the Clifford algebra associated to the Lorentz group $\mrm{SO}(1,10)$.

Similarly all the superstring theories in ten dimensions have an effective limit described by a $D=10$ supergravity theory. One has {\itshape type IIA} and {\itshape type IIB supergravities} whose field contents are determined by the 
zero-level of the spectra of $\mrm{IIA}$ and $\mrm{IIB}$ string theories presented in section \ref{classification}. 
These theories are defined by two Majorana-Weyl\footnote{In $D=10$ the irreducible decomposition of a Dirac spinor in Weyl components is compatible with the Majorana condition, such spinors are called Majorana-Weyl spinors.} gravitinos that have respectively the opposite and the same chirality 
for type $\mrm{IIA}$ and $\mrm{IIB}$ supergravity. They are maximal supersymmetric, i.e. their supercharges enjoy together 32 independent components, since in $D=10$ each Majorana-Weyl spinor has 16 real supercharges.

The spinors' structures of type $\mrm{IIA}$ supergravity suggests a relation with $D=11$ supergravity, 
in fact it is possibile to show that the Kaluza-Klein reduction of eleven-dimensional supergravity on a circle leads to type $\mrm{IIA}$ supergravity. Moreover, in section \ref{dualities} we discussed about dualities as transformations 
between fields of the strings' spectrum, thus it is possibile to relate type $\mrm{IIA}$ and $\mrm{IIB}$ supergravity by making use of the action of $\mrm T$-duality on the fields of the two theories.

Let's introduce the bosonic actions of these two supergravity theories. As regards type $\mrm{IIA}$ supergravity, we recall that the $\mrm{NS}$-$\mrm{NS}$ sector is composed by the dilaton $\Phi$, the gravitational field $g_{\mu\nu}$
and the 2-form $B_{\mu\nu}$, while the $\mrm{R}$-$\mrm{R}$ sector is described by a 1-form $C^{(1)}$ and a 3-form $C^{(3)}$. The bosonic action is given by
\begin{equation}
 \begin{split}
  S_{IIA}&=\frac{1}{2\kappa_{10}^2}\,\biggl [\int \D^{10}x\,\sqrt{-g}\,e^{-2\Phi}\left(R+4\,\partial_{\mu}\,\Phi\,\partial^{\mu}\,\Phi-\frac12\,|H_3|^2   \right)\\
  &-\frac{1}{2}\,\int \D^{10}x\,\sqrt{-g}\,\left(|F_2|^2+|\tilde{F}_4|^2\right)-\frac12\,\int B\wedge F_4\wedge F_4\biggr]\,,
 \end{split}
\end{equation}
where $\kappa_{10}^2=8\pi G_{10}$ is related to the Newton constant $G_{10}$.
The field strengths are defined as $H_3=\D B$, $F_2=\D  C^{(1)}$ and $\tilde{F}_4=\D C^{(3)}+C^{(1)}\wedge H_3$.

The vacuum expectation value of $e^{\Phi}$ is the type $\mrm{IIA}$ superstring coupling constant $g_s$, thus if one introduce the reduction ansatz for the metric of $D=11$ supergravity \cite{0264-9381-2-3-007} on a circle given by
\begin{equation}
 \D \,s^2=e^{-2\Phi/3}\,g_{\mu\nu}\D x^{\mu}\D x^{\nu}+e^{4\Phi/3}\,\left (\D\,x^{11}+C^{(1)}_{\mu}\,\D\,x^{\mu} \right)\,,
 \label{reduction}
\end{equation}
it is easy to deduce the relation $l_P= g_s^{1/3}\,l_s$ between the Plank scale $l_P$ defining the scales of energies of $\mrm M$-theory and the fundamental string's length $l_s$ characterizing the scale of the $\mrm{IIA}$ theory.

Unlike type $\mrm{IIA}$ supergravity, it is not possibile to derive an action for type $\mrm{IIB}$ supergravity by dimensional reduction from eleven-dimensional supergravity. Moreover the presence of a self-dual 5-form field strength gives an obstruction to formulate the action 
in a covariant form \cite{Schwarz:1983wa,Schwarz:1983qr,Howe:1983sra}. What is usually done is inducing an action from the equations of motion and supplementing it with the self-duality constraint on the 5-form. The field content is that one of the ground state of 
$\mrm{IIB}$ string theory; hence from the $\mrm{NS}$-$\mrm{NS}$ sector ona gains the dilaton $\Phi$, the gravitational field $g_{\mu\nu}$ and the 2-form $B_{\mu\nu}$ (with $H_3=\D B$), while the $\mrm{R}$-$\mrm{R}$ sector is composed by 
the forms $C^{(0)}, C^{(2)}, C^{(4)}$. The equations of motions can be deduced from the action
\begin{equation}
 \begin{split}
  S_{IIB}&=\frac{1}{2\kappa_{10}^2}\,\biggl [\int \D^{10}x\,\sqrt{-g}\,e^{-2\Phi}\left(R+4\,\partial_{\mu}\,\Phi\,\partial^{\mu}\,\Phi-\frac12\,|H_3|^2   \right)\\
  &-\frac{1}{2}\,\int \D^{10}x\,\sqrt{-g}\,\left(|F_1|^2+|\tilde{F}_3|^2+\frac12|\tilde{F}_5|^2\right)-\frac12\,\int C^{(4)}\wedge H_3\wedge F_3\biggr]\,,
 \end{split}
\end{equation}
where $F_{n+1}=\D C^{(n)}$, $\tilde{F}_3=\D C^{(3)}-C^{(0)}\, H_3$ and $\tilde{F}_5=\D C^{(4)}-\frac12 C^{(2)}\wedge H_3-\frac12 B\wedge F_3$. The term in $|\tilde{F}_5|^2$ does not include the self-duality condition, thus the number of degrees of freedom is doubled. The only possibility is to add ``by hand" this relation as a supplement of the equations of motion,
\begin{equation}
 \tilde{F}_5=\star \tilde{F}_5\,.
\end{equation}
In section \ref{dualities} we mention that $\mrm S$-duality maps type $\mrm{IIB}$ string theory in itself and this action can be extended to the action of $\mrm{SL}(2,\mathbb{Z})$. If one restricts the action only on the massless spectrum, namely 
on the field content of type $\mrm{IIB}$ supergravity, this symmetry is enhanced to $\mrm{SL}(2,\mathbb{R})$. Given a transformation 
\begin{equation}
R=  \left( \begin{array}{ccc}
a & b  \\
c & d \\
 \end{array} \right) \in \mrm{SL}(2,\mathbb{R})\,,
\end{equation}
then the 2-forms $B$ and $C^{(2)}$ transform as a doublet under $R$, the graviton and the 4-form are invariant while the scalars can be organized in an {\itshape axio-dilaton} 
\begin{equation}
\tau=C^{(0)}+i\,e^{-\Phi}
\end{equation}
transforming as
\begin{equation}
 \tau \rightarrow \frac{a\tau+b}{c\tau+d}\,.
\end{equation}
Finally we mention the low-energy limit of the less supersymmetric superstring theories given by {\itshape type I supergravity} and {\itshape heterotic supergravity}, that are of two types respectively belonging to $\mrm{SO}(32)$ and 
$\mrm{E}_8 \times \mrm{E}_8$ heterotic string theory.

We conclude this section with a brief discussion on the issues arising when including the $\mrm{D}8$ brane in this picture.
In sections \ref{branes1}, \ref{branes} and \ref{mtheory} we introduced the various $p$-branes in type $\mrm{II}$ string theories and we explained their eleven-dimensional origin. In the low-energy limit these are described by classical solutions in type $\mrm{II}$ and $D=11$ supergravities presented above. However there is a particular case that does not fit in this scenario. This is the case of the $\mrm{D}8$ brane. The low energy limit of this non-perturbative state cannot be described by a solution in type $\mrm{IIA}$ supergravity and, thus, escapes an $\mrm{M}$-theory completion.

More precisely the $\mrm{D}8$ requires a non-dynamical $\mrm{R}$-$\mrm{R}$ 10-form field strength $F_{10}$ that does not give any degree of freedom to $\mrm{IIA}$ supergravity. The presence of the $\mrm{D}8$ can be explained in a unique way: by considering a deformation of type $\mrm{IIA}$ supergravity called {\itshape massive type IIA supergravity} \cite{Romans:1985tz}. This deformation is unique and consists in the introduction of a mass parameter $m$ called {\itshape Romans mass} preserving formally all the supersymmetries and the gauge symmetries of type $\mrm{IIA}$ supergravity. More precisely, in this theory only the linear action of $\mrm{SUSY}$ on the supermultiplets is preserved, while the local supersymmetry is broken\footnote{This can be seen, for example, by the existence of a $\mrm{SUSY}$ transformation that gauges away the dilatino and gives mass to the gravitinos.}.

The key point is the interpretation of the Romans mass $m$ as the zero-dimensional Hodge dual $F_0$ of the non-dynamical $\mrm{R}$-$\mrm{R}$ 10-form field strength $F_{10}$ associated to the $\mrm{D}8$, i.e. $m=F_0$. The presence of $F_{10}$ implies the existence of a 9-form $C^{(9)}=\D F_{10}$ that can be interpreted as the gauge potential associate to the $\mrm{D}8$. Thus massive type $\mrm{IIA}$ supergravity describes the low-energy limit of type $\mrm{IIA}$ string theory in presence of the $\mrm{D}8$ branes.
Equivalently, it can be interpreted as the effective theory of type $\mrm{IIA}$ string theory including the open string sector of the $\mrm{D}8$ brane. 
As we said above, this inclusion breaks dynamically SUSY and the $D=10$ Poincar\'e invariance, and, for this reason, it is not known how to derive massive type $\mrm{IIA}$ supergravity from $\mrm{M}$-theory. In other words it is not possible to deform eleven-dimensional supergravity to include Romans mass and, at the same time, preserve the eleven-dimensional Poncar\'e invariance \cite{Bergshoeff:1997ak}. 

It follows that also the $\mrm{M}$-theory interpretation of the $\mrm{D}8$ brane is unknown and its presence in a consistent deformation of type $\mrm{IIA}$ supergravity hints at the existence of an eleven-dimensional $\mrm{M}9$ brane\footnote{The $\mrm{M}9$ could be related to the end-of-the-world 9-branes that we mentioned in section \ref{mtheory}.}, that describes the strong-coupling regime of the $\mrm{D}8$ and evades the low-energy regime of $\mrm{M}$-theory given by $D=11$ supergravity \cite{Bergshoeff:1998re}.

We can now introduce the bosonic action of massive type $\mrm{IIA}$ supergravity, 
\begin{equation}
  S_{mIIA}=\frac{1}{2\kappa_{10}^2}\,\biggl [\int \D^{10}x\,\sqrt{-g}\,e^{-2\Phi}\left(R+4\,\partial_{\mu}\,\Phi\,\partial^{\mu}\,\Phi-\frac12\,|H_3|^2   \right)-\frac12 \sum_{p=0,2,4} |F_p|^2  \biggr]+S_{\text{top}}\,,
  \label{massiveIIAaction}
\end{equation}
where $S_{\text{top}}$ is a topological term given by
\begin{equation}
\begin{split}
S_{\text{top}}&=-\frac{1}{4\kappa_{10}^2} \int ( B\wedge F_4\wedge F_4-\frac13  F_0\wedge B\wedge B\wedge B\wedge F_4\\
&+\frac{1}{20}F_0\wedge F_0\wedge B\wedge B\wedge B\wedge B\wedge B)\,.
\end{split}
\end{equation}
All the equations of motion can be derived consistently from \eqref{massiveIIAaction}. In particular we recall the Bianchi identity for $F_0$ which is
\begin{equation}
\D F_0=0\,,
\label{bianchiromans}
\end{equation}
implying that $F_0$ has to be constant. The relation \eqref{bianchiromans} is the equation of motion for the 10-form field strength $F_{10}=\star F_0$, then implying the existence of the gauge potential $C^{(9)}$ associated to the $\mrm{D}8$ brane.

\section{Lower-Dimensional Supergravities}

Since the observed world is four-dimensional, one has to provide a mechanism describing how the lower-dimensional physics is related to the higer-dimensional one. The core of this mechanism is represented by {\itshape compactifications} or
{\itshape dimensional reductions} and involves the propagation of strings on partially compact manifolds. More precisely, if some directions of a higher-dimensional background wrap a compact manifold, a new scale parameter associated to the size of this manifold appears. In the limit in which the size of the manifold becomes small, a lower-dimensional effective description of the theory emerges.

The propagation of strings on a background with compact coordinates implies peculiar consequences at the level of the their spectrum. The modes are defined on curved manifolds with some compact coordinates and from this it follows that, sometimes, the degrees of freedom associated to the compact directions decouple leaving intact a finite set of modes defined on the lower-dimensional background.
In these particular cases the dimensional reduction defines a {\itshape consistent truncation} and physics of high dimensions is captured by a finite set of fields defining a lower-dimensional supergravity theory whose solutions can be consistently uplifted to higher-dimensional supergravities.

A crucial ingredient is of course the geometry describing these compact manifolds. It determines the lagrangian and the main properties of lower-dimensional theories that are, by construction, classical and non-renomalizable. In this section we will present a brief 
discussion on strategies of dimensional reductions and we will roughly classify the main geometries used in compactifications. Then we will discuss about a possible classification of lower-dimensional supergravities based on the dimension and on the amount of supercharges preserved.

\subsection{Dimensional Reductions and Internal Geometries}
\label{geometry}

We already introduced an example of reduction Ansatz in (\ref{reduction}) when we considered the relation between the metric respectively of eleven-dimensional and type $\mrm{IIA}$ supergravity. That was a relatively simple example of dimensional reduction
on a $S^1$. In this section we want to introduce the general aspects relating dimensional reduction of $D=11$ and type $\mrm{II}$ supergravities.

The mechanism of dimensional reduction can be analyzed under two general aspects. The first is geometrical in the sense that it basically consists in the study of the dynamics of strings on backgrounds of the type
\begin{equation}
 M_{D}=M_{d}\times X_{D-d}
 \label{compact}
\end{equation}
where $X_{D-d}$ is a $(D-d)$-dimensional compact manifold. More precisely the $D$-dimensional background is interpreted as an $X_{D-d}$-fibration over the lower-dimensional space $M_{d}$. 
The second aspect characterizes dimensional reductions as scale effects producing lower-dimensional effective theories. In particular the couplings and the main observables respectively in $D$ and $d$ dimensions are linked by simple relations, as in the example of $\mrm{M}$-theory and $\mrm{IIA}$ string theory, depending on properties of the compact manifold $X_{D-d}$. Both these aspects are kept in a single shot in a reduction Ansatz on the higher-dimensional fields (see for example (\ref{reduction})) locally realizing a background of the type \eqref{compact}.
There are many ways to compactify a theory, we cite the two main ones:
\begin{itemize}

 \item {\itshape Kaluza-Klein (KK) reductions} \cite{Kaluza:1921tu, Klein:1926tv}: Since some coordinates are compact, the fields defined on the $D$-dimensional background can be expanded in Fourier modes along the compact coordinates of $X_{D-d}$. The modes of the zeroth level of the expansion are massless while the excited modes are massive and constitute a tower of infinite states describing the compactification. The KK reduction of gravity theories implies the appearance of gauge bosons in $d$ dimensions gauging the group $G$ of internal symmetries of the lower-dimensional theory. A consistent truncation occurs when all the $D$-dimensional equations of motion are satisfied 
 if and only if the $d$-dimensional equations of motion of a finite set of modes, belonging to the Fourier expansion, are satisfied and all the gauge fields of the isometry group $G$ are retained in the truncation \cite{Duff:1984hn}. 

 \item {\itshape Twisted reductions} \cite{Scherk:1979zr}: These are reductions in which $X_{D-d}$ is a group manifold $G$. The general strategy consists in parametrizing the reduction Ansatz of the metric in terms of the left-invariant forms on the group manifold. The Ansatz breaks the invariance of diffeomorphisms of the $D$-dimensional background in the $d$-dimensional diffeomorphisms plus a group associated to the internal symmetry. These reductions are interesting since they are examples of compactifications to lower-dimensional theories that are deformed by a potential for the scalars, appearing in the reduction, and characterized by some charged fields under the gauge group producing the potential. Furthermore they always define consistent truncations\footnote{The consistency of twisted reductions is guaranteed by the fact that the lower-dimensional fields retained by these compactifications are exactly those left-invariant.}.
 
\end{itemize}
The most interesting compactifications in string theory are those defining a consistent truncations. In this case the higher-dimensional physics is captured by a finite set of fields defining a lower-dimensional supergravity whose solutions can be consistently uplifted to higher-dimensional supergravities.

As we said, many are the geometries that can be chosen for the internal manifold $X_{D-d}$ and a general classification of them represents a very difficult task. We can generically distinguish between those reductions that can be performed directly from the lagrangian of the higer-dimensional theory and those defined only at the level of the equations of motion. The main examples of compactifications that are used in string theory are:
\begin{itemize}

 \item {\itshape Toroidal reductions}: The flatness of the torus $T^{(D-d)}$ allows the $\mrm{KK}$ reduction at the level of the action. These reductions preserve the total amount of supersymmetry of the higer-dimensional theories. For this reason, if one considers the consistent truncation on a torus of type $\mrm{II}$ or $D=11$ supergravities, the resulting supergravities in low dimensions will be maximally supersymmetric. Theese theories are called {\itshape maximal supergravities}.
 
 \item {\itshape Calabi-Yau $(\mrm{CY}_n)$ reductions}: Calabi-Yau manifolds are K{\"a}hler manifolds in $n$-dimensions with $\mrm{SU}(N)$ holonomy. The main property following by this definition is the Ricci-flatness\footnote{This holds for  
 compact Calabi-Yau manifolds. The reductions on non-compact Calabi-Yau manifolds are also possible but there are many restrictions since the relation between $\mrm{SU}(N)$ holonomy and Ricci-flatness is not automatically given.}. The lower-dimensional theories that are usually obtained when the reduction defines a consistent truncation are determined by the {\itshape deformations} admitted by these manifold. These reductions are more complicated with respect to toroidal ones and tipically are needed in order to obtain supergravities with less supersymmetry.

\end{itemize}

The reductions over torii and Calabi-Yau manifolds have been the firsts studied in literature, but unfortunately they are characterized by an intrinsic issue: they produce many massless scalar fields, or {\itshape moduli}, in the lower-dimensional theory. These scalars are free to run on the background without a well-defined expectation value. Moreover their degrees of freedom are not confirmed by particle physics and by General Relativity. 
The set of strategies used to approach this problem is called {\itshape moduli stabilization}. The aim is to find those consistent truncations leading to lower-dimensional supergravities deformed by a {\itshape scalar potential}. This is defined by the scalar fields coming from the reduction and it has to reproduce in its critical points the ground state of the lower-dimensional theory. In this way the vacuum expectation values of the scalars are then stabilized at the values extremizing the potential.

In this context an important example of moduli stabilization strategy is given by {\itshape flux compactifications} \cite{Grana:2005jc}. This consists in the inclusion of some suitable background fluxes wrapping $p$-cylces in the internal manifold $X_{D-d}$. These running fluxes deform the reduction Ansatz of the field strengths associated to the $p$-form gauge potentials and give rise to a scalar potential in the $d$-dimensional theory. This strategy is particularly suitable in Calabi-Yau reductions since their internal manifold is generally characterized by a rich plethora of non-contractible $p$-cycles.
\begin{table}[htbp]
\label{spherereduction}
\begin{center}
\begin{tabular}{c|c|c}
$d$        &   {\itshape Gaugings}   &  {\itshape Truncation}        \\
\hline
8            &  $ \mrm{SO(3)}$ & $\mrm{IIA}$ on $ S^2 $  \cite{Salam:1984ft}   \\
\hline
7            &   $  \mrm{SO(4)}    $     &  $\mrm{mIIA}$ on $ \tilde{S}^3 $  \cite{Passias:2015gya} \\
\hline
7           &   $  \mrm{SO(5)} $        &  $ D=11$ on $ S^4 $  \cite{Pilch:1984xy,Nastase:1999cb}    \\
\hline
6           &    $   \mrm{SO(5)} $    &  $\mrm{mIIA}$ on  $ \tilde{S}^4   $  \cite{Brandhuber:1999np}    \\
\hline
6           &    $   \mrm{SO(5)} $    &  $\mrm{IIA}$ on $ S^4 $  \cite{Cvetic:2000ah}    \\
\hline
5            &  $   \mrm{SO(6)} $        & $\mrm{IIB}$ on $ S^5  $  \cite{Cvetic:2000nc}  \\
\hline
4           &   $  \mrm{ISO(7)} $        &  $\mrm{mIIA} $ on $ \tilde{S}^6 $  \cite{Guarino:2015vca}  \\
\hline
4           &   $  \mrm{SO(8)} $        &  $D=11$ on $ S^7 $  \cite{deWit:1986oxb}  \\
\hline
\end{tabular}
\caption[Gauge groups of supergravities from various consistent truncations on spheres.]{\it Examples of gauge groups of supergravities obtained from various consistent truncations on spheres. The truncations of massive $\mrm{IIA}$ are realized by squashed spheres $\tilde{S}^{D-d}$ with the same topologies of the spheres, but different metrics.}
\end{center}
\end{table}

In addition to flux compactification in Calabi-Yau reductions, there are other ways to find consistent truncations producing lower-dimensional supergravities driven by a scalar potential.

The most relevant example are the truncations on {\itshape spheres} $S^{D-d}$. These truncations typically produce maximal theories in low dimension that, unlike toroidal reductions, are characterized by a scalar potential. Moreover also covariant derivatives arise together with the potential and they are associated to a set of vector fields gauging the isometry group of the sphere. The maximal supergravities obtained from toroidal reductions are thus deformed in a spherical reduction by the presence of some gauged symmetries associated to the isometries of the sphere. 

Many other compactifications of $D=11$ and type $\mrm{II}$ supergravities can be constructed. In particular we mention those on {\itshape Sasaki-Einstein manifolds} $SE_n$. These manifolds are defined as Ricci-flat manifolds with a K{\"a}hler cone structure and the truncation of higher-dimensional supergravities on them are of great relevance for $\mrm{AdS/CFT}$ correspondence.

\subsection{A Survey of Supergravities}
\label{survey}

Since there are many ways to compactify higer-dimensional supergravities, it follows that there exist many different theories in low dimensions. The simplest approach to classify them is based on the dimension $d$ of the reduced background and on the 
number of supercharges characterizing the lower-dimensional theory.

The number of supercharges is generically a multiple $\ma N$ of the number of the real components of a $d$-dimensional irreducible spinor and it defines the irreducible
representation of the super-Poincar{\'e} group. It follows that, as in any supersymmetric theory, the fields are organized in supermultiplets. Clearly there are some upper bounds on the number of supersymmetries preserved since we are considering 
fields whose maximum value of spin is two.
We already said in section \ref{highersugras} that the maximal dimension admitted is $D=11$. If we consider the number of supercharges preserved by a given theory thus, by studying the irreducible representations of the supersymmetry algebra, one conclude that the maximum amount is given by 32. The theories with 32 supercharges are exactly the maximal supergravities
introduced in section \ref{geometry} and their presence allows the classification of all supergravities from $d=9$ to $d=3$.
This classification can be performed without considering the stringy origin of supergravities and just asking a maximum number of 32 supercharges and maximum dimension 11. The remarkable fact is the perfect agreement between the higer-dimensional theories and superstring theories. 

 Imposing the bounds on $d$ and $\ma N$ discussed above, one can study the irreducible representations of the super-Poincar{\'e} algebra and finding four types of supermultiplets for a supergravity theory: 
\begin{itemize}
 
 \item {\itshape Supergravity multiplet} ($s_{max}=2$): It contains at least the graviton and $\ma N$ gravitinos. Depending on $\ma N$ it can include also vectors, forms and scalars.
 
 \item {\itshape Vector multiplets} ($s_{max}=1$): They exist in $\ma N \leq 4$ theories\footnote{The $s_{max}=3/2$ multiplet does not define an independent interacting field theory.}. Often the vectors in these multiplets gauge an extra internal symmetry group. 
 
 \item {\itshape Chiral multiplets} ($s_{max}=1/2$): They contain only $s=1/2$ fields and scalars.
 
  \item {\itshape Hypermultiplets} ($s_{max}=1/2$): They exist only in $\ma N=2$ theories and have the same role of chiral multiplets. They may transform under the internal group defined by the vector multiplets.
  
\end{itemize}
The last three multiplets define the coupling to the matter of the supergravity multiplet, thus a theory containing only the supergravity multiplet is called {\itshape minimal supergravity} otherwise it is called {\itshape 
matter-coupled supergravity}. The principal lower-dimensional supergravities are classified\footnote{Theories characterized by Symplectic-Majorana spinors (see Appendix \ref{app:SM_spinors} for an example in $d=7$) are sometimes labelled by an $\ma N$ doubled with respect to that defined by the value $\ma N$ associated to the corresponding irreducible Dirac spinor written in table \ref{sugras}. For example in chapter \ref{gaugedsugras}, following the most of literature, we will refer to the $d=5$ theory with 8 real supercharges as $\ma N=2$ supergravity.} in table \ref{sugras}.
\begin{table}[htbp]
\begin{center}
\begin{tabular}{c|c|cl}
$d$        &   $\ma N$   &  {\itshape Supercharges}        \\
\hline
11            &  1 & 32   \\
\hline
10            &   $(1,0)=\mrm{I}$, \, $(1,1)=\mrm{IIA}$, \,  $(2,0)=\mrm{IIB}$     & 16, 32, 32 \\
\hline
9,8,7           &   1, 2      &    16, 32  \\
\hline
6           &    $(1,0)=\mrm{i}$, $(1,1)=\mrm{iia}$, $(2,0)=\mrm{iib}$,  $(2,1)$,  $(2,2)$   &  8, 16, 16, 24, 32    \\
\hline
5           &    1, 2, 3, 4    & 8, 16, 24, 32     \\
\hline
4            &  1, 2, 3, 4, 5, 6, 8       & 4, 8, 12, 16, 20, 24, 32  \\
\hline
\end{tabular}
\caption[Supergravities in different dimensions. Table from \cite{Dibitetto:2012xd}.]{\it Supergravities in different dimensions. Maximal supergravities are those with 32 supercharges. In $D=10$ and $d=6$ we indicated also chiralities, for example type IIA supergravity has two gravitinos with opposite chirality, i.e. $\ma N=(1,1)$. In theories with $d=6$ also tensor multiplets are possible. Table from \cite{Dibitetto:2012xd}.} \label{sugras}
\end{center}
\end{table}

Coming back to dimensional reductions, the deeper one goes with dimensional reduction, the more becomes intricated the physics described by the lower-dimensional supergravity. As we will see, in many cases, the field content is not enough to fix the theory and strings' compactifications define not only the set of lower-dimensional fields but also the main features of the lower-dimensional theories. For example the scalar fields in the lower-dimensional supergravity often enter in the action through suitable non-linear sigma models defining a {\itshape scalar geometry}. So the moduli often parametrize some target manifolds defining a particular model that, in turn, can be derived by dimensional reduction.

The global symmetry group of a supergravity theory is associated to the set of transformations leaving invariant the scalar geometry. The gauging of some of these isometries produces the breaking of the global symmetry since a scalar potential and covariant derivatives appear in order to keep the theory invariant under the gauged symmetries. These theories are called {\itshape gauged supergravities} while if the global symmetry group is preserved and there are not gauged isometries they are called {\itshape ungauged supergravities}. 

In section \ref{geometry} we cited some examples of moduli stabilization in dimensional reductions, like flux compatification strategies in Calabi-Yau reductions or truncations on spheres.
When these examples define consistent truncations of higher-dimensional supergravites, they usually produce lower-dimensional gauged supergravities. The converse is not an obvious fact since many lower-dimensional supergravities could be described by very intricate scalar geometries admitting various gaugings and one should provide an higher-dimensional picture case by case.

\section{Solutions in Supergravity}

The solutions of equations of motion of a given supergravity theory 
capture the main information about the dynamics of strings, branes and other intrinsically stringy objects. Moreover starting from particular solutions of higher-dimensional supergravity, describing for example the modifications of a background in presence of a 
$\mrm D p$-brane, one can consider various consistent truncations giving rise to solutions of lower-dimensional supergravities. In other words the macroscopic systems strongly coupled to gravity, like black holes and domain walls,
can be explained in terms of microscopic configurations of branes and other non-perturbative objects whose physics is captured by higer-dimensional supergravities.

The compactifications of the extra dimensions can be understood as in the case of type $\mrm{IIA}$ string theory and $\mrm M$-theory: since renormalizability in theories of gravity is related to the presence of extra dimensions, the scale 
used in studying a macroscopic system tells us how large is the contribution of the higer-dimensional physics. This means for example that
the more closer we will study a black hole in $d=4$, the more it will manifest its stringy nature in terms of a more fundamental higer-dimensional solution, as the study of classical electromagnetic phenomena at short scales
manifests the existence of photons and electrons. It is clear that the supergravity regime remains a classical picture, but, anyway, with the help of supergravity theories we can understand which particular quantum configuration
gives rise to the macroscopic system and then study many of its properties.

In this section we will introduce the concept of Killing Spinor in relation to supersymmetric ($\mrm{BPS}$) solutions and we will make the distinction between extremal and non-extremal solutions. We will present some celebrated solutions in higer-dimensional
supergravities and we will briefly discuss the relations between solutions respectively in higer- and lower-dimensional supergravity

\subsection{Killing Spinors and Supersymmetry of Classical Solutions}
\label{killingspinor}

The lagrangians defining supergravity theories are invariant under local supersymmetry transformations. These are determined by a fermionic parameter $\epsilon(x)$ depending on the background. This spinor is called {\itshape Killing spinor} when it is
referred to a particular solution of the equations of motion. More precisely a set of Killing spinors is a set of $\mrm{SUSY}$ invariant spinors for which the supersymmetry transformations leaves a particular solution invariant\footnote{This section is partially based on \cite{VanProeyen:1999ni,Freedman:2012zz}.}. 

Equivalently we can say that the residual global supersymmetry of a solution is generated by a set of constant parameters contained in the Killing spinors and, thus they define the number of supercharges preserved by the solution.
In this context, a supersymmetric solution is called $\mrm{BPS}/n$ where $n$ is the fraction of supercharges contained in the Killing spinor characterizing the solution. 
The maximally supersymmetric solutions are those solutions associated to a set of Killing spinor generating the complete $\mrm{SUSY}$ algebra of the theory.

Reversing this discussion, Killing spinors constitute a powerful tool in the research of particular solutions of the field equations. Supergravities are classical supersymmetric theories and this would be quite contraddictory since classical backgrounds 
are purely bosonic. The only possibility for having a purely bosonic solutions is to consider the fermions vanishing. This condition leads to an important set of differential relations called {\itshape BPS equations}\footnote{
In this thesis we will refer to them also as {\itshape first-order flow equations} or simply {\itshape flow equations}.} that are a set of first-order equations
whose solutions solve also the equations of motion. More generally they give a set of consistent conditions that have to be respected by a $\mrm{SUSY}$ solution.

If we denote the bosonic fields of a supergravity theory collectively by $B(x)$ and the fermions by $F(x)$, the local $\mrm{SUSY}$ transformations of the theory in the limit of fermions vanishing may be written schematically in this form \cite{Freedman:2012zz}
\begin{equation}
\delta B(x)=\bar{\epsilon}(x)f_1(B(x))F(x)+o(F^3)\,, \quad \delta F(x)=f_2(B(x))\epsilon(x)+o(F^2)\,,
\label{bpsprototype}
\end{equation}
where the functions $f_1(B)$ and $f_2(B)$ are determined by the specific theory considered.

For each preserved supersymmetry there is a non-trivial configuration of the Killing spinor $\epsilon(x)$ such that both $\delta B(x)$ and $\delta F(x)$ vanish when a classical solution is substituted in (\ref{bpsprototype}).
Since the fermions vanish classically\footnote{A non-zero vacuum expectation value for fermions would violate the Lorentz symmetry.}, the condition on $\delta B(x)$ is trivially satisfied, so we need to study only the linearized fermionic variations,
\begin{equation}
\delta F(x)|_{\text{lin}}= f_2(B(x))\epsilon(x)=0\,.
\label{bpsequationsgeneral}
\end{equation}
The conditions (\ref{bpsequationsgeneral}) are a system of first-order differential equations involving bosons $B(x)$ and the Killing spinor $\epsilon(x)$ and are called $\mrm{BPS}$ equations.

Killing spinors can be used also to classify all the supersymmetric solutions invariant under a given set of isometries. Given a pair of Killing spinors $\epsilon$ and $\epsilon^\prime$, the bilinear $\epsilon \gamma^\mu \epsilon^\prime$, with $\gamma^{\mu}$ $d$-dimensional gamma matrices, is a Killing vector associated to the diffeomorphisms leaving invariant the solutions. The study of the complete systems of fermionic bilinears leads directly to the most general expression of the first-order equations associated to solutions with a given set of symmetries.

Since supergravities are defined by field equations that are of the second order and highly non-linear it is clear that the solution of a set of first-order equations semplifies dramatically the research of solutions. Generally we can say that $\mrm{BPS}$ 
equations furnish a solid interpretation on what supersymmetry effectively is when considering the low-energy limit of strings. The pairing between fermionic and bosonic degrees of freedom is needed at the Plank scale in order to avoid the presence of tachyons in the spectrum of 
superstrings. In the low-energy limit this symmetry reveals itself through a set of differential constraints that has to be respected by a $\mrm{BPS}$ solution.

\subsection{Vacuum Solutions and Maximal Symmetry}
\label{vacua}

Between classical solutions, the simplest are those preserving {\itshape maximal symmetry}. These solutions are called {\itshape vacua} since when considered in $D=11$ and type $\mrm{II}$ supergravities they describe configurations of minimal energy of the fields coming from the massless sector of closed strings. Moreover their truncations to low dimensions define configurations in absence of particles or other solitonic objects. Between various symmetries there are of course supersymmetries that are associated to the largest possible amount of linearly independent Killing spinors. 

If we want to relate vacua at different dimensions, we may give a definition of vacuum in $D=11$ or type $\mrm{II}$ supergravities as a solution described by an $\Omega_{D-d}$-fibration over a maximally symmetric solution of a $d$-dimensional supergravity. The link between the higher-dimensional vacuum and the $d$-dimensional maximally symmetric background must be established by a consistent truncation over $\Omega_{D-d}$. In this case the lower-dimensional solution is a vacumm of the $d$-dimensional supergravity, its uplift is well-defined and the the physics of the string vacuum is captured by a maximally symmetric background of a lower-dimensional supergravity. 

As an example we may consider a vacuum in type $\mrm{II}$ supergravity given by a fibration on a four-dimensional background described by the metric $\D s^2_4$ \cite{tomasiello},
\begin{equation}
\D s^2_{10}=e^{2W}\,\D s^2_4+\D s^2_6,
\end{equation}
where $\D s^2_6$ is the metric of the internal six-dimensional manifold, and the function $W$ depends only on the coordinates of the internal manifold and it is called {\itshape warping}.

This background enjoys maximal symmetry in $d=4$ and this can be realized only by three different groups: $\mrm{ISO}(3,1)$, $\mrm{SO}(3,2)$ and $\mrm{SO}(4,1)$. 
The first one is the four-dimensional Poincar{\'e} group associated to the symmetries of the {\itshape Minkowski} spacetime $\mathbb{R}^{1,3}$, the second is the symmetry group of the {\itshape Anti De Sitter} spacetime $\mrm{AdS}_4$ and the last one is the symmetry group of the {\itshape De Sitter} spacetime $\mrm{dS}_4$. 
These three backgrounds represent the four-dimensional empty universe since all the fields breaking maximal symmetry, as vectors, vanish and the metric is completely determined by symmetry properties except for the presence of the warping $W$. 

This classification holds also in supergravities in other dimensions, thus the solutions $\mathbb{R}^{1,d-1}$, $\mrm{AdS}_d$ and $\mrm{dS}_d$ are maximally symmetric and, in general\footnote{We recall that a consistent truncation relating the $D=11$ or type $\mrm{II}$ vacua with the $d$-dimensional must exists.}, they are associated to stringy vacua that we will call $\mathbb{R}^{1,d-1}$, $\mrm{AdS}_d$ and $\mrm{dS}_d$ vacua. 

The simplest example of vacua are not characterized by any warping and their topology can be written as a direct product. 
Two examples of them in $D=11$ supergravity are the so-called {\itshape Freund-Rubin solutions} \cite{Freund:1980xh},
\begin{equation}
\mrm{AdS}_4\times S^7 \qquad \text{and} \qquad  \mrm{AdS}_7\times S^4\,,
\label{freundrubin}
\end{equation}
while a relevant example \cite{Schwarz:1983qr} in type $\mrm{IIB}$ is the vacuum 
\begin{equation}
\mrm{AdS}_5\times S^5\,.
\end{equation}
Examples of warped solutions defining vacua in massive type $\mrm{IIA}$ supergravity \cite{Brandhuber:1999np, Apruzzi:2013yva} are given by
\begin{equation}
\mrm{AdS}_7\times_w \tilde{S}^3 \qquad \text{and} \qquad  \mrm{AdS}_6\times_w \tilde{S}^4\,,
\label{massiveIIAtruncations}
\end{equation}
where $ \tilde{S}^3$ and $ \tilde{S}^4$ are squashed spheres.

From the point of view of string theory, $\mrm{AdS}_d$ and $\mrm{dS}_d$ vacua are the most interesting. As we said in section \ref{geometry}, lower-dimensional theories that are characterized by non-stabilized moduli, i.e. ungauged supergravities, are sources of problems since the moduli are not stabilized. The absence of a scalar potential implies that in the maximally symmetric case the metric is flat, thus defining a $\mathbb{R}^{1,d-1}$ vacuum that in turn is characterized by a set of constant Killing spinors. 

In the case of gauged supergravities, the scalars can be stabilized and the vacuum is the solution such that the moduli take values extremizing the potential. When evaluated in its critical points the potential defines a {\itshape cosmological constant} $\Lambda$ for the theory. Even if both $\mrm{AdS}_d$ and $\mrm{dS}_d$ vacua are in principle possible, interestingly, one discovers that the most natural truncations with internal fluxes produce scalar potentials admitting negative critical points preserving SUSY, thus supersymmetric $\mrm{AdS}_d$ vacua.
Moreover all the attempts generating lower-dimensional potentials breaking all supersymmetries and leading to $\mrm{dS}_d$ backgrounds are not stable. As we discussed in the Introduction, this is one of the main open problems of string theory that, for sure, goes beyond the aims of this thesis.

\subsection{BPS Objects, Near-Horizon Limit and AdS Vacua}
\label{branesolutions}

Let's consider those solutions in $D=11$ or type $\mrm{II}$ supergravites describing less symmetric configurations respect to the vacua. In supergravity the space of solutions could be huge and difficult to study in full generality since the equations of motion are highly non-linear. As we discussed in section \ref{killingspinor}, for supersymmetric solutions fortunately a first-order formulation exists allowing their determination in a relatively ``simple way".
The most interesting effective description of solitonic states in higher-dimensional superegravities is given by {\itshape black brane solutions}. These solutions can be derived, in the simplest examples, directly by solving the equations of motion and physically represent the low-energy limit of branes. In general they are described by the spatial evolution of the supergravity fields from a flat $D=10, 11$ asymptotics, corresponding to great distances of the observer from the brane, to an {\itshape event-horizon} associated to the limit in which the observer is infinitely close to the brane.

Let's take under analysis eleven-dimensional supergravity introduced in section \ref{highersugras}. The simplest solutions obtainable by solving the equations of motion are two \cite{Gueven:1992hh}: the first one describing the $\mrm{M}2$-brane on a flat background and the second associated to the $\mrm{M}5$ on a flat background.

As regards the $\mrm{M}2$, the symmetry group is given by $\mrm{SO}(2,1)\times \mrm{SO}(8)$ and the solution for the metric and the 4-form field strength is given by
\begin{equation}
\begin{split}
&\D s^2=H^{-2/3}\D s^2_{\mathbb{R}^{1,2}}+H^{1/3}\D s^2_{\mathbb{R}^8}\,,\\
&F_4=\D x^0\wedge \D x^1\wedge \D x^2\wedge  \D H^{-1}\,,
\label{m2}
\end{split}
\end{equation}
where the coordinates $(x^0, x^1, x^2)$ are the coordinate on the worldvolume.
The metrics $\D s^2_{\mathbb{R}^{1,2}}$ and $\D s^2_{\mathbb{R}^8}$ are respectively the metric of the three-dimensional Minkowski space parametrized by the worldvolume coordinates and the metric of the euclidean transverse space ${\mathbb{R}^8}$. The function $H$ depends only on the transverse coordinates and solves the eight-dimensional Laplace equation. It follows that, given the transverse coordinates $(y^1,\cdots, y^8)$ and introducing a radial coordinate $r=|\vec{y}|$ in ${\mathbb{R}^8}$, the explicit form of $H$ is
\begin{equation}
H=1+\frac{R_2^6}{r^6},
\end{equation}
with $R_2^6=32\pi^2\,l_P^6$. The 4-form is {\itshape electric} in the sense that it is completely defined on the worldvolume and from this it follows that the $\mrm{M}2$-brane is the electric source of the 3-form gauge field included in the supergravity multiplet.

The $\mrm{M}5$-brane is invariant under the action of $\mrm{SO}(5,1)\times \mrm{SO}(5)$ and its supergravity solution is given by
\begin{equation}
\begin{split}
&\D s^2=H^{-1/3}\D s^2_{\mathbb{R}^{1,5}}+H^{2/3}\D s^2_{\mathbb{R}^5}\,,\\
&F_4=\star_{11} \left( \D x^0\wedge \D x^1\wedge \cdots \wedge \D x^5\wedge  \D H^{-1}\right)\,,
\label{m5}
\end{split}
\end{equation}
where $H$ is an harmonic function of the radial coordinate $r$ of the transverse space $\mathbb{R}^5$ and it solves the five-dimensional Laplace equation. In particular it has the form
\begin{equation}
H=1+\frac{R_5^3}{r^3},
\end{equation}
with $R_5^3=\pi\,l_P^3$. We point out that the 4-form is completely {\itshape magnetic} and then the $\mrm{M}5$-brane is the magnetic source of the 3-form gauge field.

These two solutions can be easily generalized to systems of $N$ coincident $\mrm{M2}$- and $\mrm{M5}$-branes. In fact the presence of $N$ branes has the only effect to rescale the radii $R_2$ and $R_5$ as 
\begin{equation}
R_2^6=32\pi^2\,l_P^6 N \qquad \text{and} \qquad R_5^3=\pi\,l_P^3 N\,.
\label{braneradii}
\end{equation}

As an other examples of black brane solutions we cite $p$-brane solutions on a flat background in type $\mrm{II}$ supergravities \cite{Horowitz:1991cd}. They can be carried out from $\mrm{BPS}$ equations or directly from the equations of motions at the same footing of (\ref{m2}) and (\ref{m5}). These solutions are characterized by a vanishing Kalb-Ramond field\footnote{Otherwise we would have other examples of type $\mrm{II}$ solutions like the $\mrm{NS5}$ brane or the fundamental string $\mrm{NS1}$.} and by the coupling to the $\mrm{R}$-$\mrm{R}$ $p$-forms $C^{(p+1)}$ with field strengths $F_{p+2}=\D C^{(p+1)}$. Moreover, in addition to the graviton and $\mrm{R}$-$\mrm{R}$ fields, the other field remaining is the dilaton $\Phi$.

The black $p$-brane solution is given by
\begin{equation}
\begin{split}
&\D s^2= H_p^{-1/2}\D s^2_{\mathbb{R}^{1,p}}+H_p^{1/2}\D s^2_{\mathbb{R}^{9-p}}\,,\\
&C^{(0\cdots p)}=H_p^{-1}-1\,,\\
&e^{\Phi}=g_sH_p^{(3-p)/4}\,,
\label{pbrane}
\end{split}
\end{equation}
where $\D s^2_{\mathbb{R}^{1,p}}$ is the $(p+1)$-dimensional metric on the worldvolume and $\D s^2_{\mathbb{R}^{9-p}}$ is the flat metric on the $(9-p)$-dimensional transverse space. The function $H_p$ is a function of the radial coordinate on the transverse space $\mathbb{R}^{9-p}$ and it is harmonic. Its general form is given by\footnote{From arguments coming from type $\mrm{II}$ string theories one can derive $(R_p/l_s)^{7-p}=c\, g_s N$ with $c=(2\sqrt{\pi})^{5-p}\Gamma \left ( \frac{7-p}{2}\right)$ and $N$ number of the coincident $p$-branes.}
\begin{equation}
H_p=1+\left ( \frac{R_p}{r}  \right)^{7-p}\,.
\end{equation}
The case of the $\mrm{D}3$-brane is of particular interest since the dilaton is constant, i.e. $e^{\Phi}=g_s$, and completely expressed in terms of $g_s$. 

The solutions (\ref{m2}), (\ref{m5}) and (\ref{pbrane}) are all preserving half of the total $\mrm{SUSY}$ preserved respectively by $D=11$ supergravity and type $\mrm{II}$ supergravities, thus they preserve 16 real supercharges. Generally it can be shown that a Killing spinor $\epsilon(r)$ of an electrically charged\footnote{In case of magnetically charged branes the gamma matrices appearing in the projector (\ref{projector}) are defined along the transverse directions.} brane solution has to satisfy the constraint
\begin{equation}
\left(\mathbb{I}\pm \gamma^{01\cdots p}\ma O  \right)\epsilon(r)=0\,,
\label{projector}
\end{equation} 
where $\ma O$ is an operator depending on the specific theory considered and the operator $\mathbb{I}\pm \gamma^{01\cdots p}\ma O $ is a projector eliminating half of the components of $\epsilon$. A much more complicated solution in presence of many branes could breaks more supersymmetry since the Killing spinors have to satisfy a set of different conditions of the type (\ref{projector}) coming from all the branes composing the system.

At the level of effective theories, the fluxes are obtained by integrating the field strengths of the gauge form potentials characterizing a given brane solution in supergravity. These charges are actually the sources of the fields describing the closed string background associate to the brane. In general a $(p+1)$-form gauge field $A^{(p+1)}$ with field strength $F_{p+2}=\D A^{(p+1)}$ can couple electrically to a $p$-brane and magnetically to a $(D-p-4)$-brane. This means that the magnetic dual associated to $F_{p+2}$ is the $(D-p-2)$-form $\star F_{p+2}$. The electric and magnetic charges are given by
\begin{equation}
e_p=\int_{S^{D-p-2}} \star F_{p+2} \qquad \text{and} \qquad m_{D-p-4}=\int_{S^{p+2}}  F_{p+2} \,,
\label{charges}
\end{equation}
where, in $D$ dimensions, $D-p-2$ and $p+2$ are the dimensions required to surround respectively a $p$-brane and a $(D-p-4)$-brane.  As it happens in $D=4$, a Dirac quantization condition of the type
\begin{equation}
e_p m_{\scriptsize{D-p-4}} \in 2 \pi \mathbb{Z}
\end{equation}
 holds in order to keep the gauge invariant fields uniquely defined at any point of the background. 

The reason why the solutions (\ref{m2}), ({\ref{m5}), (\ref{pbrane}) are called black branes is the existence of an horizon in $r=0$. If we consider the $\mrm{M}2$, writing the transverse metric in spherical coordinates as
\begin{equation}
\D s^2_{\mathbb{R}^8}=\D r^2+r^2\D \Omega_7^2\,,
\end{equation}
where $\D \Omega_7^2$ is the metric of the 7-sphere $S^7$, the near-horizon limit of (\ref{m2}) is 
\begin{equation}
\D s^2 = (r/R_2)^4 \, \D s^2_{\mathbb{R}^{1,2}}+ (R_2/r)^2\,\D r^2+R_2^2\,\D \Omega_7^2\,,
\label{nhm2}
\end{equation}
that is the metric describing the geometry $\mathrm{AdS}_4\times S^7$.

The near-horizon limit of $\mrm{M}5$ is obtained in a similar fashion. Writing the transverse metric in spherical coordinate one has 
$\D s^2_{\mathbb{R}^5}=\D r^2+r^2\D \Omega_4^2$ with $\D \Omega_4^2$ metric of the 4-sphere $S^4$. Performing the limit $r\rightarrow 0$ one gains
\begin{equation}
\D s^2 = (r/R_5) \, \D s^2_{\mathbb{R}^{1,5}}+ (R_5/r)^2\,\D r^2+R_5^2\,\D \Omega_4^2\,,
\end{equation}
that is the metric of $\mathrm{AdS}_7\times S^4$.
Finally, if we consider the particular case of the $\mrm{D}3$-brane, its near-horizon geometry is described by the type $\mrm{IIB}$ vacuum $\mathrm{AdS}_5\times S^5$.

From section \ref{vacua}, we know that $\mathrm{AdS}_4\times S^7$ and $\mathrm{AdS}_7\times S^4$ are exact solutions of eleven-dimensional supergravity and in particular we know that they are $\mathrm{AdS}$ vacua of M-theory. We thus observed, for the $\mrm{M}2$- and for the $\mrm{M}5$-brane, a phenomenon characterizing the low-energy description of many supersymmetric stringy objects: the near-horizon limit of their supergravity solutions can be described by $\mrm{SUSY}$ $\mathrm{AdS}_{p+2}$ vacua. On the converse, given a supersymmetric $\mathrm{AdS}_d$ vacuum in a $d$-dimensional gauged supergravity, the existence of a {\itshape brane picture} such that its near-horizon limit is a string vacuum of the type $\mrm{AdS}_d\times \Omega_{D-d}$ with $D=10, 11$, is not automatic and depends on the existence of a consistent truncation of the higher-dimensional brane solution to the $d$-dimensional solution described by an $\mrm{AdS}_d$ asymptotics.

This fact is highly non-trivial since the existence of $\mrm{AdS}_d$ solutions in $d$ dimensions is a priori independent of its higher-dimensional interpretation in terms of branes. Thus $\mathrm{AdS}_d$ vacua, black brane solutions and consistent truncations to lower-dimensional configurations are intimately related and constitute the basic links between the effective and quantum description of stringy objects.

As a final comment we point out that all the solutions considered in this section are regular with the horizon collapsed\footnote{This is related to a choice of coordinates comparable to that one used when writing the metric of a $d=4$ extremal Reissner-Nordstr\"om black hole in the form $\D s_4^2=-H(\rho)^{-2} \D t^2+H(\rho)^2(\D \rho^2+\rho^2\D \Omega_2^2)$ with $H=1+Q/\rho$ and the horizon infinitely close to the singularity in $\rho \rightarrow 0$.} in $r=0$. This is an evidence of the $\mrm{UV}$-completion produced by the physics of high dimensions. It is well known that there exist solutions in low dimensions, such black holes and black strings, that are singular. Hence one of the most important aims in studying lower-dimensional objects will be the research of their uplifts and the study of their brane picture. If such an uplift exists and whether it is possible to find out a brane solution describing it, thus one could in principle understand how and why the black hole singularities are resolved.

\subsection{Bound States, Extremality and Lower-Dimensional Flows}
\label{sugraflows}

Thanks to dimensional reduction strategies we have under control the relations between the high- and the lower-dimensional world. In particular, given a particular consistent truncation of an higher-dimensional supergravity,
it is possibile to investigate the microscopic origin of solutions in lower-dimensional supergravities by studying their uplift and their brane picture. 
Unfortunately the higher-dimensional interpretation of observable physics is not an easy task since the quantum configurations giving rise to macroscopic systems like black holes are complicated intersections 
of branes and other stringy objects.

In type $\mrm{II}$ and $D=11$ supergravities bound states of branes, or branes' intersections, are described by particular solutions of the equations of motion representing simultaneously branes of different types in equilibrium\footnote{Also in this case supersymmetry plays a crucial role since the SUSY conditions for a brane system can be interpreted as a particular realization of the static equilibrium of the system.}.
A first example could be given by those branes with a curved worldvolume: in some cases the deformations of the worldvolume can be seen as the result of the intersection with other branes, conversely a brane intersecting the worldvolume of
an other brane can be interpreted as a particular excitation of the worldvolume of the latter. 
Clearly not all the intersections are stable and the principal cause of instabilities is the presence of tachyons in the spectra of the open strings appearing when SUSY is broken\footnote{For a discussion on the instabilities of non-supersymmetric $\mrm{AdS}_d$ vacua see for example \cite{Ooguri:2016pdq, Danielsson:2016mtx,Danielsson:2017max}.}.

The construction of BPS bound state solutions in $D=11$ and in type $\mrm{II}$ supergravities is, in many cases, based on the harmonic\footnote{In the case of intersections in massive type $\mrm{IIA}$ supergravity the construction of these solutions is a difficult task because of the presence of the Romans mass.} properties of functions describing the different branes composing the intersection. In general one can use a {\itshape superposition rule} consisting
in the construction of the solution starting from the independent the solutions of each brane. For example, each metric component will be determined by the product of the corresponding warp factors of the single branes composing the intersections. 
Each warp factor composing these products is {\itshape smeared} in the sense that it is a function depending only on the overall transverse directions and it is elevated to a numeric power respecting the original division between worldvolume
and transverse coordinates.

The particular geometric configuration of a bound state tells us which directions can be wrapped and thus which particular solution can be produced in low dimensions. Assuming the existence of a consistent truncation relating the branes' intersection with a solution in a $d$-dimensional supergravity, there are two types of $d$-dimensional configurations that are particularly interesting:

\begin{itemize}

 \item {\itshape Black solutions}: These are singular solutions whose singularity is protected by an event-horizon. They are usually obtained by considering particular bound states with $(p+1)$-dimensional worldvolume and with some (or all) worldvolume coordinates wrapping non-trivial cylcles. By reducing all the worldvolume coordinates of the branes composing the intersection one obtains a black solution in a $d$-dimensional supergravity.
  If the worldvolume is wrapped on a compact manifold $\Sigma_{d-p-2}$, we can make a further distinction for topologies of the horizons of the type $\mrm{AdS}_{p+2}\times \Sigma_{d-p-2}$. The solutions with an horizon $\mrm{AdS}_2\times \Sigma_{d-2}$ are black holes, those with $\mrm{AdS}_3\times \Sigma_{d-3}$ black strings and those with $\mrm{AdS}_{p+2>3}$ black branes.
 
 \item {\itshape Domain wall solutions}: These are produced by reducing over all the transverse coordinates of a bound state. These solutions are smooth and they are very important for $\mrm{AdS/CFT}$ and cosmological applications. As for black solutions, the most interesting ones for string theory are those reproducing at certain values of their coordinates configurations of the type $\mrm{AdS}_{p+2}\times \Sigma_{d-p-2}$ and, in particular, those interpolating between two ``walls" of the type written above.
 
\end{itemize}
In both of these cases, the profile of the solution is tipically defined by a spatial {\itshape flow} of the fields of the $d$-dimensional supergravity between two limiting configurations. The most interesting scenario, also considering $\mrm{AdS/CFT}$ applications, is given by a $d$-dimensional flow in a gauged supergravity interpolating between an $\mrm{AdS}_d$ vacua and a geometry of the type $\mrm{AdS}_{p+2}\times \Sigma_{d-p-2}$. In this case we usually write
\begin{equation}
\mrm{AdS}_{d} \longrightarrow \mrm{AdS}_{p+2}\times \Sigma_{d-p-2}\,,
\label{supergravityflows}
\end{equation}
to represent the entire solution. This has an uplift to a higher-dimensional bound state solution with a $(p+1)$-dimensional worldvolume and it is characterized by two particular limits of its coordinates. One reproducing the closed string vacuum obtained as the uplift of $\mrm{AdS}_{d}$ and the other giving the vacuum associated to the wrapped coordinates, coming from $\mrm{AdS}_{p+2}\times \Sigma_{d-p-2}$.

An other important distinction concerning black solutions in supergravities is between {\itshape extremal} and {\itshape non-extremal}. 
Generally speaking an extremal black solution can be defined as a solution with the minimal possibile mass compatible with its charges (and angular momentum if it is different to zero). An extremal objects is purely mechanic in the sense that its energy is completely associated to the fundamental interactions of the fields. It follows that the temperature of extremal solutions is zero and their entropy is related to the index of degeneration of the ground state of the system. In this sense extremal solutions represent the limit in which all the thermodynamic contributions vanish and the solution describes a purely mechanical system. Actually only these solutions have a complete explanation in string theory and some examples are given by (\ref{m2}), (\ref{m5}) and (\ref{pbrane}).

Conversely non-extremal solution are described by a non-trivial thermodynamics and have not a clear explanation in terms of stringy objects since they tipically possess many event-horizons.  These are solutions only of the equations of motion and in general they don't have a first-order description\footnote{Except for some particular cases as in \cite{Klemm:2017pxv}.}. Moreover they break all the supersymmetries, but, in many cases, there are particular values of their parameters such that these solutions become extremal and supersymmetry is restored\footnote{As an example one can consider the Reissner-Nordstr{\"o}m black hole in four dimensions. This is a non-extremal electrically charged black hole with two horizons. The extremal limit in which the two horizons collapse into a single one is defined by the relation $M=|Q|$ where $M$ and $Q$ are respectively the mass and the charge of the black hole.}.

The distinction between extremal and non-extremal solutions is somehow related to supersymmetry since the limiting condition producing an extremal configuration from a non-extremal one often implies the saturation of the $\mrm{BPS}$ bound\footnote{The relation $M=|Q|$ defining the extremal limit of Reissner-nordstr{\"o}m black hole is the $\mrm{BPS}$ bound saturated.} (\ref{bpsbound}). 
Moreover, even also extremal non-supersymmetric solutions exist in supergravity, they could not survive in quantum gravity \cite{Ooguri:2016pdq}.

\subsection{An Example: the Bound State D1-D5-P}
\label{microstate}

In this section\footnote{We will partially refer to \cite{Ortin:2015hya}.} we consider the wrapping of a bound state $\mrm{D}1$-$\mrm{D}5$-$\mrm{P}$ on a $T^5$ in type $\mrm{IIB}$ string theory \cite{Strominger:1996sh, Callan:1996dv}. This bound state is defined by the intersection of a $\mrm{D}1$ and a $\mrm{D}5$-brane with a pp-wave\footnote{The pp-waves are gravitational waves that are exact solutions to the full Einstein equations of a $D$-dimensional (super)gravity theory. They 
are characterized by definition by a covariantly constant null Killing vector field and are described by a metric of the form
\begin{equation}
\D s^2= H_W^{-1}\D t^2-H_W\left[ \D y^1 -\alpha_W (H_W^{-1}-1)\D t \right ]^2-\D s^2_{\mathbb{R}^{D-2}}\,,
\label{ppwave}
\end{equation}
with $H_W$ harmonic function on $\mathbb{R}^{D-2}$. The parameter $\alpha_W$ can be positive or negative taking care of the two possibile directions of propagation along $y^1$. The coordinate of propagation $y^1$ parametrizes an $S^1$ and it is defined on $[0,2\pi R]$ where $R$ is the radius of the $S^1$. The mixed term in $(\ref{ppwave})$ is related to the time component of the spin connection that is associated to the momentum charge carried by the wave that is quantized. Alternatively one can interpret the momentum of the pp-wave as a $\mrm{KK}$ momentum associated to the wrapping on the $S^1$.
} with momentum $\mrm{P}$ and it preserves 4 real supercharges ($\mrm{BPS}/8$). Thus we relate it with the Reissner-Nordstr{\"o}m black hole in five dimensions. 

Recalling the $p$-brane solution given in (\ref{pbrane}), the low-energy limit of the bound state $\mrm{D}1$-$\mrm{D}5$-$\mrm{P}$ is obtained in type $\mrm{IIB}$ supergravity by the superposition of the independent solutions of the $\mrm{D}1$, $\mrm{D}5$ and the pp-wave (\ref{ppwave}). The solution is given by
\begin{equation}
\begin{split}
  \D s^2&= H_{\scriptsize{\mrm{D}1}}^{-1/2} H_{\scriptsize{\mrm{D}5}}^{-1/2}\{-H_W \D t^2+ H_W \left[\D y^1+ \alpha_W \left (H_W^{-1}-1\right )\D t \right]^2 \}\\
  &+ H_{\scriptsize{\mrm{D}1}}^{1/2} H_{\scriptsize{\mrm{D}5}}^{1/2} \left[\D r^2+r^2 \D \Omega_3^2 \right]\\
  &+H_{\scriptsize{\mrm{D}1}}^{1/2} H_{\scriptsize{\mrm{D}5}}^{-1/2}\left[(\D y^2)^2+(\D y^3)^2+(\D y^4)^2+(\D y^5)^2\right]\,,\\
  C^{(2)}_{ty^1}&=\alpha_ {\scriptsize{\mrm{D}1}}g_s\left (H_{\scriptsize{\mrm{D}1}}^{-1}-1\right )\,,\qquad C^{(6)}_{ty^1\cdots y^5}=\alpha_ {\scriptsize{\mrm{D}5}}g_s\left (H_{\scriptsize{\mrm{D}5}}^{-1}-1\right )\,,\\
  e^{\Phi}&=g_s \frac{H^{1/2}_{\scriptsize{\mrm{D}1}}}{H^{1/2}_{\scriptsize{\mrm{D}5}}}\,,\qquad  H_i=1+\frac{r_i^2}{r^2}\quad \text{with}\quad i=\mrm{D}1,\mrm{D}5,\mrm{P}\,,
  \label{boundstate}
  \end{split}
\end{equation}
where $r$ is the radial coordinate on the transverse space $\mathbb{R}^4$ of the bound state. The coordinates $(y^1 \cdots y^5)$ are unpacked since $y^1$ is compact and determine the direction of propagation of 
the pp-wave whilst the coordinates  
$(y^2,y^3,y^4,y^5)$ describe the smearing of the $\mrm{D}1$ on the $\mrm{D}5$. The transverse space of the bound state is described by the spherical coordinates $(r,\theta^1,\theta^2,\theta^3)$ with $\D \Omega_3^2$ the usual metric of a 3-sphere.
The parameters $\alpha_ {\scriptsize{\mrm{D}1}}$, $\alpha_ {\scriptsize{\mrm{D}5}}$, $\alpha_ W$ and $r_i$ are integration constants\footnote{The parameters $\alpha_i$ are such that $\alpha_i^2=1$ with $i=\mrm{D}1,\mrm{D}5,\mrm{P}$} and have to be fixed in terms of the physical parameters of the solution like charges\footnote{The formulation of (\ref{boundstate}) is called ``democratic" since the dual form
$C^{(6)}$ is considered independent.}  and units of momentum.
\begin{table}[h!]
\renewcommand{\arraystretch}{1}
\begin{center}
\scalebox{1}[1]{
\begin{tabular}{c||c c c c c c||c|c c c}
branes & $t$ & $y^{1}$ & $y^{2}$ & $y^{3}$ & $y^{4}$ & $y^{5}$ & $r$ & $\theta^1$ & $\theta^{2}$ & $\theta^{3}$ \\
\hline \hline
D5 & $\times$ & $\times$ & $\times$ & $\times$ & $\times$ & $\times$ & $-$ & $-$ & $-$ & $-$ \\
D1 & $\times$ & $\times$ & $-$ & $-$ & $-$ & $-$ & $-$ & $-$ & $-$ & $-$ \\
W & $\times$ & $\rightarrow$ & $-$ & $-$ & $-$ & $-$ & $-$ & $-$ & $-$ & $-$ \\
\end{tabular}
}
\end{center}
\caption[The brane picture of the bound state $\mrm{D}1\,$-$\,\mrm{D}5\,$-$\,\mrm{P}$. Table from \cite{Ortin:2015hya}.]{\it The brane picture describing the solution (\ref{boundstate}). The symbols $\times$ stand for worldvolume coordinates, the $-$ signes for the transverse one and $\rightarrow$ represents the direction of propagation of the pp-wave.
The double vertical line separate the overall transverse coordinates $(r,\theta^1,\theta^2,\theta^3)$ to the coordinates $(t,y^1)$ and to the smeared ones $(y^2,y^3,y^4,y^5)$. Table from \cite{Ortin:2015hya}.} 
\end{table}
Let's consider now the $\mrm{KK}$ reduction on a $T^5=S^1\times T^4$ with volume $V^5=2\pi R_1V^4 $ where $V^4=(2\pi)^4R_2R_3R_4R_5$ is the volume of the $T^4$ parametrized by the compact coordinates $(y^2,y^3,y^4,y^5)$ and $R_1$ is the radius of the $S^1$ parametrized by the coordinate $y^1$ of the propagation of the wave. One obtains a five-dimensional configuration given by
\begin{equation}
\begin{split}
&\D s^2 =-\left(H_{\scriptsize{\mrm{D}1}}H_{\scriptsize{\mrm{D}5}}H_W \right)^{-2/3}\D t^2+\left(H_{\scriptsize{\mrm{D}1}}H_{\scriptsize{\mrm{D}5}}H_W \right)^{1/3}\left[\D r^2+r^2 \D \Omega_3^2     \right]\,,\\
&A^{(i)}_{t}=\alpha_ {i}\left (H_i^{-1}-1\right )\,, \qquad H_i=1+\frac{r_i^2}{r^2}\quad \text{with}\quad i=\mrm{D}1,\mrm{D}5,\mrm{P}\,\\
&e^{\phi}=g_s \sqrt{\frac{R_1}{l_s}} \frac{H^{1/8}_{\scriptsize{\mrm{D}1}}H^{1/8}_{\scriptsize{\mrm{D}5}}}{H^{1/4}_W}\,, \quad k_{V^4}=
\frac{H_{\scriptsize{\mrm{D}1}}}{H_{\scriptsize{\mrm{D}5}}} \, ,\quad  k_1=\frac{R_1}{l_s}\frac{H^{1/2}_W}{H^{1/4}_{\scriptsize{\mrm{D}1}}H^{1/4}_{\scriptsize{\mrm{D}5}}}\,.
\label{rn}
\end{split}
\end{equation}
This is a charged black hole solution with three running scalar fields\footnote{Setting $r_{\scriptsize{\mrm{D}1}}=r_{\scriptsize{\mrm{D}5}}=r_{W}=r$ one gets $H_{\scriptsize{\mrm{D}1}}=H_{\scriptsize{\mrm{D}5}}=H_W=H$. In this limit the moduli are constants and the solution takes the usual form of the $d=5$ $\mrm{RN}$ black hole with metric $\D s^2 =H^{-2}\D t^2-H \,( \D r^2+r^2 \D \Omega_3^2 )$.} of $d=5$ maximal (ungauged) supergravity. It is characterized by three abelian vector fields $A^{(i)}$ that are related to the ten-dimensional forms $C^{(2)}$ and by three scalars $\phi$, $k_{V^4}$ and $k_1$. The moduli $k_{V^4}$ and $k_1$ are respectively\footnote{In particular one has $k_{V^4}=|g_{y^2y^2}g_{y^3y^3}g_{y^4y^4}g_{y^5y^5}|^{1/2}$ and $k_{1}=|g_{y^1y^1}|^{1/2}$ where $g_{\mu\nu}$ are the components of the metric (\ref{boundstate}).} related to to the volume of the $T^4$ parametrized by the coordinates $(y^2,y^3,y^4,y^5)$ and to the size measured by the compact direction $y^1$. Moreover this solution manifests a regular horizon in $r=0$ with a finite area given by
\begin{equation}
A=2\pi^2\,r_{\scriptsize{\mrm{D}1}}r_{\scriptsize{\mrm{D}5}}r_{W} \,.
\label{arearn}
\end{equation}
It is possibile now to relate the constants of (\ref{boundstate}) to the physical parameters of the $\mrm{RN}$ black hole.

Let's consider now the same intersection $(\ref{boundstate})$ but composed respectively by $N_{\scriptsize{\mrm{D}1}}$ coincident $\mrm{D}1$-branes, $N_{\scriptsize{\mrm{D}5}} $ coincident $\mrm{D}5$-branes and $N_W$ pp-waves along the direction $y^1$. If we wrap the worldvolume directions for $N_{\scriptsize{\mrm{D}1}}$, $N_{\scriptsize{\mrm{D}5}}$ and $N_W$ times, one has
\begin{equation}
r_{\scriptsize{\mrm{D}5}}^2=N_{\scriptsize{\mrm{D}5}} l_s^2 g_s\,, \qquad r_{\scriptsize{\mrm{D}1}}^2=\frac{N_{\scriptsize{\mrm{D}1}}l_s^6 g_s}{R_2R_3R_4R_5}\,, \qquad r_W^2=\frac{N_W l_s^8 g_s^2}{R_1^2R_2R_3R_4R_5}\,.
\label{parameterrn}
\end{equation}
Using (\ref{arearn}) and (\ref{parameterrn}), one can obtain the Bekenstein-Hawking entropy of the black hole
\begin{equation}
S_{BH}=\frac{A}{4G_N^{(5)}}=2\pi \sqrt{N_{\scriptsize{\mrm{D}1}}N_{\scriptsize{\mrm{D}5}}N_W} \qquad \text{with}\qquad G_N^{(5)}=\frac{G_N^{(10)}}{V^5}=\frac{\pi\, l_s^8g_s^2}{4\,R_1R_2R_3R_4R_5}
\label{rnentropy}
\end{equation}
where $G_N^{(d)}$ is the d-dimensional Newton constant. We immediately see that the entropy of the black hole depends only on the numbers of branes and pp-waves composing the bound state and thus it can be explained in terms of the degeneration of the stringy states associated to the branes' bound state.

The charges associated to the fluxes of the brane intersection are in relation to the abelian vector fields of the five-dimensional supergravity. In particular following the prescriptions of $(\ref{charges})$, one can fix the remaining intergration constants $\alpha_i$ and derive the charges $Q_{\scriptsize{\mrm{D}1}}$ and $ Q_{\scriptsize{\mrm{D}5}}$ of the bound state,
\begin{equation}
\begin{split}
Q_{\scriptsize{\mrm{D}1}}=\frac{ l_s^2 g_s^2}{8G_N^{(10)}}\int_{S^3\times T^4}\star F_{3}=\frac{\alpha_{\scriptsize{\mrm{D}1}}r_{\scriptsize{\mrm{D}1}}^2R_2\cdots R_5}{2\pi\, l_s^8\,g_s}\,, \quad  Q_{\scriptsize{\mrm{D}5}}=\frac{2\pi^4 l_s^6 g_s^2}{G_N^{(10)}}\int_{S^3 } \star F_{7}=\frac{\alpha_{\scriptsize{\mrm{D}5}}\,r_{\scriptsize{\mrm{D}5}}^2}{(2\pi)^5\, l_s^8\,g_s}\,,
\label{fluexsrn}
\end{split}
\end{equation}
where $F_3=\D C^{(2)}$ and $F_7=\D C^{(6)}$. The charge associated to the units of momentum of the pp-wave can be obtained as the $\mrm{KK}$ momentum along the circle with radius $R_1$ and it is given by $Q_W=\frac{\alpha_W\,r_W^2\,R_1^2R_2\cdots R_5}{l_s^8\,g_s^2}$. Given the expressions of charges (\ref{fluexsrn}), one can rewrite (\ref{rnentropy}) in the well-know form
\begin{equation}
S_{BH}=2\pi \sqrt{Q_{\scriptsize{\mrm{D}1}}Q_{\scriptsize{\mrm{D}5}}Q_W }\,.
\label{rnentropycharges}
\end{equation}
What has been presented is the extrapolation of a five-dimensional $\mrm{RN}$ black hole from the bound state $\mrm{D}1$-$\mrm{D}5$-$\mrm{P}$. In this context $g_s$ and $\alpha^\prime$ must be small at the uncompactifyied scales describing the bound of branes in order to keep valid the low-energy description. Conversely we know that dimensional reductions are always related to different scales of energy of a given gravitational system, in particular a lower-dimensional solution typically would describe a different regime\footnote{The couplings have to be considered as the asymptotic value of the string couplings.} of $g_s$ and $\alpha^\prime$. This means that both the type $\mrm{IIB}$ supergravity as well the entropy (\ref{rnentropycharges}) should take corrections due to string loops and perturbations of the string sigma model. At this point the relevance of supersymmetry comes out since the non-renormalization theorem for $\ma N=4$ theories protects the lower-dimensional theory from any kind of corrections.

More precisely the low-energy regime in which $\mrm{D}1$-$\mrm{D}5$-$\mrm{P}$ is described by (\ref{boundstate}) can be written as $r_i^2 \gg l_s^2$. In this limit the charges $Q_i$ of the bound state are large and the higer-order contributions to the type $\mrm{IIB}$ supergravity action are suppressed\footnote{The quantum corrections are related to the Schwarzschild radius of the black hole that in turn can be expressed in terms of the charges \cite{Strominger:1996sh}.}. Conversely the entropy (\ref{rnentropycharges}) of the black hole depends only on the charges and not on the values taken by the scalars at the horizon, thus its value can be obtained for small values of $g_s$ corresponding to a vanishing horizon and then extrapolated to the regime of large $g_s$ describing the black hole since, for the $\ma N=4$ non-renormalization theorem, the entropy is protected from quantum corrections \cite{Strominger:1996sh}.

Finally we point out that in order to obtain a complete low-energy interpretation of the microstates of (\ref{rn}) it is necessary to match the Bekenstein-Hawking entropy (\ref{rnentropycharges}) with the entropy of the quantum field theory coming from the open strings' spectrum and living on the worldvolume of the bound state of branes.  As we will see, $\mrm{AdS/CFT}$ correspondence will prove to be the best approach in solving this problem.

\section{The AdS/CFT Correspondence}
\label{adscftcorrespondence}

In this section we perform a further step introducing the  $\mrm{AdS/CFT}$ correspondence as the most exhaustive and complete framework describing the low-energy limit of non-perturbative states in string theory \cite{Maldacena:1997re, Witten:1998qj}.
As we observed in section \ref{decoupling}, in the low-energy limit there are two alternative descriptions of a solitonic object: one can study a brane from the point of view of the physics of the background, thus as a solution in supergravity like a black solution or a domain wall, or, alternatively it is possibile to consider the $\mrm{QFT}$ defined on the worldvolume. The existence of these two alternative descriptions hints the presence of a stronger relation between them called $\mrm{AdS/CFT}$ correspondence.

Roughly speaking, in the low-energy limit, non-perturbative states manifests two dual and equivalent descriptions: on one side one has a weakly coupled description given by the {\itshape near-horizon limit}\footnote{For a domain wall solution the near-horizon limit has to be substituted with the limits describing the boundaries reached smoothly by the solution, i.e. "the walls". In this section we will refer particularly to black brane solutions but one could formulate the correspondence analogously starting from domain walls.} of supergravity solutions describing branes (or bound states of branes), on the other side there is a strongly coupled description
in terms of a superconformal field theoriy ($\mrm{SCFT}$) in the {\itshape large N limit} living on the worldvolume of the branes at the near-horizon. This $\mrm{SCFT}$ is globally supersymmetric and constitutes a fixed point of the $\mrm{RG}$ flow of the $\mrm{QFT}$ living on the worldvolume.

In section \ref{branesolutions} we studied some examples of black brane solutions and we saw that in the near-horizon they are described by $\mrm{AdS}_{p+2}$ closed string vacua. One of the main belief in string theory consists in the general interpretation of $\mrm{AdS}_{d}$  vacua as the near-horizon limit of branes' solutions or just limiting value of domain walls. The $\mrm{AdS/CFT}$ correspondence is a strong motivation supporting this belief in fact, as we will see, the duality occurs thanks to the particular geometric properties of Anti De Sitter spaces and their matching with 
the properties of the $\mrm{SCFT}$s.

In this section\footnote{For a review on the AdS/CFT correspondence see \cite{zaffa}.}, we will briefly introduce superconformal theories and present two explicit examples given by $\ma N=4$ $\mrm{SYM}$ theory in four dimensions and $\mrm{ABJM}$ theory in three dimensions. We will introduce the concept of large $N$ limit and then we will formulate the correspondence in generality. Then, considering the examples of $\mrm{SCFT}$ presented, we will analyze two particular realizations of the correspondence $\mrm{AdS}_5/\mrm{CFT}_4$ and $\mrm{AdS}_4/\mrm{CFT}_3$. Finally we will introduce the concept of {\itshape RG flow across dimensions} as a formulation of the correspondence including into this scenario an entire flow in gauged supergravity and not only of their horizon configurations.

\subsection{Conformal Field Theories}
\label{cft}

A classical conformal field theory ($\mrm{CFT}$) is a classical field theory whose external invariance group is the conformal group \cite{DiFrancesco:1997nk}. A conformal transformation in $d$ dimensions is a diffeomorphism that rescales the metric by a function of the coordinates called conformal factor. These particular changes of coordinates enlarge the Lorentz group with two types of transformations: the {\itshape dilatations} and the {\itshape special conformal transformations}. The action of these transformations on a coordinate $x^\mu$ of the background is given by
\begin{equation}
x^\mu \longrightarrow \lambda x^\mu \qquad \text{and} \qquad x^\mu \longrightarrow \frac{x^\mu+b^\mu x^\mu x_\mu}{1+b^\mu x_\mu+(b^\mu x_\mu)^2}\,,
\end{equation}
where $\lambda$ and $b^\mu$ are respectively the parameters of dilatations and special conformal transformations. It can be proved that the conformal group is isomorphic to the isometry group of $\mrm{AdS}_{d+1}$ given by $\mrm{SO}(d,2)$ with a number of the generators given by $\frac{1}{2}(d+2)(d+1)$. Considering infinitesimal conformal transformations $x^\mu \rightarrow x^\mu+\xi^\mu$, it is possibile to show that they are defined by the equations
\begin{equation}
\partial_\mu \xi_\nu+\partial_\nu \xi_\mu=\frac{2}{d}(\partial_\sigma \xi^\sigma)\eta_{\mu\nu}\,.
\label{conformalkillingeq}
\end{equation}
Of particular interest in the study of the renormalization group ($\mrm{RG}$) flows in quantum theories are scale-invariant theories. At classical level it is possibile to show that under some conditions they are conformal. In particular assuming invariance under dilatations in a Lorentz-Poincar{\'e} invariant theory, a current for scale transformations will be defined as $J_\mu=T_{\mu\nu}x^\nu$ where $T_{\mu\nu}$ is the stress-energy tensor of the theory satisfying the conservation law $\partial_\mu T^{\mu\nu}=0\,$. The invariance under dilatations implies that $T_\mu^\mu=0$, thus the stress-energy tensor of the theory is traceless and it can be shown that this implies the conservation of the conformal current $T_{\mu\nu}\xi^\nu$ where $\xi^\mu$ satisfies (\ref{conformalkillingeq}).

We point out the $\mrm{CFT}$s in $d=2$ are peculiar since the conformal group in two dimensions is infinite-dimensional. In this case, if one parametrizes the Minkowski plane $\mathbb{R}^{1,1}$ as the complex line\footnote{Given the coordinates $(\tau, \sigma)$ of $\mathbb{R}^{1,1}$, one can introduce the holomorphic coordinate $z$ such that $z=e^{2(\tau-i\sigma)}$ and $\bar{z}=e^{2(\tau+i\sigma)}$.} $\mathbb{C}$ with the holomorphic coordinate $z$, thus the infinitesimal generators of the conformal transformations are given by
\begin{equation}
L_n=-z^{n+1}\partial_z \qquad \text{and} \qquad \bar{L}_n=-\bar{z}^{n+1}\partial_{\bar{z}}\,.
\end{equation}
These satisfy the Virasoro algebra (\ref{virasoro}) and the quantum Virasoro algebra (\ref{wick}) in the case of quantum theories. Thus the Virasoro algebra is the infinite-dimensional algebra generating the conformal transformations of the plane and from this it follows that the quantum field theory defined by the string embedding coordinates $X^\mu(\tau,\sigma)$ is a $\mrm{CFT}$.

The supersymmetric formulation of conformal field theories is given by the superconformal field theories ($\mrm{SCFT}$). These are supersymmetric field theories such that the conformal group is enlarged by the usual supercharges of the Poincar{\'e} group, by the R-symmetry and by a set of new supercharges called {\itshape conformal supercharges}.

If one quantizes a classical $\mrm{CFT}$, the conformal invariance is generally broken by a {\itshape conformal anomaly} and a renormalization scale $\mu$ for the coupling $g$ of the theory is introduced,
\begin{equation}
\mu \frac{\D}{\D\mu}g(\mu)=\beta(g)\,,
\end{equation}
where $\beta(g)$ is the $\beta$-function defining the $\mrm{RG}$ flow of the theory. Conformal invariance is broken since the stress-energy tensor is directly related to the  $\beta$-function\footnote{For example in a Yang-Mills theory the conformal anomaly is given by $T_\mu^\mu \sim \beta(g)F_{\mu\nu}F^{\mu\nu}$}.

Clearly all the couplings of the theory scale with $\mu$ and, in particular, also the dimension $d$ of the classical theory will be corrected by a factor $\gamma(g)$ called {\itshape anomalous dimension}.
Between classical $\mrm{CFT}$s there are some cases in which the conformal invariance is preserved also at the quantum level and these cases are the most interesting for the $\mrm{AdS/CFT}$ correspondence:
\begin{itemize}
\item {\itshape Fixed point of RG flows}: The $\mrm{RG}$ flows define a dynamics in the space of couplings of a $\mrm{QFT}$ and a fixed point $g^*$ is constitute by that theory with $\beta(g^*)=0$ and vanishing anomalies. In these cases a $\mrm{CFT}$ is defined as the fixed point of the $\mrm{RG}$ flow of a $\mrm{QFT}$.
\item {\itshape Finite theories}: These theories don't have an $\mrm{RG}$ flow in the sense that $\beta(g)=0$ for all values of $g$. Finite theories are thus conformal theories without divergences at all.
\end{itemize}

Let's focus on those $\mrm{SCFT}$s preserving conformal invariance at the quantum level. In these cases the quantum states are no longer labelled by their mass since it doesn't commute with the generator of scale transformations, namely the mass and the energy are physical quantities scaling under conformal transformations. Instead of the mass, a good quantum number is the {\itshape conformal dimension} $\Delta$ defined as follows
\begin{equation}
[\ma D,\ma O(x)]=i(\Delta+x_\mu \partial^{\mu})\ma O(x)\,,
\end{equation}
where $\ma D$ is the generator of dilatations and $\ma O(x)$ is an operator of the theory. Given $\Delta$ as a good quantum number, one can label the states by considering the irreducible representations coming from some non-compact subgroups of $\mrm{SO}(d,2)$.
As interesting property coming from the commutation relations involving $\ma D$ consists in the fact that the generators of translations $P^\mu$ and of special conformal transformations $K_\mu$ increase and decrease the conformal dimension. Then $P_\mu$ and $K_\mu$ can be viewed as creators and annihilators whose action on the vacuum defines all the fields included in the theory. In particular we call {\itshape primary operators} the operators annihilated by $K_\mu$ and {\itshape descendent operators} those obtained applying $P_\mu$.

To conclude, a $\mrm{CFT}$ is determined by a set $(\Delta_i,\ma O_i(x))$ composed by a basis of primary operators $\ma O_i(x)$ with their conformal dimensions $\Delta_i$. Moreover the conformal invariance implies that the one-point correlation functions vanish, i.e. $\langle \ma O(x)\rangle=0$, and fixes the form of two- and three-point correlations functions. For example the two-points correlation function is given by
\begin{equation}
\langle \ma O_i(x)\ma O_j(y) \rangle =\frac{c \,\delta_{ij}}{|x-y|^{2\Delta}}\,,
\end{equation}
where $c$ is a constant factor depending on the theory considered. Finally we mention that the $n$-point correlations functions can be calculated by making use of the {\itshape operator product expansions} ($\mrm{OPE}$) techniques.

\subsection{Fixed Points and Large $N$ Expansion}
\label{largeN}

In the first part of this section we will introduce two important examples of quantum $\mrm{SCFT}$s arising as description of the near-horizon of two particular brane solutions, while in the second part we will introduce the concept of large $N$ limit in a gauge theory in relation to the two examples we made above. 

We start with an intuitive argument to show how a $\mrm{SCFT}$ may emerge as the worldvolume theory of a brane solution.
Let's consider a black brane solution with a horizon described by a closed string vacuum of the type $\mrm{AdS}_{d+1}\times \Sigma_{D-d-1}$ with $D=10, 11$. The coordinates parametrizing the $\mrm{AdS}_{d+1}$ can be chosen as $x^M=(r, x^\mu)$ with $M=0,\cdots, d$. The radial coordinate $r$ of $\mrm{AdS_{d+1}}$ describes the spatial dynamics of the brane when going out\footnote{It could happen that outside of the horizon the spatial dynamics of the brane solution is described by many coordinates. For a sake of simplicity in this argument we consider just a radial flow.} of the horizon and the coordinates $x^\mu$ parametrize the worldvolume. It follows that the $x^\mu$s are the coordinates both for the supergravity background and for the $\mrm{QFT}$ defined on the worldvolume while the radial coordinate $r$ can be interpreted as an energy scale for the $\mrm{QFT}$. In this sense, the dynamics towards the horizon can be viewed as a renormalization flow of the $\mrm{QFT}$ on the worldvolume.

With these coordinates a scale transformations of the type $r\rightarrow r/\lambda$ and $x^\mu \rightarrow \lambda x^\mu$ preserves the metric of $\mrm{AdS}_{d+1}$ obtained in the near-horizon since its isometry group is $\mrm{SO}(d,2)$. For the identifications between the coordinates just presented also the $\mrm{QFT}$ gains scale invariance in the near-horizon. Thus the near-horizon limit of a brane solution can be interpreted as a scale invariant fixed point for the $\mrm{QFT}$ on the worldvolume and this means from the argument of section \ref{cft} that the theory obtained is a $\mrm{SCFT}$.

Let's consider two famous example of $\mrm{SCFT}$s arising in the near-horizon of brane solutions. We recall that these theories are rigid in the sense that the $\mrm{SUSY}$ is global and thus they don't include gravity.

As first example let us consider the $\ma N=4$ $\mrm{SYM}$ theory in $d=4$ with gauge group $\mrm{SU}(N)$. This theory arises as the worldvolume theory of a system of $N$ coincident $\mrm{D}3$ branes at the near-horizon limit, is superconformal and preserves 16 supercharges\footnote{Including superconformal generators it preserves 32 supercharges.}. The field content is given by a gauge field $A_\mu$, by six scalar fields $\varphi^i$ and four Weyl spinors $\psi_a$. The $R$-symmetry group is given by $\mrm{SU}(4)_R\simeq \mrm{SO}(6)$ which is the group associated to the symmetries of the transverse space of a $\mrm{D}3$ brane. The scalars transform under the fundamental representation of $\mrm{SO}(6)$ with the index $i=1,\cdots , 6$ and parametrize the six transverse directions of the $\mrm{D}3$ branes while the fermions $\psi_a$ trasform under the $\underline{4}$ representation with index $a=1,\cdots 4$. 
The action is given by
\begin{equation}
S=\frac{1}{g_{YM}^2}\int \D^4 x \mrm{Tr}\left( -\frac 12 F^{\mu\nu}F_{\mu\nu}-i \bar{\psi}^a \slashed D\psi_a-(D\varphi^1)^2+T_i^{ab}\psi_a[\varphi^i,\psi_b]+[\varphi^a,\varphi^b]^2   \right)\,,
\label{n4sym}
\end{equation}
where the $\mrm{Tr}$ is taken respect to the fundamental representation of the gauge group and $T_i^{ab}$ are the generators of $\mrm{SU}(4)_R$. 

As we discussed in section \ref{decoupling}, the gauge coupling in the SCFT is related to the string coupling constant $g_s$ of the string background describing the $\mrm{D3}$ branes. This can be seen directly by studying the decoupling of the $\mrm{DBI}$ action (\ref{dbi}) of the $\mrm{D3}$ in the low-energy limit. 
Comparing the kinetic term of the gauge fields in (\ref{n4sym}) with (\ref{gymgs}) and introducing the numerical factors coming from the tension of the D3, it follows that 
\begin{equation}
g_{YM}^2 =4\pi g_s\,.
\label{gym=gs}
\end{equation}
Thus the low-energy limit of the $\mrm{D3}$-branes corresponds to the strongly coupled regime of the $\ma N=4$ $\mrm{SYM}$ theory in four dimensions describing the open strings' degrees of freedom on the D3 branes.

A worldvolume $\mrm{SCFT}$ describes the space of all possible positions assumed by a brane. In fact, since the scalar fields defined on the worldvolume describe the transverse oscillations of the brane, the moduli space of the theory, i.e. the manifold parametrized by the eigenvalues of the scalars,  will define the independent positions of the brane. In particular, for $\ma N=4$ $\mrm{SYM}$ the vacuum expectation values of the scalars $\varphi^i$ parametrize the manifold $\mathbb{C}^3$ that is the space describing all the independent positions of $N$ $\mrm{D}3$ branes.

The second example is the so-called {\itshape ABJM theory} \cite{Aharony:2008ug}. This is a three-dimensional $\ma N=6$ $\mathrm{SU}(N)\times \mrm{SU}(N)$ Chern-Simons theory coupled to four complex scalars $C_I$ in the $(N,\underline{N})$ representation of the gauge group with $I=1\cdots4$, and the corresponding fermonic superpartners. 

The theory can be constructed in three step: 
\begin{itemize}

\item Consider an $\ma N=2$ vector multiplet $V$ in $d=3$ described by a $\mathrm{U}(N)\times \mrm{U}(N)$ Chern-Simons action with levels $(k, -k) \in \mathbb{N}$ coupled to $N_f$ chiral multiplets $\Phi_i$ in a representation $\underline R$ of the gauge group.

\item Consider the enhancement to $\ma N=3$ $\mrm{SUSY}$ coming from the inclusion of an auxiliary chiral multiplet $\phi$ defining a superpotential of the type $W=\bar{\Phi}_i\phi\,\Phi_i-\frac{k}{8\pi}\mrm{Tr}(\phi^2)$ \cite{Kao:1995gf, Gaiotto:2007qi, Benna:2008zy}.

\item Consider the truncation of the $\ma N=3$ theory to four chiral multiplets organized in two hypermultiplets in the bifundamental representation of the gauge group: given $A_1$, $A_2$ the bifundamental chiral multiplets and $B_1$, $B_2$ the anti-bifundamentals, they organize as doublets $(A_1, B^*_1)$ and  $(A_2, B^*_2)$. Once integrated out the auxiliary fields, the superpotential has the form $W=\frac{2\pi}{k}\epsilon^{ab}\epsilon^{\dot{a}\dot{b}}\mrm{Tr}(A_aB_{\dot{a}}A_bB_{\dot{b}})$ with a manifest global $\mrm{SU}(2)\times \mrm{SU}(2)$ invariance. Thus there is a further enhancement of symmetry since, in addition to the $\mrm{SU}(2)$  flavor rotating $A_a$ and $B_{\dot a}$, there is another $\mrm{SU}(2)$ symmetry rotating $A_a$ and $B_{\dot a}$ separately. 
Since the $R$-symmetry $\mrm{SO}(3)_R\simeq\mrm{SU}(2)_R$ of the $\ma N=3$ theory does not commute with $\mrm{SU}(2)\times \mrm{SU}(2)$, it follows that they together generate the symmetry group $\mrm{SU}(4)\simeq \mrm{SO}(6) $. Thus the scalars can be organized as $C_I=(A_1, A_2, B_1^*, B_2^*)$ and, since the $R$-symmetry group of a $d=3$ theory is $\mrm{SO}(\ma N)_R$, it follows that the theory obtained is at least $\ma N=6$ \cite{Aharony:2008ug}. 
\end{itemize}
 
Taking in consideration a gauge group given by $\mathrm{SU}(N)\times \mrm{SU}(N)$ doesn't change dramatically this construction\footnote{The construction of the moduli space is much more complicated respect to the case of
$\mathrm{U}(N)\times \mrm{U}(N)$ but it still remains the same.} and, in this way, one obtains $\mrm{ABJM}$ theory. In particular the gauge coupling of the theory is associated to the quantized Chern-Simons levels $(k, -k)$ and the moduli space is given by the manifold $\mathbb{C}^4/\mathbb{Z}_k$. Moreover $\mrm{ABJM}$ is superconformal preserving 12 supercharges. The full global symmetry is given\footnote{The group $\mrm{U(1)}$ is a baryonic symmetry under which $A_i$ and $B_i$ have respectively charges $\pm 1$.} by $\mrm{SO}(6)_R\times \mrm{U(1)}$. Furthermore we point out for the particular values of the Chern-Simons levels $k=1, 2$, one has a further enhancement of supersymmetry to $\ma N=8$ \cite{Bagger:2007jr}.

The brane picture of $\mrm{ABJM}$ is much more complicated respect that of the $\ma N=4$ $\mrm{SYM}$. The construction of the bound state of branes is firstly performed in type $\mrm{IIB}$ string theory with a bound state whose worldvolume theory is an $\ma N=3$ $\mrm{SYM}$ Chern-Simons theory in three dimensions \cite{Hanany:1996ie, Kitao:1998mf, Bergman:1999na}. Thus one can map the brane intersection in type $\mrm{IIA}$ by $\mrm{T}$-duality and then consider the strong-coupling regime by uplifting to $\mrm M$-theory. 

In this limit the brane construction is described by the near-horizon configuration of a set of $N$ $\mrm{M}2$ branes probing an eleven-dimensional background given by the superposition of two $\mrm{KK}$-monopoles\footnote{The two $\mrm{KK}$-monopoles are described respectively in type $\mrm{IIA}$ by $\mrm{D}6$ branes and $(1,k)5$ branes that in the strong-coupling limit are resolved in pure eleven-dimensional geometry.}. In particular the eleven-dimensional background is described by the manifold $\mathbb{R}^{1,2}\times X_8$ with the $\mrm{M}2$ branes extending along $\mathbb{R}^{1,2}$ and probing $X_8$. The latter is an example of hyper-toric manifold \cite{Gibbons:1997cm, Gauntlett:1997pk} and locally is given by the moduli space of $\mrm{ABJM}$, namely $\mathbb{C}^4/\mathbb{Z}_k$.  
The strong-coupling regime of the brane intersection corresponds to the low-energy limit of the $\ma N=3$ $\mrm{SYM}$ Chern-Simons theory on the worldvolume. 
In this limit the renormalization flow of the worldvolume theory has a fixed point consisting exactly by $\mrm{ABJM}$.

At the typical energy scale of the $\mrm{AdS/CFT}$ correspondence, the worldvolume theories appear in a particular regime called {\itshape large $N$ limit} \cite{tHooft:1973alw}. 
This particular limit is a systematic expansion of $\mrm{SU}(N)$ gauge theories in the strong-coupling regime in which the color number is large, i.e. $N\rightarrow \infty$.  

Generally, given a $\mrm{SU}(N)$ $\mrm{YM}$ theory, one can reorganize the perturbative expansion of the free energy as an expansion where the gauge coupling $g_{YM}$ is expressed in terms of $N$ and the pamareter
\begin{equation}
x=g_{YM}^2N\,,
\label{thooft}
\end{equation}
called  {\itshape 't Hooft coupling}. In terms of these new parameters the lagrangian of a pure $\mrm{YM}$ theory takes the generic form of the type
\begin{equation}
\mathcal L=\frac{N}{x}\,\mrm{Tr}\,F_{\mu\nu}F^{\mu\nu}\,.
\label{pureym}
\end{equation}
The perturbative contributions to the propagators and to the vertices are now respectively associated to powers of $x/N$ and $N/x$. Moreover, each loops gives a contribution of the order $N$.
The perturbative expansion in the large $N$ limit can be represented in terms of Riemann surfaces that, in turn, are classified only by their topological properties. In particular the Euler characteristic of each surface is given by $F+V-E$ where $F$ is the number of faces associated 
to the loops, $E$ the number of edges associated to each propagator and $V$ the number of vertices. It follows that in this reformulation each contribution to the free energy expansion will correspond to surfaces giving a contribution at the order $x^{E-V}N^{F+V-E}=O(N^{2-2\mathfrak{g}})$ where $\mathfrak{g}$ is the genus of the Riemann surface.

The large $N$ limit is defined as the following regime of the theory
\begin{equation}
N\rightarrow \infty \,,\qquad g_{YM}\rightarrow 0\, \quad \text{with}\quad x=g_{YM}^2N<\infty\,.
\label{largeNlimit}
\end{equation}
In this limit the perturbative expansion of the free energy collapses to the first non-trivial contribution given by {\itshape planar graphs} of the order $O(N^2)$ that in turn are associated to a Rieman surface with a spherical topology
$\mathfrak g=0$. The subleading 
contributions are supressed by powers of $1/N^2$. Including in (\ref{pureym}) the couplings to matter fields (like scalars or fermions) the loop expansion is enlarged by graphs going as powers of $1/N$. In such a general picture the topology of 
Riemann surfaces is changed since also boundaries and surfaces with an odd Euler characteristic are introduced. 

Regarding the examples of $\mrm{SCFT}$s presented above, the large $N$ limit of $\ma N=4$ $\mrm{SYM}$ is simply given by (\ref{largeNlimit}). In the case of $\mrm{ABJM}$, the gauge coupling is quantized, i.e. $g^2_{YM}=1/k$ and there are some subtleties. 
The 't Hooft coupling is given by $x=N/k$ and it is kept fixed in the limit of large $N$. Since 
the weakly-coupled regime of $\mrm{ABJM}$ is given by $k\rightarrow 0$ it follows that,
for a strong-coupling regime, the large $N$ limit must be such that $N \gg k $.

Many other examples of superconformal field theories can be extracted from the decoupling limit of branes. Usually the more complicated the brane intersection is, the richer the worldvolume theory.
An important example is given by {\itshape quiver theories} that are the worldvolume theories of
branes probing backgrounds with singular topologies\footnote{$\mrm{ABJM}$ is a example of ``simple" quiver theory since its moduli space is the orbifold $\mathbb{C}^4/\mathbb{Z}_k$.} (see for example \cite{Klebanov:1998hh}).

Finally it is interesting to mention that the highest dimension for a consistent $\mrm{SCFT}$ is six. In $d=6$ there are three types of $\mrm{SCFT}$s: the $\ma N=(2,0)$ and $\ma N =(1,1)$ theories with 16 supercharges and the $\ma N=(1,0)$ with 8 supercharges. These three theories are quite obscure since they are non-lagrangian and are related each others in a similar way of superstring theories in $D=10$. Later in this thesis we will focus our attention on the $\ma N=(1,0)$ theory.

\subsection{The Correspondence}
\label{corresponence}

We have now all the instruments to formulate the $\mrm{AdS/CFT}$ correspondence \cite{Maldacena:1997re, Witten:1998qj}. Let's concentrate our attention to a bound state composed by $N$ coincident branes. The branes can be free to move in the background or constrained between other heavier branes. Moreover, other solitonic objects like $\mrm{KK}$ monopoles, can interact with the system and modify the geometric properties of the background.

The low-energy limit $\alpha^\prime \rightarrow 0$ and $g_s \rightarrow 0$ of the bound state implies that:
\begin{itemize}
\item The physics of the bound state is captured by the massless modes of the closed string sector in terms of a solution in $D=11$ or type $\mrm{II}$ supergravities. This solution describes the modifications of a classical background due to the presence of the branes. Many black brane solutions\footnote{What follows holds also for domain walls but instead of the horizon the solution reaches smoothly the boundaries.} are characterized by a near-horizon limit given by a closed string vacuum of the type $\mrm{AdS}_{d+1}\times \Omega_{D-d-1}$ with $D=10, 11$. In this sense the brane solution can be interpreted as the {\itshape weakly-coupled description} of a non-perturbative effect modifying a closed string vacuum.
The number of coincident branes $N$ is related to the quantized fluxes of the system, thus the bigger is $N$, the larger the total charge of the system is. The low-energy limit corresponds to the limit of large charges (as an example of this see \eqref{fluexsrn}) and thus the limit of large $N$.

\item The bound state is described by a rigid supersymmetric $\mrm{SU}(N)$ quantum\footnote{As we saw for ABJM, the gauge group could be more complicated like, for example, $\mrm{SU}(N)\times \mrm{SU}(N)$.} theory defined on the worldvolume whose field content comes from the zero-level of the spectrum of open strings ending on the branes. In particular, a near-horizon limit given by a closed string vacuum of the type $\mrm{AdS}_{d+1}\times \Omega_{D-d-1}$ with $D=10, 11$ corresponds to a fixed point in the $\mrm{RG}$ flow of the worldvolume $\mrm{QFT}$ becoming in this limit a $d$-dimensional $\mrm{SCFT}_d$. The gauge coupling $g_{YM}$ of the $\mrm{SCFT}_d$ is related to $g_s$ so that the $\mrm{SCFT}_d$ gives a {\itshape strongly-coupled description} of the bound state. Moreover if one organizes the perturbations in the 't Hooft expansion and keeps 't Hooft coupling $x$ fixed, the limit $g_{YM}\rightarrow 0$ implies that $N$ is large thus, as we said in section \ref{largeN}, the strongly-coupled regime is well described by its large $N$ limit.

\end{itemize}
Keeping these two specular pictures, let's choose a bound state with a near-horizon limit described by the vacuum $\mrm{AdS}_{d+1}\times \Omega_{D-d-1}$ with $D=10, 11$ with $\Omega_{D-d-1}$ compact manifold. Since $\Omega_{D-d-1}$ is compact one can perform a $\mrm{KK}$ reduction on it and look at $\mrm{AdS}_{d+1}$ as a vacuum solution of a $(d+1)$-dimensional effective gravity theory with an infinite number of fields. Furthermore, it can be shown in many explicit realizations, that the masses of the $\mrm{KK}$ modes depend on the geometric quantities of $\Omega_{D-d-1}$ in such a way the massive modes decouple\footnote{In case of $\Omega_{D-d-1}=S^{D-d-1}$ the masses of the $\mrm{KK}$ fields goes like $1/R^2$ with $R$ radius of the $(D-d-1)$-sphere and thus, in the supergravity regime of large $R$, they decouple.}. The action of this effective theory will be a classical action $S_{\scriptsize{gravity}}$ typically including gauge fields and scalars on the background $\mrm{AdS}_{d+1}$.

The $\mrm{AdS/CFT}$ correspondence consists in a prescription relating the {\itshape euclidean}} partition function of the $(d+1)$-dimensional effective action evaluated on the $\mrm{AdS}_{d+1}$ background with the generator of the correlation functions of the worldvolume $\mrm{SCFT}_d$ in the large $N$ limit.
In order to formulate concretely this prescription we have to establish the general dictionary between fields characterizing the effective gravity theory and operators of the $\mrm{SCFT}_d$. Following the notation used in section \ref{largeN}, the coordinates of the $\mrm{AdS}_{d+1}$ can be organized as $(r,x)$ with $x=(x^0,\cdots,x^d)$.
Given a classical field $\varphi(r,x)$ featuring the $(d+1)$-dimensional gravity theory, we consider its boundary configuration 
\begin{equation}
\varphi(r,x)\longrightarrow \varphi_0(x)
\label{boundary}
\end{equation}
where the conformal boundary of $\mrm{AdS_{d+1}}$ is identified by the limit $r\rightarrow \infty$. The limit $(\ref{boundary})$ is unique once suitable boundary conditions have been imposed.

The key step is to consider $\varphi_0(x)$ as a classical source for an operator $\ma O(x)$ of the $\mrm{SCFT}$ .  This means that the generator of the correlation functions of $\ma O(x)$ is given by
\begin{equation}
\ma Z_{\scriptsize{\mrm{SCFT}_d}}(\varphi_0)=\langle e^{\,\int \varphi_0(x) \,\ma{O}(x)}  \rangle_{\scriptsize{\mrm{SCFT}}} \,.
\end{equation}
We can now consider the euclidean partition function associated to the gravity theory 
\begin{equation}
Z_{\scriptsize{gravity}}=e^{S_{\scriptsize{gravity}}}\,
\label{Zgravity}
\end{equation}
and evaluate (\ref{Zgravity}) on the $\mrm{AdS}_{d+1}$ background. The $\mrm{AdS/CFT}$ correspondence is realized by the identification
\begin{equation}
\ma Z_{\scriptsize{\mrm{SCFT}_d}}(\varphi_0)=Z_{\scriptsize{gravity}}(\varphi)|_{\scriptsize{\mrm{AdS}_{d+1}}}\,,
\label{adscft}
\end{equation}
where, on the right side, we have an on-shell quantity related to the $(d+1)$-dimensional action evaluated on the $\mrm{AdS}_{d+1}$ background and, on the left side, a functional depending on the arbitrary off-shell configuration of $\varphi_o$. 

Since $\ma Z_{\scriptsize{\mrm{SCFT}_d}}(\varphi_0)$ determines uniquely the $\mrm{SCFT}_d$, the identification (\ref{adscft}) defines a duality relation between a weakly-coupled description given by the classical gravity solution on the $\mrm{AdS}_{d+1}$ background and a strongly-coupled description given by the $\mrm{SCFT}_d$.
The correspondence is also called {\itshape holographic duality} or, simply, {\itshape holography} since the strongly-coupled regime of the $\mrm{SCFT}_d$ is defined at the conformal boundary of the $\mrm{AdS}_{d+1}$ background and captures the physics of the gravity theory living within the $\mrm{AdS}_{d+1}$ background. 

The two sides of the correspondence share the same symmetries with the difference that on the gravity side the symmetries are local while in the $\mrm{SCFT}_d$ global. In particular on the gravity side we have the group of diffeomorphisms. This includes the group of the isometries of $\mrm{AdS}_{d+1}$ given by $\mrm{SO}(d,2)$ whose rigid version generates the global conformal symmetry of the quantum theory on the boundary. This turns out to be a general property of holography: the boundary values of the fields of the background are associate to conserved currents of the global symmetries of the $\mrm{SCFT}_d$. For example the $(d+1)$-dimensional graviton is associated with the stress-energy tensor of the quantum theory on the boundary and a gauge field $A_\mu$ is related to a current $J^{\scriptsize{\mrm{SCFT}}}_\mu$,
\begin{equation}
g_{\mu\nu}^{(d+1)} \longleftrightarrow T_{\mu\nu}^{\,\scriptsize{\mrm{SCFT}}}\,,\quad A^{d+1}_\mu \longleftrightarrow J^{\scriptsize{\,\mrm{SCFT}}}_\mu \,.
\end{equation}
These identifications determine the couplings in the $\mrm{SCFT}_d$ in the following way
\begin{equation}
\int \D^dx \left ( g_{\mu\nu}T_{\,\scriptsize{\mrm{SCFT}}}^{\mu\nu}+A_\mu  J_{\,\scriptsize{\mrm{SCFT}}}^{\mu} +\cdots \right)\,,
\label{supercurrents}
\end{equation}
where each current in the $\mrm{SCFT}_d$ corresponds to a local symmetry of the classical solution on the $\mrm{AdS}_d$ background.

\subsection{Two Explicit Realizations: $\mrm{AdS}_5/\mrm{CFT}_4$ and $\mrm{AdS}_4/\mrm{CFT}_3$}
\label{realizations}

Starting from the two examples of $\mrm{SCFT}$s given in section \ref{largeN}, we will now consider two explicit realizations of the correspondence.

The first example is given by a set of $N$ coincident $\mrm{D}3$ branes probing a flat background \cite{Maldacena:1997re}. Following the scheme of the previous section, in the low-energy limit they are described by the solution (\ref{pbrane}) in type $\mrm{IIB}$ supergravity\footnote{The radius $R_3$ has to by multiplied by $N$.}. The near-horizon limit of the $\mrm{D}3$ brane solution is characterized by an enhancement of $\mrm{SUSY}$, thus it preserves 32 supercharges, and it is described by the closed string vacuum $\mrm{AdS}_5\times S^5$ with the same radius $R_3$ both for $\mrm{AdS}_5$ and for $S^5$. It can be shown from (\ref{pbrane}) that $R_3$ is given by 
\begin{equation}
R^2_3=\alpha^\prime \sqrt{4\pi\,g_s\,N}\,.
\end{equation}
Each brane contributes to the total charge of the system with a unit of $\mrm{R}$-$\mrm{R}$ 5-form flux,
\begin{equation}
\int_{S^5}\,\star F_5=N\,.
\end{equation}

On the other side the worldvolume theory of the $\mrm{D}3$s in their near-horizon limit is the $\ma N=4$ $\mrm{SYM}$ theory in four dimensions with gauge group $\mrm{SU}(N)$ introduced in section \ref{largeN}. This theory is invariant under global transformations of the conformal group $\mrm{SO}(4,2)$ that is exactly the group of the isometries of $\mrm{AdS}_5$ and it has an R-symmetry group $\mrm{SU(4)}_R$ that is the isometry group of the $S^5$. Furthermore the theory preserves 32 supercharges\footnote{Between these, 16 are coming from the super-Poincar{\'e} algebra and 16 from the conformal supercharges.} and, in the low-energy limit, it is strongly-coupled since $g_{YM}^2 =4\pi g_s$ as observed in (\ref{gym=gs}).
Let's introduce the coupling $x=g_{YM}^2N$ given in (\ref{thooft}) and organize the perturbations in $g_{YM}^2$ with the 't Hooft expansion. We thus have the identifications
\begin{equation}
4\pi g_s=\frac{x}{N} \qquad \text{and} \qquad \frac{R_3^2}{\alpha^\prime}=x^{1/2}\,.
\label{ads5cft4}
\end{equation}
On the left side we have the couplings describing the regimes of the classical (super)gravity theory associated to the vacuum $\mrm{AdS}_5\times S^5$. This theory is defined in the low-energy limit, i.e. $g_s\rightarrow 0$ and\footnote{These limit can be expressed also as $R_3\rightarrow \infty$.} $\alpha^\prime \rightarrow 0$, so if we consider the right side terms of (\ref{ads5cft4}), this is equivalent to the large $N$ regime with $x \rightarrow \infty$ describing\footnote{The values of $N$ are such that $N\gg x$, so that $x$ can be considered fixed respect to $N$ as prescribed by (\ref{largeNlimit}).} the strong coupling of the $\mrm{SCFT}_4$.

Finally one can perform a $\mrm{KK}$ reduction on $S^5$ and obtain an effective five-dimensional gravity theory with $\mrm{AdS}_5$ \cite{Kim:1985ez,Gunaydin:1984fk}. 
It is interesting to note that the $\mrm{KK}$ spectrum is organized in irreducible representations of $\mrm{SO(6)}$ and there is a complete correspondence between the $\mrm{KK}$ spectrum and the short multiplets of $\ma N=4$ $\mrm{SYM}$. In particular, as anticipated in (\ref{supercurrents}), the massless multiplet describing the gravity side and containing the graviton, the gravitinos and the $\mrm{SO(6)}$ gauge fields, corresponds to the multiplet of the supercurrent on the field theory side. Moreover the massless spectrum differs from the massive modes by quantities of the order $1/R_3^2$. So the massive modes decouple taking $R_3$ large that is equivalent to the low-energy limit.

The validity of the correspondence can be further checked by using (\ref{adscft}) and imposing suitable boundary conditions on five-dimensional fields. Thus one can derive the correlation functions of $\ma N=4$ $\mrm{SYM}$ by the functional derivation of the euclidean partition function of the five-dimensional supergravity evaluated on $\mrm{AdS}_5$ with respect to the boundary values of the classical fields associated to a given operator in the $\mrm{SCFT}_4$.

The other example takes in consideration a set of $N$ coincident $\mrm{M}2$ branes probing the singular manifold $\mathbb{C}^4/\mathbb{Z}_k$ \cite{Aharony:2008ug}. In (\ref{m2}) we wrote the supergravity solution for a single\footnote{As we said in section \ref{branesolutions}, the $\mrm{M}2$ solution is simply generalized to a system of $N$ $\mrm{M}2$ by multiplying the radius $R_2$ by $N$.} $\mrm{M}2$ brane on a flat background. When the background is given by $\mathbb{C}^4/\mathbb{Z}_k$, the near-horizon geometry (\ref{nhm2}) is slightly modified in the quotient geometry $\mrm{AdS}_4\times S^7/\mathbb{Z}_k$. 
The internal seven-dimensional background is naturally described as an $S^1$-fibration over $\mathbb{CP}^3$ that partially breaks the isometry group $\mrm{SO(8)}$ to $\mrm{SU(4)}\times \mrm{U(1)}$.
Globally, the effect of the action of $\mathbb{Z}_k$ is to reduce the volume of the $S^7$ by a factor of $k$ in order to produce a well-defined flux for the 4-form field strength. Let's write the metric of $S^7/\mathbb{Z}_k$ as
\begin{equation}
\D s^2_{S^7/\mathbb{Z}_k}=\frac{1}{k^2}(\D\phi+k\omega)^2+\D s^2_{\mathbb{CP}^3}\,,
\label{cp3fibration}
\end{equation}
where $\omega$ is the $\mrm{U}(1)$ connection associated to the fibration and $\phi \sim \phi+2\pi$. Using (\ref{cp3fibration}), one can derive the flux of the 4-form field strength obtaining a quantized value $N$ on the quotient such that $N^\prime=kN$ where $N^\prime$ is the quantized flux calculated on the $S^7$. 

Moreover, the manifold $\mathbb{C}^4/\mathbb{Z}_k$ preserves only 3/16 of the total 32 supercharges\footnote{Adding the branes doesn't break further supersymmetry.} and, when the near-horizon limit is considered, there is an enhancement to 12 supercharges preserved.

As we said in section \ref{largeN}, the theory living on the worldvolume of $N$ coincident $\mrm{M}2$ branes on $\mathbb{C}^4/\mathbb{Z}_k$ is $\mrm{ABJM}$. This is a $\ma N=6$ three-dimensional Chern-Simons theory with couplings to matter multiplets. R-symmetry group is $(\mrm{SU(4)}\times \mrm{U(1)})_R$ that matches with the isometry group of $S^7/\mathbb{Z}_k$, while the conformal group $\mrm{SO(3,2)}$ corresponds to the isometry group of $\mrm{AdS}_4$.

The peculiarity of this realization of the $\mrm{AdS/CFT}$ correspondence is the existence of two different regimes, corresponding respectively to the $\mrm{M}$-theory and type $\mrm{IIA}$ energy regimes of the background.

We recall that the gauge coupling of $\mrm{ABJM}$ is given by the inverse of the Chern-Simons level $k$, i.e. $g_{YM}^2=1/k$, and the large $N$ limit is well-defined when $\mrm{ABJM}$ is strongly-coupled, namely $k\rightarrow \infty$ and $N\gg k$. 
Let's consider the limit of (\ref{cp3fibration}) in which the radius $\phi$ of the circle is large, i.e. $N^\prime=k N \gg 1$. This radius has the same expression given in (\ref{braneradii}) divided by a factor $k$. Looking only at the $N$ and $k$ dependences, we have
\begin{equation}
\frac{R_2}{k\,l_P}\sim \frac{N^{1/6}k^{1/6}}{k}\,,.
\end{equation}
Then, when $k$ is large and $N\gg k$, we have two regimes. The first one, corresponding to $k^5 \gg N$, implies a large value of the radius $R_2/k$ and it is associated to the $\mrm{M}$-theory regime where the system of branes is well-described by a solution of eleven-dimensional supergravity.
The regime $k^5 \ll N$, implies that the radius of the circle $R_2/k$ becomes small and an effective description in type $\mrm{IIA}$ supergravity arises. This description is given by a background of the type $\mathrm{AdS}_4\times \mathbb{CP}^3$, the dilaton, a 2-form flux wrapping a 2-cycle $\mathbb{CP}^3$ and a 4-form flux on $\mrm{AdS}_4$.

From the type $\mrm{IIA}$ dilaton one obtains the relation between $g_s$ and the t'Hoof coupling $x$ of $\mrm{ABJM}$, while, from the reduction on a circle of $\mrm{AdS}_4\times S^7/\mathbb{Z}_k$, one derives the radius $R_{IIA}$ of the ten-dimensional background. It can be shown that the following dependences hold
\begin{equation}
g_s \sim \frac{x^{5/4}}{N}\qquad \text{and} \qquad \frac{R^2_{IIA}}{\alpha^\prime}\sim \sqrt{x}\,.
\label{ads4cft3}
\end{equation}
These expressions describe the same relation between the curvature of the background and the 't Hooft coupling of the other example presented in (\ref{ads5cft4}). Also the string coupling constant $g_s$ goes like $1/N$ as in $(\ref{ads5cft4})$. The relations (\ref{ads4cft3}) imply that the low-energy limit $g_s\rightarrow 0$ and $\alpha^\prime \rightarrow 0$ is equivalent to the large $N$ limit for $\mathrm{ABJM}$ that is well-defined for $k$ and $N$ large with $k \ll N$ and $N/k$ kept fixed. In particular, the strong-coupling regime of $\mathrm{ABJM}$ given by $x$ large, is dual to the weakly-coupled description in type $\mrm{IIA}$ supergravity when $k^5 \ll N$ thus $x^{5/2}/N^2$.

The four-dimensional gravity theory with $\mrm{AdS}_4$ background is obtained by $\mrm{KK}$ reduction as in the previous case \cite{Biran:1983iy,Castellani:1984vv}. The main difference here is that only the $\mathbb{Z}_k$-invariant states will survive in the dimensional reduction on $S^7$. This means that the $\mrm{KK}$ fields will organize themselves in representations of $\mrm{SU(4)}\times\mrm{U}(1)$ that match with the spectrum of the short multiplets of $\mrm{ABJM}$.

Many other realizations have been constructed (see for example \cite{Klebanov:1998hh,Guarino:2015jca,Gaiotto:2014lca}) and one of the most interesting examples is the holographic duality between an $\mrm{AdS}_7$ vacuum in massive type $\mrm{IIA}$ and the $\ma N =(1,0)$ $\mrm{SCFT}$ in six dimension \cite{Gaiotto:2014lca}. This particular realization will be considered later in this thesis.

\subsection{The RG Flow Across Dimensions}
\label{flowacrossdimensions}

In the most general scenario a consistent truncation to a $(d+1)$-dimensional gauged supergravity exists. In this case only the zero-modes of the $\mrm{KK}$ expansion determine the dynamics of the closed string and the whole brane solution is captured by a lower-dimensional solution of a $(d+1)$-dimensional gauged supergravity with $\mrm{AdS}_{d+1}$ emerging for example in its asymptotic limit.

As explained in section \ref{corresponence}, the $\mrm{AdS/CFT}$ correspondence works only when the supergravity solution associated to the bound state describes in the near-horizon a closed string vacuum of the type $\mrm{AdS_{d+1}}\times \Omega_{D-d-1}$ with $D=10, 11$. However, if the $\mrm{AdS}_{d+1}$ background is a critical point for a given $(d+1)$-dimensional gauged supergravity obtained with a consistent truncation around the closed string vacuum, it is possible to consider a more general picture in which the brane solution is completely captured by the gauged supergravity and the background $\mrm{AdS}_{d+1}$ describes a particular regime of the $(d+1)$-dimensional solution.

Let's recall the situation discussed in section \ref{sugraflows} in which the worldvolume of a bound state wraps an holomorphic non-trivial cycle. In these cases the $(d+1)$-dimensional solution interpolates between the $\mrm{AdS}_{d+1}$ vacuum, describing the asymptotics, and a less symmetric solution $\mrm{AdS}_{p+2}\times \Sigma_{d-p-2}$, where $\Sigma_{d-p-2}$ is the $(d-p-2)$-cycle describing the wrapping and the transverse geometry of a $(d+1)$-dimensional event-horizon. The uplift of the $\mrm{AdS}_{d+1}$ background describes the near-horizon configuration of the brane solution, while the uplift of the less symmetric $\mrm{AdS}_{p+1}\times \Sigma_{d-p}$ solution is associated to a new warped vacuum describing the wrapped geometry of the worldvolume.

For example we can consider a black hole \cite{Cacciatori:2009iz} in $\ma N=2$, $d=4$ $\mrm{U(1)}$-gauged supergravity described by an $\mrm{AdS}_4$ asymptotic and an $\mrm{AdS}_2\times S^2$  horizon or a black string \cite{Maldacena:2000mw} in $\ma N=2$, $d=5$ $\mrm{U(1)}$-gauged supergravity with a near-horizon given by $\mrm{AdS}_3\times H^2$ and $\mrm{AdS}_5$ asymptotics.
If one considers the uplifts of the above solutions, they can be respectively described by $\mrm{M}2$ and $\mrm{D}3$ branes with the worldvolume wrapping $S^2$ and $H^2$ cycles. Their asymptotics are uplifted to the $\mrm{AdS}_4\times S^7$ and $\mrm{AdS}_5\times S^5$ vacua, while the near-horizons limits are described in higher-dimensional supergravities by the geometries $\mrm{AdS}_2\times S^2\times_w S^7$ and $\mrm{AdS}_3\times H^2\times_w S^5$.

Both these configurations are examples of flows of the type introduced in \eqref{supergravityflows}. It is now interesting to wonder which is the dual description in terms of boundary $\mrm{SCFT}$s. If both the asymptotics and the near-horizon have an uplift to closed string vacua describing some particular limiting regimes of the corresponding branes' constructions then, for the discussion of section \ref{corresponence}, it follows the existence of two superconformal field theories, namely $\mrm{SCFT}_{d}$ and $\mrm{SCFT}_{p+1}$, dual to the backgrounds $\mrm{AdS}_{d+1}$ and $\mrm{AdS}_{p+2}$.

The flow of the $(d+1)$-dimensional solution is no longer $\mrm{AdS}_{d+1}$ when going outside to the asymptotic (or equivalently from the near-horizon). This implies that the conformal symmetry of the dual quantum theories is spoiled and the values taken by the classical fields outside to the $\mrm{AdS}_{d+1}$ are associated to deformations of the $\mrm{SCFT}_d$ breaking the conformal invariance. These new terms induce a running of the couplings of the $\mrm{SCFT}_d$, that in turn is associated to an $\mrm{RG}$ flow admitting the $\mrm{SCFT}_d$ as fixed point. Furthermore, this fixed point describes the $\mrm{UV}$ regime because the asymptotics regime of the supergravity flow corresponds to the high energy scales for the worldvolume theory (see the discussion at the beginning of section \ref{largeN}). 

From a general point of view, this $\mrm{RG}$ flow could not have an other fixed point in the $\mrm{IR}$ or, in other cases, it could have a $\mrm{IR}$ fixed point breaking all supersymmetries. Since in a renormalization group flow also the dimension of the theory can take corrections, the most interesting case is that of the supergravity flows \eqref{supergravityflows} dual to an {\itshape RG flow across dimensions} \cite{Maldacena:2000mw} interpolating between two fixed points given by the superconformal field theories $\mrm{SCFT}_{d}$ and $\mrm{SCFT}_{p+1}$. These correspond respectively to the $\mrm{UV}$ and $\mrm{IR}$ regime of the supergravity flow and in this case we will write $\mrm{SCFT}_{d} \longrightarrow\mrm{SCFT}_{p+1}$ in analogy with \eqref{supergravityflows}.

The mechanism inducing $\mrm{RG}$ flow across dimensions between two $\mrm{SCFT}s$ is clearly related to the wapping of the worldvolume of the bound state of branes \cite{Maldacena:2000mw}. If we look at this wrapping as the result of some brane intersection characterizing the bound state, whose effect is to curve the worldvolume, we can interpret the new vacuum associated to the wrapping as the result of a dynamical process of intersecting branes that, in a suitable regime, describes the fluctuations around a new closed string vacuum. This phenomenon triggers the $\mrm{RG}$ flow across dimensions, since the wrapped geometry of the worldvolume implies a spontaneous dimensional reduction\footnote{In order to preserve supersymmetry the $\mrm{SCFT}_d$ defined on the curved background has to be subjected to a {\itshape topological twist} \cite{Maldacena:2000mw}.} of the $\mrm{SCFT}_d$ on the wrapped coordinates as we go away from the asymptotics and this dimensional reduction is associated to the $\mrm{RG}$ flow of the couplings of $\mrm{SCFT}_d$. The $\mrm{IR}$ fixed point is associated to the limit in which the fluctuations of the branes along the wrapping manifold goes to zero and thus the $\mrm{IR}$ limit is captured by a lower-dimensional $\mrm{SCFT}_{p+1}$.

Let's consider for example the five-dimensional black string introduced above. The worldvolume of the $\mrm{D}3$ branes on a flat background is associated in the near-horizon limit to an $\mrm{AdS}_5$ vacuum. This is described by a dual field theory given by $\ma N=4$ $\mrm{SYM}$ in four dimensions defined on the Minkowski space $\mathbb{R}^{1,3}$. The $\mrm{RG}$ flow across dimensions is induced by the wrapping of the worldvolume on $H^2$ implying that $\ma N=4$ $\mrm{SYM}$ is now defined on the curved background $\mathbb{R}^{1,1}\times H^2$. The flow of the black string solution is associated in the infrared by an $\mrm{AdS}_3\times H^2$ and its dual field theory is the two-dimensional $\ma N=(4, 4)$ $\mrm{SCFT}_2$. Thus in this case we have an $\mrm{RG}$ flows across dimensions of the type $\mrm{SCFT}_{4} \longrightarrow\mrm{SCFT}_{2}$ corresponding to a supergravity flow of the type $\mrm{AdS}_{5} \longrightarrow\mrm{AdS}_{3}\times H^2$ \cite{Maldacena:2000mw}.

An other relevant example is that one of the black hole flow $\mrm{AdS}_{4} \longrightarrow\mrm{AdS}_{2}\times S^2$ introduced above \cite{Cacciatori:2009iz}. As we mention in previous section, the $\mrm{AdS}_4$ vacuum associated to the near-horizon of $\mrm{M}2$ branes probing the orbifold $\mathbb{C}^4/\mathbb{Z}_k$ is dual to $\mrm{ABJM}$ theory. The black hole solution is thus described in terms of the wrapping of the worldvolume of the $\mrm{M}2$ branes on an $S^2$.
This wrapping triggers an $\mrm{RG}$ flow across dimensions of the type $\mrm{SCFT}_{3} \longrightarrow\mrm{SCFT}_{1}$, where $\mrm{SCFT}_{1}$ is a superconformal quantum mechanics arising when $\mrm{ABJM}$ is placed on a background $S^1\times S^2$ \cite{Benini:2015eyy}.

As we already mentioned at the end of section \ref{microstate}, the $\mrm{RG}$ flows across dimensions are crucial in giving a microscopic interpretation of the observables of lower-dimensional solutions like black holes and black string. As an example, we mention the study of the microscopic origin of the Bekenstein-Hawking entropy of a black hole. The entropy is a particularly useful thermodynamic observable since it is related to the logarithm of the number of the microstates describing a macroscopic system. Thus the matching of the Bekenstein-Hawking entropy of a supersymmetric black hole with the number of microstates associated to the bound state of branes in the microscopic description becomes a powerful test of string theory.

Finally we mention a complementary description of RG flows across dimensions, that we will discuss in more detail later in this thesis, that is given by {\itshape conformal defects} \cite{Karch:2000gx,Lunin:2007ab}. For the moment, given the same configuration of wrapped branes discussed above, we introduce the conformal defects as a way to interpret the lower-dimensional $\mrm{SCFT}_p$ as a theory living on a hypersurface within the manifold on which the ``mother" $\mrm{SCFT}_d$ is defined.  The presence of this internal manifold manifests itself through the breaking of the conformal symmetry of the $\mrm{SCFT}_d$ and inducing an $\mrm{RG}$ flow across dimensions. This conformal symmetry breaking is described by non vanishing values of the 1-point functions in the higher-dimensional $\mrm{SCFT}_d$ and by the non-conservation of the stress-energy tensor.

%% file: chap3.tex

\chapter{The $\ma N=2$ Supergravities and their Microscopic Origin}
\label{gaugedsugras}
\thispagestyle{plain}


As we discussed in chapter \ref{effectivesupergravity} there are many ways to compactify higher-dimensional supergravities depending on the particular internal manifold considered.

In particular in section \ref{survey} we introduced the concept of gauged supergravity as a supergravity theory with some gauged symmetries. The invariance under these local symmetries implies the appearance of covariant derivatives and a potential for the scalar fields. As discussed in section \ref{geometry}, the scalar potential allows the stabilization of the scalars and the vacua configurations are given by the critical values of the potential that in turn fixes the vacuum expectation values of the scalars. Finally we recall that the moduli stabilization is necessary in order to get a realistic compactification to lower-dimensional theories.

In this chapter we do a further step in the study of supersymmetric objects in string theory by considering two particular gauged supergravities in low dimensions:
\begin{itemize}

\item Matter-coupled $\ma N=2$ abelian gauged supergravity in $d=4$ \cite{Ferrara:1976fu}.

\item Matter-coupled $\ma N=2$ abelian gauged supergravity in $d=5$ \cite{Chamseddine:1980sp}.

\end{itemize}

These theories are particularly interesting for two reasons. The first is related to their scalar geometries (or moduli spaces). Since the number of supercharges preserved is relatively ``small" there is a huge freedom concerning the explicit realizations of the scalar geometries. In particular in the ``$\ma N=2$ theories" the moduli parametrize two kind of geometries encoding the informations coming to higher dimensions, these are the {\itshape Special} and {\itshape Quaternionic geometries} respectively defined by the scalar fields of vector- and hypermultiplets. 

The second reason is technical. In these theories some techniques to derive and solve $\mrm{BPS}$ equations can be formulated in a relatively easy way. An example of this is given by the Hamilton-Jacobi formulation for black holes and black strings that will be consider later in this thesis.

The dependence of these theories on the scalar geometries implies that the lagrangians are not fixed by the field content of the theory but they include non-linear sigma model terms for the scalars that are determined by the metrics characterizing the target manifolds. Moreover abelian gauge fields are present and their couplings to the scalars are related to the geometric quantities of the scalar manifolds. In particular, in the four-dimensional case the relation between the non-linear sigma models and the vectors is driven by the {\itshape electromagnetic duality} \cite{Gaillard:1981rj}.

In the gauged case this picture becomes more complicated since some isometries of the scalar manifolds are local and the global symmetry is broken.

The crucial consequence of these arguments is that the $\ma N=2$ theories produce many different physical configurations determined by the field content of matter multiplets, by the scalar geometry and by the eventual gauging. All these things will fix uniquely a {\itshape model} describing a particular lower-dimensional physics. As we said in section \ref{survey} the higher-dimensional interpretation of all these configurations is not automatic and depends on the existence of a consistent truncation of eleven-dimensional or type $\mrm{II}$ supergravities on the particular model under consideration.

In this chapter we will firstly introduce these two theories in generality and we will discuss about their main features, taking in particular account the geometric properties of their moduli spaces. Then we will study their higher-dimensional origin relating their main features, like field content, moduli spaces and the gauging to the internal manifolds defining the consistent truncations of $D=11$ and type $\mrm{II}$ supegravities to the $\ma N=2$ theories in $d=4, 5$.

\section{The $\ma N=2$ Gauged Supergravities in $d=4, 5$}

As we already said in the introduction of this chapter the peculiarity of $\ma N=2$ supergravities in $d=4,5$ consists in the intrinsic richness of their scalar geometries. The two theories are quite similar both in their field content and in their scalar geometries. 

In this section we will present the multiplets' organization of $\ma N=2$ supergravities in $d=4,5$ in presence of couplings to the matter. We will introduce and classify the geometries describing their moduli spaces with particular attention on their symmetry properties. Thus we will present the lagrangians and the equations of motion in the case of abelian gauging and we will discuss about possible gaugings in relation to the {\itshape embedding tensor} that is the most powerful tool to study gauged symmetries of the moduli spaces.

\subsection{Multiplets and Moduli Spaces}
\label{multiplets}

When we consider truncations of higher-dimensional supergravities to supergravities described by non-linear sigma models we should always take in account that the lower-dimensional result represents a particular realization of a more general lower-dimensional theory. This theory can be presented per se without any reference to its stringy origin. 
In this section we are going to present $\ma N=2$ theories in $d=4, 5$ in their generality without specify the number of multiplets and their moduli spaces.

Let's consider the field content and its organization in multiplets. Recalling the general classification of the supermultiplets in supergravity theories of section \ref{survey}, the coupling to the matter of the supergravity multiplet in $d=4$ is realized by $n_V$ vector multiplets and $n_H$ hypermultiplets, as well as in the $d=5$ case\footnote{We will consider $n_V-1$ vector multiplets in $d=5$ instead of $n_V$ to make more explicit the relations between the two theories. The curved indices $\mu, \nu, \cdots$ of the background and the flat indices $a, b, \cdots$ of the vielbein go from 0 to 3 and 0 to 4 respectively for the cases $d=4,5$.} by $n_V-1$ vector multiplets and $n_H$ hypermultiplets.
The field content is organized as follows:

\begin{itemize}

\item {\itshape Supergravity multiplets in $d=4, 5$}: $(e^a_{\mu}, \psi^{A}_{\mu},A_{\mu}^{0})$. They contains the graviton, the gravitinos\footnote{They are two Majorana spinors in $d=4$ with opposite chirality and 4 real independent components each. They are indicated as $\psi^{A}_{\mu}$ and $\psi_{\mu\, A}$. Regarding the $d=5$ case, Symplectic-Majorana spinors exist and the degrees of freedom correponding to the $s=3/2$ contribution to the supergravity multiplet can be organized in a $\mrm{SU(2)}_R$ doublet $\psi^{A}_{\mu}$. In total this doublet enjoys 8 independent real components defining an irreducible $d=5$ Dirac spinor.} in the doublet representation $A=1,2$ of the R-symmetry, given by $(\mrm{U(1)}\times\mrm{SU(2)})_R$ in $d=4$ and by $\mrm{SU(2)}_R$ in $d=5$, and a vector field called {\itshape graviphoton}.

\item {\itshape Vector multiplets in $d=4$}: $(A_{\mu}^{i}, z^{i}, \chi^{A i})$ with $i=1, \cdots, n_V$. Each one is composed by a vector field, a complex scalar and by two spinors with opposite chirality, the {\itshape gauginos}. 

\item {\itshape Vector multiplets in $d=5$}: $(A_{\mu}^{i}, \phi^{i}, \chi^{A\, i})$ with $i=1, \cdots, n_V-1$. Each one is composed by a vector field, a real scalar and by two gauginos. 

\item {\itshape  Hypermultiplets in $d=4, 5$}: $(q^u,\zeta^ {\scriptsize{\ma A}})$ with $u=1, \cdots, 4n_H$ and $\ma A=1,\cdots,2n_H$. Each one contains four real scalar fields, called {\itshape hyperscalars}, and two spinors called {\itshape hyperinos}.
\end{itemize} 
Considering for simplicity the ungauged case, the scalar geometry is parametrized by the scalars of the vector multiplets respectively $z^i$ and $\phi^i$ for $d=4, 5$ and by the hyperscalars $q^u$. These defines the scalar sector of the theory through a non-linear sigma model lagrangian given by
\begin{equation}
\mathscr L_{scalar}=\frac{1}{2}\,g_{MN}(\varphi)\partial_\mu \varphi^M \partial^\mu \varphi^N\,,
\label{nonlinearsigmamodel}
\end{equation}
where $\varphi$ is a collective coordinate such that $\varphi^M=(z^i, q^u)$ or $\varphi^M=(\phi^i, q^u)$ respectively for $d=4, 5$ and $g_{MN}$ is the metric describing a product manifold $\ma{M}_{s}$ given by \cite{Andrianopoli:1996cm}
\begin{equation}
\ma{M}_{s}=\ma{SM}\otimes\ma{HM}\,,
\label{moduliN=2}
\end{equation}
where $\ma{SM}$ and $\ma{HM}$ are the manifolds parametrized respectively by the scalars belonging to the vector multiplets and by the hyperscalars. Both on $\ma{SM}$ and \ma{HM} some peculiar geometric structures are defined.
The manifold $\ma{HM}$ parametrized by the hyperscalars $q^u$ is a {\itshape quaternionic manifold} and this particular type of geometry is the same in $d=4, 5$.

The situation change when considering vector multiplets. For $d=4$ the moduli space $\ma{SM}$, parametrized by the complex scalars $z^i$, is described by a {\itshape Special K{\"a}hler manifold}. In the $d=5$ case the manifold $\ma{SM}$ is defined by the real moduli $\phi^i$ and it is represented by a {\itshape Very Special K{\"a}hler manifold}.

An $\ma N=2$ model is a concrete realization of $\ma N=2$ supergravities identified by a given number of vector- and hypermultiplets with their explicit scalar manifolds $\ma{SM}$ and $\ma{HM}$.

Usually the properties of a model\footnote{We will see some particular examples later in this thesis.} are studied by considering its symmetries. Given an ungauged model, we can consider the group of the isometries $G_0$ of $\ma{M}_{s}$. This group acts on the scalar fields as a global symmetry group leaving invariant (\ref{nonlinearsigmamodel}), furthermore, in $\ma N=2$ supergravities, $G_0$ has a non-trivial action also on other fields (namely the vectors) since, as we will see, the couplings between the moduli and the vectors are $G_0$-invariants. In general the complete $\ma N=2$ action is preserved in absence of gaugings and $G_0$ corresponds to the global symmetry group of the model.
This implies that the supergravity fields will organize themselves in certain representations of $G_0$.

An important class is given by {\itshape homogeneous models}. If $H$ is the maximal compact subgroup of $G_0$, a model is called homogeneous if its moduli space can be written as the coset $\ma{M}_s\simeq G_0/H$ . In this case the dimension of the moduli space coincides with the total number of physical scalars. On the converse the models don't satisfying this property will be called {\itshape non-homogeneous}.

The continuous global symmetry group $G_0$ is also called $\mrm{U}$-duality group since it can be related to the string discrete $\mrm{U}$-duality\footnote{We recall that (ungauged) maximal supergravities can be obtained by toroidal reductions of $D=10, 11$ supergravities (see section \ref{geometry}). The global invariance group of their moduli space is directly related to the $\mrm{U}$-duality group arising from toroidal reduction. If one consider less-supersymmetric supergravity models obtained by truncating the supermultiplets of maximal theories, the global invariance group of the moduli spaces must have an embedding into the isometry group of the maximal theory. For example by reducing eleven-dimensional supergravity on a $T^7$ one obtain $\ma N=8$, $d=4$ supergravity described by a moduli space with $G_0=\mrm{E_{7(7)}}$ and $\mrm{E_{7(7)}}$ is the $\mrm{U}$-duality group arising from the $T^7$-truncation. Moreover if one considers the truncation of the $\ma N=8$ multiplets to $\ma N=2$, one obtains a class of moduli spaces $\ma{M}_{s}=\ma{SM}\times\ma{HM}$ with global symmetry groups that have a non-trivial embedding in $\mrm{E_{7(7)}}$.} introduced in section \ref{dualities}.

\subsection{Special K{\"a}hler Geometry}
\label{specialgeometry}

As we discussed in the previous section, the manifold $\ma{SM}$ parametrized by the scalars of the vector multiplets are different in the two cases of four- and five-dimensional supergravities. In $d=4$ the complex scalars $z^i$ define a special K{\"a}hler manifold, while in $d=5$ we have a very special K{\"a}hler manifold parametrized by the real moduli $\phi^i$ \cite{Freedman:2012zz,Andrianopoli:1996cm,Gunaydin:1983bi, deWit:1991nm, deWit:1992wf, deWit:1992cr, deWit:1993rr, deWit:1995jd}.

Let's start by $n_V$ complex scalars $z^i$ in four dimensions.
An $n_V$-dimensional special K\"ahler manifold is a K\"ahler-Hodge 
manifold, with K\"ahler metric $g_{i\bar\jmath}(z,\bar z)$, which is the base of a $\mrm{Sp}(2n_V+2,\mathbb{R})$-bundle with the covariantly {\itshape symplectic sections}\footnote{In this thesis the symplectic sections will be also called {\itshape symplectic frames}.}
\begin{equation}
 \ma{V}(z, \bar{z})=\left(\begin{array}{c}
                   L^\Lambda \\
                   M_\Lambda
                  \end{array}\right) \qquad \text{with} \qquad \Lambda=0,\cdots,n_V+1\,
                  \label{specialembedding}
\end{equation}
obeying the constraint
\begin{equation}
 \left\langle\ma{V}, \ma{\bar V}\right\rangle\equiv
 \bar{L}^\Lambda M_\Lambda-L^\Lambda \bar{M}_\Lambda=-i\,, \label{eq:sympcond}
\end{equation}
where $\ma{K}$ is the K\"ahler potential, i.e. $g_{i\bar\jmath}=\partial_i\partial_{\bar{\jmath}}\,\ma K$.
The physical scalars $z^i$ are expressed as follows
\begin{equation}
z^i=\frac{L^i}{L^0} \qquad \text{with} \qquad i=1,\cdots,n_V\,.
\label{physicalscalars}
\end{equation}
From this structure it follows the existence of a line-bundle associated to the so-called {\itshape K{\"a}hler transformations} acting on the K{\"a}hler potential through the holomorphic function $f$ in the following way
\begin{equation}
 \ma{K}\longrightarrow \ma{K}+f(z)+\bar{f}(\bar{z})\,,
\end{equation}
and thus preserving the metric. The derivatives $\partial_i \ma{K}$ and $\partial_{\bar{\imath}} \ma{K}$ define the $\mrm{U(1)}$ connection of the line bundle
\begin{equation}
\masf{A}_{\mu}=\frac{i}{2}\,\left(\partial_\mu\bar{z}^{\bar{\imath}}\partial_{\bar{\imath}}\ma K- \partial_\mu z^{i}\partial_i \ma K     \right)
\label{U1connection}
\end{equation}
and the correspondent complex covariant derivatives defined as follows
\begin{equation}
 D_{ i}\ma{V}=\partial_{i}\ma{V}
                  +\frac{1}{2}\left(\partial_{i}\ma{K}\right)\ma{V}\qquad \text{and}\qquad D_{\bar \imath}\ma{V}=\partial_{\bar \imath}\ma{V}
                  -\frac{1}{2}\left(\partial_{\bar \imath}\ma{K}\right)\ma{V}=0\,.
\end{equation}
The relation (\ref{physicalscalars}) defines a projective structure associated to the symplectic one, in fact one can introduce a new set of {\itshape holomorphic sections} as follows
\begin{equation}
 v \equiv e^{-\mathcal{K}/2}\ma{V}\equiv\left(\begin{array}{c}
						  X^\Lambda\\
						  F_\Lambda
						  \end{array}\right)\,.
\end{equation}
 If one expresses the constraint (\ref{eq:sympcond}) in terms of the sections $v$, one obtains
\begin{equation}
\label{eq:sympcond2}
 \left\langle v, \bar{v}\right\rangle\equiv\bar{X}^\Lambda F_\Lambda-X^\Lambda{\bar{F}}_\Lambda=
 -i e^{-\mathcal{K}}.
\end{equation}

An important property is the existence of an appropriate symplectic frame for the sections $\ma V$ where it is possible to introduce a homogeneous function of second degree $F(X)$, called {\itshape prepotential}, such
that 
\begin{equation}
F_\Lambda=\partial_\Lambda F\,.
\end{equation}
This function can be viewed as a section on the line-bundle defined by K{\"a}hler transformations and its existence allows to derive easily all the geometric quantities characterizing the manifold. 
For example, the couplings of the vector fields to the scalars are determined by the $(n_V+1)\times(n_V+1)$ {\itshape period matrix} \ma{N} that is defined by the relations
\begin{equation}
 M_\Lambda = \ma{N}_{\Lambda\Sigma}\, L^\Sigma\,, 
 \qquad D_{\bar\imath}\bar{M}_\Lambda=\ma{N}_{\Lambda\Sigma}\,D_{\bar \imath}\bar{L}^\Sigma\,.
 \label{perioddefinition}
\end{equation}
If the theory is defined in a frame in which a prepotential exists, \ma{N} can be obtained from
\begin{equation}
  \label{eq:period_matrix_prep}
  \ma{N}_{\Lambda\Sigma}=\bar{F}_{\Lambda\Sigma}
  + 2i\frac{(N_{\Lambda\Gamma}X^\Gamma)(N_{\Sigma\Delta}X^\Delta)}{X^\Omega N_{\Omega\Psi}X^\Psi}\,,
\end{equation}
where $F_{\Lambda\Sigma}=\partial_\Lambda\partial_\Sigma F$ and $N_{\Lambda\Sigma}=\mathrm{Im}(F_{\Lambda\Sigma})$.

The presence of the symplectic structure implies that the group $G_0$ of the global symmetry group of an $\ma N=2$ model in $d=4$ introduced in section \ref{multiplets} must have an embedding in $\mrm{Sp}(2n_V+2,\mathbb{R})$. As we will see in detail later, this embedding implies the possibility to recast in a {\itshape symplectic covariant form} the model. This covariant form basically consists in an organization of the fields in symplectic vectors of the type as, for example, $\ma V^M=(L^\Lambda, M_\Lambda)$. Because of the symplectic embedding, these vectors globally transform in the fundamental representation of $\mrm{Sp}(2n_V+2,\mathbb{R})$ labelled by an index\footnote{We will usually omit the fundamental index $M$ in the symplectic covariant quantities.} $M=1,\cdots,2n_V+2$. In turn the fundamental representation is splitted in a doublet representation labelled by the upper and lower indices $\Lambda, \Sigma, \cdots=0,\cdots , n_V$.

As an example of symplectic covariant object we can define the $2(n_V+1)\times 2(n_V+1)$ matrix
\begin{equation}
 \ma{M}=\left(\begin{array}{cc}
 \left(I + 
 R  I^{-1}  R\right)_{\Lambda\Sigma} & \,\,- \left( R I ^{-1}\right)_{\,\Lambda}^{\,\,\,\,\Sigma} \\
- \left(I ^{-1}  R\right)^{\Lambda}_{\,\,\,\,\Sigma}  &  \left(I ^{-1}\right)^{\Lambda\Sigma} \\
\end{array}\right)\,,
\label{Mmatrix}
\end{equation}
where we introduced the matrices
\begin{equation}
I_{\Lambda\Sigma}=\mathrm{Im}\,\ma{N}_{\Lambda\Sigma}\,, \qquad R_{\Lambda\Sigma}=\mathrm{Re}\,\ma{N}_{\Lambda\Sigma}\,,
\qquad I^{\Lambda\Sigma} I_{\Sigma\Gamma}=\delta^\Lambda{}_\Gamma\,.
\end{equation}
The symplectic matrix \eqref{Mmatrix} respects an important covariant relation linking the sections to their derivatives,
\begin{equation}
\frac12 (\mathcal M - i\Omega) = \Omega\bar{\mathcal V}\mathcal V\Omega + \Omega D_i\mathcal V
g^{i\bar\jmath}D_{\bar\jmath}\bar{\mathcal V}\Omega\,, \label{eq:sympid}
\end{equation}
where
\begin{equation}
\Omega = \left(\begin{array}{cc} 0 & -\mathbb{I}_{n_V+1} \\ \mathbb{I}_{n_V+1} & 0 \end{array}\right)\,
\label{sympmatrix}
\end{equation}
is the standard symplectic matrix realizing the scalar product (\ref{eq:sympcond}).

We can now consider some explicit realizations of Special K{\"a}haler manifolds with an interesting physical interpretation. In particular we consider the class of models defined by a cubic prepotential of the form
\begin{equation}
F(X)= C_{ijk}\frac{X^iX^jX^k}{X^0}\,,
\label{specialprepotential}
\end{equation}
where $C_{ijk}$ is a completely symmetric real tensor indentifying a specific realization of the special geometry. As we will see these models emerge naturally from truncations of type $\mrm{II}$ superstrings and M-theory \cite{Cecotti:1988qn}. 

A famous example in this class is the so-called {\itshape $STU$} model. Given $n_V=3$ vector multiplets defined by the three scalars $z^i=(z^1, z^2, z^3)$, we consider the prepotential \eqref{specialprepotential} with a single non-vanishing component $C_{123}=\frac16\,$. In this case the scalars parametrize the homogeneous\footnote{We point out that the prepotential \eqref{specialprepotential} includes also non-homogeneous geometries.} special K{\"a}hler manifold
\begin{equation}
\ma{SM}_{STU}=\left(\frac{\mrm{SU(1,1)}}{\mrm{U(1)}}\right)^3\,.
\label{stumanifold}
\end{equation}
Choosing $X^0=1$, the symplectic sections of the $STU$ model are defined by the K{\"a}hler potential $\ma K=-\log[-i(z^1-\bar{z}^1)(z^2-\bar{z}^2)(z^3-\bar{z}^3)]$ and have the form
\begin{equation}
\ma V=e^{\scriptsize{\ma K}/2}(1, z^1, z^2, z^3, - z^1 z^2 z^3, z^2z^3, z^1z^3, z^1z^2)^t \,.
\end{equation}

\subsection{Very Special K{\"a}hler Geometry}
\label{veryspecialmanifold}

Let's take in exam the five-dimensional case where the $n_V$ real scalars $\phi^i$ parametrize a very special K{\"a}hler manifold \cite{Freedman:2012zz,Gunaydin:1983bi, deWit:1991nm, deWit:1992wf, deWit:1992cr, deWit:1993rr, deWit:1995jd}. These geometries are more constrained respect to the four-dimensional special K{\"a}hler manifolds and this is related to the request that the $\mrm{KK}$ reduction on a $S^1$ of $\ma N=2$ supergravity in $d=5$ has to reproduce the four-dimensional $\ma N=2$ models \eqref{specialprepotential}. In particular it is possible to formulate a precise mapping called {\itshape $r$-map} encoding the dictionaries of the two theories and realizing the circle reduction \cite{deWit:1991nm}. The crucial point is that the $r$-map is defined only for the $d=4$ models described by a prepotential of the type \eqref{specialprepotential}.

Then the main property of very special K\"ahler manifolds is that they are identified by an hypersurface in the global space parametrized by the scalars $\phi^i$ with $i=1,\cdots,n_V-1$ corresponding to the cubic models in $d=4$. The equation describing the locus of this hypersurface is an algebraic equation for a cubic polynomial that defines uniquely a $d=5$ model.
If we introduce the new set of ``homogeneous" real coordinates\footnote{There are not homogeneous functions in the sense given by a projective structure. We call them homogeneous to underline the analogy with the coordinates $X^\Lambda$ of the four-dimensional case.} $h^I(\phi^i$) with $I=1,\cdots,n_V$, the equation of the hypersurface parametrized by the physical $d=5$ scalars is given by 
\begin{equation}
\ma V\equiv\,\frac16 \,C_{IJK} h^I h^J h^K= 1\,,
\label{veryspecial}
\end{equation}
where $C_{IJK}$ is a completely symmetric tensor whose components defines a particular realization of very special K\"ahler manifold. By applying the $r$-map, the tensor $C_{IJK}$ define the $d=4$ prepotential \eqref{specialprepotential} through the identification of the index $I$ with the four-dimensional index $i$. As we will see, it is possible to give a clear higher-dimensional interpretation to the quantity $\ma V$, in the context of Calabi-Yau compactifications, as the volume modulus of the internal manifold and to the components of the tensor $C_{IJK}$ as the {\itshape intersection numbers} on the space of the deformations of the Calabi-Yau \cite{Cadavid:1995bk}. 

A $(n_V-1)$-dimensional very special K{\"a}hler manifold parametrized by the moduli $\phi^i$ is a real manifold defined by a set of continuous function $h^I(\phi^i)$ and a completely antisymmetric tensor $C_{IJK}$ satisfying the relation \eqref{veryspecial}, and with a metric $\ma G_{ij}$ given by
\begin{equation}
\ma G_{ij} = \partial_i h^I\partial_j h^J\left.G_{IJ}\right|_{\mathcal V=1}\,,
\end{equation}
with 
\begin{equation}
G_{IJ} = -\frac12\frac{\partial}{\partial h^I}\frac{\partial}{\partial h^J}\left.\log\mathcal V\right|_{\mathcal
V=1}\,.
\label{GIJ}
\end{equation}
From the relation \eqref{GIJ} it follows the possibility to derive from the tensor $C_{IJK}$ all the geometric quantities describing the manifold. In particular one can recast \eqref{GIJ} in the following form
\begin{equation}
G_{IJ} = \frac92 h_I h_J - \frac12 C_{IJK}h^K\,.
\label{veryspcialmetric}
\end{equation}
Thus if one introduces the dual coordinates $h_I$ as follows
\begin{equation}
h_I = \frac23 G_{IJ}h^J \qquad \text{and} \qquad \partial_i h_I = -\frac23 G_{IJ}\,,
\label{dualH}
\end{equation}
it is possible to derive a set of equivalent relations featuring the very special geometry
\begin{equation}
\begin{split}
\mathcal G^{ij}\partial_i h^I\partial_j h^J = G^{IJ} & -\frac23 h^I h^J\,, \qquad\mathcal G^{ij}\partial_i h_I
\partial_j h_J = \frac49 G_{IJ} - \frac23 h_I h_J\,, \\
&\mathcal G^{ij}\partial_i h^I\partial_j h_J = -\frac23\delta^I_J + \frac23 h^I h_J\,.
\label{eq:veryspecial1}
\end{split}
\end{equation}
In the special case where the tensor $T_{ijk}$ that determines the Riemann tensor of the vector multiplet
scalar manifold $\cal SM$ (see \cite{Gunaydin:1983bi} for details) is covariantly constant\footnote{This condition
implies that $\cal SM$ is a locally symmetric space.}, one has also
\begin{equation}
C_{IJK} C_{J'\left(LM\right.} C_{\left.PQ\right)K'}\delta^{JJ'}\delta^{KK'} = \frac43\delta_{I\left(L\right.}
C_{\left.MPQ\right)}\,, \label{adjoint-id}
\end{equation}
which is the {\itshape adjoint identity} of the associated Jordan algebra \cite{Gunaydin:1983bi}. Using
\eqref{adjoint-id} and defining $C^{IJK}\equiv\delta^{II'}\delta^{JJ'}\delta^{KK'}C_{I'J'K'}$, one can show
that
\begin{equation}
G^{IJ} = -6 C^{IJK} h_K + 2h^I h^J\,.
\end{equation}
As an explicit realization of these geometries we mention the model identified by a symmetric tensor given by $C_{123}=1$ that is the image under $r$-map of the $STU$ model introduced in \eqref{stumanifold}. This scalar manifold is parametrized by two real scalars $\phi_1$ and $\phi_2$ such that
\begin{equation}
h^1=e^{-\frac{\phi_1}{\sqrt{6}}-\frac{\phi_2}{\sqrt{2}}}\,, \quad h^2=e^{-\frac{\phi_1}{\sqrt{6}}+\frac{\phi_2}{\sqrt{2}}}\, \quad h^3=e^{2\,\frac{\phi_2}{\sqrt{2}}}\,.
\label{5dstu}
\end{equation}
with a metric given by $\ma G_{ij}=\frac12\, \delta_{ij}$.

\subsection{Quaternionic Geometry}
\label{quaternionic}

In this section we summarize the properties of the moduli space $\ma{HM}$ parametrized by $4 n_H$ real hyperscalars $q^u$ with $u=1,\cdots , 4n_H$. As we said the geometry of the scalar manifold of this sector remains the same when considering the four- and the five-dimensional cases and it is described by quaternionic manifolds \cite{Freedman:2012zz,Andrianopoli:1996cm,Ferrara:1989ik, deWit:2001brd, Bergshoeff:2004nf} . 

The name of these geometries belongs to their holonomy group given by $\mrm{Usp}(2n_H)\times \mrm{SU(2)}$. Since $\mrm{Usp}(2n_H)\simeq \mrm{U}(n_H, \mathbb{H})$ where $\mathbb{H}$ is the field of quaternions, each quadruple of hyperscalars $q^u$ can be regarded as a quaternion.

A quaternionic K\"ahler manifold with metric $h_{uv}(q)$ is a $4n$-dimensional Riemannian manifold admitting a locally defined triplet 
$\vec{K}_u^{\phantom{u}v}$ of almost complex structures satisfying the quaternion relation
 \begin{equation}
   \label{eq:quaternionic_kahler_complexstruct_definition}
h^{st}K^x_{\phantom{x}us}K^y_{\phantom{y}tw}=-\delta^{xy}h_{uw}+\varepsilon^{xyz}K^z_{\phantom{z}uw}\,,
 \end{equation}
and whose Levi-Civita connection preserves $\vec{K}$ up to a rotation,
 \begin{equation}
  \label{eq:quaternionic_kahler_complexstruct_rotation}
  \nabla_w {\vec K}_u^{\phantom{u}v}+\,\vec{\omega}_w\times{\vec K}_u^{\phantom{u}v}=0\,,
 \end{equation}
where $\vec\omega= \vec\omega_u (q)\, dq^u$ is the connection of a $\mathrm{SU}(2)$ bundle for which the quaternionic manifold is the base and $x=1,2,3$ the index labelling its vector representation. An important property is that the $\mathrm{SU}(2)$ curvature $\Omega^x$ is proportional to the
complex structures,
 \begin{equation}
  \label{eq:quaternionkahl_su2curv}
  \Omega^x=\, d\omega^x+\frac12\varepsilon^{xyz}\omega^y\wedge\omega^z=-\,K^x\,.
 \end{equation}
 Moreover it turns out that quaternionic manifolds are Einstein manifolds, i.e. 
 \begin{equation}
 R_{uv}=\frac{1}{4n_H}R\,,
 \end{equation}
where $ R_{uv}$ is the curvature associate to the affine connections of the manifolds. 

Finally we mention an important explicit realization of quaternionic geometry given by the so-called {\itshape universal hypermultiplet} \cite{Cecotti:1988qn}. This is given by a single quadruple of real hyperscalars $q^u=(\phi, \xi^0, \tilde{\xi}_0, a)$ parametrizing the following manifold
\begin{equation}
\ma{HM}_{\scriptsize{UHM}}=\frac{\mrm{SU(2,1)}}{\mrm{SU(2)}\times\mrm{U(1)}}\,.
\label{uhm}
\end{equation}
The metric of this manifold is given by
\begin{equation}
\label{eq:hypersol_hyper_metric}
 h_{uv}dq^u dq^v=d\phi^2+\frac14 e^{4\phi}\left(da-\frac12\langle\xi|d\xi\rangle\right)^2
 +\frac14 e^{2\phi}[(d\xi^0)^2+(d\tilde{\xi}_0)^2]\,,
\end{equation}
where $\langle\xi|d\xi\rangle = \tilde{\xi}_0 d\xi^0-\xi^0 d\tilde{\xi}_0$. By using the defining relation \eqref{eq:quaternionic_kahler_complexstruct_definition} one can derive the almost complex structures $K^x$ and thus the corresponding $\mrm{SU(2)}$ connection $\omega^x$ by using \eqref{eq:quaternionkahl_su2curv}. In particular one obtains the components
\begin{equation}
 \omega^1=e^\phi d\tilde{\xi}_0\,,\quad \omega^2=e^\phi d\xi^0\,,\quad \omega^3=\frac{e^{2\phi}}{2}\left( da-\frac12 \langle\xi|d\xi\rangle\right) \,, \label{su2connectionuhm}
\end{equation}
defining the $\mrm{SU(2)}$ connection $\omega^x$ for the universal hypermultiplet.

\subsection{Gauging and Embedding Tensor}
\label{gaugingsembedding}

As we discussed in sections \ref{geometry} and \ref{survey} the realistic compactifications are characterized by the stabilization of the moduli by a suitable scalar potential whose critical points are associated to the vacua. From the point of view of supergravity theories the procedure including the scalar potential is called {\itshape gauging}.

Let's consider a particular model\footnote{Since the $\ma N=2$ theories in $d=4,5$ considered preserve the same amount of SUSY given by a quarter of the supercharges of the correspondent maximal theories, the arguments included in this section are analogue for both cases. For this reason we will index the quantities transforming under gauge transformations using the four-dimensional index $\Lambda$. The $d=5$ case will be obtained simply by substituting the index $\Lambda$ with the $d=5$ index $I$. Moreover when some differences will arise, we will discuss the cases separately.} in $\ma N=2$ ungauged supergravity in $d=4, 5$ characterized by a moduli space \eqref{moduliN=2} enjoying a group $G_0$ of isometries. 
The associated algebra $\mathfrak{g}_0$ will be spanned by the set of generators $\{ t^a\}$ with $a=1,\cdots,\text{dim}(\mathfrak{g_0})$ such that $[t_a, t_b]=f^{\,\,\,\,\,\,c}_{ab}\,t_c$ with structure constants $f^{\,\,\,\,\,\,c}_{ab}$.  

A gauging of an $\ma N=2$ model is a {\itshape deformation} of the model performed by promoting a subalgebra $\mathfrak{g}\subset\mathfrak{g}_0$ to a gauge algebra. 

In this context a crucial tool is the so-called {\itshape embedding tensor}\footnote{This has been developed in maximal supergravities \cite{Nicolai:2000sc, deWit:2002vt, deWit:2005hv} and secondly in less supersymmetric theories (for a review in half-maximal theories see \cite{Weidner:2006rp}). In this section we mainly follow \cite{deWit:2005ub}.}. If we introduce the generators $k_\Lambda$ of the gauged isometries or {\itshape Killing vectors}, the embedding tensor is a map $\Theta_\Lambda^a$ realizes the embedding of the local subalgebra $\mathfrak{g}$ in the global one $\mathfrak{g}_0$ in the following way,
\begin{equation}
 k_\Lambda=\Theta_\Lambda^a t_a\,.
\end{equation}
We point out that not all the vector fields are involved in the gauging and the effect of the embedding tensor is to project out these fields. The remaining vectors are the gauge fields for the local isometries of $G$ and organize themselves in a representation (not necessary irreducible) of the gauge group as $A_\mu^\Lambda$ described by the index\footnote{We excluded from the counting of multiplets those whose vector fields have been projected away.} $\Lambda=0,\cdots, n_V$. Moreover they continue to transform linearly under global transformations $(t_a)^{\,\,\,\, \Gamma}_\Sigma$.

As in any gauge theory, the ordinary derivatives must be substituted by the covariant ones. These are defined as follows
\begin{equation}
\hat{\partial}_\mu = \partial_\mu-  A_\mu^\Lambda \,\Theta_\Lambda^a\, t_a=\partial_\mu-  A_\mu^\Lambda\,k_\Lambda\,,
\end{equation}
implying that some of the gauge fields will become massive.
Furthermore, in order to preserve the gauge symmetry, a scalar potential arises. 

Finally we mention the constraints that have to be satisfied by the embedding tensor in order to produce a consistent gauging. The gauge invariance of the embedding tensor implies two quadratic constraints,
\begin{equation}
\begin{split}
&\Theta^{\Lambda\,[a}\,\Theta_\Lambda^{\,b]}=0\,,\\
&f^{\,\,\,\,\,\,c}_{ab}\,\Theta_\Lambda^a\Theta_\Sigma^b+(t_a)^{\,\,\,\, \Gamma}_\Sigma\,\Theta_\Lambda^a\Theta_\Gamma^c=0\,,
\label{quadraticconstraints}
\end{split}
\end{equation}
The second constraint of (\ref{quadraticconstraints}) is needed for the closure of the gauge algebra, i.e. $[k_\Lambda, k_\Sigma]=-X^{\quad \,\,\, \Gamma}_{\Lambda\Sigma}\, k_\Gamma$ with $X^{\quad  \Gamma}_{\Lambda\Sigma}= \Theta_\Lambda^a (t_a)^{\,\,\,\, \Gamma}_\Sigma$.
Furthermore a set of linear constraint have to be imposed in order to keep the action supersymmetric. 

In this thesis we will consider only {\itshape abelian gaugings} on the scalar manifold \eqref{moduliN=2}. This implies that the subalgebra of local isometries of $G_0$ is such that $[k_\Lambda, k_\Sigma]=0$.

Since the scalar manifold in $\ma N=2$ supergravity is the product manifold \eqref{moduliN=2}, it follows that the group of local abelian isometries $G$ of $\ma M_{s}$ must have an independent action respectively on the special geometry $\ma{SM}$ and on the quaternionic sector $\ma{HM}$. Hence, for $\ma N=2$ gauged supergravity in $d=4, 5$, the embedding tensor is splitted as $\Theta^a=(\Theta_{\scriptsize{\ma{SM}}}^a, \Theta_{\scriptsize{\ma{HM}}}^a)$.
The means that the Killing vectors on the scalar manifold will respect the same splitting and we can study the local symmetries associated to a given gauge group $G$ on $\ma M_{s}$ separately.

Let's consider firstly local symmetries on special geometries. On the special K{\"a}hler manifold they are generated by a set of holomorphic Killling vectors $k_\Lambda^i(z)$ associated to the gauge fields $A_\mu^\Lambda$ . These gauged symmetries must have an embedding in $\mrm{Sp}(2n_V+2,\mathbb{R})$ in order to preserve the symplectic structure of the manifold.
By applying the $r$-map to the geometries \eqref{specialprepotential}, one can pass to the five-dimensional very special K{\"a}hler manifolds whose local isometries are described by real Killing vectors $k_I^i(\phi)$ whose properties are inherited from the symmetry features of the correspondent $d=4$ cubic model, taking in account that the symplectic structure is completely spoiled under the $r$-map.

We won't rest in the analysis of properties of these local symmetries since in the abelian case their are trivial. The reason is that the scalars $z^i$ and $\phi^i$ belongs to the vector multiplets including also the gauge fields. This implies that they transform under gauge transformations in the adjoint representation of $G$ that in the abelian case is trivial. In fact if we express the $d=4, 5$ K\"ahler moduli spaces respectively in the coordinates $L^\Lambda$ and $h^I$, one can write the infinitesimal action of the local isometries in special and very special K{\"a}hler manifolds as
\begin{equation}
\delta L^\Lambda=L^{\Gamma}\, X^{\quad \Lambda}_{\Sigma \Gamma}\qquad \text{and}\qquad \delta h^I=L^{K}\, X^{\quad I}_{JK}\,,
\end{equation}
where $X^{\quad \Lambda}_{\Sigma \Gamma}$ represent in the first relation the structure constants of the gauge algebra in $d=4$ and in second those ones on $d=5$.

On the contrary local abelian isometries on quaternionic manifolds are not trivialized. A Killing vector on a quaternionic manifold\footnote{Quaternionic geometries maintain the same structure in $d=4, 5$, thus as in other part of this section we will use the four-dimensional index $\Lambda$. The five-dimensional analysis will be gained using the $d=5$ index $I$ instead of $\Lambda$.}  $k_\Lambda^u (q)$ must be {\itshape triholomorphic} in the sense that it must preserve the quaternionic structure, i.e. it must commute with the hyper-complex stuctures $K^x$ defined by the relations \eqref{eq:quaternionic_kahler_complexstruct_definition} and \eqref{eq:quaternionic_kahler_complexstruct_rotation}.

The requirement that the quaternionic K\"ahler structure must be preserved implies the existence, for each
Killing vector, of a triplet of Killing potentials, or {\itshape moment maps} $P_\Lambda^x$, such that
\begin{equation}
\label{eq:mommaps}
\masf{D}_u P_\Lambda^x = \partial_u P_\Lambda^x + \varepsilon^{xyz}\omega_{\phantom{y} u}^y
P_\Lambda^z = -2\Omega^x{}_{uv} k_\Lambda^v\,.
\end{equation}
The set of moment maps can be used to define a gauging on the quaternionic manifold at the same footing of the Killing vectors.
One of the most important relations satisfied by the moment maps is the so-called equivariance relation that, for abelian gaugings, has the form
\begin{equation}
\frac12\epsilon^{xyz} P^x_\Lambda P^y_\Sigma - \Omega^x_{uv} k^u_\Lambda k^v_\Sigma =  0\,.
\label{eq:equivariance}
\end{equation}

Finally we conclude this section by considering the unique realization of abelian gauging in absence of hypermultiplets, i.e. $n_H=0$. A property of the moments maps $P_\Lambda^x$ is to be defined up to an additive real constant. This shift corresponds to a further $\mrm{U}(1)$ symmetry belonging to the R-symmetry group. The gauging of this abelian\footnote{We mention that there exists also an $\mrm{SU(2)}$ Fayet-Iliopoulos gauging corresponding to the gauging of the entire $\mrm{SU(2)}_R$.} group is not trivial in the sense that it produce a scalar potential both in four- and five-dimensions and it is called {\itshape Fayet-Iliopoulos (FI) gauging}. 

More precisely, in $d=4$ the $\mrm{FI}$ gauging involves a $\mrm{U(1)}$ subgroup of the $\mrm{SU(2)}_R$ component of the R-symmetry that, we recall, is given by $(\mrm{U(1)}\times\mrm{SU(2)})_R$. The moment maps associated to the gauging are constant of the form
\begin{equation}
P^x_\Lambda=g_\Lambda\, e^x\,
\label{fi4d}
\end{equation}
where the $g_\Lambda$ are called {\itshape FI parameters} and define the particular realization of the $\mrm{FI}$ gauging while $e^x$ is an arbitrary constant $\mrm{SU(2)}$ vector.
 In $d=5$ there is no possibility of confusion since the R-symmetry group is $\mrm{SU(2)}_R$. The moment maps corresponding to an abelian $\mrm{FI}$ gauging are given by
 \begin{equation}
P^x_I=V_I\, e^x\,
\end{equation}
where the $V_I$ are the $\mrm{FI}$ parameters realizing the gauging in five-dimensions.

\subsection{Electromagnetic Duality and Symplectic Covariance}
\label{symplecticcovariance}

The gauging of global symmetries in four-dimensional $\ma N=2$ supergravity exhibits some peculiarities that are in relation to the $d=4$ electromagnetic duality for vector fields.  As we briefly observed in section \ref{specialgeometry}, from the presence of the symplectic structure on special K{\"a}hler manifolds, it follows that the global symmetry group $G_0$ must have an embedding as
\begin{equation}
G_0 \subset \mrm{Sp}(2n_V+2)\,.
\label{symplecticembedding}
\end{equation}
In other words it is always possible to find a certain representation (not necessarily irreducible) of the global symmetry group $G_0$ coinciding with the fundamental representation of $\mrm{Sp}(2n_V+2,\mathbb{R})$.

Let's introduce an infinitesimal\footnote{An infinitesimal transformation is defined as a matrix $\ma T$ such that $\ma S=\mathbb{I}+\ma T \in \mrm{Sp}(2n_V+2,\mathbb{R})$. If $\ma S=\left(\begin{array}{cc}
 U&  Z \\
W  &  V  \\
\end{array}\right)$, it follows that the relations $U\simeq \mathbb{I}-b^t$, $V\simeq \mathbb{I}+b$, $W\simeq -c$ and $Z\simeq-d$ define the matrix $\ma T$. }
 global symplectic transformation as a $2(n_V+1)\times 2(n_V+1)$ matrix of the type
\begin{equation}
 \ma{T}_{\,\,\,\,\,\,N}^{M}=\left(\begin{array}{cc}
b^\Lambda_{\,\,\,\,\Sigma} &  c^{\Lambda\Sigma} \\
d_{\Lambda\Sigma}  &  -(b^t)_\Lambda^{\,\,\,\,\Sigma}  \\
\end{array}\right)_{\,\,\,\,\,\,N}^{M} \,\in \mathfrak{sp}(2n_V+2, \mathbb{R})\,,
\label{symptransformation}
\end{equation}
with $c^{\Lambda\Sigma}=c^{\Sigma\Lambda}$ and $d_{\Lambda\Sigma}=d_{\Sigma\Lambda}$. The indices $M, N, \cdots=0,\cdots,2(n_V+1)$ are associated to the fundamental representation of $\mrm{Sp}(2n_V+2,\mathbb{R})$ while the indices $\Lambda, \Sigma,\cdots=0,\cdots n_V+1$ define the splitting in symplectic doublets as in \eqref{specialembedding}.

In the $\ma N=2$ vector multiplets are included also vector fields $A_\mu^i$ and, considering also the graviphoton, they can be organized as $A_\mu^\Lambda=(A_\mu^0, A_\mu^i)$ with field-strength $F^\Lambda_{\mu\nu}$. The Maxwell fields in $d=4$ enjoy\footnote{In the five-dimensional case, the electromagnetic duality takes the form of a duality between vectors and 2-forms \cite{Ceresole:2000jd}.} the electromagnetic duality \cite{Gaillard:1981rj} that is exactly realized by the symplectic group. It follows that the symplectic structure of special K{\"a}hler manifold in $\ma N=2$ supergravity realizes also the electromagnetic\footnote{For an extensive analysis on the electromagnetic duality in $\ma N=2$ supergravity in $d=4$ we refer to \cite{Freedman:2012zz}.} duality and viceversa.

To make this argument more explicit, we consider the ungauged lagrangian sector describing the whole contribution of vectors to $\ma N=2$ ungauged supergravity in $d=4$,
\begin{equation}
\sqrt{-g}^{-1}\mathscr{L}_{vector}=\frac14 \,\mrm{Im}\ma N_{\Lambda\Sigma}F^{\Lambda\mu\nu}F^{\Sigma}{}_{\mu\nu} 
  + \frac14 \, \mrm{Re}\ma N_{\Lambda\Sigma}F^{\Lambda\mu\nu}\star\! F^{\Sigma}{}_{\mu\nu}\,,
  \label{dualityn2lagrangian}
  \end{equation}
where the couplings between vector fields and scalars $z^i$ are manifestly realized through the period matrix $\ma N_{\Lambda \Sigma}$ introduced in \eqref{perioddefinition}. The Maxwell equations and the Bianchi identities following from \eqref{dualityn2lagrangian} are given by
\begin{equation}
\partial_{\mu}\left(\sqrt{-g} \star_4 G_{\Lambda}^{\mu\nu}  \right)=0 \qquad \text{and} \qquad \partial_{\mu}\left(\sqrt{-g} \star_4 F^{\Lambda\,\mu\nu}  \right)=0 \,.
\label{ungaugedmaxwell}
\end{equation}
These equations are covariant under global symplectic transformations once the symplectic vector
\begin{equation}
\ma F_{\mu\nu}=\left(\begin{array}{c}
                   F^\Lambda_{\mu\nu} \\
                 G_{\Lambda\,\mu\nu}
                  \end{array}\right) \qquad \text{with} \qquad G_{\Lambda} = -\frac2{\sqrt{-g}}\star_4\ \frac{\delta\mathscr{L}}{\delta F^{\Lambda}}\,,
                  \label{eq:Gdual}
\end{equation}
has been defined. The equations \eqref{ungaugedmaxwell} are over-determined since the physical degrees of freedoms of $A^\Lambda_\mu$ come from Maxwell equations while Bianchi identities furnish a set of consistency conditions realizing the electromagnetic duality. This abundance of degrees of freedoms is manifest also at the level of special geometries. In fact, in order to make explicit the symplectic structure in section \ref{specialgeometry}, we introduced the homogeneous coordinates $L^\Lambda$ defining the doublet representation \eqref{specialembedding} of $\ma V$ that have $2(n_V+2)$ components. The scalar physical degrees of freedom are given only by the $n_V$ scalars $z^i$ defined as \eqref{physicalscalars}. 

From this argument it follows that it is always possible to find a representation for a global symplectic transformation $\ma S$, given in \eqref{symptransformation} in its infinitesimal form, characterized by a smaller number of independent parameters. There are two possibilities: $\ma T$ takes a completely lower-triangular form and thus $c^{\Lambda\Sigma}=0$ or, viceversa, it has an upper-triangular form with $d_{\Lambda\Sigma}=0$. In the first case $\ma S$ is called {\itshape electric transformation} and its action on a symplectic vector does not mix the upper part with the lower one. The symplectic sections $\ma V$ transforming under electric transformations constitute the {\itshape electric frames}. In the opposite case of upper-triangular transformations $\ma S$, the action on a symplectic vector does not mix the lower part. These are called {\itshape magnetic transformations} and the correspondent symplectic sections form the {\itshape magnetic frames}. 

Let's consider the gaugings. First of all we point out that the vector fields $A_\mu^\Lambda$ are objects transforming under lower-triangular symplectic transformations (they carry an upper $\Lambda$ index) and this implies that the gauging described by the $d=4$ gauge fields $A_\mu^\Lambda$ introduced in section \ref{gaugingsembedding} are necessarily associated to lower-triangular local transformations. A gauging of this type is called {\itshape purely electric gauging} and produces the breaking of the symplectic invariance, since the covariant derivatives and the potential will be expressed in a preferred symplectic frame. This global symmetry breaking appears dynamically through new couplings between scalars and giving mass to some gauge fields. This means that the Maxwell equations are non-longer homogeneous and thus the duality symmetry with Bianchi identities is broken. Moreover, since the global symplectic structure is broken, all the gauge degrees of freedom involved are physical and they are determined by the couplings between gauge fields and scalars. 

The breaking of the global symplectic invariance due to a purely electric gaugings is related to the particular choice of local lower-triangular transformations, thus to an electric formulation of the symplectic frames.
It follows that, by considering different symplectic frames, one can obtain gaugings of more general symplectic transformations provided that one includes some auxiliary degrees of freedom. Let's introduce a set of {\itshape magnetic gauge fields} $A_{\mu\,\Lambda}$ and define a symplectic vector 
\begin{equation}
\ma A_\mu=\left(\begin{array}{c}
                   A^\Lambda_{\mu} \\
                 A_{\Lambda\,\mu}
                  \end{array}\right)\,,
                  \label{symvector}
\end{equation}
where the $ A_{\Lambda\,\mu}$ fields are interpreted as the gauge fields associated to the dual field-strengths $G_{\Lambda\,\mu\nu}$. 
 The symplectic vector \eqref{symvector} is physically over-determined but the symplectic embedding ensures the possibility to kill the lower component by a suitable lower-triangular global symplectic transformation. It can be shown that the $\ma N=2$ theory formulated in terms of \eqref{symvector} is manifestly covariant under global symplectic transformations and its action is dependent on auxiliary fields \cite{deWit:2005ub}. As we will see, its formulation has various complications respect to the purely electric case, but it is covariant under global symplectic transformations. Anyway we point out that this is a reformulation without any new physical degree of freedom and this can be verified directly since the consistency of the gauging implies that it is always possible to ``rotate" the theory in a purely electric frame.

The nice consequence following from this covariant formulation is the possibility to consider more general gaugings respect to the purely electric one. In particular if we choose a specular frame such that all the gauged isometries are upper-triangular thus all the gauge degrees of freedom will be included in $A_{\Lambda\,\mu}$ and we will have a {\itshape purely magnetic gauging}. The most general framework is that one with both electric and magnetic gauged isometries. This particular type of gauging is called {\itshape dyonic gauging}.

This analysis of generic gaugings in $\ma N=2$ four-dimensional supergravity in relation to the electromagnetic duality is automatically encoded by the properties of the embedding tensor \cite{deWit:2005ub}. In particular the quadratic constraints \eqref{quadraticconstraints} guarantees the possibility to recast the theory in a completely electric frame by a suitable global symplectic transformation.

If we consider a dyonic gauging, the symplectic structure is respected by the embedding tensor. In fact it will transform in general in the fundamental representation of the symplectic group labelled by $M=0,\cdots,2(n_V+2)$ and, then, it will be splitted in a doublet as $\Theta_M^a=(\Theta^{\Lambda \, a}, \Theta_\Lambda^a)^t$. It follows that a purely magnetic gauging is realized by the particular case $\Theta_M^a=(\Theta^{\Lambda \, a}, 0)^t$ and the purely electric one, introduced in section \ref{gaugingsembedding}, by a zero in the upper component.
Hence, if we consider abelian gaugings, the local isometries on the special K{\"a}hler manifold remain trivial while, for quaternionic gauged symmetries, we can introduce magnetic Killing vectors $k^{\Lambda u}$ and magnetic moment maps $P^{x\Lambda}$ that are both associated to magnetic gauged isometries. These quantities obey the same relations of quaternionic isometries presented in section \ref{gaugingsembedding} and, together with electric Killing vectors and moment maps, are organized in simplectic vectors of the form
\begin{equation}
\mathcal K^u = \left(\begin{array}{c}
k^{\Lambda u} \\ k_\Lambda^u \end{array}\right)\,, \qquad \text{and} \qquad
\mathcal P^x = \left(\begin{array}{c}
P^{x\Lambda} \\ P^x_\Lambda \end{array}\right)\,. \label{eq:magneticgaugings}
\end{equation}
As was shown in \cite{Erbin:2014hsa}, the quadratic constraint for the embedding tensor introduced in \eqref{quadraticconstraints} can be rewritten as
\begin{equation}
\langle\ma K^u,\ma P^x\rangle = 0\,.
\label{eq:cons}
\end{equation}
This relation implies the possibility to rotate any gauging to a frame with purely electric components.

Also the $\mrm{FI}$ abelian gauging introduced at the end of section \ref{gaugingsembedding} can be magnetic or dyonic. In particular if one consider a symplectic covariant moment map given by
\begin{equation}
\ma P^x= \ma G \, e^x \qquad \text{with} \qquad \ma G=\left(\begin{array}{c}
g^{\Lambda} \\ g_\Lambda \end{array}\right)\,,
\label{4dFImagnetic}
\end{equation}
the parameters $g^\Lambda$ are associated to the $\mrm{FI}$ magnetic gauging.

 \subsection{The $d=4$ Lagrangians and the Equations of Motion}
 \label{lagrangians4d}
 
In this section we are going to present the bosonic lagrangians and the equations of motion for $\ma N=2$ abelian gauged supergravites in $d=4$ with general couplings to vector- and hypermultiplets. Moreover we will refer to the discussion of section \ref{symplecticcovariance} and we will present a covariant formulation of the theory including gauged magnetic symmetries.

Let's start by $\ma N=2$ gauged supergravity in $d=4$ coupled to $n_V$ vector multiplets and $n_H$ hypermultiplets characterized by an abelian purely electric gauging on the quaternionic moduli space.
In this case the bosonic Lagrangian \cite{Andrianopoli:1996cm} reads\footnote{We recall the notation introduced in section \ref{specialgeometry} given by the two matrices $R_{\Lambda\Sigma}=\mathrm{Re}\,\ma N_{\Lambda\Sigma}$ and
$I_{\Lambda\Sigma}=\mathrm{Im}\,\ma N_{\Lambda\Sigma}$.}
\begin{equation}
\begin{split}
\label{eq:mainaction}
\sqrt{-g}^{-1}\!\mathscr{L} &= \frac R2 - g_{i\bar\jmath}\,\partial_{\mu}z^i\partial^{\mu}
\bar z^{\bar\jmath} - h_{uv}\hat{\partial}_{\mu} q^u\hat{\partial}^{\mu} q^v\\[2mm]
   &+\frac14 I_{\Lambda\Sigma}F^{\Lambda\mu\nu}F^{\Sigma}{}_{\mu\nu} 
  + \frac14 R_{\Lambda\Sigma}F^{\Lambda\mu\nu}\star_4\! F^{\Sigma}{}_{\mu\nu} - V_g(z,\bar z,q)\,,
\end{split}
\end{equation}
where the scalar potential has the form
\begin{equation}
\label{eq:scal_pot}
V_g = 4h_{uv} k^u_{\Lambda} k^v_\Sigma L^\Lambda\bar{L}^\Sigma + (g^{i\bar\jmath}D_i L^{\Lambda}
D_{\bar\jmath}\bar{L}^{\Sigma} - 3 L^\Lambda\bar{L}^{\Sigma})P^x_\Lambda P^x_\Sigma\,, 
\end{equation}
and the covariant derivatives acting on the hyperscalars are
\begin{equation}
\hat{\partial}_\mu q^u = \partial_\mu q^u + A^\Lambda_\mu k_\Lambda^u\,.
\label{eq:hypercov4d}
\end{equation}
 
Furthermore\footnote{The $\mrm{SUSY}$ transformations for fermions can be found for example in \cite{Andrianopoli:1996cm}. We won't write them since they we will not explicitly used in this thesis.} we note that the scalar potential \eqref{eq:scal_pot} can be written in terms of a $\mrm{SU(2)}$ triplet $\ma W^x$ of complex functions given by
\begin{equation}
 \mathcal W^x =  L^\Lambda P^x_\Lambda \,.
\label{electricWx}
\end{equation}
In particular, using the quaternionic relations (\ref{eq:quaternionic_kahler_complexstruct_definition}),
(\ref{eq:quaternionkahl_su2curv}), (\ref{eq:mommaps}), the scalar potential (\ref{eq:scal_pot}) can be
rewritten in the form \cite{Klemm:2016wng}
\begin{equation}
\label{eq:superV}
V_g = \tilde{\mathbb G}^{AB}\mathbb D_A\mathcal W^x\mathbb D_B\bar{\mathcal W}^x -
3|\mathcal W^x|^2\,,
\end{equation}
where we introduced
\begin{equation}
\tilde{\mathbb G}^{AB} = \left(\begin{array}{cc}
g^{i\bar\jmath} & 0 \\
0 & \frac13 h^{uv}\\
\end{array}\right)\,, \qquad \mathbb D_A = \left(\begin{array}{c} D_i \\ \masf{D}_u\end{array}\right)\,.
\end{equation}

By taking the variation with respect to the bosonic fields $g_{\mu\nu}$, $z^i$, $q^u$ and $A_\mu^\Lambda$, one obtains respectively the Einstein\footnote{We splitted the traceless and trace part of Einstein equations. The covariant derivative $\nabla_\mu$ is associated to the Levi-Civita connections of the four-dimensional background.} equations, the equations for the scalars and the Maxwell equations,
\begin{equation}
\begin{split} 
\label{electriceom4d}
& R_{\mu\nu}- 2 g_{i \bar{\jmath}} \partial_{\mu}z^i \partial_{\nu}\bar{z}^{\bar{\jmath}} -2 h_{uv} \hat\partial_{\mu} q^u \hat\partial_{\nu} q^v
- I_{\Lambda\Sigma}( F^{\Lambda}{}_{\mu\rho} F^{\Sigma}{}_{\nu}^{\,\,\rho} - 
\frac14 g_{\mu\nu} F^{\Lambda}{}_{\sigma\rho} F^{\Sigma\sigma\rho})- g_{\mu\nu} V_g =0\,,\\[2mm]
& R  - 2 h_{uv} \hat\partial_{\mu} q^u \hat\partial^{\mu} q^v - 2 g_{i \bar{\jmath}} \partial_{\mu}z^i \partial^{\mu}z^{\bar{\jmath}} - 4 V_g =0\,, \\[2mm]
&\nabla_{\mu}(g_{i \bar{\jmath}}\,\partial^{\mu}\bar{z}^{\bar{\jmath}}) -\partial_i g_{j \bar{k}} \partial_{\mu} z^j \,\partial^{\mu}\bar{z}^{\bar{k}}+  \frac14 \partial_i I_{\Lambda\Sigma} F^{\Lambda\mu\nu} F^{\Sigma}{}_{\mu\nu} 
+   \frac14 \partial_i I_{\Lambda\Sigma}  F^{\Lambda\mu\nu}\star_4\! F^{\Sigma}{}_{\mu\nu} - \partial_i V_g=0\,,\\[2mm]
& 2 \nabla_{\mu}( h_{uv} \hat{\partial}^{\mu}q^v)- 2 h_{vw}\partial_u k_\Lambda^v \, A^\Lambda_{\mu}\,  \hat{\partial}^{\mu}q^w
- \partial_u h_{vw}\hat{\partial}_{\mu}q^v \hat{\partial}^{\mu}q^w - \partial_u V_g=0\,,\\[2mm]
&\partial_\mu(\sqrt{-g} I_{\Lambda\Sigma} F^{\Sigma\mu\nu} + \frac12
\epsilon^{\mu\nu\rho\sigma} R_{\Lambda\Sigma} F^\Sigma{}_{\rho\sigma}) = 2\sqrt{-g}\, h_{uv}
k^u_\Lambda\,\hat{\partial}^\nu q^v\,.
\end{split} 
\end{equation}
Let's include in the formulation also magnetic gaugings \eqref{eq:magneticgaugings}.
In this more general picture the general action (\ref{eq:mainaction}) is modified in a non-trivial way
by some topological terms \cite{deWit:2005ub} in order to reconstruct the symplectic covariance. The consistency of the theory requires the introduction of the auxiliary 2-forms $B_a=\frac12 B_{a\mu\nu}\D x^\mu\wedge\D x^\nu$ that do not change
the number of degrees of freedom. The action has the form \cite{deWit:2005ub,Samtleben:2008pe}
\begin{equation}
\begin{split}
\sqrt{-g}^{-1}\!\mathscr{L} &= \frac R2 - g_{i\bar\jmath}\,\partial_\mu z^i\partial^\mu
\bar z^{\bar\jmath} - h_{uv}\hat{\partial}_\mu q^u\hat{\partial}^\mu q^v + \frac14 I_{\Lambda\Sigma}
H^{\Lambda\mu\nu} H^{\Sigma}{}_{\mu\nu} + \\[2mm]
& \frac14 R_{\Lambda\Sigma} H^{\Lambda\mu\nu}\star_4\! H^{\Sigma}{}_{\mu\nu} -
\frac{\epsilon^{\mu\nu\rho\sigma}}{4\sqrt{-g}}\Theta^{a\Lambda} B_{a\mu\nu}\partial_\rho
A_{\Lambda\sigma} + \\[2mm]
& \frac1{32\sqrt{-g}}\Theta^{\Lambda a}\Theta_\Lambda^b\epsilon^{\mu\nu\rho\sigma} B_{a\mu\nu}
B_{b\rho\sigma} - V_g\,, \label{actionmag}
\end{split}
\end{equation}
where the modified field strength $H^\Lambda{}_{\mu\nu}=F^\Lambda{}_{\mu\nu}+\frac12
\Theta^{\Lambda a} B_{a\mu\nu}$ was introduced. The co\-var\-iant derivatives of the hyperscalars and the
scalar potential read respectively \cite{Samtleben:2008pe,Michelson:1996pn,deWit:2005ub}
\begin{equation}
\hat{\partial}_\mu q^u = \partial_\mu q^u - A_\mu^\Lambda \Theta_\Lambda^a k_a^u -
A_{\Lambda\mu}\Theta^{\Lambda a} k_a^u = \partial_\mu q^u - \langle\ma A_\mu,\ma K^u\rangle\,,
\label{eq:covmag}
\end{equation}
\begin{equation}
V_g = 4 h_{uv} \langle\mathcal K^u,\mathcal V\rangle\langle\mathcal K^v,\bar{\mathcal V}\rangle
+ g^{i\bar{\jmath}}\langle\mathcal P^x, D_i\mathcal V\rangle\langle\mathcal P^x,\bar {D}_{\bar{\jmath}}
\bar{\mathcal V}\rangle - 3\langle\mathcal P^x,\mathcal V\rangle\langle\mathcal P^x,
\bar{\mathcal V}\rangle\,. \label{eq:Vg-sympl-cov}
\end{equation}
Note that also in the covariant case it is possible to introduce a triplet of functions $\ma W^x$ such that the potential \eqref{eq:Vg-sympl-cov} can be rearranged in the form \eqref{eq:superV}. This can be obtained simply by applying a global symplectic rotation on \eqref{eq:superV} and, in this case, the $\ma W^x$ introduced in \eqref{electricWx} take the form
\begin{equation}
\ma W^x=\langle \ma P^x, \ma V \rangle \,,
\label{tripletsuperpotential}
\end{equation}
where now $\ma P^x$ are given by \eqref{eq:magneticgaugings}.

The equations of motion for $A_{\Lambda\mu}$, $A_\mu^\Lambda$ and $B_{a\mu\nu}$ following from
(\ref{actionmag}) are
\begin{equation}
\begin{split}
&\frac14 \epsilon^{\mu\nu\rho\sigma}\partial_\mu B_{a\nu\rho}\Theta^{\Lambda a} = -2\sqrt{-g}
h_{uv}\Theta^{\Lambda a} k_a^u\hat{\partial}^\sigma q^v\,, \\[2mm]
&G_{\Lambda\mu\nu}\Theta^{\Lambda a} = \Theta^{\Lambda a}(F_{\Lambda\mu\nu} - \frac12
\Theta_\Lambda^b B_{b\mu\nu})\,, \\[2mm]
&\partial_\mu\Bigl(\sqrt{-g} I_{\Lambda\Sigma} H^{\Sigma\mu\nu} + \frac12
\epsilon^{\mu\nu\rho\sigma} R_{\Lambda\Sigma} H^\Sigma{}_{\rho\sigma}\Bigl) = 2\sqrt{-g} h_{uv}
\Theta^a _\Lambda k^u_a\hat{\partial}^\nu q^v\,,
\label{eq:eommag1}
\end{split}
 \end{equation}
where $G_{\Lambda\mu\nu}$ is defined by \eqref{eq:Gdual}. The equations \eqref{eq:eommag1} can be rewritten in a completely symplectically covariant form in terms of the fundamental representation of $\mrm{Sp}(2n_V+2, \mathbb{R})$ labelled by the indices $M,N=0,\cdots,2n_V+2$, in the following way
\begin{equation}
\frac12\epsilon^{\mu\nu\rho\sigma}\partial_\nu G_{\rho\sigma}^M = \Omega^{MN} J^\mu_N\,, \qquad
\Theta^{a M} (H-G)_M = 0\,, 
\end{equation}
where
\begin{equation}
H^M_{\mu\nu} = F^M_{\mu\nu} + \frac12\Omega^{MN}\Theta^a_N B_{a\mu\nu}\,, \qquad
G_{\mu\nu}^M = (H^\Lambda{}_{\mu\nu},G_{\Lambda\mu\nu})\,,
\end{equation}
and $J^\mu _M$ are the currents coming from the coupling to the matter. As we can see also from the lagrangian \eqref{actionmag}, the modified field strength $H^\Lambda$ substitutes the normal one $F^\Lambda$. This can be observed also by deriving the other equations of motion,
\begin{equation}
\begin{split} 
& R_{\mu\nu}- 2 g_{i \bar{\jmath}} \partial_{\mu}z^i \partial_{\nu}\bar{z}^{\bar{\jmath}} -2 h_{uv} \hat\partial_{\mu} q^u \hat\partial_{\nu} q^v
- I_{\Lambda\Sigma}( H^{\Lambda}{}_{\mu\rho} H^{\Sigma}{}_{\nu}^{\,\,\rho} - 
\frac14 g_{\mu\nu} H^{\Lambda}{}_{\sigma\rho} H^{\Sigma\sigma\rho})- g_{\mu\nu} V_g =0\,,\\[2mm]
& R  - 2 h_{uv} \hat\partial_{\mu} q^u \hat\partial^{\mu} q^v - 2 g_{i \bar{\jmath}} \partial_{\mu}z^i \partial^{\mu}z^{\bar{\jmath}} - 4 V_g =0\,, \\[2mm]
&\nabla_{\mu}(g_{i \bar{\jmath}}\,\partial^{\mu}\bar{z}^{\bar{\jmath}}) -\partial_i g_{j \bar{k}} \partial_{\mu} z^j \,\partial^{\mu}\bar{z}^{\bar{k}}+  \frac14 \partial_i I_{\Lambda\Sigma} H^{\Lambda\mu\nu} H^{\Sigma}{}_{\mu\nu} 
+   \frac14 \partial_i I_{\Lambda\Sigma}  H^{\Lambda\mu\nu}\star_4\! H^{\Sigma}{}_{\mu\nu} - \partial_i V_g=0\,,\\[2mm]
& 2 \nabla_{\mu}( h_{uv} \hat{\partial}^{\mu}q^v)- 2 h_{vw}\langle \partial_u \ma K^v, \mathcal A _{\mu} \rangle \, \hat{\partial}^{\mu}q^w- \partial_u h_{vw}\hat{\partial}_{\mu}q^v \hat{\partial}^{\mu}q^w - \partial_u V_g=0\,,
\end{split} 
\end{equation}
where the symplectic vectors $\ma A_\mu$ and $\ma K^u$ have been respectively introduced in \eqref{symvector} and in \eqref{eq:magneticgaugings}.

 \subsection{The $d=5$ Lagrangian and the $r$-map}
 \label{lagrangians5d}
 
In this section we will present the bosonic lagrangian and the equations of motion for $\ma N=2$ abelian gauged supergravity in $d=5$ with couplings to $n_V-1$ vector multiplets and $n_H$ hypermultiplets. Moreover we will construct explicitly the $r$-map to the $d=4$ theory \eqref{eq:mainaction}.
 
The bosonic Lagrangian \cite{Gunaydin:1983bi,Cadavid:1995bk,Lukas:1998yy} is given by\footnote{The lagrangian \eqref{eq:genlaghyper} has been obtained from the \cite{Ceresole:2001wi} by rescaling $a_{IJ}\to\frac23 G_{IJ}$, $C_{IJK}\to\frac16 C_{IJK}$,
$k_I\to 2k_I$, $A^I\to\sqrt{\frac32}A^I$, $\mrm{g}\to\sqrt{\frac32}\mrm{g}$. In this case $\sqrt{-g}^{-1}$ and $\nabla_\mu$ are referred to the five-dimensional background.}
\begin{eqnarray}
\sqrt{-g}^{-1}\mathscr L &=&\frac12 R - \frac12\mathcal G_{ij}\partial_{\mu}\phi^i\partial^{\mu}\phi^j - 
h _{uv}\hat{\partial}_{\mu}q^u\hat{\partial}^{\mu} q^v - \frac14 G_{IJ} F^I_{\mu\nu} F^{J\mu\nu}
\nonumber \\[2mm]
&&+ \frac {e^{-1}}{48} C_{IJK}\epsilon^{\mu\nu\rho\sigma\lambda} F^I_{\mu\nu} F^J_{\rho\sigma}
A^K_{\lambda} - \mrm{g} ^2 V_{5d}\,, \label{eq:genlaghyper}
\end{eqnarray}
with covariant derivatives\footnote{The gauge coupling $\mrm{g}$ could be absorbed into the Killing vectors $k^u_I$ as in the four-dimensional case. We kept it explicit since its presence will be useful when we will consider the $r$-map to $d=4$.}
\begin{equation}
\hat{\partial}_{\mu}q^u = \partial_{\mu}q^u + 3 \mrm{g} A_{\mu}^I k_I^u\,
\label{eq:covhyper}
\end{equation}
and the scalar potential
\begin{equation}
V_{5d} = P_I^x P_J^x\left(\frac92\mathcal G^{ij}\partial_i h^I\partial_j h^J - 6 h^I h^J\right) + 9 h_{uv} k_I^u
k_J^v h^I h^J\,. \label{eq:pothyper}
\end{equation}
From the variations with respect to  $g_{\mu\nu}$, $\phi^i$, $q^u$ and $A_\mu^I$, one finds the following set of equations of motion,
\begin{equation}
\begin{split} 
\label{eq:eom5d}
& R_{\mu\nu}- \mathcal G_{ij} \partial_{\mu}\phi^i \partial_{\nu}\phi^j - 2 h_{uv}\hat\partial_{\mu} q^u \hat\partial_{\nu}q^v 
- G_{IJ} (F^I_{\mu\sigma} F^J_{\nu}{}^{\sigma} - \frac16 g_{\mu\nu} F^I_{\sigma\rho} F^{J\sigma\rho}) + \frac23 \,\mrm{g} ^2\,g_{\mu\nu}V_{5d}=0\,,\\[2mm]
& R  - \mathcal G_{ij} \partial_{\mu}\phi^i \partial^{\mu}\phi^j - 2 h_{uv}\hat\partial_{\mu} q^u \hat\partial^{\mu}q^v
-\frac16 G_{IJ} F^I_{\mu\nu} F^{J\mu\nu} + \frac{10}{3}\, \mrm{g} ^2 V_{5d} =0\,,\\\nonumber
\end{split}
\end{equation}
\begin{equation}
 \begin{split}
&\nabla_{\mu}(\mathcal G_{ij}\partial^{\mu}\phi^j) -\frac12\partial_i \mathcal G_{kj}\partial_{\mu} \phi^k \partial^{\mu}\phi^j
-  \frac14 \partial_i G_{IJ}F^{I\mu\nu} F^{J}{}_{\mu\nu} - \mrm{g} ^2 \partial_i  V_{5d}=0\,,\\[2mm]
& 2 \nabla_{\mu}( h_{uv} \hat{\partial}^{\mu}q^v)- 2\, \mrm{g}\, h_{vw} \partial_s k^u_I A^I_{\mu} \hat{\partial}^{\mu}q^w
- \partial_u h_{vw}\hat{\partial}_{\mu}q^v \hat{\partial}^{\mu}q^w - \mrm{g} ^2\partial_u V_{5d}=0\,,\\[2mm]
&\nabla_{\mu}(G_{IK} F^{I\mu\nu}) + \frac14 C_{IJK}F^I_{\mu\sigma}F^J_{\rho\lambda}\epsilon^{\mu\sigma\rho\lambda\nu}=6\,\mrm{g}\, k_{K}^{u}h_{uv}\hat\partial^{\nu}q^{v}\,.
\end{split} 
\end{equation}

Given the five-dimensional $\ma N=2$ lagrangian \eqref{eq:genlaghyper} and the equations of motion \eqref{eq:eom5d}, we want to construct\footnote{We will follow the procedure presented in \cite{Klemm:2016kxw}.} their mapping to the class of $\ma N=2$ models in four dimensions defined by cubic prepotentials \eqref{specialprepotential}.
To this aim we introduce a Kaluza-Klein Ansatz realizing the dimensional reduction along a compact spatial direction, namely the $z$-direction\footnote{In this $\mrm{KK}$ analysis we will consider the indices $\mu,\nu,\ldots$ as curved indices for the four-dimensional background.}. The Ansatz is the following
\begin{equation}
\D s^2_5 = e^{\frac{\phi}{\sqrt3}}\D s^2_4 + e^{-\frac2{\sqrt3}\phi}(\D z + K_\mu\D x^\mu)^2\,,
\qquad A^I = B^I\D z + C^I_\mu\D x^\mu + B^I K_\mu\D x^\mu\,. \label{KK5to4}
\end{equation}
Defining $K_{\mu\nu}=\partial_\mu K_\nu-\partial_\nu K_\mu$ and $C^I_{\mu\nu}=\partial_\mu
C^I_\nu-\partial_\nu C^I_\mu$, the five-dimensional Lagrangian \eqref{eq:genlaghyper} reduces
to\footnote{We choose $\epsilon^{\mu\nu\rho\sigma z}_5=-\epsilon^{\mu\nu\rho\sigma}_4$.}
\begin{equation}
\begin{split}
\sqrt{-g_4}^{-1}\mathscr L_{4d} &= \, \frac{R^{(4)}}2 - \frac18 e^{-\sqrt3\phi} K^{\mu\nu} K_{\mu\nu} - \frac14
G_{IJ} e^{-\frac{\phi}{\sqrt3}}(C^{I\mu\nu} + B^I K^{\mu\nu})(C^J_{\mu\nu} + B^J K_{\mu\nu})\\[2mm]
& -\frac12 e^{\frac{2\phi}{\sqrt3}} G_{IJ}\partial_{\mu} B^I\partial^{\mu} B^J - \frac12 G_{IJ}\partial_{\mu}
h^I\partial^{\mu} h^J - \frac14\partial_{\mu}\phi\partial^{\mu}\phi - h_{uv}\hat{\partial}_{\mu} q^u \hat{\partial}^{\mu} q^v\\[2mm]
& -\frac{e_4^{-1}}{16}\epsilon^{\mu\nu\rho\sigma} C_{IJK}\big(C^I_{\mu\nu} C^J_{\rho\sigma} B^K +
\frac13 K_{\mu\nu} K_{\rho\sigma} B^I B^J B^K + C^I_{\mu\nu} K_{\rho\sigma} B^J B^K\big)\\[2mm]
& -e^{\sqrt3\phi}\mathrm{g}^2 B^I k_I^u B^J k_J^v h_{uv} - \mathrm{g}^2 e^{\frac{\phi}{\sqrt3} }V_{5d}\,.
\label{eq: lag4d}
\end{split}
\end{equation}
Now we want to rewrite $\mathscr L_{4d}$ in the dictionary of $\ma N=2$, $d=4$ supergravity, by using
the identifications of the ungauged case \cite{Ceresole:2007rq}. The coordinates of the special
K\"ahler manifold, the K\"ahler potential, the K\"ahler metric and electromagnetic field strengths are
given in terms of five-dimensional data respectively by
\begin{equation}
\begin{split}
&z^I = -B^I - i e^{-\frac{\phi}{\sqrt3}} h^I\,, \qquad e^{\cal K} = \frac18 e^{\sqrt3\phi}\,, \\
& g_{I\bar J} = \frac12 e^{\frac{2\phi}{\sqrt3}} G_{IJ}\,, \qquad F_{\mu\nu}^{\Lambda} =
\frac1{\sqrt2} (K_{\mu\nu}, C^I_{\mu\nu})\,,
\end{split}
\label{eq: rmapping}
\end{equation}
where capital greek indices $\Lambda,\Sigma,\ldots$ range from $0$ to $n_{V}$. If we introduce
the matrices
\begin{equation}
R_{\Lambda\Sigma} =
-\begin{pmatrix}
\frac 13 B & \frac 12 B_J \\
\frac 12 B_I & B_{IJ}
\end{pmatrix}\,,
\qquad
I_{\Lambda\Sigma} = -e^{-\sqrt3\phi}
\begin{pmatrix}
1 + 4 g & 4 g_ {\bar J} \\
4 g_I & 4 g_{I\bar J}
\end{pmatrix}\,,
\end{equation} 
where we defined
\begin{equation}
\begin{split}
B_{IJ} = & C_{IJK} B^K\,, \qquad B_I = C_{IJK} B^J B^K\,, \qquad B = C_{IJK} B^I B^J B^K\,, \\
& g = g_{I\bar J} B^I B^J\,, \qquad g_{I\bar J} B^J = g_I = g_{\bar I} = g_{\bar I J} B^J\,,
\end{split}
\end{equation}
the Lagrangian \eqref{eq: lag4d} can be cast into the form
\begin{equation}
\begin{split}
\sqrt{-g_4}^{-1}\mathscr L_{4d} = &\frac R2 - g_{I\bar J}\partial_{\mu} z^I\partial^{\mu}\bar z^{\bar J}
- h_{uv}\hat{\partial}_{\mu} q^u\hat{\partial}^{\mu} q^v \\
& +\frac14 I_{\Lambda\Sigma} F^{\Lambda\mu\nu} F^{\Sigma}_{\mu\nu}
+ \frac18 e_4^{-1}\epsilon^{\mu\nu\rho\sigma} R_{\Lambda\Sigma} F^{\Lambda}_{\mu\nu}
F^{\Sigma}_{\rho\sigma} - V_{4d}\,,
\label{{eq: lag4d}}
\end{split}
\end{equation}
with the four-dimensional potential given by
\begin{equation}
V_{4d} = \mrm{g}^2e^{\frac{\phi}{\sqrt3}} V_{5d} + e^{\sqrt3\phi}\,\mrm{g}^2 h_{uv} k_I^u k_J^v
B^I B^J\,. \label{eq:potn}
\end{equation}
The underlying prepotential of the special K\"ahler manifold turns out to be
\begin{equation}
F = \frac16\frac{C_{IJK} X^I X^J X^K}{X^0}\,, \label{eq:prepo}
\end{equation}
chosen the parametrization $X^I/X^0=z^I=-B^I-ie^{-\phi/\sqrt3}h^I$ \cite{Ceresole:2007rq}.

The four-dimensional scalar potential \eqref{eq:potn} reads
\begin{eqnarray}
\frac{V_{4d}}{\mrm{g}}&=& -9 e^{\frac{\phi}{\sqrt3}} P^x_I P^x_J\left(h^I h^J - \frac12 G^{IJ}\right) 
+ 9 e^{\frac{\phi}{\sqrt3}} h_{uv} k^u_I k^v_J h^I h^J + 9 e^{\sqrt3\phi} h_{uv} k_I^u k_J^v B^I B^J
\nonumber \\
&=& 18 P^x_I P^x_J\left(\frac14 e^{\frac{\phi}{\sqrt3}} G^{IJ} + \frac12 e^{\sqrt3\phi} B^I B^J
- 4\frac{e^{\sqrt3\phi}}8 (e^{-\frac{\phi}{\sqrt3}} h^I)(e^{-\frac{\phi}{\sqrt3}} h^J)
- \frac12 e^{\sqrt3\phi} B^I B^J\right) \nonumber \\
&& + 72\frac{e^{\sqrt3\phi}}8 h_{uv} k_I^u k_J^v (e^{-\frac{2\phi}{\sqrt3}} h^I h^J + B^I B^J)\,.
\label{eq:potrmap} 
\end{eqnarray}
The first two terms in the second line of \eqref{eq:potrmap} combine to give
$-\frac12 I^{\Lambda\Sigma}$, while the last two
terms yield $-4X^I\bar X^J$. Finally by fixing $\mrm{g}=\frac{1}{3\sqrt{2}}$ one obtains
\begin{equation}
\begin{split}
V_{4d} = &\left[P^x_\Lambda P^x_\Sigma\left(-\frac12 I^{\Lambda \Sigma} -
4 X^{\Lambda}\bar{X}^{\Sigma}\right) + 4 h_{uv} k_\Lambda^u k_\Sigma^v X^\Lambda\bar{X}^\Sigma
\right]\Big|_{P^x_0 = 0, k^u_0 = 0} \\
= & V_g \Big|_{P^x_0 = 0, k^u_0 = 0}\,,
\label{eq:trunpot}
\end{split}
\end{equation}
which is precisely a truncated form of the potential of the four-dimensional theory given in \eqref{eq:scal_pot}.
The final step is the matching of the covariant derivative of the hyperscalars,
\begin{equation}
\hat{\partial}_{\mu} q^u = \partial_{\mu} q^u + 3\mrm{g}\, C^I_\mu k^u_I = \partial_{\mu} q^u + 
A^I_{\mu} k^u_I\,,
\end{equation}
where $A^I_{\mu}$ are the four-dimensional gauge fields.
We conclude that, once the five-dimensional index $I=1,\cdots, n_V$ has been substituted to the four-dimensional one $i=1,\cdots, n_V$, the lagrangian \eqref{eq: lag4d} coincides with the lagrangian of $\ma N=2$ gauged supergravity in $d=4$ with special K\"ahler moduli space described by the cubic models \eqref{specialprepotential} and electric local isometries on the quaternionic manifold.

\section{The Higher-Dimensional Origin of the $\ma N=2$ Theories}

Even if $\ma N=2$ gauged supergravities in $d=4, 5$ can be formulated independently from their relations with the microscopic theory, it is clear that their relevance comes from their higher-dimensional origin. 

In what follows we will consider those $\ma N=2$ supergravity models whose fields have an higher-dimensional interpretation (as for example \eqref{specialprepotential} and \eqref{uhm}). To this aim we will study some compactifications of $D=11$ and type $\mrm{II}$ supergravities admitting consistent truncations to $\ma N=2$ supergravities. In particular we are interested in those truncations including maximally symmetric backgrounds like $\mathbb{R}^{1,3}$ and $\mathbb{R}^{1,4}$ or $\mrm{AdS}_4$ and $\mrm{AdS}_5$, whose uplifts are consistently described by closed string vacua.

The first examples considered historically are based on Calabi-Yau compactifications. In this approach the scalars of the lower-dimensional supergravity parametrize the deformations of the Calabi-Yau and then the construction of the $\ma N=2$ theory mainly involves the analysis of the geometric properties of the internal manifold. In this context in order to produce gauged models with $\mrm{AdS}_d$ vacua one has to include in the compactification background fluxes turned on in the internal manifold.

An other possibility is based on the truncations on spheres or more general geometries like Sasaki-Einstein manifolds. In these cases one obtains directly various examples of $\ma N=2$ gauged models with $\mrm{AdS}_d$ vacua.

In this section we will consider firstly the truncations of $D=11$ and type $\mrm{II}$ supergravities on Calabi-Yau threefolds producing ungauged $\ma N=2$ models in $d=4, 5$. Thus we will make some explicit examples of reductions on a Calabi-Yau threefold including backgroung fluxes and producing a gauging in the lower-dimensional theories. Finally we will move to some more complicated examples of truncations on Sasaki-Einstein manifolds leading to models with $\mrm{AdS}_d$ vacua.

\subsection{$\mrm{SU(3)}$-Structures and $\ma N=2$ Moduli Spaces}
\label{modulispace}

Historically the first examples of dimensional reductions producing $\ma N=2$ supergravity models in four and five dimensions have been the consistent truncations to the massless $\mrm{KK}$ modes respectively of type $\mrm{II}$ and eleven-dimensional supergravities on a suitable Calabi-Yau threefold $\mrm{CY}_3$ without internal background fluxes (see for example \cite{Cecotti:1988qn, Bodner:1990zm,Cadavid:1995bk}). In this case one obtains ungauged $\ma N=2$ supergravities with vacua of the type $\mathbb{R}^{1,3}$ and $\mathbb{R}^{1,4}$.

One of the main properties of Calabi-Yau manifolds is the existence of a class of continuous deformations of the parameters describing their size and their shape that preserve the global properties of the manifold, i.e. the manifold remains Ricci-flat and the Hodge numbers\footnote{The Hodge number $h^{p, q}$ of a manifold $\ma M$ is the dimension of the $(p, q)$ Dolbeault cohomology group $H^{p, q}(\ma M)$ of the Calabi-Yau. The Hodge numbers are topological invariants and thus they label the complex cohomology classes of the manifold.} $h^{p, q}$ are preserved. Furthermore, the parameters of these deformations are directly related to the {\itshape $G$-structure}\footnote{Calabi-Yau reductions preserve some supersymmetry. From this it follows that the internal manifold $\mrm{CY}_n$ admits at least a covariantly constant Killing spinor. A way to characterize Calabi-Yau manifolds is through a set of differential closed forms defined by the non-trivial fermonic bilinears of the Killing spinors and respecting a set of conditions equivalent to the $\mrm{BPS}$ conditions. More precisely the set of these forms defines a $G$-structure composed by the forms on $\mrm{CY}_n$ that are stabilized by the subgroup $G=\mrm{SU}(n)$ of the holonomy group that preserves the Killing vectors.} associated to the Calabi-Yau and to the $\mrm{SUSY}$ preserved by the theory reduced on the Calabi-Yau.

The massless scalar fields in the lower-dimensional theories are directly related to the parameters of these deformations. In particular the total number of the moduli, the organization in multiplets and the scalar geometries come from the study of the topological invariants of the Calabi-Yau and from the analysis of the deformations of the $G$-structures.

Coming back to $\ma N=2$ supergravities in $d=4, 5$, they both preserve 8 real supercharges and this means that a covariant Killing spinor is defined on the six-dimensional manifold $\mrm{CY}_3$.
In this particular case on can consider a six-dimensional $\mrm{SU(3)}$-structure\footnote{The holonomy group associated to the parallell transports of a six-dimensional spinor is $\mrm{Spin(6)}\simeq \mrm{SU(4)}$. The presence of 8 preserved supercharges implies that the holonomy group is broken to $\mrm{SU(3)}$. In particular $J$ and $\omega$ are defined by the relations
\begin{equation}
 \D J=0\,,\qquad \D \omega=0\,,\qquad J\wedge \omega=0\,,\qquad \frac16 J^3=\frac{i}{8}\omega\wedge \bar{\omega}\neq 0\,.
 \label{6dstructure}
 \end{equation}
 } defined by a (1,1) closed form $J \in H^{1,1}(\mrm{CY}_3)$ and a (2,1) closed form $\omega \in H^{2,1}(\mrm{CY}_3)$. In particular it can be shown that $J$ is related to the K{\"a}hler form of the $\mrm{CY}_3$, while the presence of $\omega$ is needed for a vanishing Chern class.

A general deformation of the Calabi-Yau can be written as a transformation\footnote{For many Calabi-Yau manifolds it is not possible to construct an explicit form of the metric. In this discussion the metrics $g_{mn}$ and $\delta g_{mn}$ will be considered ``formally".} on its six-dimensional metric $g_{mn}$ of the following form
\begin{equation}
g_{mn}\rightarrow g_{mn}+\delta g_{mn} \qquad \text{with}\qquad R_{mn}(g+\delta g)=0\,,
\end{equation}
where $\delta g_{mn}$ represents the metric on the manifold spanned by the parameters of the deformations, i.e. the metric of the moduli space.
All the geometric quantities of the Calabi-Yau can be expressed in terms of $J$ and $\omega$. From this fact it can be shown that the metric $\delta g_{mn}$ splits into two pieces corresponding respectively to the independent variations of $\omega$ and $J$ namely $\delta \omega$ and $\delta J$. Thus the geometry of the moduli space $\ma{M}_{mod}$ has a product structure of the type 
\begin{equation}
\ma{M}_{mod}=\ma{M}^{2,1}\times\ma{M}^{1,1}\,,
\label{modulisplitting}
\end{equation}
where $\ma{M}^{2,1}$ is called {\itshape complex-structure moduli space} and $\ma{M}^{1,1}$ {\itshape K{\"a}hler-structure moduli space}. 

The variations of $\omega$ are associated to the shape of the Calabi-Yau and parametrize the manifold $\ma{M}^{2,1}$. It turns out that $\ma{M}^{2,1}$ is a special K{\"ahler} manifold \cite{Bodner:1990zm} parametrized by a set of $h^{2,1}$ complex moduli $z^a$ with $a=1,\cdots,h^{2,1}$. The special geometry is firstly realized by the K{\"a}hler potential given by
\begin{equation}
\ma K^{2,1}(z, \bar{z})=-\log \left(i \int \omega \wedge \omega     \right)\,
\end{equation}
and defining the metric $g_{a\bar{b}}=\partial_a\partial_{\bar{b}}\,\ma K^{2,1}$. Secondly by the definition of a $\mrm{Sp}(2h^{2,1}+2)$-structure through the embedding coordinates $Z^A(z)$ defined as 
\begin{equation}
Z^A(z)=\int_{\Sigma^{A}} \omega \qquad \text{with} \qquad A=0,\cdots, h^{2,1} \,,
\end{equation}
where $\Sigma^{A}$ is a non-trivial 3-cycle in $\ma{M}^{2,1}$. In this picture the physical moduli describing the complex deformations of $\mrm{CY}_3$ are given by 
\begin{equation}
z^a=\frac{Z^a}{Z^0}\,.
\end{equation}
The $\mrm{Sp}(2h^{2,1}+2)$-structure implies the existence of a set of dual coordinates $\ma G_A$ functions of the coordinates $Z^A$
\begin{equation}
\ma G_A=\int_{\Gamma_{A}} \omega\,,
\end{equation}
where $\Gamma_{A}$ together with $\Sigma^A$ form an homology basis of 3-cycles\footnote{In the sense that their intersections are given by $\Sigma^A \cap \Gamma_{B}=\delta^A_B$ and $\Sigma^A \cap\Sigma^B=\Gamma_{A} \cap \Gamma_{B}=0$.}. Given $(\alpha_A, \beta^A) \in H^{2,1}(\ma M^{2,1})$, the cohomology basis of (2,1)-forms dual of $(\Sigma^A, \Gamma_{A})$, one has the $\mrm{Sp}(2h^{2,1}+2)$-invariant decomposition of $\omega$ one the basis
\begin{equation}
\omega=Z^A \alpha_A-\ma G_A \beta^A\,.
\label{cyvolumeomega}
\end{equation}
It follows that one can organize the quantities $Z^A$ and $\ma G_A$ in symplectic sections 
\begin{equation}
 V=\left(\begin{array}{c}
                   Z^A\\
                   \ma G^A
                 \end{array}\right),
                  \label{symsections}
\end{equation}
obeying the constraint
\begin{equation}
 \left\langle V| \bar V\right\rangle=
 \bar{Z}^A \ma G_A-Z^A \bar{\ma G}_A=-i\,,
 \end{equation}
 where $\langle \,,\, \rangle$ is the inner product defined by the standard symplectic matrix \eqref{sympmatrix}.
The (2,1)-form $\omega$ is defined up to rescaling by an holomorphic function $f_{\omega}(Z)$ of the coordinates as $\omega \rightarrow e^{f_{\omega}(Z)}\,\omega$. These are the K{\"a}hler tranformations of the potential $\ma{K}^{2,1}$.
It follows that in this frame a prepotential $F_{\omega}(Z)$ such that $\ma G_A=\partial_A F_{\omega}$ exists. It is important to note that $F_{\omega}$ is protected by corrections in $\alpha^\prime$. 

Keeping $\omega$ fixed and varying $J$ one obtains $\ma{M}^{1,1}$ parametrized by $h^{1,1}$ real moduli $v^i$ with $i=1, \cdots, h^{1,1}$. These variations are relatated to the size of the Calabi-Yau. These moduli are called {\itshape K{\"a}hler moduli} and are related to certain geometric properties of the Calabi-Yau, in particular to volumes contained in non-trivial cycles, thus this sector of the moduli space is subjected to the $\alpha^\prime$-corrections.

Given a basis of harmonic (1,1)-forms $ E_i \in H^{1,1}(\ma M^{1,1})$, as in (\ref{cyvolumeomega}) it is always possible to
expand $J$ as $J=v^i E_i$ and construct a K{\"a}hler metric as
\begin{equation}
g_{i\bar{\jmath}}=\frac12 g(E_i, E_j)=\frac{\partial}{\partial v^i}\frac{\partial}{\partial v^j}\, \ma K^{1,1}\,,
\end{equation}
where $\ma K^{1,1}$ is a K{\"a}her potential related to the volume modulus\footnote{The volume modulus can be expressed in terms of $\omega$ using the $\mrm{SU(3)}$-structure conditions as $\ma V=\frac 32i\,\int \omega \wedge \bar{\omega}$.} $\ma V$ of the Calabi-Yau
\begin{equation}
\ma V\equiv \frac{1}{6} \int J \wedge J\wedge J=\frac18 e^{-\scriptsize{\ma K^{1,1}}} \,.
\label{cyvolume}
\end{equation}
Also in this case it is possible to define an homogeneous function of degree-two with the role of prepotential. In particular one has
\begin{equation}
F(X)=\frac{1}{6}\,C_{ijk}\frac{X^iX^jX^k}{X^0}=\frac{1}{6X^0}\int \ma J \wedge \ma J \wedge \ma J \,,
\label{leadingprepotential}
\end{equation}
where we introduced a set of projective coordinates $X^\Lambda=(X^0, X^i)=(X^0,X^0\,z^i)$ where $X^0$ is an auxiliary coordinate. The (1,1)-form $\ma J=B+iJ$ is defined as the complexification of the K{\"a}hler form with the Kalb-Ramond field\footnote{In the case of $\mrm{CY}_3$ reductions from eleven to five dimensions the $B$-field is not present and thus one has only the real scalars $v^i$.} while the scalars 
\begin{equation}
z^i= b^i+i v^i
\label{axioscalars}
\end{equation}
are the components of $\ma J$ expanded on the basis of the (1,1)-forms $ E_i$. The tensor $C_{ijk}$ is a completely symmetric tensor whose components define the {\itshape intersection numbers} introduced in \eqref{specialprepotential} and are given by
\begin{equation}
C_{ijk}=\,\int E_i \wedge E_j \wedge E_k\,.
\end{equation}
It can be shown that the complexified scalars \eqref{axioscalars} parametrize a special K{\"a}hler manifold governed by the prepotential \eqref{leadingprepotential}.

Thus we explained the microscopic origin of the class of the $d=4$ special K{\"a}hler manifolds with cubic prepotential\footnote{Up to a rescaling $C_{ijk}\rightarrow \frac16 C_{ijk}$.} presented in \eqref{specialprepotential}: given the massless truncation of type $\mrm{II}$ supergravities on a $\mrm{CY}_3$, \eqref{specialprepotential} arise from the moduli space of a Calabi-Yau threefold.
Furthermore we point out that the sector $\ma{M}^{1,1}$ and its prepotential are subjected to $\alpha^\prime$ corrections\footnote{This means that (\ref{leadingprepotential}) represents only the leading result respect the perturbative expansion in $\alpha^\prime$. In particular since $F(X)$ must be homogeneous of degree two, the perturbative corrections will be given by 
\begin{equation}
F_{pert}=i \ma Y (X^0)^2\,
\end{equation}
where the parameter $\ma Y$ is expressed in terms of the Euler characteristic of the manifold $\ma Y=\frac{\zeta(3)}{2(2\pi)^3}\chi(\scriptsize{\mrm{CY}_3})$. Later in the thesis we will consider the case of a non-perturbative deformation of the prepotential.}.

The K{\"a}hler moduli space $\ma M^{1,1}$ coming from the reduction of $D=11$ supergravity on a $\mrm{CY_3}$ is peculiar\footnote{We redefine the coordinates as $X^\Lambda=(X^0, X^i)\rightarrow h^I=(h^1, h^I)$, the indices $\Lambda=(0,\cdots, h^{1,1}+1)\rightarrow I=(1,\cdots, h^{1,1})$ and  $i=(1,\cdots, h^{1,1}) \rightarrow i=(1,\cdots, h^{1,1}-1)$ and the scalars $v^i\rightarrow \phi^i$. With this redefinition the analysis becomes compatible with section \ref{veryspecialmanifold} on very special K{\"a}hler manifolds in $d=5$.}.
The absence of the $B$-field in eleven-dimensional supergravity implies that the compactification identifies an hypersurface within the K{\"a}her structure moduli space $\ma{M}^{1,1}$ where the moduli are constrained. This hypersurface is defined by the equation
\begin{equation}
\ma V=1\,,
\label{hypersurface5d}
\end{equation}
where $\ma V$ is the volume modulus defined in (\ref{cyvolume}). Since $B=0$ the relation (\ref{leadingprepotential}) turns out to be expressed only in terms of $J$ and on the real moduli $\phi^i$.
Given the coordinates  $h^I=(1, \phi^i)$ on $\ma M^{1,1}$, where we have chosen $h^0=1$, one can recast the equation (\ref{hypersurface5d}) as a condition on the intersection numbers $C_{IJK}$ and on the $h^I$s given by
\begin{equation}
\frac16\,C_{IJK}h^Ih^Jh^K=1\,,
\label{hypersurface5d2}
\end{equation}
where we used the relations (\ref{cyvolume}) and (\ref{leadingprepotential}).

The result \eqref{hypersurface5d} matches exactly with the defining condition \eqref{veryspecial} for $d=5$ very special K{\"a}hler manifolds describing the moduli space of vector multiplets in five-dimensional $\ma N=2$ supergravity.
Thus reducing eleven-dimensional supergravity on a Calabi-Yau threefold and truncating to massless modes, we find that the K{\"a}hler-structure moduli space $\ma M^{1,1}$ defines a very special K{\"a}hler manifold in five dimensions governed by the relation \eqref{hypersurface5d}. Furthermore \eqref{hypersurface5d} implies that, if only variations of $J$ are considered, i.e. no hyperscalars running, the volume of the Calabi-Yau is fixed, otherwise the condition have to be satisfied ``dynamically" with the hyperscalars parametrizing the variations of the volume.

\subsection{The Calabi-Yau Origin of the $\ma N=2$ Multiplets}
\label{cytruncation}

In this section we give an explanation of the higher-dimensional origin of the $\ma N=2$ multiplets in four and five dimensions introduced in section \ref{multiplets} based on the massless truncations on Calabi-Yau threefolds.

Let's start from type $\mrm{IIA}$ supergravity and an $\ma N=2$ vacuum $\mathbb{R}^{1,3}\times \mrm{CY}_3$ where $\mrm{CY}_3$ is a Calabi-Yau threefold preserving 8 real supercharges \cite{Bodner:1990zm}. We want to present the result of the $\mrm{KK}$ reduction on the Calabi-Yau and show that, after the truncation to the massless modes, they are organized in the supermultiplets of $\ma N=2$ supergravity in $d=4$ given in section \ref{multiplets}.

If one considers the type $\mrm{IIA}$ fields $(G_{MN}, B_{MN}, \Phi, C^{(1)}_M, C^{(3)}_{MNP})$ defined on a background of the type $M_4\times \mrm{CY}_3$, one can require that the components of the $\mrm{IIA}$ fluxes vanish along the $\mrm{CY}_3$. Then one can split the ten-dimensional spacetime index as ${M}=(\mu, m, \bar{m})$, where $\mu$ is the index along the four-dimensional spacetime and $(m, \bar{m})$ are the indices associated to the $\mrm{SU(3)}$-structure.
Giving a suitable reduction Ansatz \cite{Bodner:1990zm} on the metric and on the gauge potentials and truncating to the massless modes, one obtains the following four-dimensional massless scalars:
\begin{itemize}
\item$h^{2,1}$ complex scalars\footnote{We use the same notation of section \ref{modulispace} for scalars $v^i$ and $z^a$ even if they now describe the four-dimensional $\mrm{KK}$ zero-modes of ten-dimensional fields.} $z^a$ coming from the components $G_{mn}$ along the complex-structure moduli space $\ma M^{2,1}$.
\item $h^{1,1}$ real scalars $v^i$ corresponding to the components $G_{m\bar{n}}$ along the K{\"a}hler-structure moduli space $\ma M^{1,1}$.
\item $h^{1,1}$ real scalars $b^i$ coming from the internal directions of the Kalb-Ramond field $B_{m\bar{n}}$. The scalars can be extracted by expanding $B$ on the basis of the harmonic (1,1)-forms $E_i$ as $B=b^i \,E_i$ with $i=1,\cdots, h^{1,1}$. These fields are called {\itshape Betti moduli}.
\item A real scalar $\phi$ related to the type $\mrm{IIA}$ dilaton $\Phi$.
\item Two sets of $h^{2,1}+1$ real scalars $\xi^A$ and $\tilde{\xi}_A$ with $A=0, \cdots, h^{2,1}$. These come from the components of the $\mrm{R}$-$\mrm{R}$ 3-form on the internal manifold $C^{(3)}_{mn\bar{p}}$. The scalars can be extracted by expanding on the basis $(\alpha_A, \beta^A)$ of the harmonic 3-forms as $C^{(3)}=\xi^A\alpha_A-\tilde{\xi}_A\beta^A$.
\item A real scalar $a$ coming from the dualization of the four-dimensional directions of the Kalb-Ramond field $B$, i.e. $\star_{4}\,\D B=\D a$\,.
\end{itemize}
The vector fields come from the mixed components $C^{(3)}_{\mu\bar{n}}$ and can be extracted by expanding on the basis of harmonic (1,1)-forms $E_i$. In this way one obtains a set of $h^{1,1}$ four-dimensional vector fields defined by the expansion
\begin{equation}
C^{(3)}=A^i_\mu\D x^\mu \wedge E_i\,.
\end{equation}
Moreover an other vector field, the graviphoton, is produced by the reduction along the spacetime directions of the $\mrm{R}$-$\mrm{R}$ 1-form, i.e. $C^{(1)}_\mu=A^0_\mu$. It remains the four-dimensional components $C^{(3)}_{\mu\nu\rho}$ that are dual to a free parameter $e_0$ in the four-dimensional theory.

If one organizes the scalars $v^i$ and $b^i$ in $h^{1,1}$ complex scalars $z^i$ as in (\ref{axioscalars}), the above four-dimensional fields form the $\ma N=2$ supermultiplets. The supergravity multiplet is given by the four-dimensional components of the metric $g_{\mu\nu}$ and by the graviphoton $A^0_\mu$. Then one has $h^{1,1}$ abelian vector multiplets, each one of these is composed by a complex scalar $z^i$ and a vector $A^i_\mu$. Finally we obtain $h^{2,1}$ hypermultiplets composed by the scalars $(z^a, \xi^a, \tilde{\xi}_a)$ that we can rearrange in quadruples $q^u$ of real hyperscalars with $u=1,\cdots, 4h^{2,1}$. 

Moreover there is a further hypemultiplet composed by $(\phi,\, a,\, \xi^0,\, \tilde{\xi}_0)$ that is the universal hypermultiplet introduced in section \ref{quaternionic}, describing the geometry \eqref{uhm} with the metric \eqref{eq:quaternionic_kahler_complexstruct_definition}. Its name is due to the fact the it appears in any $\ma N=2$ Calabi-Yau compactification. 

The truncation of type $\mrm{IIB}$ on a Calabi-Yau threefold $\mrm{\widetilde{CY}}_3$ follows the same procedure adopted above and produces the same result \cite{Bohm:1999uk}. This fact is related to a {\itshape mirror symmetry} between the two Calabi-Yau threefolds $\mrm{CY}_3$ and $\mrm{\widetilde{CY}}_3$. In fact it can be shown that $h^{1,1}(\mrm{CY}_3)=h^{2,1}(\mrm{\widetilde{CY}}_3)$ and $h^{2,1}(\mrm{CY}_3)=h^{1,1}(\mrm{\widetilde{CY}}_3)$ and this implies that $\ma M^{2,1}=\widetilde{\ma M}^{1,1}$ and $\ma M^{1,1}=\widetilde{\ma M}^{2,1}$.

In this case one obtains $h^{2,1}$ real massless scalars $v^a$ from $\widetilde{\ma M}^{2,1}$ and $h^{1,1}$ complex massless scalars $z^i$ from $\widetilde{\ma M}^{1,1}$. Thus a complex field comes from the type $\mrm{IIB}$ axio-dilaton $\tau=C_0+i\,e^{\Phi}$. The $B$-field is decomposed like in type $\mrm{IIA}$ producing the real $b^a$ moduli.
Furthermore expanding respectively the 2-form $C^{(2)}$ and the 4-form $C^{(4)}$ on the basis of the harmonic (1,1)-forms and (2,1)-forms, one obtains the $2h^{2,1}$ real scalars $c^a$ and $\tilde{c}^a$. Moreover\footnote{The mixed components of the 4-form given by $C^{(4)}=C^a_{\mu\nu}E_a$ are associated to the scalars $\tilde{c}^a$ for the self-duality condition $F_5=\star F_5$.} two further real scalars $a$ and $C$ come from the dualization of the spacetime components $B_{\mu\nu}$ and $C^{(2)}_{\mu\nu}$.

Hence we obtain $h^{2,1}$ vector multiplets composed by the complex scalars $z^i$ and by the vectors $A^i_\mu$ extracted from the 4-form\footnote{We point out that the further contribution coming from $A_{A\,\mu}\D x^\mu \wedge \beta^A$ does not give new degrees of freedom for the self-duality condition on $F_5$. This gives a higher-dimensional explanation of the discussion on magnetic gaugings and symplectic covariance of section \ref{symplecticcovariance}.} as
\begin{equation}
C^{(4)}=A^A_\mu\D x^\mu \wedge \alpha_A\,,
\end{equation}
where $A_{\mu}^0$ is the graviphoton.
The $h^{1,1}$ hypermultiplets are composed by the reals scalars $(v^a, b^a, c^a, \tilde{c^a})$ with the universal hypermultiplet given by $(\tau, a, C)$. 

Let's now discuss the higher-dimensional origin of $\ma N=2$ multiplets in five-dimensional supergravity\footnote{Also in this case we redefine the coordinates as $X^\Lambda=(X^0, X^i)\rightarrow h^I=(h^1, h^I)$, the indices $\Lambda=(0,\cdots, h^{1,1}+1)\rightarrow I=(1,\cdots, h^{1,1})$ and  $i=(1,\cdots, h^{1,1}) \rightarrow i=(1,\cdots, h^{1,1}-1)$, and the scalars $v^i\rightarrow \phi^i$. With these redefinitions the analysis becomes compatible with section \ref{veryspecialmanifold}}. Starting from a Minkowski vacuum in $D=11$ supergravity of the type $\mathbb{R}^{1,4}\times \mrm{CY}_3$, we consider the $\mrm{KK}$ reduction on the Calabi-Yau and the truncation to the massless spectrum. Given a suitable Ansatz \cite{Cadavid:1995bk} on the fields $(G_{MN}, A^{(3)}_{MNP})$, the eleven-dimensional indices $M, \cdots$ are broken as in the case studied above, i.e. ${M}=(\mu, m, \bar{m})$, where $\mu$ is the index along the five-dimensional spacetime and $(m, \bar{m})$ are the $\mrm{SU(3)}$ indices along the Calabi-Yau. Performing the $\mrm{KK}$ reduction, the massless scalars are given by
\begin{itemize}
\item $h^{2,1}$ complex scalars $z^a$ associate to the components $G_{mn}$ parametrizing the complex-structure moduli space $\ma M^{2,1}$.
\item $h^{1,1}-1$ real scalars $\phi^i$ coming from the K{\"a}hler moduli space $\ma M^{1,1}$ and associated to the metric components $G_{m\bar{n}}$.
\item $2\,h^{2,1}$ complex scalars belonging to the internal components of the 3-form $A^{(3)}_{mn\bar{p}}$.
\item One complex scalar associated to the completely antisymmetric internal components $A^{(3)}_{mnp}=\varepsilon_{mnp} C$.
\item One real scalar $a$ associated to the five-dimensional components $A^{(3)}_{\mu\nu\rho}$.
\item One real scalar given by the volume modulus $\ma V$. 
\end{itemize}
Moreover from the components $A^{(3)}_{\mu m \bar{n}}$ we obtain $h^{1,1}-1$ vectors that can be extracted by expanding the 3-form on the basis of the harmonic (1, 1)-forms $E_i$.

The massless $\mrm{KK}$ spectrum can be organized in $\ma N=2$ five-dimensional supermultiplets and the final result is similar to the four-dimensional case derived above. The supergravity multiplet is composed by the $d=5$ components of the metric $g_{\mu\nu}$ and the graviphoton given by the model dependent combination $A^0_\mu = \phi^i A^i_\mu$.

We have $h^{1,1}-1$ vector multiplets each one of these is composed by a vector $A^i_\mu$ and a real scalar $\phi^i$. The reality of the $h^{1,1}$ scalars $\phi^i$ is the main difference respect to the four-dimensional case and, as we said, it is related to the absence of the $B$-field in eleven-dimensional supergravity.

Finally we have $h^{2,1}$ hypermultiplets remaining unchanged respect those in $d=4$. In fact each one of these is composed by the two degrees of freedom of the complex scalar $z^a$ together with the degrees of freedom coming from the complex scalar $A^{(3)}_{mn\bar{p}}$. As in the four-dimensional case, we organize the $h^{2,1}$ scalars composing the hypermultiplets in hyperscalars $q^u$ with $u=1\cdots 4\,h^{1,1}$. 
We note that also in this case the universal hypermultiplet is included in the field content and it is composed by the scalars $(C, a, \ma V)$. 

\subsection{Flux Compactifications on Calabi-Yau Threefolds}
\label{cyfluxes}



The truncations of higher-dimensional supergravities on Calabi-Yau threefolds presented in section \ref{cytruncation} produce the $\ma N=2$ theories in $d=4, 5$, respectively given in \eqref{eq:mainaction} and \eqref{eq:genlaghyper}, without any gauging and thus without the scalar potentials \eqref{eq:scal_pot}, \eqref{eq:pothyper} and covariant derivatives \eqref{eq:hypercov4d}, \eqref{eq:covhyper}. 

Following the ideas discussed in section \ref{geometry}, in order to derive gauged $\ma N=2$ models, we want now to deform the reduction Ansatz for the background fluxes including some new components wrapping non-trivial cycles within the Calabi-Yau. These contributions have to define new quantized values for the fluxes that, in turn, will be related to the gauging of the lower-dimensional theory. 

Many complications immediately arise in this more general context . The quantized values of a given background flux $\ma F$ identify a disconnected component in the moduli space $\ma M_{\scriptsize{\ma F}} \subset \ma M_{mod}$ corresponding to the values of the moduli leading to that particular compactification with the running flux $\ma F$. It follows that many new vacua can be defined in relation to the particular background flux chosen and the existence of a consistent truncation have to be demonstrated case by case. Anyway this procedure allows to generate in some cases a non-trivial scalar potential that stabilize the moduli and then to construct consistent truncations with $\mrm{AdS}_d$ vacua.

Let's consider $\ma N=2$ abelian gauged supergravity in $d=4$ with a purely electric gauging whose action is written in \eqref{eq:mainaction}. 
We want to endow the Calabi-Yau reduction of type $\mrm{IIA}$ string theory on a $\mrm{CY}_3$ with some non-zero internal values for the $\mrm{R}$-$\mrm{R}$ background fluxes $F_2$ and $F_4$. We start from the truncation Ansatz for the gauge potentials mentioned in section \ref{cytruncation}. This has the following explicit form \cite{Bodner:1990zm}
\begin{equation}
\begin{split}
&C^{(1)}=A^0\,, \qquad B=a+b^i E^i\,,\\
&C^{(3)}=C^{(3)}+A^i\wedge E_i+\xi^A \alpha_A+\xi_A \beta^A\,,
\label{ungaugedfluxes}
\end{split}
\end{equation}
where $E_i$ and $(\alpha_A, \beta^A)$ had been defined as the basis respectively of harmonic (1,1)- and (2,1)-forms.
We can now introduce a deformation of the Ansatz \eqref{ungaugedfluxes} given by the follwing new expressions for the field strengths $F_2$ and $F_4$ associated to the R-R fields $C^{(1)}$ and $C^{(3)}$ \cite{Michelson:1996pn, Taylor:1999ii, DallAgata:2001brr},
\begin{equation}
\begin{split}
F_2&=\D A^0\, , \qquad H=\D B+\D b_i E_i \,,\\
F_4&=\D C^{(3)}-B\wedge \D A^0+(\D A^i-b^i\D A^0)\wedge E_i\\
&+(\D \xi^A \alpha_A+\D \tilde{\xi}_A \beta^A)+e_i \tilde{E}^i\,,
\label{fluxes}
\end{split}
\end{equation}
where the forms $\tilde{E}^i$ define the dual basis respect to that of the forms $E_i$ and the new  $h^{1,1}$ parameters\footnote{As we already noted in section \ref{cytruncation}, the parameter $e_0$, obtained by dualizing the 3-form, was also present in the compactification without fluxes. In this case, it gives the charge to the scalar $a$ of the universal hypermultiplet.} $e_i$. These parameters are associated to the embedding tensor of the four-dimensional $\ma N=2$ theory.
Plugging the Ansatz \eqref{fluxes} into the type $\mrm{IIA}$ action, one obtains the lagrangian \eqref{eq:mainaction} for cubic models \eqref{specialprepotential} coupled to hypermultiplets with quaternionic Killing vectors given by
\begin{equation}
k^u_\Lambda=-2 e_\Lambda \delta^{ua}\,,
\end{equation}
where $\Lambda={0,\cdots,h^{1,1}}$ and $a$ is the scalar of the universal hypermultiplet becoming charged with the gauging. 

This procedure cannot be easily generalized to the symplectic covariant action with magnetic gaugings \eqref{actionmag}. In order to produce the auxaliary 2-form in four dimensions allowing the symplectic covariant formulation of the theory, one has to consider the compactification on a Calabi-Yau threefold of massive type \mrm{IIA} supergravity. The Ansatz for the background fluxes $\hat{F}_2$ and $\hat{F}_4$ in massive type $\mrm{IIA}$ is formulated as the deformation of \eqref{fluxes} given by \cite{Louis:2002ny}
\begin{equation}
\begin{split}
&\hat{F}_2= F_2+mB-(m^i-m b^i)E_i\,,\\
&\hat{F}_4= F_4-\frac{m}{2}B\wedge B+(m^i B-m B b^i)\wedge E_i+C_{ijk}\,(b^i \,m^j-\frac{1}{2}m\,b^i\,b^j) \tilde{E}^k\,.
\label{magneticfluxes}
\end{split}
\end{equation}
In this case one has $2h^{1,1}+2$ flux parameters given by $(e_0, e_i, m, m^i)$ where $m=\hat{F_0}$ is Romans mass. 
It can be shown that, plugging this Ansatz in the action of massive type $\mrm{IIA}$, one obtains a $\mrm{Sp}(2h^{1,1}+2,\mathbb{R})$-structure on the moduli space and the symplectic covariant lagrangian \eqref{actionmag} with electric and magnetic gaugings determined by $(e_\Lambda, m^\Lambda)$. In this compactification the four-dimensional auxiliary 2-form $B_a$ comes from the $B$-field and the modified field strengths are given by $H^\Lambda=\D A^\Lambda+m^\Lambda B$.

Finally between the compactifications with fluxes producing $\ma N=2$ gauged models, we mention also those including the $\mrm{NS}$-$\mrm{NS}$ flux $H$ \cite{Michelson:1996pn, Taylor:1999ii, DallAgata:2001brr, Louis:2002ny}. 

An analogous procedure can be applied to the flux $F_4$ in the compactification of eleven-dimensional supergravity on a Calabi-Yau threefold. For example in \cite{Behrndt:2000zh}, five-dimensional $\ma N=2$ gauged supergravity \eqref{eq:genlaghyper} coupled to the universal hypermultiplet is obtained with the following Ansatz on the 4-form field strength,
\begin{equation}
\hat{F}_4=F_4+\frac{e_i}{\ma V}\,\tilde{E}^i\,,
\end{equation}
where $F_4$ is the flux associated to the Calabi-Yau reduction of $D=11$ supergravity presented in \ref{cytruncation} and $e_i$ are the parameter of the gauging on the quaternionic moduli space \eqref{uhm}. We note that for large volume and keeping fixed the parameters $e_i$, we find the ungauged limit.

\subsection{M-Theory Truncations on Sasaki-Einstein Manifolds}
\label{Mtheorytruncation}

In this section we change a little the strategy in obtaining $\ma N=2$ gauged supergravity models by considering the truncations of $\mrm{M}$-theory and type \mrm{II} string theories on different manifolds like spheres and Sasaki-Einstein manifolds. These truncations are particularly important for the $\mrm{AdS/CFT}$ correspondence since they define $\ma N=2$ models including many examples of $\mrm{AdS}_5$ and $\mrm{AdS}_4$ vacua.

There are many examples of these truncations both to $d=4$ \cite{Gauntlett:2007ma, Gauntlett:2009zw, Cassani:2009ck, Cassani:2012pj} and to $d=5$ models \cite{Buchel:2006gb, Gauntlett:2007ma, Gauntlett:2007sm, Skenderis:2010vz, Liu:2010sa}. In this section we will take in consideration, as an example, the $\mrm{M}$-theory truncations on seven-dimensional Sasaki-Einstein manifolds $M_7$ preserving 8 supercharges and including $\mrm{AdS}_4$ vacua \cite{Cassani:2012pj}. 

In section \ref{geometry} we classified the compactifications on spheres as the truncations on gauged supergravities preserving maximal supersymmetry. If we consider in particular the consistent truncation of eleven-dimensional supergravity on a $S^7$, we obtain $\ma N=8$ supergravity in $d=4$ with a $\mrm{SO(8)}$ gauging \cite{deWit:1981sst}. This theory admits an $\mrm{AdS}_4$ vacuum whose uplift is given by the $\mrm{M}$-theory vacumm $\mrm{AdS}_4\times S^7$ given in \eqref{freundrubin}.

Since we are interested in $\ma N=2$ truncations, it is possible to show that the same $\mrm{M}$-theory vacuum can be obtained by considering a suitable truncation of the supermultiplets of the maximal supergravity where only massless gauge fields are kept in the description. This leads to the $STU$ model \eqref{stumanifold} with a purely electric $\mrm{FI}$ gauging \eqref{fi4d}.

If one now considers truncations on more complicated seven-dimensional geometries \cite{Cassani:2012pj}, many new $\mrm{M}$-theory vacua arise. In particular we want to study the consistent truncations of eleven-dimensional supergravity on manifolds of the type
\begin{equation}
M_7=G/H\,,
\label{sasakieinstein}
\end{equation}
where only Kaluza-Klein modes that are left-invariant under the action of $G$ are kept in the reduction. Moreover, since we are truncating to the $\ma N=2$ theory in $d=4$, one has to require that \eqref{sasakieinstein} is endowed by a seven-dimensional $\mrm{SU(3)}$-structure. This is quite similar respect to the six-dimensional one: there is a (1,1)-form $J$ and a (2,1)-form $\omega$ and they respect the same relations of the six-dimensional case written in \eqref{6dstructure}. A further real 1-form $\eta$ is present defining a K\"ahler cone in $M_7$ with metric given by
\begin{equation}
\D s^2(M_7)=\eta^2+\D s^2(B_6)\,,
\label{kahlercone}
\end{equation}
where $B_6$ is a six-dimensional subspace orthogonal to $\eta$ with an induced $\mrm{SU(3)}$-structure generated by $J$ and $\omega$.
The reduction procedure is analogous to that one sketched in section \ref{cyfluxes} of massive type \mrm{IIA} on a Calabi-Yau threefold in presence of background fluxes and one verifies that there are many explicit realizations of internal geometries of the type \eqref{kahlercone}. In particular if one searches for truncations admitting $\mrm{AdS}_4$ vacua, one finds an entire class of $\ma N=2$ models with vector- and hypermultiplets with both electric and magnetic gaugings.
\begin{table}[htbp]
\label{spherereduction}
\begin{center}
\begin{tabular}{c|ccccc}
\hline
$M_7$        &   $n_V$  &   $n_H$   &  $\ma{SM}$  & $F(X)$ &      $\ma{HM}$   \\
\hline
$S^7$=$\frac{\mrm{SU(4)}}{\mrm{SU(3)}}$ & 1 & 1  &   $\frac{\mrm{SU(1,1)}}{\mrm{U(1)}}$  & $-\frac{(X^1)^3}{X^0}$ &  $\frac{\mrm{SU(2,1)}}{\mrm{SU(2)}\times\mrm{U(1)}}$       \\
$M^{110}$   & 2 & 1 &  $\left(\frac{\mrm{SU(1,1)}}{\mrm{U(1)}}\right)^2$ & $-\frac{(X^1)^2X^2}{X^0}$  &  $\frac{\mrm{SU(2,1)}}{\mrm{SU(2)}\times\mrm{U(1)}}$    \\
$Q^{111}$   & 3 & 1 &   $\left(\frac{\mrm{SU(1,1)}}{\mrm{U(1)}}\right)^3$ & $-\frac{X^1X^2X^3}{X^0}$  &  $\frac{\mrm{SU(2,1)}}{\mrm{SU(2)}\times\mrm{U(1)}}$    \\
$V_{5,2}$   & 1 & 2 &    $\frac{\mrm{SU(1,1)}}{\mrm{U(1)}}$  & $-\frac{(X^1)^3}{X^0}$ & $\frac{G_{2(2)}}{\mrm{SO(4)}}$ \\
\hline
\end{tabular}
\caption[Examples of $\ma N=2$ models from Sasaki-Einstein truncations of M-theory. Table from \cite{Cassani:2012pj}.]{\it Four examples of $\ma N=2$ gauged models with $\mrm{AdS}_4$ vacua describing the M-theory truncations on a Sasaki-Einstein manifold $M_7$. Table from \cite{Cassani:2012pj}.} \label{truncations}
\end{center}
\end{table}
Between these models there is, of course, the maximally symmetric one corresponding to the truncation on the $S^7$ mentioned above. The other gauged models involves hypermultiplets and are in the most cases described by an $M_7$ manifold with a Sasaki-Einstein structure\footnote{There are two examples of internal spaces, called $N(k, l)$ and $N(1, 1)$, described by a tri-Sasaki-Einstein manifold \cite{Cassani:2012pj}.}. For example we mention the truncation on $Q^{111}$ corresponding to the $STU$ model coupled to the universal hypermultiplet \eqref{uhm}. In this truncation two vector fields acquire mass for a dyonic gauging of the non-compact abelian subgroup\footnote{These gauged symmetries correspond to the translations of $a$ and to the rotations between $\xi^0$ and $\tilde{\xi}_0$.} $\mathbb{R}\times \mrm{U}(1)$ of the isometries of the universal hypermultiplet. 
More generally, we point out that all these models are characterized by a prepotential of the type \eqref{specialprepotential} for the scalars of the vector multiplets

Let's follow the same notation used in the above sections on type $\mrm{IIA}$ truncations and have a look on the structure of the moduli spaces characterizing these truncations.
Thanks to the cone structure \eqref{kahlercone}, the moduli space splitting remain the same of \eqref{modulisplitting}, thus from the deformations of $\ma M^{1,1}$ we have $h^{1,1}$ complex scalars $z^i$ defined as \eqref{axioscalars}, where in this case the Betti moduli $b_i$ comes from the expansion of the 3-form gauge potential $A^{(3)}=b^i \,E_i\wedge \theta$ with $\theta$ harmonic real 1-form.

Moreover there are $h^{2,1}$ complex scalars $z^a$ describing the deformations of $\ma M^{2,1}$ and the real scalars $\xi^A$, $\tilde{\xi}_A$, $\phi$, $a$, respectively resulting from the expansion of the 3-form $A^{(3)}$ on the basis of the harmonic (2,1)-forms, from the volume modulus of the internal manifold and from the dualization of a 2-form $B$ involved\footnote{This 2-form determines the $d=4$ auxiliary 2-form $B_a$ associated to the symplectic covariant formulation of the $d=4$ theory given by the action \eqref{actionmag}.} in the Ansatz of the 3-form. These moduli organize themselves in $h^{2,1}$ hypermultiplets $(z^a, \xi^a, \tilde{\xi}_a)$ plus the universal hypermultiplet $(\phi, a, \xi^0, \tilde{\xi}_0)$.
Moreover we note that the hypermultiplets define a class of quaternionic manifolds whose metrics was classified in \cite{Ferrara:1989ik}.

Finally, as far as concerns the gauging on quaternionic manifolds, as we said above, they can be both electric and magnetic. In general they will be determined by the non-compact translations along the hyperscalars $(a, \xi^A, \tilde{\xi}_A)$ and by the $\mrm{U(1)}$ rotations of the $(z^a, \bar{z}^{\bar{a}})$ and $(\xi^A, \tilde{\xi}_A)$.

%% file: chap4.tex

\chapter{Extremal Flows in $\ma N=2$ Supergravities}
\label{flows}
\thispagestyle{plain}


In chapter \ref{gaugedsugras} the properties of the scalar geometries characterizing $\ma N=2$ gauged theories in $d=4, 5$ and their stringy origin have been introduced. In particular we presented some of the many possible consistent truncations of eleven-dimensional and type $\mrm{II}$ supergravities to $\ma N=2$ gauged supergravity models in $d=4, 5$ and we discussed about the large plethora of closed string vacua that are well described as $\mrm{AdS}_d$ backgrounds in some of these $\ma N=2$ models. 

Furthermore, in section \ref{flowacrossdimensions} we introduced the concept of $\mrm{RG}$ flow across dimensions as a general realization of the $\mrm{AdS/CFT}$ correspondence consisting in the description of a flow in a gauged supergravity interpolating between two different vacua in terms of the renormalization flow across dimensions between the two $\mrm{SCFT}$s dual to the vacua. In this framework it is often possible to extrapolate the classical observables describing the flow from string theory and the most important example is certainly given by the {\itshape microstates counting} of black holes. This basically consists in the derivation of the Bekenstein-Hawking entropy from the counting of the quantum states of the $\mrm{SCFT}$ on the worldvolume of the correspondent branes' system \cite{Strominger:1996sh}. Thanks to the $\mrm{AdS/CFT}$ techniques, many important results have been obtained in the study of microstate structure of black holes and nowadays this represents one of the most productive research directions\footnote{For a non-exhaustive bibliography on $\mrm{RG}$ flows and black holes in $\ma N=2$ gauged supergravity we refer to \cite{Maldacena:2000mw,Benini:2015eyy,Benini:2016rke,Hosseini:2016cyf,Hosseini:2016ume,Hosseini:2017mds,Azzurli:2017kxo,Amariti:2015ybz,Amariti:2016mnz,Benini:2013cda,Benini:2012cz,Benini:2015bwz,Guarino:2015jca,Guarino:2015qaa,Guarino:2015vca,Guarino:2016err,Guarino:2017eag,Guarino:2017pkw,Guarino:2017jly,Bobev:2018uxk}.}. 

As a relevant example of a microscopic entropy calculation we recall the example of $\mrm{RG}$ flow across dimensions mentioned in section \ref{flowacrossdimensions} given by the black hole solution to $\ma N=2$, $d=4$ $\mrm{FI}$ gauged supergravity constructed in \cite{Cacciatori:2009iz} and interpolating behavior $\mrm{AdS}_4\rightarrow \mrm{AdS}_2\times S^2$. Holographically this solution is interpreted as an $\mrm{RG}$ flow across dimensions between $\mrm{ABJM}$ on $S^1\times S^2$ and a superconformal quantum mechanics. In particular the partition function of $\mrm{ABJM}$ on $S^1\times S^2$ can be interpreted as the {\itshape Witten index} of the superconformal quantum mechanics. The derivation of the index in the large $N$ limit implies that, when evaluated in its critical points, it is directly related to the entropy of the black hole \cite{Benini:2015eyy}.

Guided by this general framework, in this chapter we will study two examples of extremal black solutions in $\ma N=2$ gauged supergravities: the black hole in $d=4$ and the black string in $d=5$. We will consider those flows that are static and spherically/hyperbolically symmetric and we will derive their general first-order equations using the {\itshape Hamilton-Jacobi approach} that is slightly more general with respect to the Killing spinor analysis presented in section \ref{killingspinor} since it allows also extremal flows breaking all the supersymmetries.

Furthermore their near-horizon properties will be studied in relation to the so-called {\itshape attractor mechanism} \cite{Ferrara:1995ih, Strominger:1996kf, Ferrara:1996dd, Ferrara:1996um, Ferrara:1997tw}, consisting in a dynamical process of stabilization of the scalars at the horizon to charge-dependent values independently to they asymptotic values. These  black solutions are also called {\itshape attractors} and their entropy is strictly dependent on the charges and on the gauging parameters of the $\ma N=2$ model considered. As we explained in detail in section \ref{microstate} where we described the higher-dimensional interpretation of a $\mrm{RN}$ black hole in $d=5$ in terms of the bound state $\mrm{D}1$-$\mrm{D}5$-$\mrm{P}$, the strict dependence on the fluxes of the Bekenstein-Hawking entropy allows its extrapolation from a higher-dimensional solution describing the bound state of brane. In other words the attractor mechanism manifests that the entropy of the black hole is a moduli-independent quantity and then its value is the same also in the regime in which the interpretation of the black hole in terms of branes' intersection holds.

The entropy of the attractors manifests many properties characterizing the black solution. Among these, if we refer to the case of $d=4$ black holes, there is a peculiar non-linear symmetry of the Bekenstein-Hawking entropy. This symmetry is called {\itshape Freudenthal duality} \cite{Borsten:2009zy,Ferrara:2011gv} and its stringy origin is still unknown. It can be defined as an anti-involutive, non-linear map acting on the electromagnetic charges of the black holes and leaving invariant the entropy.

To sum up, in this chapter we will construct the $\mrm{BPS}$ first-order equations for the $d=4$ black hole \cite{Klemm:2016wng} and for the $d=5$ black string \cite{Klemm:2016kxw} in $\ma N=2$ gauged supergravities with general matter couplings by making use of the Hamilton-Jacobi approach. Then the attractor mechanism will be introduced in $d=4$ and formulated by solving the general $\ma N=2$ equations of motion at the horizon \cite{Chimento:2015rra}. The attractors will be also studied from the point of view of the first-order formulation introduced in the first part of the chapter and the general expressions of the {\itshape attractor equations} in $d=4, 5$ will be derived \cite{Klemm:2016wng,Klemm:2016kxw}. Finally the $\mrm{AdS}_3$ central charge of the black string will be obtained in complete generality by solving explicitly the attractor equations in $d=5$ \cite{Klemm:2016kxw}.

We will conclude the chapter formulating in generality Freudenthal duality for those four-dimensional attractors belonging to $\ma N=2$ gauged models, firstly considering general $\mrm{FI}$ gauging and then extending to the hypermultiplets \cite{Klemm:2017xxk}.

\section{First-Order Description and Hamilton-Jacobi Flows}

In section \ref{killingspinor} we introduced the Killing spinor analysis as a crucial tool in deriving and studying supersymmetric solutions in supergravities. This formulation of the first-order equations \eqref{bpsequationsgeneral} is equivalent to the determination of the set of spinors such that the $\mrm{SUSY}$ variations of fermions are vanishing.

A more general possibility, that still produces a system of first-order equations, is the {\itshape Hamilton-Jacobi
approach}. This includes the Killing spinor equations as a special subcase, but is
quite easily generalizable to extremal non-$\mrm{BPS}$- or even non-extremal black holes. Using Hamilton-Jacobi
theory is essentially\footnote{``Essentially" means that in many flow equations obtained in the literature
by squaring an action, the rhs of \eqref{flow-intro} is not a gradient, or, in other words, the flow is
not driven by a (fake) superpotential (no gradient flow).}
equivalent to writing the action as a sum of squares such that their vanishing correspond to the extremization of the action.
Suppose that, owing to various symmetries (like for example staticity and spherical symmetry), the supergravity
action can be dimensionally reduced to just one-dimensional\footnote{When there is less symmetry, e.g.~for
rotating black holes, one obtains a field theory living in two or more dimensions, instead of a mechanical
system \cite{Andrianopoli:2012ee,Hristov:2012nu}. In this case, the Hamilton-Jacobi formalism has to be
generalized to the so-called De Donder-Weyl-Hamilton-Jacobi theory \cite{Andrianopoli:2012ee}.} {\itshape effective action} given by
\begin{equation}
S = \int\D r\left[\frac12{\mathscr G}_{\Lambda\Sigma}\dot q^\Lambda\dot q^\Sigma - U(q)\right]\,,
\label{action-intro}
\end{equation}
where $r$ is a radial variable (the flow direction), the $q^\Lambda(r)$ denote collectively the dynamical
variables, $U(q)$ is the potential and ${\mathscr G}_{\Lambda\Sigma}(q)$ the metric on the
target space parametrized by the $q^\Lambda$, with inverse ${\mathscr G}^{\Lambda\Sigma}$.
Now suppose that $U$ can be expressed in terms of a (fake) {\itshape superpotential} $W$ as
\begin{equation}
U = E - \frac12{\mathscr G}^{\Lambda\Sigma}\frac{\partial W}{\partial q^\Lambda}
\frac{\partial W}{\partial q^\Sigma}\,, \label{Ham-Jac-intro}
\end{equation}
where $E$ is a constant. Then, the action \eqref{action-intro} becomes
\begin{equation}
S = \int\D r\left[\frac12{\mathscr G}_{\Lambda\Sigma}\left(\dot q^\Lambda - {\mathscr G}^{\Lambda
\Omega}\frac{\partial W}{\partial q^\Omega}\right)\left(\dot q^\Sigma - {\mathscr G}^{\Sigma
\Delta}\frac{\partial W}{\partial q^\Delta}\right) + \frac{\D}{\D r}(W - Er)\right]\,,
\end{equation}
which is up to a total derivative equal to
\begin{equation}
S = \int\D r\,\frac12{\mathscr G}_{\Lambda\Sigma}\left(\dot q^\Lambda - {\mathscr G}^{\Lambda
\Omega}\frac{\partial W}{\partial q^\Omega}\right)\left(\dot q^\Sigma - {\mathscr G}^{\Sigma
\Delta}\frac{\partial W}{\partial q^\Delta}\right)\,.
\end{equation}
The latter is obviously stationary if the first-order flow equations
\begin{equation}
\dot q^\Lambda = {\mathscr G}^{\Lambda\Omega}\frac{\partial W}{\partial q^\Omega} \label{flow-intro}
\end{equation}
hold. But \eqref{Ham-Jac-intro} is nothing else than the reduced Hamilton-Jacobi equation, with
$W$ Hamilton's characteristic function, while \eqref{flow-intro} represents the expression for the
conjugate momenta
$p_\Lambda=\partial{\mathscr L}/\partial\dot q^\Lambda={\mathscr G}_{\Lambda\Sigma}\dot q^\Sigma$
in Hamilton-Jacobi theory\footnote{For further discussions of the relationship between the Hamilton-Jacobi
formalism and the first-order equations derived from a (fake) superpotential see \cite{Andrianopoli:2009je,Trigiante:2012eb}.}.

First-order flow equations, derived either by writing the dimensionally reduced action as
a sum of squares or from the Hamilton-Jacobi formalism, appear for many different settings in the
literature, both in ungauged and gauged supergravity, and for BPS-, extremal non-BPS- and even 
non-extremal black holes\footnote{See for example \cite{Miller:2006ay,Ceresole:2007wx,LopesCardoso:2007qid,
Cardoso:2008gm,Andrianopoli:2009je,Dall'Agata:2010gj,Galli:2011fq,Barisch:2011ui,Trigiante:2012eb,
BarischDick:2012gj,Klemm:2012vm,Gnecchi:2012kb,Gnecchi:2014cqa,Cardoso:2015wcf,Klemm:2016wng,Klemm:2017pxv}.}. In particular, in \cite{Ceresole:2001wi} the general
properties of supersymmetric flow equations for domain walls in five-dimensional $\ma N=2$ gauged
supergravity coupled to vector- and hypermultiplets are established.

In this section we want to apply this analysis to $\ma N=2$ gauged supergravities in $d=4, 5$ with general matter couplings. We will formulate two general static Ans{\"a}tze respectively in $d=4, 5$ with spherical/hyperbolic symmetry including in their possible concrete realizations black hole and black string solutions. Thus, we will use them to write the effective one-dimensional actions describing the correspondent flows and we will derive the first-order equations by using the Hamilton-Jacobi approach introduced above. In particular, in the $d=4$ case, both the situations with pure electric and dyonic gauging will be treated.

\subsection{The Effective Action of the Black Hole}
\label{effectivebh}

Let's construct the most general static field configuration with spherical or hyperbolic symmetry in four-dimensional $\ma N=2$ gauged supergravity \eqref{eq:mainaction} with general couplings to $n_V$ vector multiplets and $n_H$ hypermultiplets. For simplicity we will consider only purely electric gaugings on the quaternionic moduli space $\ma{HM}$. This defines a scalar potential of the form given in \eqref{eq:scal_pot} and the covariant derivatives \eqref{eq:hypercov4d}. 

Let's introduce the following Ansatz for the metric
\begin{equation}
\D s^2 = - e^{2U(r)} \D t^2 + e^{-2U(r)} \D r^2 + e^{2(\psi(r) - U(r))}\D\Omega_{\kappa}^2\,,\label{eq:Ansatzmet} 
\end{equation}
where $\D\Omega_{\kappa}^2=\D\theta^2+f_{\kappa}^2(\theta)\D\varphi^2$ is the metric on the two-dimensional surfaces $\Sigma_\kappa=\{S^2,H^2\}$ of constant scalar curvature
$R=2\kappa$, with $\kappa\in\{1,-1\}$, and
\begin{equation}
f_\kappa(\theta) = \frac{1}{\sqrt{\kappa}} \sin(\sqrt{\kappa}\theta) = 
\left\{\begin{array}{c@{\quad}l} \sin\theta\, & \kappa=1\,, \\                                             
                                             \sinh\theta\, & \kappa=-1\,. \end{array}\right.
\end{equation} 
The scalar fields depend only on the radial coordinate,
\begin{equation}
z^i = z^i(r)\,, \qquad q^u = q^u(r)\,,
\label{eq:Ansatzscalars}
\end{equation}
while the abelian gauge fields $A^\Lambda$ are given by
\begin{equation}
\label{eq:Ansatzgaugefields}
A^\Lambda = A^\Lambda_t(r)\D t - \kappa p^\Lambda f^{\prime}_{\kappa}(\theta)\D\phi\,.
\end{equation}
Their field strengths $F^\Lambda=\D A^\Lambda$ must have the form
\begin{equation}
F^\Lambda_{tr} = e^{2(U - \psi)} I^{\Lambda\Sigma}\left(R_{\Sigma\Gamma} p^\Gamma -
e_\Sigma(r)\right)\,, \qquad F^\Lambda_{\theta\phi} = p^\Lambda f_{\kappa}(\theta)\,, \label{eq:AnsatzF}
\end{equation}
where the magnetic and electric charges $(p^\Lambda, e_\Lambda)$ are defined as
\begin{equation}
p^{\Lambda} = \frac1{\mbox{vol}(\Sigma_{\kappa})}\int_{\Sigma_{\kappa}} F^{\Lambda}\,, \quad   e_{\Lambda}(r) = \frac1{\mbox{vol}(\Sigma_\kappa)}\int_{\Sigma_\kappa} G_{\Lambda}\,, \quad
\mbox{vol}(\Sigma_\kappa) = \int f_\kappa(\theta)\D\theta\wedge\D\phi\,,
\label{eq:charges}
\end{equation}
with $G_{\Lambda}$ given by \eqref{eq:Gdual}.
The metric \eqref{eq:Ansatzmet} is quite general and admits various particular realizations. As we said, we are interested in those describing static extremal four-dimensional black holes with spherical/hyperbolic symmetry, thus solutions presenting a singularity protected by an event-horizon and with zero temperature. Moreover we note that these black holes can be both supersymmetric or non-supersymmetric depending on their configuration of charges \cite{Ortin:1996bz}.

We point out that the electric charges can depend on the radial coordinate. This can be easily understood,
since the running hyperscalars are electrically charged, and thus contribute to the total electric charge
inside the 2-surfaces $\Sigma_\kappa(r)$ of constant $r$ and $t$.
In fact, the Maxwell equations written in \eqref{electriceom4d} can be written as
\begin{equation}
\partial_\mu(\sqrt{-g}\star_4\! G_\Lambda^{\phantom{\Lambda}\mu\nu})  = -2\sqrt{-g}\, h_{uv}
k^u_\Lambda\hat{\partial}^\nu q^v\,. \label{eq:max}
\end{equation}
Imposing the Ansatz (\ref{eq:Ansatzmet}), (\ref{eq:Ansatzscalars}) and (\ref{eq:Ansatzgaugefields}) on the
$t$-component, one obtains the radial variation of the electric charges,
\begin{equation}
e^\prime_{\Lambda} = -2e^{2\psi - 4U} h_{uv} k^u_\Lambda k^v_\Sigma A^\Sigma_t\,.
\label{eq:maxAnsatz-t}
\end{equation}
On the other hand, the magnetic charges are always constant as a consequence of the Bianchi identities
$\nabla_\nu\star_4 F^{\Lambda\mu\nu}=0$. Thus we explicitly verified that the gauging on the quaternioninc manifold breaks the electromagnetic duality as we explained in section \ref{symplecticcovariance}.

The equations of motion following from (\ref{eq:mainaction}) with the Ansatz (\ref{eq:Ansatzmet}), (\ref{eq:Ansatzscalars}) and (\ref{eq:Ansatzgaugefields}) can also be obtained from the effective action
\begin{equation}
S = \int\D r L = \int\D r\left[e^{2\psi}\left(U'^2 - \psi'^2 + h_{u v} q^{\prime\,u} q^{\prime\,v} +
g_{i\bar{\jmath}} z^{\prime\,i}\bar{z}^{\prime\,\bar{\jmath}}\right) + e_\Lambda A^{\prime\,\Lambda}_t
- V\right]\,, \label{eq:redlagrangian} 
\end{equation}
where $V$ is given by
\begin{equation}
V = - e^{2(U-\psi)}V_{\text{BH}} + e^{2\psi-4U} h_{uv} k^u_\Lambda k^v_\Sigma A^\Lambda_t
A^\Sigma_t + \kappa - e^{2(\psi-U)} V_g\,, \label{eq:V1d}
\end{equation}
with $V_{\text{BH}}$ to be defined below.
In addition to the equations of motion following from (\ref{eq:redlagrangian}), one has to impose the
Hamiltonian constraint
\begin{equation}
H = L - e_\Lambda A^{\prime\,\Lambda}_t + 2V = 0\,, \label{eq:Hamcons} 
\end{equation}
the $\varphi$-component of the Maxwell equations\footnote{Plugging the
spherical/hyperbolic Ansatz into the $\varphi$-component of the Maxwell equations, one obtains
$p^\Lambda k_\Lambda^u k_{u\Sigma}=0$, which implies (\ref{eq:maxAnsatz-phi}). The
$\theta$-component is trivial.} (\ref{eq:max}),
\begin{equation}
p^\Lambda k_\Lambda^u = 0\,, \label{eq:maxAnsatz-phi}
\end{equation}
as well as the $r$-component
\begin{equation}
k_{\Lambda u} q'^u=0\,. \label{eq:maxAnsatz-r}
\end{equation}
The effective potential $V$ is determined by the scalar potential $V_g$, the charge-dependent {\itshape black hole
potential} $V_{\text{BH}}$, and by a contribution coming from the covariant derivatives of the hyperscalars
plus a constant term depending on the scalar curvature $\kappa$. In particular, $V_{\text{BH}}$ can be
written in the symplectically covariant form
\begin{equation}
V_{\text{BH}} = -\frac12\ma{Q}^T\ma{M}\ma{Q}\,, \qquad \ma{Q} = \left(\begin{array}{c}
p^\Lambda \\ e_\Lambda \end{array}\right)\,.
\label{Vbh}
\end{equation}
Notice that the effective action (\ref{eq:redlagrangian}) does not result by merely substituting the Ansatz
(\ref{eq:Ansatzmet}), (\ref{eq:Ansatzscalars}), (\ref{eq:Ansatzgaugefields}) into the general action
(\ref{eq:mainaction}). This can be seen from $V_{\text{BH}}$ in (\ref{eq:V1d}), that does not arise by
rewriting the kinetic terms of the gauge fields. In fact it is easy to see that the gauge fields enter the
equations of motion of the whole system via their stress-energy tensor, whose components are expressed
in terms of $V_{\text{BH}}$ \cite{Ferrara:1996dd,Bellucci:2008cb,Chimento:2015rra}.

In this sense, the presence of the term $e_\Lambda A_t^{\prime\,\Lambda}$ is necessary for having the right dynamics of the variables $e_\Lambda$ and $A_t^\Lambda$. Indeed, varying the effective action
(\ref{eq:redlagrangian}) with respect to $A_t^\Lambda$, one obtains exactly (\ref{eq:maxAnsatz-t}). Variation
with respect to $e_\Lambda$ yields
\begin{equation}
A^{\prime\,\Lambda}_t = - e^{2(U-\psi)} I^{\Lambda\Sigma}(R_{\Sigma\Gamma}p^\Gamma - e_\Sigma(r))\,,
\label{eq:max1}
\end{equation}
which is exactly the expression \eqref{eq:AnsatzF} for the $(t,r)$-component of $F^{\Lambda\mu\nu}$. 

Introducing
\begin{equation}
\ma{H}_{\Lambda\Sigma} = k^u_\Lambda h_{uv} k^v_\Sigma\,, \label{eq:matriceH}
\end{equation}
(\ref{eq:maxAnsatz-t}) becomes 
\begin{equation}
e^{\prime}_{\Lambda} = - 2 e^{2\psi-4U}\ma{H}_{\Lambda\Sigma}A^\Sigma_t\,,
\label{eq:max2}
\end{equation}
which allows to express $A^\Sigma_t$ in terms of the other fields as follows. Since
$\ma{H}_{\Lambda\Sigma}$ is real and symmetric, there exists a matrix $O\in\text{O}(n_V+1)$ such that
\begin{equation}
\ma{H}_{\Lambda\Sigma} = (O^TDO)_{\Lambda\Sigma} = O^\Omega{}_\Lambda O^\Gamma{}_\Sigma
D_{\Omega\Gamma}\,,
\end{equation}
with $D$ diagonal. Without loss of generality, suppose that the first $n$ eigenvalues of $D$ are
nonvanishing ($0\le n\le n_V+1$), while the remaining ones are zero. Let hatted indices
$\hat\Lambda,\hat\Sigma,\ldots$ range from $0$ to $n-1$, and define
\begin{equation}
{\hat A}^\Gamma_t = O^\Gamma{}_\Sigma A^\Sigma_t\,.
\end{equation}
\eqref{eq:max2} yields then
\begin{equation}
O_{\hat\Psi}{}^\Lambda e^{\prime}_\Lambda = -2 e^{2\psi-4U}D_{\hat\Psi\hat\Gamma}
{\hat A}^{\hat\Gamma}_t\,, \label{eq:Oeprime}
\end{equation}
where indices are raised and lowered with the flat metric, i.e.
$O_\Psi{}^\Lambda=\delta_{\Psi\Omega}\delta^{\Lambda\Gamma}O^\Omega{}_\Gamma$.
We also get
\begin{equation}
O^{\Psi\Lambda} e^{\prime}_{\Lambda} = 0 \quad \text{for} \quad \Psi\ge n\,. \label{eq:Oeprime2}
\end{equation}
\eqref{eq:Oeprime} gives
\begin{equation}
{\hat A}^{\hat\Lambda}_t = -\frac12 e^{4U-2\psi} {(D^{-1})}^{\hat\Lambda\hat\Psi}
O_{\hat\Psi}{}^\Lambda e^{\prime}_{\Lambda}\,. \label{eq:A}
\end{equation}
Using these relations in the effective action (\ref{eq:redlagrangian}) to eliminate $A^\Sigma_t$, one obtains
\begin{equation}
S = \int\D r\left[e^{2\psi}(U^{\prime\,2} - \psi^{\prime\,2} + h_{u v} q^{\prime\,u} q^{\prime\,v} +
g_{i\bar{\jmath}} z^{\prime\,i}\bar{z}^{\prime\,\bar{\jmath}} + \frac14 e^{4(U-\psi)}
\ma{H}^{\Lambda\Sigma} e'_{\Lambda} e'_{\Sigma}) - \tilde{V}\right]\,, \label{eq:Seff}
\end{equation}
where we defined the effective potential
\begin{equation}
\tilde{V} = -e^{2(U-\psi)} V_{\text{BH}} + \kappa - e^{2(\psi-U)} V_{g}\,, \label{eq:effpot}
\end{equation}
as well as
\begin{equation}
\ma{H}^{\Lambda\Sigma} = O_{\hat\Lambda}{}^\Lambda (D^{-1})^{\hat\Lambda\hat\Sigma}
O_{\hat\Sigma}{}^\Sigma\,. \label{eq:H-upper}
\end{equation}
Note that, unless $n=n_V+1$, $\ma{H}^{\Lambda\Sigma}$ is not the inverse of
$\ma{H}_{\Lambda\Sigma}$ (which is not invertible), but we have the weaker relation
\begin{equation}
\ma{H}^{\Lambda\Gamma}\ma{H}_{\Lambda\Sigma}\ma{H}_{\Gamma\Omega} = \ma{H}_{\Sigma\Omega}\,,
\label{eq:HHH}
\end{equation}
that will be used below to square the action.

One can then rewrite the constraint \eqref{eq:Hamcons} in terms of the effective Hamiltonian
\begin{equation}
H = \frac14 e^{-2\psi} p_U^2 - \frac14 e^{-2\psi} p_{\psi}^2+ \frac14 e^{-2\psi}h^{u v} p_{q^u} p_{q^v}
+ e^{-2\psi}g^{i\bar{\jmath}} p_{z^i} p_{\bar{z}^{\bar{\jmath}}} + e^{4(U-\psi)}\ma{H}^{\Lambda\Sigma}
p_{e_{\Lambda}} p_{e_{\Sigma}} + \tilde{V}\,,
\label{eq:Heff}
\end{equation}
where the canonical momenta $p_U$, $p_\psi$, $p_{q^u}$, $p_{z^i}$, $p_{\bar z^{\bar\jmath}}$
and $p_{e_\Lambda}$ are defined in the usual way.
The effective action (\ref{eq:Seff}), together with the relations \eqref{eq:Hamcons}, \eqref{eq:maxAnsatz-phi},
\eqref{eq:maxAnsatz-r}, reproduce the complete set of equations of motion for the spherical/hyperbolic
Ansatz (\ref{eq:Ansatzmet}), (\ref{eq:Ansatzscalars}) and (\ref{eq:Ansatzgaugefields}).

\subsection{Extension to Dyonic Gaugings}
\label{magneticAnsatz}

Let's extend the analysis presented in section \ref{effectivebh} to the case of general dyonic gauging on the quaternionic moduli space where the action of the general $\ma N=2$ theory is now given by \eqref{actionmag}. This leads to an effective action enjoying the global symplectic invariance.

Keeping valid the Ans{\"a}tze on the metric \eqref{eq:Ansatzmet} and on the scalars \eqref{eq:Ansatzscalars}, we have to modify the form of the gauge fields since in the action \eqref{actionmag} now appear the auxiliary 2-form $B_a$ and the magnetic gauge fields $A_{\Lambda}$. In particular the most general Ans{\"a}tze for the gauge fields and for the 2-form $B_a$, compatible with their equations of motion \eqref{eq:eommag1}, have the form
\begin{equation}
A^\Lambda = A_t^\Lambda\D t- \kappa p^\Lambda f^\prime_\kappa(\theta)\D\phi\,, \qquad
A_\Lambda = A_{\Lambda t}\D t - \kappa e_\Lambda f^\prime_\kappa(\theta)\D\phi\,,
\label{Ansatzmag1}
\end{equation}
\begin{equation}
B^\Lambda = 2\kappa p^{\prime\,\Lambda} f^\prime_\kappa(\theta)\D r\wedge\D\phi\,, \qquad 
B_\Lambda = -2\kappa e^{\prime}_\Lambda f^\prime_\kappa(\theta)\D r\wedge\D\phi\,,
\label{Ansatzmag2}
\end{equation}
which implies for the field strengths
\begin{equation}
H^\Lambda{}_{tr} = e^{2(U - \psi)} I^{\Lambda\Sigma}(R_{\Sigma\Gamma} p^\Gamma - e_\Sigma)\,,
\qquad H^\Lambda{}_{\theta\phi} = p^\Lambda f_\kappa(\theta)\,,
\end{equation}
\begin{equation}
G_{\Lambda tr} = e^{2(U - \psi)}\left(I_{\Lambda\Sigma} p^\Sigma + R_{\Lambda\Gamma}
I^{\Gamma\Omega} R_{\Omega\Sigma} p^\Sigma - R_{\Lambda\Gamma} I^{\Gamma\Omega}
e_\Omega\right)\,, \qquad G_{\Lambda\theta\phi} = e_\Lambda f_\kappa(\theta)\,.
\label{eq:AnsatzH}
\end{equation}
Introducing the symplectic matrix
\begin{equation}
\ma H = (\mathcal K^u)^T h_{uv}\mathcal K^v\,,
\end{equation}
and plugging the above Ansatz into (\ref{eq:eommag1}), one obtains
\begin{equation}
\mathcal A^\prime_t = - e^{2(U - \psi)}\Omega\mathcal M\mathcal Q\,, \qquad
\mathcal Q^\prime = -2 e^{2\psi - 4U}\mathcal H\Omega\mathcal A_t\,, \label{eq:maxsim} 
\end{equation}
where the constraints
\begin{equation}
\ma H\Omega\ma Q = 0\,, \qquad \mathcal K_u q^{\prime u} = 0
\label{eq:symcon}
\end{equation}
have been imposed. 

Following the same procedure used previously for pure electric gaugings, one finds the effective
action that generalizes (\ref{eq:Seff}),
\begin{equation}
\begin{split}
&S = \int\D r\left[e^{2\psi}(U^{\prime\,2} - \psi^{\prime\,2} + h_{u v} q^{\prime\,u} q^{\prime \, v} +
g_{i\bar{\jmath}}\,z^{\prime\, i}\bar{z}^{\prime\,\bar{\jmath}} + \frac14 e^{4(U - \psi)}
{\ma Q^\prime}^T\ma{H}^{-1}\ma Q^\prime) - \tilde{V}\right]\,, \\
&\tilde{V} = -e^{ 2(U - \psi)} V_{\text{BH}}  + \kappa - e^{2(\psi - U)} V_{g}\,,
\label{eq:Seffmag}
\end{split}
\end{equation}
where, in a slight abuse of notation, $\ma H^{-1}$ denotes the symplectically covariant generalization
of the matrix $\ma H^{\Lambda\Sigma}$ defined by \eqref{eq:H-upper}. (Note that one has not
necessarily $\ma H^{-1}\ma H=\mathbb{I}$, as we explained in section \ref{effectivebh}, but
$\ma H^{-1}$ in \eqref{eq:Seffmag} can be defined in a way similar to \eqref{eq:H-upper}).

We conclude this section by deriving a special form for the symplectic invariant scalar potential $V_g$ written in \eqref{eq:Vg-sympl-cov}. In section \ref{lagrangians4d} we have shown that it can be recasted in the particular form given in \eqref{eq:superV} in terms of the $\mrm{SU(2)}$ triplet of functions $\ma W^x$ defined in \eqref{tripletsuperpotential}.

Once spherical or hyperbolic symmetry has been imposed on the $\ma N=2$ fields, it is possible to show that \eqref{eq:superV} can be further re-expressed in terms of a particular symplectic invariant complex function $\ma L$ entering in the potential only through its square and the square of its derivatives. This function is crucial for the Hamilton-Jacobi formulation of first order flows since it is involved in the definition of the superpotential $W$.

Let's define the symplectic invariant function $\ma L$ as
\begin{equation}
\ma L = \ma Q^x\ma W^x = \langle \ma Q^x \ma P^x, \ma V \rangle 
\label{Lsuperpotential}
\end{equation}
where
\begin{equation}
Q^x= \langle \ma P^x, \ma Q \rangle\,.
\end{equation}
If we require spherical/hyperbolic invariance, the scalar potential (\ref{eq:superV}) can be rewritten in a
way analogous to \cite{Dall'Agata:2010gj}. Namely, using (\ref{eq:maxAnsatz-phi}), the quaternionic
relations (\ref{eq:quaternionic_kahler_complexstruct_definition}), (\ref{eq:quaternionkahl_su2curv}),
(\ref{eq:mommaps}), (\ref{eq:equivariance}) and imposing\footnote{Notice that
$\partial_\mu(\ma Q^x\ma Q^x)=\partial_u(\ma Q^x\ma Q^x)\partial_\mu q^u$, and
$\partial_u(\ma Q^x\ma Q^x)=\masf{D}_u(\ma Q^x\ma Q^x)=2\ma{Q}^x\masf{D}_u\ma{Q}^x$. Using
the definition of $\ma Q^x$ together with \eqref{eq:mommaps}, this is equal to
$-4\ma{Q}^x p^\Lambda\Omega^x{}_{uv}k^v_\Lambda$, which vanishes by virtue
of \eqref{eq:maxAnsatz-phi}. $\ma Q^x\ma Q^x$ is thus a constant of motion, that we choose to be one.} a sort of quantization condition given by
\begin{equation}
\ma Q^x\ma Q^x =1\,,
\label{eq:quantization}
\end{equation}
one can show that the scalar potential (\ref{eq:superV}) can be expressed in terms of $\ma L$ as
\begin{equation}
V_g = \mathbb G^{AB}\mathbb D_A\ma L\,\mathbb D_B\bar{\ma L} - 3|\ma L|^2\,, \label{eq:superV1}
\end{equation}
where
\begin{equation}
\mathbb G^{AB} = \left(\begin{array}{cc}
g^{i\bar\jmath} & 0 \\ 0 & h^{uv} \end{array}\right)\,, \qquad
\mathbb D_A = \left(\begin{array}{l} D_i \\ \masf{D}_u \end{array}\right)\,.
\label{eq:globalmetric}
\end{equation}
We note that if the magnetic Killing vectors and moment maps are zero, the expression \eqref{eq:superV1} becomes the scalar potential in the case of pure electric gaugings and spherical/hyperbolic symmetry.

Furthermore it is interesting to consider the case of $\mrm{FI}$ gauging \eqref{4dFImagnetic}. If one chooses the moment maps\footnote{This choice can be made since the theory is invariant under $\mrm{SU(2)}$.} as 
\begin{equation}
\ma P^1=0\,, \qquad \ma P^2=0\,, \qquad \ma P^3= \ma G\,.
\label{FIchoice}
\end{equation}
With these particular moment maps, the function \eqref{Lsuperpotential} becomes
\begin{equation}
\ma L=-\kappa\,\langle \ma G, \ma V \rangle\,
\label{FIsuperpotential}
\end{equation}
and the quantization condition \eqref{eq:quantization} collapses to
\begin{equation}
\langle \ma G, \ma Q \rangle=-\kappa\,,
\end{equation}
that is the well-know symplectic covariant quantization condition with $\mrm{FI}$ gauging \cite{Dall'Agata:2010gj}.

Finally, considering a generic $\mrm{SU(2)}$ frame, one can show that the general prescription  
\begin{equation}
-\kappa \, Q^x \ma P^x \longrightarrow \ma G
\label{FIprescription}
\end{equation}
reproduces exactly the $\mrm{FI}$ limit \cite{Dall'Agata:2010gj} in case of abelian gauging, spherical/hyperbolic symmetry and constant hyperscalars.

\subsection{Flow Equations for the BPS Black Hole}
\label{Flow Equations}

We want now to find the first-order equations for the effective action
(\ref{eq:Seff}) by using the Hamilton-Jacobi approach \cite{Andrianopoli:2009je,Trigiante:2012eb} in the case of a pure electric gauging. So we come back to section \ref{effectivebh} and we consider the effective action \eqref{eq:Seff}.
Introducing a superpotential $W$ associated to (\ref{eq:Seff}), we are going to write the action as a sum of squares from which one can derive the flow equations.

The first observation is that the effective potential \eqref{eq:effpot} contains not only $V_g$, and thus the superopotential $W$ (that solves the ``time" (i.e. $r$)-independent Hamilton-Jacobi equation) must contain
also other contributions in addition to $\ma L$ given in \eqref{Lsuperpotential}. This happens both in the case without hypermultiplets
and $\text{U}(1)$ Fayet-Iliopoulos gauging \cite{Dall'Agata:2010gj,Klemm:2012vm}, and with hyperscalars running.

In the last case the general structure of the effective action remains essentially
the same except for the presence of some new kinetic terms. The main difference is the internal form of the scalar
potential $V_g$,  which is now governed by the gauging on the quaternionic manifold and thus on the moment maps. Guided by these observations we
introduce the real function
\begin{equation}
W = e^U |\mathcal Z  + i\kappa e^{2\psi - 2U} \ma L|\,,
\label{eq:superpotential}
\end{equation}
and a phase $\alpha$ defined by
\begin{equation}
e^{2i \alpha} = \frac{\mathcal Z  + i\kappa e^{2(\psi - U)}\mathcal L}{\bar{\mathcal Z} - i\kappa
e^{2(\psi - U)}\bar{\mathcal L}}\,, \qquad \mbox{or} \qquad \mathrm{Im}(e^{-i\alpha}\mathcal Z) =
- \kappa e^{2(\psi - U)}\mathrm{Re}(e^{-i\alpha}\mathcal L)\,, \label{eq:defal}
\end{equation}
where $\ma Z= \langle\ma Q,\ma V\rangle$ is the central charge and $\ma L$ has the form given in \eqref{Lsuperpotential} in terms of only electric moment maps $P^x_\Lambda$. Defining ``tilded" variables by
$\tilde{\ma X}=e^{-i\alpha}\ma X$ etc., we can rewrite $W$ as
\begin{equation}
W = e^U\mathrm{Re}\tilde{\ma Z} - \kappa e^{2\psi - U}\mathrm{Im}\tilde{\ma L}\,.
\label{eq:Wnd}
\end{equation}
Using (\ref{eq:sympid}), (\ref{eq:mommaps}) and (\ref{eq:quantization}), it is possible to show that
\begin{eqnarray}
&& e^{-2\psi}\left((\partial_U W)^2 - (\partial_\psi W)^2 + 4 g^{i\bar{\jmath}}\partial_i W
\partial_{\bar{\jmath}} W +  h^{uv}\partial_u W\partial_v W + 4 e^{4(\psi - U)}\ma H_{\Lambda\Sigma}
\partial_ {e_\Lambda} W\partial_{e_\Sigma} W\right) \nonumber \\
&& \qquad -e^{2(\psi - U)}  V_g -  e ^{2(U - \psi)} V_{\text{BH}} + \kappa=0\,, \label{eq:sqpot}
\end{eqnarray}
or, in other words, that $2W$ solves the Hamilton-Jacobi equation associated to the Hamiltonian (\ref{eq:Heff}) with zero energy. By virtue of \eqref{eq:sqpot}, up to a total derivative, the action
(\ref{eq:Seff}) can be written as
\begin{equation}
\begin{split}
S &= \int\D r \Bigl[ e^{2\psi} \bigl(U^\prime + e^{-2 \psi }\partial_U W\bigl)^2 
- e^{2\psi} \bigl(\psi^\prime - e^{-2\psi}\partial_\psi W\bigl)^2 + \\
&e^{2\psi }g_{i\bar\jmath}\bigl(z^{\prime\,i} + 2e^ {-2\psi} g^{i\bar{k}}\partial_{\bar{k}} W\bigl) 
\bigl(\bar{z}^{\prime\,\bar{\jmath}} + 2e^ {-2\psi} g^{\bar{\jmath}\,l} \partial_l W\bigl) + \\
& e^{2\psi} h_{uv} \bigl(q^{\prime\,u} + e^{-2\psi} h^{us}\partial_s W\bigl)
\bigl(q^{\prime\,v} + e^{-2\psi} h^{vt}\partial_t W\bigl) + \\
&\frac14 e^{4U - 2\psi}\ma H^{\Lambda\Gamma}\bigl(e^\prime_\Lambda + 4e^{2\psi- 4U}
\ma H_{\Lambda\Sigma} \partial_{e_{\Sigma}} W\bigl)\bigl(e^\prime_{\Gamma} + 4e^{2\psi- 4U}
\ma H_{\Gamma\Omega}\partial_{e_{\Omega}} W\bigl)\Bigl]\,,
\end{split} \label{eq:BPSaction}
\end{equation}
where we used also \eqref{eq:HHH} and the fact that \eqref{eq:Oeprime2} implies
\begin{equation}
\ma H^{\Lambda\Gamma}\ma H_{\Lambda\Sigma}\partial_{e_\Sigma}W e^{\prime}_\Gamma
= \partial_{e_\Gamma} W e^{\prime}_\Gamma\,.
\end{equation}
The BPS-rewriting \eqref{eq:BPSaction} guarantees that the solutions of the first-order
equations obtained by setting each quadratic term to zero do indeed extremize the action.
If one explicitly computes the derivatives of $W$,
these first-order flow equations become
\begin{equation}
\begin{split}
&U^\prime = -e^{U-2\psi}\mathrm{Re}\tilde{\mathcal Z} - \kappa e^{-U}\mathrm{Im}\tilde{\mathcal L}\,, \\
&\psi^\prime = -2\kappa e^{-U}\mathrm{Im}\tilde{\mathcal L}\,, \\
&z^{\prime\,i} = -e^{i\alpha} g^{i\bar{\jmath}}\left(e^{U-2\psi} D_{\bar\jmath}\bar{\mathcal Z} - i\kappa
e^{-U} D_{\bar\jmath}\bar{\mathcal L} \right)\,, \\
&q^{\prime\,u} = \kappa e^{-U} h^{uv}\mathrm{Im}(e^{-i\alpha}\partial_v\mathcal L)\,, \\
&e^\prime_\Lambda = -4 e^{2\psi-3U}\ma H_{\Lambda\Sigma}\mathrm{Re}\tilde{L}^\Sigma\,.
\end{split} \label{eq:flow-equ-electr}
\end{equation}
These relations, plus the constraints that we had to impose, are equivalent to those obtained in
\cite{Halmagyi:2013sla}\footnote{\eqref{eq:flow-equ-electr} corrects some sign errors in appendix B of
\cite{Halmagyi:2013sla}.} from the Killing spinor equations. To see this, note that comparing
the expression for $e^{\prime}_\Lambda$ in \eqref{eq:flow-equ-electr} with (\ref{eq:max2}) yields
the additional condition
\begin{equation}
2e^U\ma H_{\Lambda\Sigma}\mathrm{Re}\tilde{L}^\Sigma\ = \ma{H}_{\Lambda\Sigma}
A^\Sigma_t\,, \label{eq:ReL-A}
\end{equation}
which is just (B.44) of \cite{Halmagyi:2013sla} contracted with $h_{uv}k^v_\Sigma$. To be precise,
\eqref{eq:ReL-A} is equivalent to
\begin{equation}
2 e^U k^u_\Lambda\mathrm{Re}\tilde{L}^\Lambda = k^u_\Lambda A^\Lambda_t + m^u\,,
\end{equation}
where $m^u$ must satisfy $k^v_\Sigma h_{uv}m^u = 0\,\,\forall\,\Sigma$. If $n^\Sigma$ is an
eigenvector of $\ma H_{\Lambda\Sigma}$ with zero eigenvalue, i.e.,
$\ma H_{\Lambda\Sigma}n^\Sigma=0$, then we can take the linear combination
$m^u=k^u_\Lambda n^\Lambda$. (B.44) of \cite{Halmagyi:2013sla} has $m^u=0$, and is thus
slightly stronger than \eqref{eq:ReL-A}. Notice also that the number of independent constraints
coming from \eqref{eq:ReL-A} is equal to $n$, where $n$ denotes the number of nonvanishing
eigenvalues of $\ma H_{\Lambda\Sigma}$. This becomes evident by casting \eqref{eq:ReL-A} into
the form
\begin{equation}
2 e^U D_{\Omega\Gamma}O^\Gamma{}_\Sigma\mathrm{Re}\tilde{L}^\Sigma =
D_{\Omega\Gamma}\hat A^\Gamma_t\,.
\end{equation}
The auxiliary field $\alpha$ is related to the phase of the Killing spinor associated to the BPS solution,
as was shown for the case without hypers and $\text{U}(1)$ Fayet-Iliopoulos gauging
in \cite{Dall'Agata:2010gj}\footnote{Without hypermultiplets and for $\text{U}(1)$ FI gauging
one can always make the choice \eqref{FIchoice}. The superpotential $W$ boils down to equ.~(2.40) of \cite{Dall'Agata:2010gj} for $\kappa=1$.}, and for
the case including hypermultiplets in \cite{Halmagyi:2013sla}.

Finally, since the equations \eqref{eq:flow-equ-electr} describe extremal configurations, there exists an
additional constant of motion $\mathbb Q$ \cite{Trigiante:2012eb} such that
\begin{equation}
\frac {\D\mathbb Q}{\D r} = H = 0\,.
\end{equation}
Using the first order equations for $U$ and $\psi$, one gets from \eqref{eq:sqpot}
\begin{equation}
\mathbb Q = e^{2 \psi }(U' - \psi') + W\,.
\end{equation}

\subsection{Hamilton-Jacobi and Symplectic Covariance}
\label{symplecticallyinvariant}

Let's extend the Hamilton-Jacobi analysis to magnetic gaugings. We want to obtain a set of $\mrm{BPS}$ first-order equations that are invariant under global symplectic transformations and that generalize \eqref{eq:flow-equ-electr}. Furthermore we will see how to produce a system of non-$\mrm{BPS}$ flow equations through the action of a global symplectic transformations on the charges $\ma Q$ imposed at the level of the $\mrm{BPS}$ flow equations obtained. 

We start by repeating the procedure that has been performed in section \ref{Flow Equations} by considering the Ans{\"a}tze on the gauge fields given in \eqref{Ansatzmag1} and \eqref{Ansatzmag2} and the covariant effective action \eqref{eq:Seffmag}. 

Introducing the function $W$ and the phase $\alpha$ as in (\ref{eq:superpotential}) and (\ref{eq:defal}),
with $\mathcal L$ given by \eqref{Lsuperpotential}, it is straightforward to show that $W$
satisfies the Hamilton-Jacobi equation for the action (\ref{eq:Seffmag}),
\begin{eqnarray}
&& e^{-2\psi}\left((\partial_U W)^2 - (\partial_\psi W)^2 + 4 g^{i\bar{\jmath}}\partial_i W
\partial_{\bar{\jmath}} W +  h^{uv}\partial_u W \partial_v W + 4 e^{4(\psi - U)}({\partial_ {\ma Q}
W})^T\ma H\partial_{\ma Q }W\right) \nonumber \\
&& \qquad -e^{2(\psi - U)}  V_g - e^{2(U - \psi)} V_{\text{BH}} + \kappa = 0\,, \label{eq:sqpotmag}
\end{eqnarray}
provided the charge-quantization condition \eqref{eq:quantization} holds, with
$\mathcal Q^x = \langle\mathcal P^x, \mathcal Q\rangle$. Using \eqref{eq:sqpotmag} as well as
\eqref{eq:cons} and discarding total derivatives, the action (\ref{eq:Seffmag}) can be cast into the form
\begin{equation}
\begin{split}
S &= \int\D r\Bigl[e^{2\psi}\bigl(U^\prime + e^{-2\psi }\partial_U W\bigl)^2
- e^{2\psi}\bigl(\psi^\prime - e^{-2\psi}\partial_\psi W\bigl)^2 + \\
&e^{2\psi} g_{i\bar{\jmath}}\bigl(z^{\prime\,i} + 2 e^ {-2\psi} g^{i\bar{k}}\partial_{\bar{k}} W\bigl)
\bigl(\bar{z}^{\prime\,\bar{\jmath}} + 2 e^ {-2\psi} g^{\bar{\jmath}l}\partial_l W\bigl) + \\
& e^{2\psi} h_{uv}\bigl(q^{\prime\,u} + e^{-2\psi} h^{us}\partial_s W\bigl)
\bigl(q^{\prime\,v} + e^{-2\psi} h^{vt}\partial_t W\bigl) + \\
&\frac14 e^{4U - 2\psi}\bigl(\ma Q^\prime + 4 e^{2\psi- 4U}\ma H\partial_{\ma Q} W\bigl)^T
\ma H^{-1}\bigl(\ma Q^\prime + 4 e^{2\psi- 4U}\ma H\partial_{\ma Q} W\bigl)\Bigl]\,.
\label{eq:magsquaring}
\end{split}
\end{equation}
All first-order equations following from \eqref{eq:magsquaring} except the one for $z^i$ are 
symplectically covariant. Computing explicitely $\partial_{\bar k}W$, the latter reads
\begin{equation}
z^{\prime\,i} = - e^{i\alpha} g^{i\bar{\jmath}}\left(e^{U-2\psi} D_{\bar\jmath}\bar{\mathcal Z}
- i\kappa e^{-U} D_{\bar\jmath}\bar{\mathcal L}\right)\,.
\end{equation}
Contracting this with $D_i\ma V$ and using \eqref{eq:sympid}, one obtains a symplectically
covariant equation for the section $\ma V$,
\begin{eqnarray}
\mathcal V^\prime + i \masf A_r\mathcal V &=& e^{i\alpha} e^{U - 2\psi}\left(-\frac12\Omega\mathcal M
\mathcal Q - \frac i2\mathcal Q + \bar{\mathcal V}\mathcal Z\right) \nonumber \\
&& - i\kappa e ^{i\alpha} e^{-U}\left(-\frac12\Omega\mathcal M\mathcal P^x\mathcal Q ^x -
\frac i2\mathcal P^x\mathcal Q^x + \bar{\mathcal V}\mathcal L\right)\,,
\end{eqnarray}
where $\masf A_r=\mathrm{Im}(z^{\prime\,i}\partial_i\mathcal K)$ is the K\"ahler connection \eqref{U1connection}.
Calculating the remaining derivatives of $W$, the first-order flow equations become
\begin{equation}
\begin{split}
&U^\prime = -e^{U-2\psi}\mathrm{Re}\tilde{\mathcal Z} - \kappa e^{-U}\mathrm{Im}\tilde{\mathcal L}\,, \\
&\psi^\prime = -2\kappa e^{-U}\mathrm{Im}\tilde{\mathcal L}\,, \\
&q^{\prime\,u} = \kappa e^{-U} h^{uv}\mathrm{Im}(e^{-i\alpha}\partial_v \mathcal L)\,, \\
&\ma Q^\prime = -4 e^{2\psi - 3U}\mathcal H\Omega\mathrm{Re}\tilde{\mathcal V}\,, \\
&\mathcal V^\prime  = e^{i\alpha} e^{U - 2\psi}\left(-\frac12\Omega\mathcal M\mathcal Q -
\frac i2\mathcal Q + \bar{\mathcal V}\mathcal Z\right) \\
&- i\kappa e^{i\alpha} e^{-U}\left(-\frac12\Omega\mathcal M\mathcal P^x\mathcal Q^x -
\frac i2\mathcal P^x\mathcal Q^x + \bar{\mathcal V}\mathcal L\right) - i \masf A_r\mathcal V\,.
\label{eq:flowmag}
\end{split}
\end{equation}
These equations have a more useful form if one consider the phase $\alpha$ as a dynamical variable. 
Introducing the quantity $\mathcal S=\mathcal Z + i\kappa e^{2(\psi - U)}\mathcal L$, the relations (\ref{eq:defal}) and (\ref{eq:Wnd}) can be rewritten as
\begin{equation}
e^{2i\alpha} = \frac{\mathcal S}{\bar{\mathcal S}}\,, \qquad \mathrm{Im}(e^{-i\alpha}\mathcal S) = 0\,,
\qquad W = e^U\mathrm{Re}(e^{-i\alpha}\mathcal S)\,, \qquad W^2 = e^{2U}\mathcal S\bar{\mathcal S}\,.
\end{equation} 
One has thus
\begin{equation}
\alpha^\prime = \frac{\mathrm{Im}(e^{-i\alpha}\mathcal S^\prime)}{e^{-U} W}\,, \qquad \mathcal S^\prime
= U^\prime\partial_U\mathcal S + \psi^\prime\partial_\psi\mathcal S + \mathcal V^\prime
\partial_{\mathcal V}\mathcal S + q^{\prime\,u}\partial_u\mathcal S + \mathcal Q^{\prime\,T}
\partial_{\mathcal Q}\mathcal S\,.
\end{equation}
Inserting (\ref{eq:flowmag}) and the derivatives of $\ma S$ in this last expression, one gets
\begin{equation}
\alpha' + \masf A_r = 2\kappa e^{-U}\mathrm{Re}(e^{-i\alpha}\mathcal L)\,.
\end{equation}
Finally, plugging the equation for $U$ into the expression of $\mathrm{Im}\tilde{\ma V}^\prime$, one
can write the first-order flow equations in the form\footnote{Some comments on the limit of flat horizons ($\kappa=0$) are in order.
This case was not considered above, where we took $\kappa=\pm 1$ only. For $\kappa=0$, taking
(as in \cite{Cardoso:2015wcf}) $\ma P^1=\ma P^2=\ma Q^3=0$, one can again write the action
as a sum of squares, now with the Hamilton-Jacobi function $W=e^U|\mathcal Z-ie^{2(\psi-U)}\ma W^3|$.
The resulting first-order equations agree then, for pure electric gauging, precisely with those derived
in \cite{Cardoso:2015wcf}. (Note that in \cite{Cardoso:2015wcf} only electric gaugings are considered, and it is not identified the ``superpotential" that drives their first-order flow)}
\begin{eqnarray}
&& 2 e^{2\psi}\left(e^{-U}\mathrm{Im}(e^{-i\alpha}\mathcal{V})\right)^{\prime} - \kappa e^{2(\psi -
U)}\Omega\mathcal{M}\ma Q^x\mathcal{P}^x + 4 e^{2\psi-U}(\alpha^{\prime} + \masf A_r)\mathrm{Re}
(e^{-i\alpha}\mathcal{V}) + \mathcal{Q} = 0\,, \nonumber \\
&& \psi^{\prime} = -2\kappa e^{-U}\mathrm{Im}(e^{-i\alpha}\mathcal{L})\,, \nonumber \\
&& \alpha^{\prime} + \masf A_r = 2\kappa e^{-U}\mathrm{Re}(e^{-i\alpha}\mathcal{L})\,, \nonumber \\
&& q^{\prime\,u} = \kappa e^{-U} h^{uv}\mathrm{Im}(e^{-i\alpha}\partial_v\mathcal L)\,, \nonumber \\
&& \ma Q^\prime = -4 e^{2\psi - 3U}\mathcal H\Omega\mathrm{Re}\tilde{\mathcal V}\,,
\label{eq:flowequations}
\end{eqnarray}
where also (\ref{eq:symcon}) and (\ref{eq:quantization}) must hold together with
\begin{equation}
2 e^U\mathcal H\Omega\mathrm{Re}\tilde{\mathcal V} = \ma H\Omega\ma A_t\,,
\label{eq:relalg}
\end{equation} 
since the last equation of \eqref{eq:flowequations} has to coincide with \eqref{eq:maxsim}. Then \eqref{eq:relalg}
is the symplectically covariant generalization of the constraint \eqref{eq:ReL-A}.

A consequence of the flow equations in the Hamilton-Jacobi formalism is that the squaring of
the action is not unique; one can find another flow that squares the effective action in a similar way. This
was done for the ungauged case in \cite{Ceresole:2007wx} and for gauged supergravity with FI terms
in \cite{Gnecchi:2012kb}. We shall now generalize this procedure to the presence of hypermultiplets.

By repeating essentially the same computations as in the preceding subsection, one can show
that there is an alternative set of first-order equations that comes from the Hamilton-Jacobi function
\begin{equation}
W = e^U\left|\langle\tilde{\mathcal Q},\mathcal V\rangle + i\kappa e^{2(\psi - U)} \langle\ma W^x
\tilde{\ma Q^x},\mathcal V\rangle\right|\,, \label{eq:prW}
\end{equation}
with the associated constraints
\begin{equation}
\ma H\Omega\ma Q = 0\,, \qquad 2 e^U\mathcal H\Omega\mathrm{Re}\tilde{\mathcal V} = S\ma H
\Omega\ma A_t\,,
\label{eq:constraints}
\end{equation}
where we introduced a ``field rotation matrix" $S\in\text{Sp}(2n_v+2,\mathbb R)$ that rotates the
charges as $\tilde{\ma Q} = S\ma Q$ and that has to satisfy the compatibility conditions
\begin{equation}
S\mathcal H S^T = \mathcal H\,, \qquad S^T\mathcal M S = \mathcal M\,. \label{eq:comp-cond}
\end{equation}
Moreover, the rotated charges must obey the analogue of \eqref{eq:quantization}, namely
\begin{equation}
\tilde{\ma Q^x}\tilde{\ma Q^x} =1\,.
\end{equation}
The first equ.~of \eqref{eq:constraints} is a consequence of spherical/hyperbolic symmetry, and implies,
together with $S\mathcal H S^T = \mathcal H$ and the fact that $S$ is symplectic,
the additional condition $\ma H\Omega\tilde{\ma Q}=0$. The latter and the equation
$\ma H\Omega\ma Q = 0$ lead respectively to
\begin{equation}
\langle\ma K^u,\tilde{\ma Q}\rangle = \langle\ma K^u,\ma Q\rangle = 0\,,
\label{eq:algebraic}
\end{equation} 
which are quite restrictive constraints on the possible gaugings. Moreover, in general it is not guaranteed
that a nontrivial solution to \eqref{eq:comp-cond} exists. Note that the technique of ``rotating charges"
was first introduced in \cite{Ceresole:2007wx,LopesCardoso:2007qid}, and generalizes the sign-flipping
procedure of \cite{Ortin:1996bz}. It was applied to abelian $\mrm{FI}$ gauged supergravity
in \cite{Klemm:2012vm,Gnecchi:2012kb}. As we discussed in the Introduction of the thesis, we recall that the UV completion of effective charged solutions breaking all supersymmetries could manifest instabilities that are invisible at the supergravity level \cite{Ooguri:2016pdq}.

\subsection{The Flow of the Black String}

Following the arguments of previous sections on the four-dimensional case, we want now to consider an Ansatz on the fields of $\ma N=2$ gauged supergravity in $d=5$ with general couplings to the matter introduced in \eqref{eq:genlaghyper} reproducing a static and extremal black string with spherical/hyperbolic symmetry and derive its general system of $\mrm{BPS}$ first-order equations in the Hamilton-Jacobi approach. Also in this case we will then see how to produce a set on non-$\mrm{BPS}$ first-order equations by rotating the charges of the black string. Finally, by recalling the dictionary of the $r$-map given in section \ref{lagrangians5d}, we will discuss the relations between the four-dimensional flow \eqref{eq:flowmag} and that one of the black string.

From the analysis of sections \ref{lagrangians5d} and \ref{modulispace} it follows that the very special real K\"ahler moduli spaces can be viewed as the pre-image of the $r$-map \cite{Berkooz:2008rj,Cortes:2011aj}. This suggests to consider the five-dimensional
background as a Kaluza-Klein uplift of the usual static black hole in four dimensions described by the Ansatz \eqref{eq:Ansatzmet}. Moreover, a static string solution in $d=5$ supports only magnetic charges, thus the field configuration reads
\begin{equation}
\begin{split}
&\D s^2 = e^{2T(r)}\D z^2 + e^{-T(r)}\left(-e^{2U(r)}\D t^2 + e^{-2U(r)}\D r^2 + e^{2\psi(r) - 2U(r)}
\D\Omega^2_\kappa\right)\,, \\
& F^I = p^I f_\kappa (\theta)\,\D\theta\wedge\D\phi\,, \qquad\phi^i = \phi^i(r)\,, \label{eq:ans}
\end{split}
\end{equation}
where $\D\Omega_{\kappa}^2=\D\theta^2+f_{\kappa}^2(\theta)\,\D\varphi^2$ is the metric on the
two-dimensional\footnote{The following analysis can be extended to the case of the black brane described by $\Sigma_\kappa=\mathbb{R}^2$ (see \cite{LopesCardoso:2007qid, Barisch:2011ui, BarischDick:2012gj, Cardoso:2015wcf}).} surfaces $\Sigma_\kappa=\{S^2, H^2\}$ of constant scalar
curvature $R=2\kappa$, with $\kappa\in\{1,-1\}$, and
\begin{equation}
f_\kappa(\theta) = \frac{1}{\sqrt{\kappa}} \sin(\sqrt{\kappa}\theta) = 
\left\{\begin{array}{c@{\quad}l} \sin\theta\, & \kappa=1\,, \\
                                                 \sinh\theta\, & \kappa=-1\,. \end{array}\right.
\end{equation}
Let's start by imposing \eqref{eq:ans} on Maxwell equations given in \eqref{eq:eom5d}. The $t$-, $\theta$- and $z$-components are automatically satisfied, while the $r$- and $\varphi$-components become respectively
\begin{equation}
h_{uv} k_I^u q^{\prime\,v} = 0\,, \qquad k_I^u p^I = 0\,.
\end{equation}
The remaining equations of motion for the Ansatz \eqref{eq:ans} can be derived from the effective action
\begin{equation}
\begin{split}
&S = \int\D r\left[e^{2\psi}\left(U^{\prime 2} + \frac34 T^{\prime 2} - \psi^{\prime 2} +
\frac12\mathcal G_{ij}\phi^{\prime\,i}\phi^{\prime\,j} +  h_{uv} q^{\prime\,u} q^{\prime\,v}\right) -
\tilde{V}\right]\,, \\
&\tilde{V} = \kappa - e^{2\psi - 2U - T}\mathrm{g}^2 V - \frac12 e^{2U + T - 2\psi}G_{IJ} p^I p^J\,,
\end{split}
\end{equation}
supplemented by the Hamiltonian constraint on the effective hamiltonian associated, i.e. $H=0$. The latter leads to the
Hamilton-Jacobi equation
\begin{equation}
e^{-2\psi}\left((\partial_U W)^2 - (\partial_\psi W)^2 + \frac43(\partial_T W)^2 + 2\mathcal G^{ij}\partial_i
W\partial_j W +  h^{uv}\partial_u W\partial_v W\right) + \tilde{V} = 0\,.
 \label{eq:hjbshyper}
\end{equation}
Guided by the four-dimensional expression of the superpotential given in \eqref{eq:superpotential}, we formulate the following prescription on the five-dimensional superpotential,
\begin{equation}
W = c e^{U + \frac T2}\ma Z + d e^{2\psi - U - \frac T2}\ma L\,, \label{eq:mainHJhyper}
\end{equation}
where
\begin{equation}
\ma Z = p^I h_I\,, \qquad \ma L = \ma Q^x\ma W^x\,, \qquad \ma Q^x = p^I P^x_I\,, \qquad
\ma W^x = h^J P_J^x\,. \label{def-ZLQxWx}
\end{equation}
Using the relations of very special geometry \eqref{veryspcialmetric}, \eqref{dualH} and \eqref{eq:veryspecial1} as well as the quaternionic ones
\eqref{eq:quaternionic_kahler_complexstruct_definition}, \eqref{eq:quaternionkahl_su2curv} and
\eqref{eq:equivariance}, one can show that \eqref{eq:mainHJhyper} solves indeed \eqref{eq:hjbshyper}
provided that
\begin{equation}
c = -\frac34\,, \qquad d = -\frac92\kappa\mathrm{g}^2\,, \qquad \ma Q^x\ma Q^x =
\frac1{9\mathrm{g}^2}\,.
\end{equation}
The solution \eqref{eq:mainHJhyper} leads then to the first-order equations for the black string
\begin{equation}
\begin{split}
& U^{\prime} = -\frac34 e^{U + \frac T2 - 2\psi}\ma Z + \frac92\kappa\mathrm{g}^2 e^{-U -\frac T2}
\ma L\,, \qquad T^{\prime} = \frac23 U^{\prime}\,, \\
&\psi^{\prime} = 9\kappa\mathrm{g}^2 e^{-U - \frac T2}\ma L\,, \\
&\phi^{\prime\,i} = \mathcal G^{ij}\left(-\frac32 e^{U + \frac T2 - 2\psi}\partial_j\ma Z - 9\kappa
\mathrm{g}^2 e^{-U - \frac T2}\partial_j\ma L\right)\,, \\
& q^{\prime\,u} = -\frac 92 \kappa\mathrm{g}^2 e^{-U - \frac T2} h^{uv}\partial_v\ma L\,.
\label{eq:hyperflow}
\end{split}
\end{equation}
It is interesting to note that it is possible recast \eqref{eq:hyperflow} into a form very similar to \eqref{eq:flowequations}. Integrating $T^{\prime} = \frac23 U^{\prime}$
and plugging this into the remaining equations of  \eqref{eq:hyperflow}, one gets
\begin{equation}
\begin{split}
& T^{\prime} = -\frac12 e^{2T - 2\psi}\ma Z + 3\kappa\mathrm{g}^2 e^{- 2T}\ma L\,, \\
&\psi^{\prime} = 9\kappa\mathrm{g}^2 e^{-2T}\ma L\,, \\
&\phi^{\prime\,i} = \mathcal G^{ij}\left(-\frac32 e^{2T - 2\psi}\partial_j\ma Z - 9\kappa\mathrm{g}^2
e^{-2T}\partial_j\ma L\right)\,, \\
& q^{\prime\,u} = -\frac 92 \kappa\mathrm{g}^2 e^{-2T} h^{uv}\partial_v\ma L\,. \label{eq:hyperflow1}
\end{split}
\end{equation}
Using the equation for $\phi^{\prime\,i}$ together with $(h^I)^\prime=\phi^{\prime\,i}\partial_i h^I$ and
\eqref{eq:veryspecial1}, the equations for $T$ and $\phi^i$ can be rewritten as
\begin{equation}
e^{2\psi}\left(e^{-2T} h^I\right)^\prime + 9\mathrm{g}^2\kappa e^{2\psi-4T}\ma Q^x P^x_J G^{IJ}-p^I
= 0\,. \label{eq:hyperflow2}
\end{equation}

Note that the case of $\mrm{FI}$ gauging can be recovered imposing a prescription of the type \eqref{FIchoice}. We choose $P^1_I=P^2_I=0$ and $P^3_I=V_I$ with $V_I$ constant $\mrm{FI}$ parameters. Then the charge quantization condition $\ma Q^x\ma Q^x=1/(9\mathrm{g}^2)$ boils down to $\ma Q^3=p^IV_I=\pm\kappa/(3\mathrm{g})$ (use $\kappa^2=1$), while $\ma L$ in \eqref{def-ZLQxWx} becomes $\ma L=\pm\frac{\kappa}{3\mathrm{g}} h^JV_J$. 
In absence of hypers and with $\mrm{FI}$ gauging, one can check that \eqref{eq:hyperflow} coincides with the system obtained in \cite{Cacciatori:2003kv} from the Killing spinor equations. In particular, it is easy to verify that the supersymmetric magnetic black string solution of \cite{Klemm:2000nj} satisfies \eqref{eq:hyperflow}. Moreover, requiring $\mrm{FI}$ gauging, introducing a new radial coordinate $R$ and the warp factors $f$ and $\rho$ such that
\begin{equation}
U = \frac32 f\,, \qquad \psi = 2f + \rho\,, \qquad T = f\,, \qquad \frac{\D R}{\D r} = e^{-3f}\,,
\label{eq:maldaansatz}
\end{equation}
and specifying to the model \eqref{5dstu}, one shows that \eqref{eq:hyperflow} is precisely the system of equations derived in \cite{Maldacena:2000mw} from the Killing spinor equations.

As we saw in the four-dimensional case, also in this case the Hamilton-Jacobi formalism allows for a simple generalization of the first-order flow driven by \eqref{eq:mainHJhyper} to a non-BPS one. In analogy to the final discussion of section \ref{symplecticallyinvariant},  one can introduce
a ``field rotation matrix" ${S^I}_J$ such that
\begin{equation}
G_{LK} {S^L}_I {S^K}_J = G_{IJ}\,. \label{eq:rotmat}
\end{equation}
A non-trivial $S$ (different from $\pm\mathbb{I}$) allows to generate new solutions from known ones by
``rotating charges". This technique was first introduced in \cite{Ceresole:2007wx,LopesCardoso:2007qid},
and generalizes the sign-flipping procedure of \cite{Ortin:1996bz}. Let's consider for simplicity an $\mrm{SU(2)}$ frame such that $P^1_I=P^2_I=0$, $P^3_I=P_I$ and $\ma Q^3=p^IP_I=-\kappa/(3\mathrm{g})$.
Using \eqref{eq:rotmat}, one easily verifies that
\begin{equation}
\tilde W = -\frac34 e^{U + \frac T2}  h_I {S^I}_J p^J + \frac32\mathrm{g} e^{2\psi - U - \frac T2} P_I h^I
\end{equation}
satisfies again the Hamilton-Jacobi equation \eqref{eq:hjbshyper}, provided the modified quantization condition
\begin{equation}
P_I {S^I}_J p^J = -\frac{\kappa}{3\mathrm{g}}
\end{equation}
holds. This leads to the first-order flow driven by $\tilde W$,
\begin{equation}
\begin{split}
& U^{\prime} = -\frac34 e^{U + \frac T2 - 2\psi} h_I {S^I}_J p^J - \frac32\mathrm{g}\, e^{-U - \frac T2}
P_I h^I\,,\\
& \psi^{\prime} = -3\mathrm{g}\, e^{-U - \frac T2} P_I h^I\,, \qquad T^{\prime} = \frac23 U^{\prime}\,, \\
& \phi^{\prime\,i} = 3\mathcal G^{ij}\left(-\frac12 e^{U + \frac T2 - 2\psi}\partial_j h_I  {S^I}_J p^J + \mathrm{g}\, e^{-U - \frac T2} P_I \partial_j h^I\right)\,,\\
& q^{\prime\,u} = \frac 32 \,\mathrm{g}\, e^{-U-T/2} h^{uv} h^I \partial_v P_I\,. \label{eq:hyperflow1}
\end{split}
\end{equation} 
An interesting example for which \eqref{eq:rotmat} admits non-trivial solutions is the model
$\mathcal V = h^1 h^2 h^3 = 1$ introduced in \eqref{5dstu}, for which (see for example \cite{Cacciatori:2003kv}).
In this case a particular solution of \eqref{eq:rotmat} is given by
\begin{equation}
{S^I}_J =
\begin{pmatrix}
\epsilon_1   &  0  &  0 \\
0  &  \epsilon_2 &  0 \\
0  &  0  &  \epsilon_3
\end{pmatrix}\,,
\end{equation}
with $\epsilon_I=\pm 1$. These matrices form a discrete subgroup
$D=(\mathbb Z_2)^3\subset\text{GL}(3,\mathbb R)$. Since there are two equivalent BPS branches,
the independent solutions correspond to elements of the quotient group $D/\mathbb Z_2$.

The $r$-map constructed in section \ref{lagrangians5d} has been formulated is generality. Furthermore it is interesting to use it to compare the flow \eqref{eq:hyperflow} for a black string in $d= 5$ with the
flow equations for black holes in four dimensions obtained in \eqref{eq:flow-equ-electr}.
The latter is driven by the four-dimensional superpotential written in \eqref{eq:superpotential} given by
\begin{equation}
W_{4d} = e^U\mathrm{Re}(e^{-i\alpha}\mathcal Z_{4d}) - \kappa e^{2\psi - U} \mathrm{Im}
(e^{-i\alpha}\mathcal L_{4d})\,, \label{eq:Wnd}
\end{equation}
where the phase $\alpha$ is defined by
\begin{equation}
e^{2i\alpha} = \frac{\mathcal Z_{4d} + i\kappa e^{2(\psi - U)}\mathcal L_{4d}}{\bar{\mathcal Z}_{4d}
- i\kappa  e^{2(\psi - U)}\bar{\mathcal L}_{4d}}\,.
\label{eq:alpha}
\end{equation}
Specifying to a pure magnetic charge configuration $\hat{p}^\Lambda=(0,p^I/\sqrt2)$, pure electric 
couplings with $P^x_0=0,k^u_0=0$, and restricting to imaginary scalars, $z^I=-ie^{-\phi/\sqrt3}h^I$,
the quantities defining \eqref{eq:Wnd} become
\begin{equation}
\begin{split}
&\mathcal Z_{4d} = \frac{3}{2\sqrt2} e^{T/2}\hat{p}^I h_I = \frac34 e^{T/2}\ma Z\,, \qquad
\ma Q^x_{4d} = P^x_I\hat{p}^I = \frac1{\sqrt2}\ma Q^x\,, \\
&\ma W^x_{4d} = -\frac i{2\sqrt2} e^{-T/2} P^x_I h^I = -\frac i{2\sqrt2} e^{-T/2}\ma W^x\,, \qquad
\ma L_{4d} = \ma Q^x_{4d}\ma W^x_{4d}\,,
\end{split}
\end{equation}
where the quantities $\ma Z$, $\ma Q^x$ and $\ma W^x$ were defined in section \eqref{def-ZLQxWx}.
Note that the axions are absent, since for magnetically charged black strings the $z$-components $B^I$ of the five-dimensional gauge potentials vanish.
For this choice, \eqref{eq:alpha} becomes $e^{2 i\alpha}=1$. Moreover, taking in account that
$\mathrm{g}=\frac{1}{3\sqrt2}$ and choosing $e^{i\alpha}=-1$, the function \eqref{eq:Wnd} boils 
down to \eqref{eq:mainHJhyper}. On the other hand, the Hamilton-Jacobi equation satisfied by
\eqref{eq:Wnd}, namely \eqref{eq:sqpot}, becomes \eqref{eq:hjbshyper} once the dictionary
is imposed. This proves the expected equivalence between the flows in five and four dimensions.

\section{The Attractor Mechanism}

The {\itshape attractor mechanism} is a phenomenon discovered in $\ma N=2$ supergravities characterizing some charged supersymmetric extremal black solutions. It has been formulated firstly in \cite{Ferrara:1995ih, Strominger:1996kf, Ferrara:1996dd, Ferrara:1996um, Ferrara:1997tw} for four-dimensional black holes in ungauged supergravities and then extended to the case of gaugings \cite{Bellucci:2008cb, Chimento:2015rra}.

As we mentioned at the beginning of this chapter, an {\itshape attractor} is a black solution basically enjoying two related features:

\begin{itemize}

\item The flow of the moduli fields is subjected to a dynamical stabilization at the horizon values. The scalars, independently to their asymptotic values, take fixed values at the horizon that are pure charge-dependent, or also gauging-dependent\footnote{In the sense that they depend on the set of independent parameters of the embedding tensor.} if the model is gauged.

\item The entropy is strictly charge-dependent or also gauging-dependent if the model is gauged. This implies that the entropy does not take stringy corrections (see section \ref{microstate}) and this allows the extrapolation of the entropy from a correspondent higher-dimensional bound state of branes and the microstate counting.

\end{itemize}

These two aspect characterizing the attractors are kept in a single shot by a particular function of the scalars, that we will call $V_{eff}$, defined by a process of extremization at  the horizon. Furthermore the critical value of $V_{eff}$ at the horizon is given exactly by the Bekenstein-Hawking entropy, that is strictly charge-dependent since the near-horizon values of the moduli are charge-dependent\footnote{It could happen that a scalar is not stabilized at the horizon, but the entropy remains moduli-independent. In this case the scalar field is called {\itshape flat direction} \cite{Nampuri:2007gv,Ferrara:2007tu} and the attractor mechanism is preserved.}.

In this section we will present the general formulation of the attractor mechanism for $d=4$ black holes in presence of hypermultiplets extending the formulation of \cite{Bellucci:2008cb} for $\mrm{FI}$ gauged supergravity. Thus we will consider the first-order formulation for black holes and black strings at their near-horizon limits and we will relate it with the attractor mechanism.

\subsection{Attractors and Hypermultiplets}
\label{attractorhyper}

In this section we will limit to consider the four dimensional black hole since, thanks to dictionary of the $r$-map formulated in section \ref{lagrangians5d}, one can easily uplift\footnote{See \cite{Hristov:2014eba, Hristov:2014eza} where the uplifts of various near-horizon geometries in $\ma N=2$ supergravities are performed.} the results to the case of the five-dimensional static black string .

In \cite{Bellucci:2008cb} the authors present a generalization of the well-known black hole attractor mechanism in $\ma N=2$ ungauged supergravity introduced in \cite{Ferrara:1995ih, Strominger:1996kf, Ferrara:1996dd, Ferrara:1996um, Ferrara:1997tw} to extremal static black holes in four-dimensional $\ma N=2$ $\mrm{FI}$ supergravity coupled to abelian vector multiplets. In this section we closely follow their argument, generalizing it by taking into account the presence of gauged hypermultiplets. For simplicity we will consider the case of electric abelian gaugings and, as in \cite{Bellucci:2008cb}, we will not make any assumption on the form of the scalar potential \eqref{eq:scal_pot} and on the moduli spaces.

We start by considering the Ansatz for a static extremal black hole with spherical/hyperbolic symmetry\footnote{As in the other sections, the case of flat surface $\Sigma_0=\mathbb{R}^2$ can be easily included in this picture.} given by \eqref{eq:Ansatzmet}, \eqref{eq:Ansatzscalars} and \eqref{eq:Ansatzgaugefields} and evaluating it into the equations of motion \eqref{electriceom4d}. In this way we obtain the system of the equations of motion describing the black hole,
\begin{equation}
\begin{split}
\label{sphericaleoms}
 &e^{2 U}\left( 2 U^\prime \psi^\prime+U^{\prime\prime}  \right)-e^{4(U-\psi)}V_{\text{BH}}-2 e^{-2U}A^\Lambda_t\,k_{\Lambda\,u}A^\Sigma_t\,k_\Sigma^u
  +V_g=0\,, \\
  &e^{2 U}\left(U'^2+ \psi^{\prime 2}+ \psi^{\prime\prime} \right)- e^{4(U-\psi)} V_{\text{BH}}+ e^{2U}g_{i\bar\jmath}z^{i\prime}{\bar z}^{\bar\jmath\prime}
  + e^{2U}h_{uv}q^{u\prime}q^{v\prime}\\
 & \quad - e^{-2U}A^\Lambda_t \,k_{\Lambda\,u} \,A^\Sigma_t \,k_\Sigma^u+V_g=0\,, \\
& e^{2 U}\left(2 \psi^{\prime 2}+\psi^{\prime\prime}-\kappa e^{-2 \psi}  \right)-2 e^{-2U}A^\Lambda_t\, k_{\Lambda\,u} \,A^\Sigma_t \,k_\Sigma^u+2\,V_g=0\,, \\
& e^{2 U}\left(z^{i\,\prime\prime}+2 \psi^\prime z^{i\,\prime}+ g^{i\bar\jmath}\partial_l g_{k\bar\jmath}z^{l\,\prime}z^{k\,\prime}\right) -e^{4(U-\psi)}\partial^i  V_{\text{BH}}-\partial^i V_g=0\,, \\
&2\,e^{2 U}\left(q^{u\,\prime\prime}+2 \psi^\prime q^{u\prime}+ \Gamma^u_{vz}q^{v\prime}q^{z\prime}\right)-
  2\,e^{-2U}A^\Lambda_t\,k_\Lambda^v \,A^\Sigma_t\, \nabla_v\,k_\Sigma^u  - \partial^u V_g=0\,, \\
 \end{split}
\end{equation}
where the prime denotes, as usual, the derivative with respect to $r$ and $\nabla_u$ the covariant derivative with respect to the Levi-Civita connection on the quaternionic manifold.

Suppose that the horizon is identified in the limit $r\rightarrow 0$. We want to consider now a near-horizon limit of the type $\mrm{AdS}_2\times \Sigma_\kappa$. If we consider the general form of the metric \eqref{eq:Ansatzmet}, the explicit realization of this near-horizon configuration is given by the following expression for the warp factors $U$ and $\psi$,
\begin{equation}
\label{eq:att_metric_limits}
 U= \log \frac{r}{R_{\mathrm{AdS}_2}}\,,\qquad \psi=\log\left( \frac{R_H}{R_{\mathrm{AdS}_2}}\, r \right)\,,
\end{equation}
where $R_{\mathrm{AdS}_2}$ and $R_H$ are respectively the curvature radii\footnote{In case of $\Sigma_\kappa$ non-compact, it has to be considered compactified on a Riemann surface of genus $\mathfrak{g}$.} of $\mrm{AdS}_2$ and of $\Sigma_\kappa$.
We require all the fields, their derivatives, the scalar potential and the couplings to be regular on the horizon. Then we can choose a gauge such that
\begin{equation}
\label{eq:att_At_gauge}
 \left.A^\Lambda_t\right|_{r=0}=0\qquad\Longrightarrow\qquad A^\Lambda{}_t\stackrel{r\to 0}{\sim}\left.F^\Lambda{}_{rt}\right|_{r=0} r\,.
\end{equation}
It is also reasonable to assume that the derivative of the electric charges $e_\Lambda^\prime$ remains finite on the horizon. In this case, Maxwell equations \eqref{eq:max2} implies that in the near-horizon limit the quantity $A^\Sigma_t k_{\Sigma\,u} k_\Lambda^u$ is at least of order $r^2$. If we expand in powers of $r$, in the gauge \eqref{eq:att_At_gauge} the order zero term automatically vanishes, while for the 
order one term we have
\begin{equation}
\label{eq:att_Fk}
 0=\left.\partial_r \left(A^\Sigma_t\,k_{\Sigma\,u}\,k_\Lambda^u\right)\right|_{r=0}=
 \left.-F^\Sigma_{tr}\,k_{\Sigma\,u}\,k_\Lambda^u\right|_{r=0}
 \qquad\Longrightarrow\qquad \left.F^\Lambda_{tr}\,k_\Lambda^u\right|_{r=0}=0\,.
\end{equation}
Using \eqref{eq:att_At_gauge} and \eqref{eq:att_Fk} one can see that the terms with $A^\Lambda_t$ in the equations of motion, 
$e^{-2U}A^\Lambda_t\,k_{\Lambda\,u}A^\Sigma_t\,k_\Sigma^u$  and 
$e^{-2U}A^\Lambda_t\,k_\Lambda^v\, A^\Sigma_t\, \nabla_v\,k_\Sigma^u$, go to zero in the
near-horizon limit.
In this limit the equations of motion \eqref{sphericaleoms} thus reduce to
\begin{equation}
\begin{split}
& \frac{1}{R_{\mathrm{AdS}_2}^2}=\frac{V_{\text{BH}}}{R_H^4}-V_g\,, \\
& \frac{\kappa}{R_H^2}=\frac{1}{R_{\mathrm{AdS}_2}^2}+2 V_g\,, \\
& \partial_i\left(\frac{V_{\text{BH}}}{R_H^4}+V_g \right)=0\,, \\
 &\partial_u V=0\,,
 \label{nheom}
 \end{split}
\end{equation}
where $V_{\text{BH}}$ has to be considered evaluated in the the limit ${e_\Lambda(r)\to e_\Lambda(0)}$. Solving the first two
equations for $R_H^2$ and $R_{\mathrm{AdS}_2}^2$ one gets
\begin{gather}
 R_H^2=\left.\frac{\kappa\pm\sqrt{\kappa^2-4 V_{\text{BH}} V_g }}{2\,V_g}\right|_{r=0}\,, \\
 \nonumber\\
 R_{\scriptsize{\mathrm{AdS}_2}}^2=\left.\mp \frac{R_H^2}{\sqrt{\kappa^2-4 V_{\text{BH}} V_g }}\right|_{r=0}\,,
\end{gather}
and since of course $R_{\mathrm{AdS}_2}^2>0$ we have to choose the lower sign. We also have to require
$R_H^2>0$, which means that flat or hyperbolic
geometries, $\kappa=0,-1$, are only possible if the scalar potential takes negative values on the horizon, $V_g|_{r=0}<0$. Spherical 
geometry ($\kappa=1$), on the other hand, is compatible with both positive or negative values of $V_g$ on the horizon, but for $V_g|_{r=0}>0$ 
there is the restriction $V_{\text{BH}} V_g|_{r=0}<\frac{1}{4}$, since $V_{\text{BH}}$ is always positive.

We can define the function $V_{eff}$ driving the attractor mechanism as a function of the scalars,
\begin{equation}
 V_{eff}(z,\bar z, q)= \frac{\kappa-\sqrt{\kappa^2-4 V_{\text{BH}} V_g }}{2\,V_g}\,,
 \label{effective}
\end{equation}
defined for $V_{\text{BH}} V_g<\frac{1}{4}$, and write
\begin{gather}
 R_H^2=\left.V_{eff}\right|_{z_H,q_H}\,,\label{rH} \\
 \nonumber\\
R_{\mathrm{AdS}_2}^2=\left.\frac{V_{eff}}{\sqrt{\kappa^2-4 V_{\text{BH}} V_g}}\right|_{z_H,q_H}\,,\label{rA}
\end{gather}
with $z^i_H=\lim_{r\to 0} z^i$, $q^u_H=\lim_{r\to 0} q^u$.
Because of the last two equations of \eqref{nheom}, $V_{eff}$ is extremized on the horizon by all the scalar fields of the theory,
\begin{equation}
\label{eq:att_Veff_extr}
 \left.\partial_i V_{eff}\right|_{z_H,q_H}=0\,,\qquad \left.\partial_u V_{eff}\right|_{z_H,q_H}=0\,.
\end{equation}

The values $z^i_H, q^u_H$ of the scalars on the horizon are then determined by the extremization conditions \eqref{eq:att_Veff_extr},
and the Bekenstein--Hawking entropy density is given by the critical value of $V_{eff}$,
\begin{equation}
 s_{BH}=\frac{S_{BH}}{\mrm{vol}({\Sigma_\kappa})}=\frac{A}{4 \,\mrm{vol}({\Sigma_\kappa})}=\frac{R_H^2}{4}=\frac14\,V_{eff}(z_H,\bar z_H, q_H)\,, \label{Seffective}
\end{equation}
where $A$ is the area of the portion of spacetime enclosed\footnote{When $\Sigma_\kappa$ is non-compact, the area is taken by considering compactifications of $\Sigma_\kappa$ on Riemann surfaces of genus $\mathfrak{g}$.} by $\Sigma_\kappa$.
For a given theory this critical value, and thus also the entropy, depend only on the charges (on the horizon) $p^\Lambda$ and 
$e_\Lambda (0)$, so that the attractor mechanism still works. On the other hand $z^i_H$ and $q^u_H$ may not be uniquely determined, since
in general $V_{eff}$ may have flat directions.

The limit for $V_g\to 0$ of $V_{eff}$ only exists for $\kappa=1$, in which case
$V_{eff}\to V_{\text{BH}}$ and one recovers the attractor
mechanism for ungauged supergravity. The fact that this limit does not exist for $\kappa=0,-1$ is not surprising since flat or hyperbolic
horizon geometries are incompatible with vanishing cosmological constant.

\subsection{Attractor Equations for BPS Black Holes}
\label{BPSattractor}

In order to analyze the attractor mechanism from the point of view of the first-order formulation for the $d=4$ black hole, one has to derive the near-horizon limit of the symplectically invariant first-order equations \eqref{eq:flowequations}. If the charge-dependent stabilization at the horizon of the moduli is imposed, one obtains a set of algebraic equations, called {\itshape attractor equations}, depending on all the parameters characterizing the horizon of the black hole, namely the charges, the gauging parameters and the radii of the horizon. By solving these equations one can determine the near-horizon configuration of the black hole including the expression for the entropy that turns out to be strictly dependent on the charges and on the gauging parameters. This configuration will be supersymmetric since the attractor equations have been derived from the $\mrm{BPS}$ first-order flow equations \eqref{eq:flowequations}.

Let's recall the $\mrm{AdS}_2\times \Sigma_\kappa$ Ansatz written in \eqref{eq:att_metric_limits} and suppose that the horizon is located in $r=0$, thus it is easy to show that superpotential\footnote{We are considering the most general case with dyonic gaugings, thus the superpotential \eqref{eq:superpotential} will be expressed in terms as $\ma L=\langle \ma Q^x\ma P^x, \ma V \rangle$ with $\ma P^x$ including magnetic gaugings.} \eqref{eq:superpotential} vanishes at the horizon, i.e. $W=0$ at $r=0$. In fact the flow equations for $U$ and $\psi$ in \eqref{eq:flowmag} can be recasted as
\begin{equation}
U^\prime = -e^{-2(A+U)}(W - \partial_A W)\,, \qquad A^\prime = e^{-2(A+U)}W\,,
\end{equation}
where $A=\psi-U$ and $A\to\log(R_H)$ for $r\to 0$.
The relation $W=0$ immediately implies
\begin{equation}
\ma Z = -i\kappa R_H^2\ma L\,.
\label{attclass1}
\end{equation}
Assuming $z^{\prime\,i}=0$ and $q^{\prime\,u}=0$ at the horizon, it follows that
\begin{equation}
D_i\ma Z = -i\kappa R_H^2 D_i\ma L\,, \qquad \mathrm D_u\ma L = 0\,,
\label{attclass2}
\end{equation}
and $\alpha^\prime=0$. From $\mathrm{D}_u\ma L=0$ we get
\begin{equation}
\langle\ma K^v,\ma V\rangle = 0\,,
\label{eq:eqhyprNH}
\end{equation}
if we use also the algebraic relation $\langle\ma K^v,\ma Q\rangle=0$ written in \eqref{eq:algebraic}
together with \eqref{eq:quaternionic_kahler_complexstruct_definition}, \eqref{eq:quaternionkahl_su2curv}
and \eqref{eq:mommaps}.
As in section \ref{attractorhyper}, we can choose the gauge $\ma A_t=0$ at the horizon. Then, from
\eqref{eq:relalg} and the last equation of \eqref{eq:flowequations}, one obtains $\ma Q^\prime=0$.

With these assumptions, the BPS flow equations \eqref{eq:flowequations} become
\begin{equation}
\begin{split}
&4\,\mathrm{Im}(\bar{\ma Z}\ma V) - \kappa\, R_H^2\,\Omega\ma M\ma Q^x\ma P^x + \ma Q = 0\,, \\
&\ma Z = -\frac{R_H^2}{2 R_{\scriptsize{\mathrm{AdS}_2}}} e^{i\alpha}\,, \\
&\langle\ma K^v,\ma V\rangle = 0\,,
\label{eq:attractor4d}
\end{split}
\end{equation}
that must be supplemented by the constraints $\ma Q^x\ma Q^x=1$ and $\ma H\Omega\ma Q=0$.
If one rotates to a frame with pure electric gauging, $\ma Q^x$ boils down to
$p^\Lambda P^x_\Lambda$, and the magnetic charges $p^\Lambda$ become constant. One can then
use a local (on the quaternionic K\"ahler manifold) $\text{SU}(2)$ transformation to set
$\ma Q^1=\ma Q^2=0$, and the equations \eqref{eq:attractor} reduce to the ones obtained
in \cite{Erbin:2014hsa}.

The solutions of \eqref{eq:attractor4d} are the near-horizon values of the scalars and the radii of the horizon in terms of the charges and the gauging parameter. In particular from the solution for $R_H^2$ one derives the Bekenstein-Hawking entropy. This has been obtained, in general homogeneous models in \cite{Erbin:2014hsa,Halmagyi:2013qoa} for a gauging of the type $\ma P^1=\ma P^2=0$ and $\ma P^3=\ma P$. In our case one can obtain the same result but with the $\mrm{SU(2)}$ covariance restored, i.e. $\ma P^3\to -\kappa\ma Q^x\ma P^x$. The main difference with respect to the $\mrm{FI}$ case consists obviously in the hypers-dependence, whose near-horizon values are fixed by
\eqref{eq:eqhyprNH} and by $\ma H\Omega\ma Q=0$. Furthermore it is easy to verify that our general expression \eqref{Seffective} matches with the result of \cite{Erbin:2014hsa,Halmagyi:2013qoa} if one imposes the conditions \eqref{attclass1}, \eqref{attclass2} and \eqref{eq:eqhyprNH}. Clearly the expression of \eqref{Seffective} is completely general and thus, imposing different conditions $\ma Z$, $\ma L$, $D_i \ma Z$ and $D_i\ma L$ one can obtain also the entropy of non-supersymmetric attractors.

We conclude by a underlying the relevance of \eqref{eq:attractor4d} in microstate countings in $\mrm{AdS/CFT}$. In fact the resolution of the attractor equation furnishes directly the expression of the entropy without knowing the analytical flow on the whole background. Moreover the conditions \eqref{attclass1}, \eqref{attclass2} and \eqref{eq:eqhyprNH} hint the possibility of a classification of all the possible attractor configurations in terms of $\ma Z$, $\ma L$, $D_i \ma Z$ and $D_i\ma L$ for a given $\ma N=2$ model including those breaking $\mrm{SUSY}$. It follows that if we consider, for example, the models belonging the $\mrm{M}$-theory truncations with $\mrm{AdS}_4$ vacua  presented in section \ref{Mtheorytruncation}, the conditions \eqref{attclass1}, \eqref{attclass2} and \eqref{eq:eqhyprNH} tell us if attractor configurations interpolating between an $\mrm{AdS}_4$ vacuum and $\mrm{AdS}_2\times \Sigma_\kappa$ exist and thus if new examples of $\mrm{RG}$ flows across dimensions can be realized in a particular string truncation.

\subsection{Black String and Central Charges}
\label{centralchargebs}

In this section we investigate the near-horizon configurations of the extremal black string in $d=5$. To keep
things simple, we shall first concentrate on the $\mrm{FI}$ gauged case\footnote{We recall that the $\mrm{FI}$ gauging is realized by a prescription of the type \eqref{FIchoice}, i.e. $P^1_I=P^2_I=0$ and $P^3_I=V_I$ with $V_I$ $\mrm{FI}$ parameters. In this case the charge quantization condition $\ma Q^x\ma Q^x=1/(9\mathrm{g}^2)$ boils down to $\ma Q^3=p^IV_I=\pm\kappa/(3\mathrm{g})$ and $\ma L$ defined in \eqref{def-ZLQxWx} becomes $\ma L=\pm\frac{\kappa}{3\mathrm{g}} h^JV_J$. Moreover in this section we will set the parameter $\mathrm{g}=1$ to keep lighter the notation.} and secondly we will extend to the case of running hyperscalars. 

For the near-horizon analysis it is more suitable to consider the coordinates $(t,R,z,\theta,\phi)$ introduced in \eqref{eq:maldaansatz}. In this new frame the metric \eqref{eq:ans} takes the form
\begin{equation}
\D s^2 = e^{2f}(-\D t^2 + \D R^2 + \D z^2) + e^{2\rho}\,\D\Omega^2_{\kappa}\,,
\end{equation}
and the first-order flow equations \eqref{eq:hyperflow} become
\begin{equation}
\begin{split}
& f^\prime = -e^f(h^I V_I + \frac12 e^{-2\rho}\ma Z )\,, \\
&\rho^\prime = -e^f(h^I V_I - e^{-2\rho}\ma Z)\,, \\
&\phi^{\prime\,i} = 3\mathcal G^{ij} e^f (\partial_j h^I V_I - \frac12 e^{-2\rho}\partial_j\ma Z)\,,
\label{eq:flowequation2}
\end{split}
\end{equation}
where the primes now denote derivatives with respect to $R$.
Supposing that the horizon is located in $R=0$, the near-horizon geometry of a static extremal black string is of the type $\mathrm{AdS}_3\times\Sigma_\kappa$ with $\Sigma_\kappa=\{S^2, H^2\}$, and we assume that the scalars stabilize regularly at the horizon, i.e. $\phi^{\prime\,i}=0$. 
For a product space $\mathrm{AdS}_3\times\Sigma_\kappa$ we have
\begin{equation}
e^{2f} = \frac{R_{\mathrm{AdS}_3}^2}{R^2}\,, \qquad e^{2\rho} = R_{ H}^2\,. \label{eq:nhans}
\end{equation}
Plugging this together with $\phi^{\prime\,i}=0$ into \eqref{eq:flowequation2}, one obtains a system of
algebraic equations whose solution fixes the near-horizon values of the scalars in terms of the charges and
the gauging parameters,
\begin{equation}
h^I V_I = \frac2{3R_{\mathrm{AdS}_3}}\,, \qquad \ma Z = R_{ H}^2 h^I V_I\,, \qquad
\partial_i\ma Z = 2R_{ H}^2\partial_i h^I V_I\,. \label{eq:attractor}
\end{equation}
For the Ansatz \eqref{eq:nhans}, the FI-version of \eqref{eq:hyperflow2} (obtained by taking
$\ma Q^xP^x_J=\ma Q^3P^3_J=-\kappa V_J/(3\mathrm{g})$) reduces to
\begin{equation}
e^{f + 2\rho}(e^{-2f} h^I)^\prime - 3 e^{2\rho} G^{IJ}V_J - p^I = 0\,.
\end{equation}
Using \eqref{eq:nhans} and \eqref{eq:attractor}, this can be rewritten as
\begin{equation}
p^I + 3 R_{ H}^2 G^{IJ} V_J = 3\ma Z h^I\,. \label{eq:attractor2}
\end{equation}
The equations \eqref{eq:attractor} and \eqref{eq:attractor2} constitutes the five-dimensional attractor equations for the black string and correspond, via $r$-map, to those derived in $d=4$ given in \eqref{attclass1}, \eqref{attclass2} and \eqref{eq:attractor4d} and written for a $\mrm{FI}$ gauging.

We want now to solve the attractor equations \eqref{eq:attractor} (or equivalently \eqref{eq:attractor2}) in
order to express $R_{\mathrm{AdS}_3}$, $R_{ H}$ and $h^I$ in terms of $p^I$ and $V_I$. To this end, we
contract the third relation of \eqref{eq:veryspecial1} with $V_I$ to get
\begin{equation}
\mathcal G^{ij}\partial_i h^I V_I\partial_j h_J = -\frac23 V_J + \frac23 h^I V_I h_J\,. \label{eq:veryspecialL}
\end{equation}
With \eqref{eq:attractor}, the above relation becomes
\begin{equation}
R_{ H}^2 V_J = -\frac34\mathcal G^{ij}\partial_i\ma Z\partial_j h_J + \ma Z h_J\,.
\end{equation}
Using $h_I=\frac16 C_{IJK}h^Jh^K$ and \eqref{eq:veryspecial1}, one obtains
\begin{equation}
R_{ H}^2 V_J = \frac16 C_{JKL} p^K h^L\,. \label{RH^2V_J}
\end{equation}
Let us define the charge-dependent matrix
\begin{equation}
C_{p\,IJ}= C_{IJK}p^K\,.
\end{equation}
Using the adjoint identity \eqref{adjoint-id}, one easily shows that $C_{p\,IJ}$ is invertible, with inverse
\begin{equation}
C^{IJ}_p = 3\,\frac{C^{IJK} C_{KMN} p^M p^N - p^I p^J}{C_p}\,, \label{eq:inverseC}
\end{equation}
where $C_p=C_{IJK}p^Ip^Jp^K$. The relation \eqref{RH^2V_J} implies then
\begin{equation}
h^I = 6 R_{ H}^2 C^{IJ}_p V_J\,. \label{eq:nhrelation}
\end{equation}
Plugging \eqref{eq:nhrelation} into \eqref{veryspecial}, one can derive a general expression for
$R_{ H}$ in terms of the intersection numbers, the charges and the FI parameters,
\begin{equation}
R_{ H}^2 = (36\, C_{IJK} C_p^{IM} C_p^{JN} C_p^{KP} V_M V_N V_P)^{-\frac13}\,. \label{eq:entropy}
\end{equation}
Using this in \eqref{eq:nhrelation} gives the values of the scalars at the horizon,
\begin{equation}
h^I = \frac{6 C^{IJ}_p V_J}{(36\, C_{KLM} C_p^{KN} C_p^{LP} C_p^{MR} V_N V_P V_R)^{\frac13}}\,.
\label{eq:scalarsnh}
\end{equation}
Contracting \eqref{eq:nhrelation} with $V_I$ and using the first equation of \eqref{eq:attractor} as well as
\eqref{eq:entropy}, we obtain an expression for the $\text{AdS}_3$ curvature radius $R_{\mathrm{AdS}_3}$,
\begin{equation}
R_{\mathrm{AdS}_3} = \frac{(36 \,C_{IJK} C_p^{IM} C_p^{JN} C_p^{KP} V_M V_N V_P)^{\frac13}}{9C^{RS}_p
V_R V_S}\,. \label{eq:Rads}
\end{equation}
Finally, one can plug \eqref{eq:inverseC} into \eqref{eq:entropy}, \eqref{eq:scalarsnh} and \eqref{eq:Rads},
and use \eqref{adjoint-id} to write the solutions of \eqref{eq:attractor} and \eqref{eq:attractor2} as
\begin{equation}
\begin{split}
& R_{ H}^2 = (\mas C^{IJK}(p) \,V_I V_J V_K)^{-\frac13}\,, \\
& h^I = \frac{6\kappa}{C_p}\frac{p^I + 3\kappa\, C^{IJK} C_{KLM} p^L p^M V_J}{(\mas C^{NPR}(p) V_N V_P
V_R)^{\frac13}}\,, \\
& R_{\mathrm{AdS}_3} = \frac{C_p}{27}\frac{(\mas C^{IJK}(p) V_I V_J V_K)^{\frac13}}
{C^{LMN} C_{NRS}\, p^R p^S V_L V_M - \frac19}\,, \label{eq:solutions}
\end{split}
\end{equation}
where
\begin{equation}
\mas C^{IJK}(p) = -\frac{108}{C_p}\left[2 C^{IJK} - \frac{9}{C_p} p^{(I} C^{JK)M} C_{MNP} p^N p^P +
\frac{9}{C_p} p^I p^J p^K\right]\,.
\end{equation}
The central charge of the two-dimensional conformal field theory that describes the black strings
in the infrared \cite{Maldacena:2000mw,Benini:2013cda,Hristov:2014eza}, is given
by \cite{Brown:1986nw}
\begin{equation}
c = \frac{3R_{\text{AdS}_3}}{2G_3}\,, \label{central}
\end{equation}
where $G_3$ denotes the effective Newton constant in three dimensions, related to $G_5$ by
\begin{equation}
\frac1{G_3} = \frac{R_{ H}^2\mathrm{vol}(\Sigma_\kappa)}{G_5}\,. \label{G_3}
\end{equation}
In what follows, we assume $\Sigma_\kappa$ to be compactified to a Riemann surface of genus $\mathfrak{g}$,
with $\mathfrak{g}$. The unit $\Sigma_\kappa$ has Gaussian curvature $K=\kappa$, and thus
the Gauss-Bonnet theorem gives
\begin{equation}
\mathrm{vol}(\Sigma_\kappa) = \frac{4\pi(1 - \mathfrak{g})}{\kappa}\,. \label{GB}
\end{equation}
Using \eqref{G_3} and \eqref{GB} in \eqref{central} yields for the central charge
\begin{equation}
c = \frac{6\pi(1 - \mathfrak{g}) R_{\text{AdS}_3} R_{ H}^2}{\kappa \, G_5}\,.
\end{equation}
The curvature radii $R_{\text{AdS}_3}$ and $R_{ H}$ can be expressed in terms of the constants
$C_{IJK}$, the magnetic charges $p^I$ and the FI parameters $V_I$ by means of \eqref{eq:solutions}.
This leads to
\begin{equation}
c = \frac{2\pi(1 - \mathfrak{g})\, C_p}{\kappa \, G_5 \,(9 \, C^{IJK} C_{KMN} p^M p^N V_I V_J - 1)}\,. \label{c-final}
\end{equation}

If the hyperscalars are running, one has to consider also the near-horizon limit of the last equation of \eqref{eq:hyperflow}. Assuming $q^{\prime\,u}=0$ at the horizon and using \eqref{eq:mommaps},
one easily derives the algebraic condition
\begin{equation}
k^u_I h^I=0\,.
\end{equation}
As far as the remaining equations of \eqref{eq:hyperflow} are concerned, one can follow the same steps
as in this section, with the only difference that $V_I$ has to be replaced everywhere by
$-3\kappa \ma Q^x P_I^x$.

\section{Freudenthal Duality and Attractors}

Between the various properties of attractors, one of the most intriguing is represented by {\itshape Freudenthal duality}. This has been discovered originally in \cite{Borsten:2009zy} for black holes in four-dimensional Einstein-Maxwell systems coupled to non-linear sigma models of scalar fields as a particular transformation of the charges of the black hole leaving invariant the Bekenstein-Hawking entropy. 
Generally it can be defined as an anti-involutive, non-linear map, acting on symplectic manifolds and thus in particular on the representation space in which the electromagnetic charges of the black hole sit.

In the context of ungauged supergravity, Freudenthal duality was proved to be a symmetry not only of the classical Bekenstein-Hawking entropy, but also of the critical points of the black hole potential \eqref{Vbh} \cite{Ferrara:2011gv} . Moreover, it was consistently extended to any generalized special geometry, thus encompassing all $\ma N>2$ supergravities, as well as $\ma N=2$ generic special geometry, not necessarily having a coset space structure.

The existence of this further symmetry enlarges the set of invariance symmetries of the entropy function, thereby setting up the challenging question of its realization and interpretation in string theory and $\mrm{M}$-theory once the attractor solution has been consistently uplifted \cite{Mandal:2017ioi}.

In this section we will present the consistent formulation of Freudenthal duality in the context of abelian gauging of four-dimensional $\ma N=2$ supergravity; this will be performed both for $\mrm{U(1)}$ $\mrm{FI}$ gauging and then for models with couplings to the hypermultiplets \cite{Klemm:2017xxk}. Finally we will see the non-linear symmetry in action by considering the explicit example of the model $F=-i X^0 X^1$.

\subsection{\label{sec:ungauged}The Original Formulation}

In this section we are going to formulate Freudenthal duality in $\ma N=2$ ungauged supergravity \cite{Borsten:2009zy,Ferrara:2011gv,Ferrara:2013zga}.
Let's start by recalling the attractor mechanism in ungauged supergravity \cite{Ferrara:1995ih,Strominger:1996kf,Ferrara:1996dd,Ferrara:1996um,Ferrara:1997tw}. When there is no gauging, the hypermultiplets decouple from the theory and the whole moduli space is special K\"ahler and parametrized by the complex scalars $z^i$. In section \ref{attractorhyper} we showed that, in the ungauged case, the function driving the attractor mechanism \eqref{effective} is the black hole potential $V_{\text{BH}}(\mathcal Q, z^i)$ defined in \eqref{Vbh}, where $\ma Q$ is the symplectic vector of the charges\footnote{Since Freudenthal duality will be introduced as a non-linear transformation of the charges $\ma Q$, in this section we will make explicit the charge- and the moduli-dependence as for example $V_{\text{BH}}(\mathcal Q, z^i)$ or $z_{H}^i(\mathcal Q)$.}.

Let's recall the expression for the black hole potential given in \eqref{Vbh} as follows
\begin{equation}
V_{BH}(\mathcal Q, z^i) = -\frac12{\mathcal Q}^t{\mathcal M}{\mathcal Q}\,.
\label{eq:vbh}
\end{equation}
The horizon values $z_{H}^i(\mathcal Q)=\lim_{r\rightarrow 0}\,z^i$ are thus determined by the criticality conditions
\begin{equation}
\partial _i V_{\text{BH}}(\mathcal Q, z^i)|_{z_H^i(\mathcal Q)} = 0\,, \label{eq:critun}
\end{equation}
and the Bekenstein-Hawking entropy is given by
\begin{equation}
S_{BH} = \pi V_{\text{BH}}(\mathcal Q, z^i)|_{z_H^i(\mathcal Q)}\,. \label{SBH}
\end{equation}
Following \cite{Ferrara:2011gv}, we introduce the scalar-field dependent {\itshape Freudenthal duality operator}
$\mathfrak{F}_z$ by
\begin{equation}
\mathfrak{F}_z(\mathcal Q) = \hat{\mathcal Q} = -\Omega\mathcal{M}\mathcal{Q}\,, \qquad
\mathfrak{F}_z(\mathcal V) = \mathcal{V}\,, \label{eq:actF}
\end{equation}
where the symplectic section $\mathcal V$ and the matrices $\mathcal M$, $\Omega$ were defined in section \ref{specialgeometry}. It is possible to show that they satisfy the relations
\begin{equation}
{\mathcal M}^t = {\mathcal M}\,, \qquad {\mathcal M}\Omega{\mathcal M} = \Omega\,, \qquad
{\mathcal M}{\mathcal V} = i\Omega{\mathcal V}\,, \qquad {\mathcal M}D_i{\mathcal V} = -i\Omega 
D_i{\mathcal V}\,. \label{eq:Mprop}
\end{equation}
As a consequence of \eqref{eq:Mprop}, it follows that the action of $\mathfrak{F}_z$ on $\mathcal Q$ is 
anti-involutive, $\mathfrak{F}_z^2(\mathcal Q) = -\mathcal{Q}$. Using again \eqref{eq:Mprop}, one
shows that
\begin{equation}
\mathfrak{F}_z (V_{\text{BH}}(\mathcal Q, z^i)) = -\frac12\hat{\mathcal{Q}}^t\mathcal{M}
\hat{\mathcal{Q}} = V_{\text{BH}}(\mathcal Q, z^i)\,, \label{eq:FV}
\end{equation}
implying that the black hole potential is invariant under Freudenthal duality.
Moreover, the second of \eqref{eq:Mprop} yields
\begin{equation}
\partial_i\mathcal M = \mathcal{M}\,\Omega\,(\partial_i\mathcal{M})\,\Omega\,\mathcal M\,. \label{eq:Id}
\end{equation}
The direct application of this identity implies that under $\mathfrak{F}_z$, $\partial_i V_{\text{BH}}$ flips
sign\footnote{Since the operator $\mathfrak F_z$ does not commute with $\partial_i$, it is
important to specify that $\mathfrak F_z$ acts always after the action of $\partial_i$. Notice that
\eqref{eq:FVd} corrects equation (3.11) of \cite{Ferrara:2011gv}.},
\begin{equation}
\mathfrak F_z (\partial_i V_{\text{BH}}(\mathcal Q, z^i)) = -\frac12\hat{\mathcal Q}^t(\partial_i\mathcal M)
\hat{\mathcal Q} = -\partial_i V_{\text{BH}}(\mathcal Q, z^i)\,. \label{eq:FVd}
\end{equation}
Since the $z_H^i(\mathcal Q)$ are the critical points of $V_{\text{BH}}$, one has
\begin{equation}
0 = \partial_i V_{\text{BH}}|_{z_H^i(\mathcal Q)} = -\mathfrak F_z(\partial_i
V_{\text{BH}})|_{z_H^i(\mathcal Q)} = \frac12\hat{\mathcal Q}^t (\partial_i\mathcal M)
\hat{\mathcal Q}|_{z_H^i(\mathcal Q)} = \frac12\hat{\mathcal Q}_{H}^t\partial_i
\mathcal M (z_H^i(\mathcal Q))\hat{\mathcal Q}_{H}\,, \label{eq:criticalungauged}
\end{equation}
where we introduced {\itshape Freudenthal duality} $\mathfrak F$ {\itshape at the horizon} as
\begin{equation}
\mathfrak F(\mathcal Q) = \mathfrak F_z(\mathcal Q)|_{z_H^i(\mathcal Q)} = -\Omega
\mathcal M_{H}\mathcal Q = \hat{\mathcal Q}_{H}\,. \label{hor-limit}
\end{equation}
On the other hand, applying \eqref{eq:critun} to the charge configuration $ \hat{\mathcal Q}_{H}$
leads to
\begin{equation}
0 = -\partial_i V_{\text{BH}}(\hat{\mathcal Q }_H, z^i)|_{z_H^i( \hat{\mathcal Q}_{H})}
= \frac12\hat{\mathcal Q }_{\text h}^t\partial_i\mathcal M (z_H^i( \hat{\mathcal Q}_{H}))
 \hat{\mathcal Q}_{H}\,. \label{eq:dualcriticalungauged}
\end{equation}
Comparing \eqref{eq:criticalungauged} and \eqref{eq:dualcriticalungauged}, one can conclude that the
attractor configuration
\begin{equation}
z_H^i( \hat{\mathcal Q}_{H}) = z_H^i(\mathcal Q)\,,  \label{eq:stabsca}
\end{equation}
is a solution also for \eqref{eq:dualcriticalungauged} \cite{Ferrara:2011gv}. The equation \eqref{eq:stabsca} can be interpreted as the stabilization of the near horizon configuration under Freudenthal duality, but an explicit
verification of this claim is possible only if all the charges are different from zero. In any case one can always
verify that $z^i_{H}$ is critical point for both $V_{\text{BH}}(\ma Q, z^i)$ and
$V_{\text{BH}}(\hat{\ma Q}_{H}, z^i)$.

This fact turns out to be crucial in order to extend \eqref{eq:actF} to a symmetry of the black hole entropy
$S_{BH}$. In fact, using \eqref{SBH}, \eqref{eq:FV} and \eqref{eq:stabsca}, one obtains
\begin{eqnarray}
\frac1{\pi}\mathfrak F (S_{BH}) &=& \mathfrak F\left(-\frac12\mathcal Q^t\mathcal M (z_H^i
(\mathcal Q))\mathcal Q\right) = -\frac12 \hat{\mathcal Q}_{H}^t\mathcal M (z_H^i
( \hat{\mathcal Q}_{H})) \hat{\mathcal Q}_{H} \nonumber \\
&=& -\frac12\mathcal Q^t\mathcal M_{H}\mathcal Q = \frac{S_{BH}}\pi\,. \label{eq:Sinv}
\end{eqnarray}
Thus, the entropy pertaining to the charge configuration $\mathcal Q$ is the same as the one pertaining
to the Freudenthal dual configuration $\mathfrak F(\mathcal Q)$. Since $\mathfrak F(\mathcal Q)$
in \eqref{hor-limit} is homogeneous of degree one (but generally non-linear) in $\mathcal Q$,
\eqref{eq:Sinv} results in the quite remarkable fact that the Bekenstein-Hawking entropy of a black hole in
ungauged supergravity is invariant under an intrinsically non-linear map acting on charge configurations.
Note that no assumption has been made on the underlying special K\"ahler geometry, nor did we use
supersymmetry.

It is worthwhile to note that the action of Freudenthal duality on the charges takes the form of the action of a {\itshape symplectic gradient} on the Bekenstein-Hawking entropy \cite{Ferrara:2011gv}.
Taking (\ref{eq:vbh}) into account, one can show that
\begin{equation}
\mbox{$\mathfrak{F}$}_{z}(\mbox{$\mathcal{Q}$})=\Omega \frac{\partial V_{%
\text{BH}}}{\partial \mbox{$\mathcal{Q}$}}\, , \label{eq:Fsymgra}
\end{equation}
with the near-horizon limit given by
\begin{equation}
\mbox{$\mathfrak{F}$}(\mbox{$\mathcal{Q}$})=\frac{1}{\pi }\Omega \frac{%
\partial S_{BH}}{\partial \mbox{$\mathcal{Q}$}}\,.
\end{equation}%
The invariance (\ref{eq:Sinv}) of $S_{BH}$ can be recasted as
\begin{equation}
S_{BH}\left( \mathcal{Q}\right) =S_{BH}\left( \frac{1}{\pi }%
\Omega \frac{\partial S_{BH}}{\partial \mbox{$\mathcal{Q}$}}\right)\, ,
\label{ung-inv}
\end{equation}
 where the criticality of $V_{\text{BH}}$ in the near-horizon limit has been used.

\subsection{\label{freudenthalgauged}Freudenthal Duality in Gauged Supergravity}

The definition of Freudenthal duality in gauged supergravity is based on the symplectic covariant formulation with general gaugings. In section \ref{attractorhyper} we presented the attractor mechanism only for the case of electric gaugings on the quaternionic moduli space. The extension to the case of magnetic gaugings is almost automatic since the covariant formulation does not add any new term to the near-horizon expression of the equations of motion \eqref{nheom}. The only difference is in the expression of $V_g$, entering in the function $V_{eff}(\mathcal Q, q^u, z^i)$ in \eqref{effective}, that now depends also on magnetic Killing vectors and moment maps.

In this section we will firstly formulate the action of Freudenthal duality in the case of a general $\mrm{FI}$ gauging, thus we will extend the duality to the case of couplings to the hypermultiplets. 

Let's recall the results of section \ref{attractorhyper} by taking into account that in this case $V_g$ includes magnetic gaugings. The function $V_{eff}$ driving the attractor mechanism is given by
\begin{equation}
V_{eff}(\mathcal Q, q^u, z^i) = \frac{\kappa - \sqrt{\kappa^2 - 4V_{\text{BH}} V_g}}{2V_g}\,, \label{eq:effective}
\end{equation}
and its critical points determine the horizon values through the equations
\begin{equation}
\partial_i V_{eff}(\mathcal Q, q^u, z^i)|_{z_H^i,\, q_H^u} = 0\,, \qquad
\partial_u V_{eff}(\mathcal Q, q^u, z^i)|_{z_H^i,\, q_H^u} = 0\,. \label{eq: crithyp}
\end{equation}
Thus the entropy density reads
\begin{equation}
s_{BH} =\frac{S_{BH}}{\text{vol}(\Sigma_\kappa)} = \frac{V_{eff}(\mathcal Q, q^u.
z^i)|_{z_H^i,\,q_H^u}}4\,. \label{entr-density}
\end{equation}

The case of $\mrm{FI}$ gaugings is obtained by requiring constant hyperscalars and choosing the moment maps as in \eqref{FIchoice}. Thus the complex function $\ma L$ entering in the covariant scalar potential $V_g$ written in \eqref{eq:superV1} takes the form \eqref{FIsuperpotential}. 
It follows that the parameters describing a black hole horizon will be doubled by the $\mrm{FI}$
couplings $\mathcal G$, thus we write $z^i_H(\ma Q, \ma G)=\lim_{r\rightarrow 0}z^i $.

As a first step, we extend the action of the field-dependent Freudenthal duality $\mathfrak F_z$ by
acting on both $\mathcal Q$ and $\mathcal G$ according to
\begin{equation}
\mathfrak F_z (\mathcal Q) = \hat{\mathcal Q} = -\Omega\mathcal M\mathcal Q\,, \qquad
\mathfrak F_z (\mathcal G) = \hat{\mathcal G} = -\Omega\mathcal M\mathcal G\,, \label{eq:Fimages}
\end{equation}
while, by definition, $\mathfrak F_z$ leaves the symplectic section $\mathcal V$ (and its covariant derivatives) invariant. Now we can recast the scalar potential \eqref{eq:superV1} in the following form
\begin{equation}
V_g(\mathcal G, z^i) = g^{i\bar\jmath} D_i\mathcal L\bar D_{\bar\jmath}\bar{\mathcal L} - 3\vert\mathcal L\vert^2
= -\frac12\mathcal G^t\mathcal M\mathcal G - 4\vert\mathcal L\vert^2 \,, \label{Vg}
\end{equation}
where we used \eqref{eq:Mprop}, \eqref{eq:Id} and we remind that
\begin{equation}
\mathcal L=-\kappa \, \mathcal G\Omega\mathcal V = -\kappa \langle\mathcal G,\mathcal V\rangle\,. \label{L-call}
\end{equation}
Thus one obtains
\begin{equation}
\begin{split}
&\mathfrak F_z (V_g(\mathcal G, z^i)) = -\frac12\hat{\mathcal G}^t\mathcal M\hat{\mathcal G}
- 4\hat{\mathcal L}\hat{\bar{\mathcal L}} = V_g(\mathcal G, z^i)\,, \\
&\mathfrak F_z (\partial_i V_g(\mathcal G, z^i)) = -\frac12\hat{\mathcal G}^t (\partial_i\mathcal M) 
\hat{\mathcal G} - 4(D_i\hat{\mathcal L})\hat{\bar{\mathcal L}} = -\partial_i V_g(\mathcal G, z^i)\,.
\end{split} \label{July}
\end{equation}
Since $V_{eff}$ and $\partial_i V_{eff}$ can be written
as functions of $V_{\text{BH}}$, $V_g$, $\partial_i V_{\text{BH}}$ and $\partial_i V_g$, the relation \eqref{July}
together with \eqref{eq:FV} and \eqref{eq:FVd} implies
\begin{equation}
\mathfrak F_z (V_{eff}(\mathcal Q, \mathcal G, z^i)) = V_{eff}(\mathcal Q, \mathcal G, z^i)\,,
\qquad\mathfrak F_z (\partial_i V_{eff}(\mathcal Q, \mathcal G, z^i)) = -\partial_i V_{eff}
(\mathcal Q, \mathcal G, z^i)\,. \label{eq:FVgge}
\end{equation}
Using the second relation of \eqref{eq:FVgge}, one has then
\begin{eqnarray}
0& =& -\partial_i V_{eff}|_{z_H^i(\mathcal Q, \mathcal G)} = \mathfrak F_z (\partial_i
V_{eff})|_{z_H^i(\mathcal Q, \mathcal G)} \nonumber \\
&=&\partial_i V_{eff}(\hat{\mathcal Q}, \hat{\mathcal G}, z^i)|_{z_H^i (\mathcal Q,
\mathcal G)} = \partial_i V_{eff}( \hat{\mathcal Q}_{H}, \hat{\mathcal G}_{H}, z_H^i
(\mathcal Q, \mathcal G))\,. \label{eq:Feffective}
\end{eqnarray}
Let us define Freudenthal duality at the horizon by
\begin{eqnarray}
\mathfrak F (\mathcal Q) &=& \mathfrak F_z (\mathcal Q)|_{z_H^i(\mathcal Q, \mathcal G)}
= -\Omega\mathcal M_{H}\mathcal Q = \hat{\mathcal Q}_{H}\,, \nonumber \\
\mathfrak F (\mathcal G) &=& \mathfrak F_z (\mathcal G)|_{z_H^i(\mathcal Q, \mathcal G)}
= -\Omega\mathcal M_{H}\mathcal G =\hat{\mathcal G}_{H}\,. \label{eq:hF}
\end{eqnarray}
From the comparison of \eqref{eq:Feffective} with the definition
\begin{equation}
0 = \partial_i V_{eff}( \hat{\mathcal Q}_{H}, \hat{\mathcal G}_{H}, z^i)|_{z_H^i
( \hat{\mathcal Q}_{H}, \hat{\mathcal G}_{H})} = \partial_i V_{eff}( \hat{\mathcal Q}_{H},
\hat{\mathcal G}_{H}, z_H^i ( \hat{\mathcal Q}_{H}, \hat{\mathcal G}_{H}))\,,
\label{eq:dualderveff}
\end{equation}
it follows that
\begin{equation}
z_H^i ( \hat{\mathcal Q}_{H}, \hat{\mathcal G}_{H}) = z_H^i (\mathcal Q,
\mathcal G) \label{eq:stabscag}
\end{equation}
is a solution also for \eqref{eq:dualderveff}, thus it is a critical point for both $V_{eff}$ and
$\maf F_z (V_{eff})$.

The equations \eqref{entr-density}, \eqref{eq:FVgge} and \eqref{eq:stabscag} imply that $S_{BH}$ is
invariant under Freudenthal duality,
\begin{eqnarray}
4\mathfrak F (s_{BH}) &=& V_{eff}( \hat{\mathcal Q}_{H}, \hat{\mathcal G}_{H},
z_H^i ( \hat{\mathcal Q}_{H}, \hat{\mathcal G}_{H})) =
V_{eff}( \hat{\mathcal Q}_{H}, \hat{\mathcal G}_{H}, z_H^i (\mathcal Q,\mathcal G))
\nonumber \\
&=& V_{eff}(\mathcal Q, \mathcal G, z_H^i (\mathcal Q, \mathcal G)) =
4 s_{BH}\,. \label{eq:Sinvg}
\end{eqnarray}
It is immediate to see that in the limit $\mathcal G\to 0$, one recovers the results of the ungauged case.
Notice that the origin of Freudenthal duality is firmly rooted into the properties \eqref{eq:Mprop}. The
action of $\mathfrak F$ yields a new attractor-supporting configuration $( \hat{\mathcal Q}_{H},
\hat{\mathcal G}_{H})$ that, in general, belongs to a physically different theory, specified by a
different choice of gauge couplings.
It is worthwhile to note that no assumption has been made on the special K\"ahler geometry in this formulation of the duality.

As in the ungauged case, it is possibile to write the action of Freudenthal duality as the symplectic gradient of the entropy density. 
Taking into account that (\ref{eq:effective}) is critical at the horizon, one can derive its derivative respect to the charges and 
obtain\footnote{If one tries to apply the same procedure taking the derivative respect to $\ma G$ one discovers that it reproduces a symplectic gradient only imposing $\ma L_H=0$.}
\begin{equation}
\mathfrak{F}(\ma Q)=2\left( \frac{R_{H}}{R_{\scriptsize{\mrm{AdS}_2}}}\right)^{2}\Omega \, \frac{\partial\, V_{eff}( q_H, \ma G_{H}, z_H^i (\mathcal Q,\mathcal G))}{\partial \ma Q}=
8\left( \frac{R_{H}}{R_{\scriptsize{\mrm{AdS}_2}}}\right)^{2}\Omega \, \frac{\partial S_{BH}}{\partial \mbox{$\mathcal{Q}$}},
\end{equation}
where we used (\ref{entr-density}) and the expressions of the radii $R_H$ and $R_{\scriptsize{\mrm{AdS}_2}}$ of the horizon in terms of $V_{eff}$, $V_g$ and $V_{\text{BH}}$ given in \eqref{rH} and \eqref{rA}.
In this way, the invariance of the entropy density under
Freudenthal duality can be recasted in the form
\begin{equation}
s_{BH}\left(\ma Q, \ma G \right)
\,=s_{BH}\left( 8 \left( \frac{R_{H}}{R_{AdS_{2}}}\right)^{2}\Omega \frac{\partial S_{BH}}{\partial \ma Q},\,
\mathfrak{F}(\ma G)\right) \, .
\end{equation}

Let's generalize our analysis to include also hypermultiplets. The dynamics of the attractor mechanism is now governed by the potentials $V_{\text{BH}}(\ma Q, z^i)$ and $V_g(\ma P^x(q^u),{\mathcal K}^u,z^i)$ given in \eqref{eq:superV1}, where $\ma P^x$ and $\mathcal K^u$ are given in \eqref{eq:magneticgaugings}.

The field-dependent Freudenthal duality is again defined by \eqref{eq:actF}, supplemented with
\begin{equation}
\mathfrak F_z (\mathcal P^x)=\hat{\mathcal P}^x = -\Omega\mathcal M\mathcal P^x\,,
\qquad\mathfrak F_z (\mathcal K^u )=\hat{\mathcal K}^u = -\Omega\mathcal M\mathcal K^u\,.
\end{equation}
One easily shows that $\mathfrak F_z(\ma Q^x)=\ma Q^x$ and, with slightly more effort, that
\begin{equation}
\begin{split}
&\mathfrak F_{z} (V_{eff}(\ma Q,\ma P^x (q^u), \mathcal K^u (q^u), z^i)) = V_{eff}
(\ma Q,\ma P^x (q^u), \mathcal K^u (q^u), z^i)\,, \\
&\mathfrak F_{z} (\partial_A V_{eff}(\ma Q, \ma P^x (q^u), \mathcal K^u (q^u), z^i)) =
-\partial_A V_{eff}(\ma Q, \ma P^x (q^u), \mathcal K^u (q^u), z^i)\,.
\label{eq:invarianceVeffhyper}
\end{split}
\end{equation}
Thus, in analogy to the $\mrm{FI}$ case, one has to consider the criticality conditions\footnote{For simplicity of notation we introduced a collective variable for the attractor points $\phi^A_H=(z^i_H,q^u_H)$.} \eqref{eq: crithyp}
and apply the second relation of \eqref{eq:invarianceVeffhyper},
\begin{equation}
\begin{split}
0 & = -\partial_A V_{eff}(\ma Q, \mathcal P^x, \mathcal K^u, z^i)|_{\phi_{H}^A} =
\maf F_z(\partial_A V_{eff}(\ma Q, \mathcal P^x, \mathcal K^u, z^i))|_{\phi_{H}^A} = \\
& = \partial_A V_{eff}(\hat{\ma Q}, \hat{\mathcal P}^x, \hat{\mathcal K}^u, z^i)|_{\phi_{H}^A}
= \partial_A V_{eff}(\hat{\ma Q}_{H}, \hat{\mathcal P}^x_{H} (q^u_{H}),
\hat{\mathcal K}^u (q^u_{H}), z^i_{H})\,, \label{eq:criticalhyper}
\end{split}
\end{equation}
where
\begin{equation}
\hat{\mathcal P}^x_{H} (q^u)  = -\Omega\mathcal M_{H}\mathcal P^x( q^u)
\end{equation}
is the dual expression for the moment maps that depends on the scalar fields, the charges and the 
parameters contained in the quaternionic Killing vectors.
Defining $\hat{\ma Q}_{H}$ as in \eqref{eq:hF}, the criticality condition of the attractor points
$\hat{\phi}_{H}^A$ for the dual configuration of $(\ma Q, \ma P^x(q^u))$, namely for
$(\hat{\ma Q}_{H},\hat{\mathcal P}_{H}^x(q^u))$, reads
\begin{equation}
0 = \partial_A V_{eff}(\hat{\ma Q}_{H}, \hat{\mathcal P}_{H}^x, \hat{\mathcal K}^u,
z^i)|_{\hat{\phi}_{H}^A} = \partial_A V_{eff}(\hat{\ma Q}_{H}, \hat{\mathcal P}_{H}^x (\hat q^u_{H}), \hat{\mathcal K}^u (\hat q^u_{H}), \hat z^i_{H})\,.
\label{eq:dualcriticalhyper}
\end{equation}
Thus a comparison between \eqref{eq:criticalhyper} and \eqref{eq:dualcriticalhyper} shows that the
configuration
\begin{equation}
\phi_{H}^A = \hat{\phi}_{H}^A
\end{equation}
is a solution for both criticality conditions. It follows that
\begin{equation}
\begin{split}
4\,\maf F (s_{BH}) = & V_{eff}(\hat{\ma Q}_{H}, \hat{\mathcal P}^x_{H}
(\hat{q}^u_{H}), \hat{z}_{H}^i) = V_{eff}(\hat{\ma Q}_{H},\hat{\mathcal P}^x_{H}
(q^u_{H}), z^i_{H}) \\
& = V_{eff}(\ma Q, \mathcal P^x_{H} (q^u_{H}), z^i_{H}) = 4 s_{BH}\,,
\label{eq:Sinvhyper}
\end{split}
\end{equation}
namely the entropy density of the two configurations related by the Freudenthal operator is the same.

\subsection{The Example of $F=-iX^0X^1$}

As an illustrative example, let us check the action of Freudenthal duality for the simple model with
prepotential\footnote{The main properties of this model will be introduced in section \ref{X0X1UHM}. Alternatively see \cite{Cacciatori:2009iz} for more details.} $F=-i X^0 X^1$ and purely electric FI gauging.
To keep things simple, we assume that the electric charges vanish. One has thus
\begin{equation}
\ma Q = \left(\begin{array}{c}
p^0 \\
p^1\\
0\\
0
\end{array}\right)\,, \qquad \ma G = \left(\begin{array}{c}
0 \\
0 \\
g_0 \\
g_1
\end{array}\right)\,.
\end{equation}
This model has just one complex scalar $z=x+iy$, and the matrix $\mathcal M$ is given by
\begin{equation}
\mathcal M = \left(\begin{array}{cccc} -\frac{x^2 + y^2}x & 0 & \frac yx & 0 \\
0 & -\frac1x & 0 & -\frac yx \\
\frac yx & 0 & -\frac1x & 0 \\
0 & -\frac yx & 0 & -\frac{x^2 + y^2}x
\end{array}\right)\,.
\end{equation}
The black hole- and scalar potential read respectively
\begin{eqnarray}
V_{\text{BH}} &=& -\frac12\mathcal Q^t\mathcal M\mathcal Q = \frac{x^2 + y^2}{2x}(p^0)^2 +
\frac{(p^1)^2}{2x}\,, \nonumber \\
V_g &=& -\frac1{2x} (g_0^2 + 4g_0 g_1 x + g_1^2 (x^2 + y^2))\,. \label{VVBHX0X1}
\end{eqnarray}
Plugging this into the effective potential \eqref{eq:effective}, one shows that the latter is extremized for
\begin{equation}
x = x_{H} = \frac{u g_0}{g_1}\,, \qquad y = y_{H} = 0\,, \label{eq:X0X1}
\end{equation}
where $u$ is a solution of the quartic equation
\begin{equation}
\left[(1 - \nu^2) u + 2(u^2 - \nu^2)\right]^2 = k(1 - u^2)(\nu^2 - u^2)\,, \label{eq:u-quartic}
\end{equation}
with
\begin{equation}
\nu=\frac{g_1 p^1}{g_0 p^0}\,, \qquad k=\frac{\kappa^2}{(g_0 p^0)^2}\,.
\end{equation}
Note that positivity of the kinetic terms in the action requires $x>0$. Depending on the sign of
$g_0/g_1$, this means that either only negative or only positive roots of \eqref{eq:u-quartic} are
allowed, and such roots may not exist for all values of $\nu$ and $k$. Notice also that in the special
case where
\begin{equation}
(2g_0 p^0)^2 = (2g_1 p^1)^2 = \kappa^2\,, \label{eq:cond-BPS-sign}
\end{equation}
the effective potential \eqref{eq:effective} becomes completely flat,
\begin{equation}
V_{eff} = -\frac{\kappa}{2 g_0 g_1}\,,
\end{equation}
and the scalar $z$ is thus not stabilized at the horizon, a fact first noted in \cite{Cacciatori:2009iz}.
(Nonetheless, the entropy is still independent of the arbitrary value $z_H$, in agreement with
the attractor mechanism, i.e. it constitutes a flat direction). The relation \eqref{eq:cond-BPS-sign} corresponds to the BPS conditions
found in \cite{Cacciatori:2009iz}, or to a sign-flipped modification of them\footnote{In the BPS case,
$g_0p^0$ and $g_1p^1$ must have the same sign. It would be interesting to see whether the
appearance of flat directions is a generic feature of the BPS case, or just a consequence of the
simplicity of the model under consideration.}.

Using \eqref{eq:u-quartic}, one can derive the near-horizon value of $V_{eff}$, and thus the entropy
density \eqref{entr-density},
\begin{equation}
s_{BH} = \frac{V_{eff}(\mathcal Q, \mathcal G, z^i)|_{z_H^i(\mathcal Q, \mathcal G)}}4
= \frac{g_0 {p^0}^2 [(1 - \nu^2) u + 2(u^2 - \nu^2)]}{4\kappa g_1 (1 - u^2)}\,.
\label{eq:entropyX0X1}
\end{equation}
We now determine the action of Freudenthal duality on the charges and the FI parameters.
The definitions \eqref{eq:hF} yield
\begin{equation}
\maf F(\ma Q)=\left(\begin{array}{c} 0 \\ 0 \\ \hat{q}_0 \\ \hat{q}_1\end{array}\right) =
\left(\begin{array}{c} 0 \\ 0 \\ p^0 x_H \\ p^1/x_H\end{array}\right)\,, \qquad
\maf F(\ma G)=\left(\begin{array}{c} \hat{g}^0 \\ \hat{g}^1 \\ 0 \\ 0\end{array}\right) =
\left(\begin{array}{c} -g_0/x_H \\ -g_1 x_H \\ 0 \\ 0\end{array}\right)\,.
\label{eq:dualchargesX0X1}
\end{equation}
The dual configuration is thus electrically charged and has purely magnetic gaugings.
For the transformed potentials one gets
\begin{eqnarray}
\maf F (V_{\text{BH}}) &=& -\frac12\hat{\mathcal Q}^t_{H}\mathcal M \hat{\mathcal Q}_{H} =
\frac{x^2 + y^2}{2x}\hat{q}_1^2 + \frac{\hat{q}_0^2}{2x}\,, \\
\maf F (V) &=& -\frac12\hat{\mathcal G}^t_{H}\mathcal M\hat{\mathcal G}_{H} - 4|\langle
\hat{\mathcal G}_{H}, \mathcal V\rangle|^2
= -\frac1{2x}\left((\hat{g}^1)^2 + 4\hat{g}^0\hat{g}^1 x + (\hat{g}^0)^2 (x^2 + y^2)\right)\,. \nonumber
\end{eqnarray}
These are identical to \eqref{VVBHX0X1}, except for the replacements
\begin{displaymath}
(p^0)^2 \to \hat{q}_1^2\,, \qquad (p^1)^2 \to \hat{q}_0^2\,, \qquad g_0^2 \to (\hat{g}^1)^2\,,
\qquad g_1^2 \to (\hat{g}^0)^2\,, \qquad g_0 g_1 \to \hat{g}^0\hat{g}^1\,.
\end{displaymath}
The critical points of $\maf F(V_{eff})$ are thus $\hat{x}_{H}=\hat{g}^1\hat{u}/\hat{g}^0$
and $\hat{y}_{H}=0$, where $\hat u$ satisfies
\begin{equation}
\left[(1 - \hat{\nu}^2)\hat u + 2(\hat{u}^2 - \hat{\nu}^2)\right]^2 = \hat k(1 - \hat{u}^2)(\hat{\nu}^2
- \hat{u}^2)\,, \label{eq:hatu-quartic}
\end{equation}
with
\begin{equation}
\hat\nu=\frac{\hat{g}^0\hat{q}_0}{\hat{g}^1\hat{q}_1}\,, \qquad\hat k=
\frac{\kappa^2}{(\hat{g}^1\hat{q}_1)^2}\,.
\end{equation}
Now, using \eqref{eq:dualchargesX0X1}, one easily shows that
\begin{displaymath}
\hat{\nu}^2 = \frac1{\nu^2}\,, \qquad \hat k = \frac k{\nu^2}\,.
\end{displaymath}
Plugging this into \eqref{eq:hatu-quartic} and multiplying with $\nu^4/\hat{u}^4$ yields
\begin{equation}
\left[(1 - \nu^2)\hat{u}^{-1} + 2(\hat{u}^{-2} - \nu^2)\right]^2 = k(1 - \hat{u}^{-2})(\nu^2 - \hat{u}^{-2})\,.
\end{equation}
Comparing with \eqref{eq:u-quartic}, we see that $u$ and $\hat{u}^{-1}$ satisfy the same equation,
and have thus the same set of solutions. Hence, up to permutations of possible multiple roots, one gets
$u=\hat{u}^{-1}$, which, by means of \eqref{eq:dualchargesX0X1}, leads to $\hat x_H=x_H$,
and therefore $V_{eff}$ and $\maf F(V_{eff})$ share the same critical points.

The transformed entropy density is given by
\begin{equation}
\maf F(s_{BH}) = \frac{V_{eff}(\maf F(\mathcal Q), \maf F(\mathcal G),z^i)|_{\hat{
z}_{H}^i(\maf F(\mathcal Q), \maf F(\mathcal G))}}4 = \frac{\hat{g}^1\hat{q}_1^2 [(1 - \hat{\nu}^2)
\hat u + 2(\hat{u}^2 - \hat{\nu}^2)]}{4\kappa\hat{g}^0 (1 - \hat{u}^2)}\,.
\end{equation}
Using again \eqref{eq:dualchargesX0X1}, it is easy to see that this coincides with \eqref{eq:entropyX0X1},
so that the entropy is indeed invariant under Freudenthal duality.

%% file: chap5.tex

\chapter{BPS Black Holes in  $\ma N=2$, $d=4$ Supergravity}
\label{blackholes}
\thispagestyle{plain}


In this chapter we focus our attention on extremal black hole solutions in four-dimensional $\ma N=2$ gauged supergravity. In particular we will present two examples of supersymmetric static black holes enjoying spherical/hyperbolic symmetry obtained by using the first-order formulation presented in sections \ref{Flow Equations} and \ref{symplecticallyinvariant}.
These two solutions are analytical and they can be schematically introduced as follows:
\begin{itemize}

\item A dyonic black hole \cite{Klemm:2015xda} describing a flow of the type $\mrm{AdS}_4\longrightarrow \mrm{AdS}_2\times \Sigma_\kappa$
obtained in a non-homogeneous deformation of the $STU$ model \eqref{stumanifold}. This deformation is defined by the prepotential $F=\frac{X^1X^2X^3}{X^0}-\frac{A}{3}\frac{(X^3)^3}{X^0}$ and by a dyonic $\mrm{FI}$ gauging with three magnetic parameters $g^1, g^2, g^3$ and an electric parameter $g_0$. The flow is characterized by two real scalars given by the imaginary parts (thus with vanishing Betti moduli) of two of the three special K\"ahler moduli $z^i$, and by one magnetic and two electric charges.

\item A magnetic black hole \cite{Chimento:2015rra} describing a flow interpolating between an $\mrm{AdS}_2\times H^2$ horizon and an asymptotics conformally equivalent to $\mrm{AdS}_2\times H^2$. This black hole has been obtained in a model defined by a special K\"ahler moduli space associated to the prepotential $F=-iX^0X^1$ and by the universal hypermultiplet \eqref{uhm}. The solution is completely characterized by the flow of the dilaton coming from the universal hypermultiplet and by a pure electric gauging on the quaternionic manifold realized by the non-compact gauge group $\mathbb{R}\times \mrm{U(1)}$.

\end{itemize}
These solutions are interesting for various reasons. The first one constitute the only example of black hole solution with an $\mrm{AdS}_4$ vacuum in the asymptotics, obtained in a non-homogeneous special geometry. This particular scalar geometry is associated to the non-perturbative corrections to the prepotential of the $STU$ model due to worldsheet istantons \cite{Louis:1996mt}. As far as we know the brane picture associated to this $\mrm{AdS}_4$ vacuum is still unknown as well as its holographic interpretation.

The second solution represents the unique known example in gauged supergravity of exact $d=4$ black hole flow with a charged hyperscalar running. The study of solutions including couplings to the hypermultiplets constitutes one of the most iteresting research directions in gauged supergravity. As we saw in section \ref{Mtheorytruncation}, except to that one on the $S^7$, all the $\mrm{M}$-theory truncations with $\ma N=2$ $\mrm{AdS}_4$ vacua on seven-dimensional Sasaki-Einsteins include hypermultiplets, thus the discovery of $\mrm{AdS}_4$ flows in this context is crucial for their implication in $\mrm{AdS/CFT}$. Some numerical results have been obtained in \cite{Halmagyi:2013sla,Hosseini:2016ume} for models discussed in section \ref{Mtheorytruncation} and others in the $\ma N=2$ truncations of massive type $\mrm{IIA}$  \cite{Guarino:2015jca,Guarino:2015qaa,Guarino:2015vca,Guarino:2016err,Guarino:2017eag,Guarino:2017pkw,Guarino:2017jly}.


\section{A Black Hole in the Non-Homogeneous $STU$ Model}

As we already mention in section \ref{modulispace}, the inclusion in the scenario of supergravity of the non-perturbative effects, typically coming from the istantonic corrections to the worldsheet theory \cite{Becker:1995kb,Witten:1999eg}, implies the appearance of non-perturbative corrections to the prepotential of the K\"ahler moduli space of the Calabi-Yau. 
To explain in particular the deformations of the cubic models \eqref{specialprepotential}, one has to take in account the duality relating the strong coupling regime of $\mrm{E}_8\times \mrm{E}_8$ heterotic string theory on the orbifold $(K3\times T^2)/\mathbb{Z}_2$ with the weak coupling of type $\mrm{IIA}$ string theory on a suitable Calabi-Yau threefold \cite{Hull:1994ys,Witten:1995ex}. 
This duality relation implies that the respective moduli spaces are identical and thus the weak-coupling computations in a type $\mrm{IIA}$ setting give non-perturbative insights about the dual heterotic vacua
and viceversa.

In particular, if we consider the heterotic $\ma N=2$ vacua in $d=4$, these have a dual description in terms of vacua in type $\mrm{IIA}$ arising from a K\"ahler moduli space with prepotential corrected by non-perturbative effects typically described as worldsheet istantons \cite{Ferrara:1995yx, Kachru:1995wm}. 
In this section we will consider an example of these truncations given by a four-dimensional K\"ahler moduli space described by $n_V=3$ vector multiplets and a quaternionic moduli space defined by $n_H=273$ hypermultiplets\footnote{The number of hypermultiplets $n_H=273$ is required for anomaly cancellation in heterotic string theory \cite{Louis:1996mt}.} \cite{Louis:1996mt}. For this particular truncation the prepotential is given by
\begin{equation}
F=\frac{X^1X^2X^3}{X^0}-\frac{A}{3}\frac{(X^3)^3}{X^0}\,,
\label{deformedstu}
\end{equation}
with $A=-1$. This prepotential is a deformation of the prepotential of the $STU$ model \eqref{stumanifold} and it describes a special K\"ahler manifold that in non-symmetric nor homogeneous.

In particular, we will consider a $\mrm{FI}$ gauging on \eqref{deformedstu} and we will explicitly solve the $\mrm{BPS}$ first-order equations to construct a static supersymmetric black hole with radial symmetry and an $\mrm{AdS}_4$ asymptotics. Thus we will study the main properties of this black hole solution.

\subsection{\label{Setup}Non-Homogeneous Special Geometry}

The general framework is given by $\ma N=2$, $d=4$ gauged supergravity \eqref{actionmag} coupled to $n_{V}$ vector multiplets and a dyonic $\mrm{FI}$ abelian gauging realized by the symplectic vector $\ma G$ of $\mrm{FI}$ parameters as in \eqref{FIchoice}.

As is the case for many other known solutions \cite{Cacciatori:2009iz,Mohaupt:2011aa, Gnecchi:2013mta,Errington:2014bta}, we shall assume vanishing Betti moduli. This implies that, in this particular setting, the moduli space of the the model is realized by purely imaginary scalars $\lambda^i$ (with $\lambda^{i}>0$) coming from the Calabi-Yau complex structure moduli space,
\begin{equation}
z^{i}=b^{i}-i\lambda ^{i}\,,\qquad b^{i}=0\qquad \text{with}\qquad i=1,2,3\,.  \label{eq:vanishing_axions}
\end{equation}
This choice is also called {\itshape axions vanishing condition} and it implies that, for some values of the $\mrm{FI}$ parameters in $\mathcal{G}$, the first-order equations  (\ref{eq:BPS}) are further simplified.

Let's construct the main quantities of the special geometry, introduced in section \ref{specialgeometry}, associated to the deformation \eqref{deformedstu}. The moduli $\lambda^i$ can be expressed in terms of the homogeneous coordinates $X^\Lambda$ as
\begin{equation}
\frac{X^{1}}{X^{0}}=-i\lambda ^{1}\,,\qquad \frac{X^{2}}{X^{0}}=-i\lambda
^{2}\,,\qquad \frac{X^{3}}{X^{0}}=-i\lambda ^{3}\,.
\end{equation}
Thus, the symplectic sections (\ref{specialembedding}) become
\begin{equation}
\begin{split}
& L^{\Lambda }=e^{\mathcal{K}/2}\left( 1,-i\lambda ^{1},-i\lambda
^{2},-i\lambda ^{3}\right) ^{t}\,, \\
& M_{\Lambda }=e^{\mathcal{K}/2}\left( -i\left( \lambda ^{1}\lambda
^{2}\lambda ^{3}-\frac{A}{3}(\lambda ^{3})\mbox{}^{3}\right) ,-\lambda
^{2}\lambda ^{3},-\lambda ^{1}\lambda ^{3},-\lambda ^{1}\lambda
^{2}+A(\lambda ^{3})\mbox{}^{2}\right) ^{t}\,,
\end{split}
\end{equation}
while the K\"{a}hler potential reads
\begin{equation}
e^{-\mathcal{K}}=8\left( \lambda ^{1}\lambda ^{2}\lambda ^{3}-\frac{A}{3}
(\lambda ^{3})\mbox{}^{3}\right)\,.
\end{equation}
For vanishing axions, the special K\"{a}hler metric takes the form
\begin{equation}
g_{i\bar{\jmath}}=\frac{1}{4\left( \lambda ^{1}\lambda ^{2}\lambda ^{3}-
\frac{A}{3}(\lambda ^{3})\mbox{}^{3}\right) ^{2}}\left(
\begin{array}{ccc}
(\lambda ^{2})\mbox{}^{2}(\lambda ^{3})\mbox{}^{2} & \frac{A}{3}(\lambda
^{3})\mbox{}^{4} & -\frac{2}{3}A\lambda ^{2}(\lambda ^{3})\mbox{}^{3} \\
&  &  \\
\frac{A}{3}(\lambda ^{3})\mbox{}^{4}\; & (\lambda ^{1})\mbox{}^{2}(\lambda
^{3})\mbox{}^{2} & -\frac{2}{3}A\lambda ^{1}(\lambda ^{3})\mbox{}^{3} \\
&  &  \\
-\frac{2}{3}A\lambda ^{2}(\lambda ^{3})\mbox{}^{3}\; & -\frac{2}{3}A\lambda
^{1}(\lambda ^{3})\mbox{}^{3} & \;\;(\lambda ^{1})\mbox{}^{2}(\lambda ^{2})%
\mbox{}^{2}+\frac{A^{2}}{3}(\lambda ^{3})\mbox{}^{4}%
\end{array}%
\right) \,.  \label{eq:Kaehler_metric}
\end{equation}
The symplectic matrix $\mathcal{N}_{\Lambda \Sigma }$ has, in the axion-free
case under consideration, vanishing real part $R_{\Lambda \Sigma }$, while
$I_{\Lambda \Sigma }$ is given by
\begin{equation}
I_{\Lambda \Sigma }=-\frac{1}{8}e^{-\mathcal{K}}\left(
\begin{array}{cc}
1 & 0 \\
0 & 4g_{i\bar{\jmath}}
\end{array}
\right)\,,
\end{equation}
which is thus consistently negative definite.

When dyonic $\mrm{FI}$ gauging is imposed, the effective action \eqref{eq:Seffmag} associated to the black hole Ansatz \eqref{eq:Ansatzmet}, \eqref{eq:Ansatzscalars} and \eqref{Ansatzmag1}  boils down to 
the effective action derived in \cite{Dall'Agata:2010gj} given by
\begin{equation}
\begin{split}
S=& \int \mathrm{d}r\left\{ e^{2\psi }\left[U^{\prime 2}-\psi
^{\prime 2}+g_{i\bar{\jmath}}z^{i\,^{\prime }}\bar{z}^{\bar{\jmath}%
\,^{\prime }}+e^{2U-4\psi }V_{\text{BH}}+e^{-2U}V_{g}\right] -1\right\} \,,
\end{split}
\label{eq:S_eff}
\end{equation}
where $V_{\text{BH}}$ has been written in \eqref{Vbh} and $V_g$ corresponds to \eqref{eq:superV1} where the $\mrm{FI}$ prescriptions \eqref{FIchoice} has been imposed on the quaternionic moment maps. This has the form 
\begin{equation}
V_{g}=g^{i\bar{\jmath}}D_{i}\mathcal{L}{\bar{D}}_{\bar{\jmath}}\bar{\mathcal{L}}-3|\mathcal{L}|^{2}\,,  \label{eq:scalar_pot}
\end{equation}
with $\ma L=-\kappa \, \langle \ma G, \ma V \rangle$.
Furthermore the quantization condition \eqref{eq:quantization} becomes in this case 
\begin{equation}
\langle \ma G, \ma Q \rangle=-\kappa\,,
\label{eq:kappa}
\end{equation}
as we already shown in \eqref{FIchoice}.

As far as concerns to the symplectic covariant flow equations \eqref{eq:flowequations}, in the $\mrm{FI}$ case they reduce to
\begin{align}
& 2e^{2\psi }\left( e^{-U}\mathrm{Im}(e^{-i\alpha }\mathcal{V})\right)
^{\prime}+ e^{2(\psi -U)}\Omega \mathcal{M}\mathcal{G}+4e^{-U}(\alpha
^{\prime }+\masf{A}_{r})\mathrm{Re}(e^{-i\alpha }\mathcal{V})+\mathcal{Q}%
=0\,,  \nonumber  \label{eq:BPS} \\
& \psi ^{\prime}=2e^{-U}\mathrm{Im}(e^{-i\alpha }\,\ma G^t \Omega \ma V)\,, \\
& \alpha ^{\prime }+\masf{A}_{r}=-2e^{-U}\mathrm{Re}(e^{-i\alpha }\,
\ma G^t \Omega \ma V)\,, \nonumber
\end{align}
where $\masf{A}_\mu$ is the connection associated to the K\"{a}hler transformations introduced in \eqref{U1connection}, while the phase $\alpha $ written in \eqref{eq:defal} can be expressed as
\begin{equation}
e^{2i\alpha }=\frac{\mathcal{Z}-ie^{2(\psi -U)}\,\ma G^t \Omega \ma V}{\bar{\mathcal{Z}}
+ie^{2(\psi -U)}\,\ma G^t \Omega \bar{\ma V}}\,.  \label{eq:alpha}
\end{equation}

\subsection{\label{FI-Gauging}Gauging, $\text{AdS}$ Vacua and Near-Horizon Analysis}

To proceed further, we specify a dyonic gauging by choosing the $\mrm{FI}$ parameters $\mathcal{G}$. In particular we choose
\begin{equation}
\mathcal{G}^{t}=(0,g^{1},g^{2},g^{3},g_{0},0,0,0)^{t}\,. \label{eq:G}
\end{equation}
Together with the vanishing axion condition (\ref{eq:vanishing_axions}), this choice fixes the phase $\alpha $ in (\ref{eq:alpha}) to the constant value\footnote{Another possible choice
yielding the same constant value for $\alpha $ is $\mathcal{G}^{t}=(g^0,0,0,0,0,g_1,g_2,g_3)^t$,
which would in turn
require $\mathcal{Q}$ to assume the (magnetic) form $\mathcal{Q}^t=(0,p^1,p^2,p^3,q_0,0,0,0)^t$.} $\alpha =\pm \pi /2$. This
can be checked from the explicit expressions of the symplectic invariants $%
\mathcal{Z}$ and $\mathcal{L}$,
\begin{equation}
\begin{split}
& \mathcal{Z}=ie^{\mathcal{K}/2}\left( p^{0}\left( \lambda ^{1}\lambda
^{2}\lambda ^{3}-\frac{A}{3}\lambda ^{3}\right) -q_{1}\lambda
^{1}-q_{2}\lambda ^{2}-q_{3}\lambda ^{3}\right) \,, \\
& \mathcal{L}=-\kappa \,e^{\mathcal{K}/2}\left( g_{0}+g^{1}\lambda ^{2}\lambda
^{3}+g^{2}\lambda ^{1}\lambda ^{3}+g^{3}(\lambda ^{1}\lambda ^{2}-A(\lambda
^{3})\mbox{}^{2})\right) \,.
\end{split}
\end{equation}
As can be inferred from the BPS equations (\ref{eq:BPS}), the choice (\ref{eq:G}) requires some charges
to vanish, so that the vector $\mathcal{Q}$ takes the form
\begin{equation}
\mathcal{Q}^{t}=(p^{0},0,0,0,0,q_{1},q_{2},q_{3})^{t}\,. \label{eq:Q}
\end{equation}
With the choice \eqref{eq:G}, the scalar potential (\ref{eq:scalar_pot})
becomes
\begin{eqnarray}
V_{g} &=&-g^{2}g^{3}\lambda ^{1}-g^{1}g^{3}\lambda ^{2}-\left(
g^{1}g^{2}-A(g^{3})^{2}\right) \lambda ^{3}  \nonumber \\
&&-\frac{g_{0}}{\lambda ^{1}\lambda ^{2}\lambda ^{3}-\frac{A}{3}(\lambda
^{3})\mbox{}^{3}}\left( g^{2}\lambda ^{1}\lambda ^{3}+g^{1}\lambda
^{2}\lambda ^{3}+g^{3}\left( \lambda ^{1}\lambda ^{2}-A(\lambda ^{3})\mbox{}%
^{2}\right)\right)\,, \label{pot-axionfree}
\end{eqnarray}
which matches the known expression for the $STU$ model \cite{Ceresole:2007rq,Bellucci:2008sv,Cacciatori:2009iz}
for $A=0$. In what follows we shall assume that all gauge coupling constants $g_0,g^i$ are positive.
Then the potential \eqref{pot-axionfree} has two critical points, namely one for
\begin{equation}
\lambda^1 = \frac{g^1}{g^3}\lambda^3\,, \qquad \lambda^2 = \frac{g^2}{g^3}\lambda^3\,, \qquad
\lambda^3 = \sqrt{\frac{g_0g^3}{g^1g^2 - \frac A3(g^3)^2}}\,, \label{usual-crit-pt}
\end{equation}
and the other for
\begin{equation}
\lambda^1 = \frac{g^1}{g^3}\lambda^3\,, \qquad \lambda^2 = -\frac1{g^1g^3}\left(g^1g^2 -
\frac23 A(g^3)^2\right)\lambda^3\,, \qquad \lambda^3 = \sqrt{\frac{g_0g^3}{g^1g^2 -
\frac A3(g^3)^2}}\,. \label{new-crit-pt}
\end{equation}
The first has $V_g=-3\ R_{\scriptsize{\mrm{AdS}_4}}^{-2}$, and the second $V_g=-R_{\scriptsize{\mrm{AdS}_4}}^{-2}$, where $R_{\scriptsize{\mrm{AdS}_4}}$ is the curvature radius of $\mrm{AdS}_4$, so both correspond to $\mrm{AdS}_4$ vacua. One easily shows that \eqref{usual-crit-pt} is also a critical point of $\ma L$, while \eqref{new-crit-pt} is not. The vacuum \eqref{usual-crit-pt}
is thus supersymmetric, whereas \eqref{new-crit-pt} breaks supersymmetry. Moreover, reality and
positivity of the scalars $\lambda^i$ implies that the second vacuum exists only in the range
\begin{equation}
\frac32\frac{g^1g^2}{(g^3)^2} < A < 3\frac{g^1g^2}{(g^3)^2}\,,
\end{equation}
in particular it is not present for zero deformation parameter $A$.

As far as concerns to the first-order equations when this particular gauging is performed, from the constancy of $\alpha$, it follows that the $\mrm{BPS}$ flow equations (\ref{eq:BPS}) boil down to
\begin{equation}
\begin{split}  \label{eq:to_solve}
& 2 e^{2\psi}\left(e^{-U}\mathrm{Re}\mathcal{V}\right)^{\prime}+e^{2(\psi-U)}\Omega\mathcal{M}\mathcal{G} + \mathcal{Q} = 0\,, \\
& (e^\psi)^{\prime}=2e^{\psi-U } \mathrm{Re}\,(\,\ma G^t \Omega \ma V)\,.
\end{split}
\end{equation}
The near-horizon geometry is required to be $\text{AdS}_2\times\Sigma_\kappa $, i.e. the warp factors enter in the near-horizon Ansatz as in \eqref{eq:att_metric_limits} while the scalar fields $z^{i}=-i\lambda ^{i}$ assume constant values on the horizon.
Under this assumption, the $\mrm{BPS}$ attractor equations \eqref{eq:attractor4d} simplify to
\begin{equation}
\begin{split}
& \mathcal{Q}+R_{H}^{2}\Omega \mathcal{M}\mathcal{G}=-4\mathrm{Im}\left(\overline{\mathcal{Z}}\mathcal{V}\right) \,, \\
& \mathcal{Z}=i\,\frac{R_{H}^{2}}{2R_{\mathrm{AdS}_2}}\,.
\end{split}
\label{eq:nearhor}
\end{equation}
In addition, one has to impose the constraint \eqref{eq:kappa}. As we said in section \ref{BPSattractor}, the $\mrm{BPS}$ attractor equations \eqref{eq:nearhor} provide a set of equations for the variables $\{R_H, R_{\mathrm{AdS}_2},\lambda^i\} $ as functions of the gaugings $g_0,g^i$ and the charges $p^0,q_i$.
Unfortunately the attractor equations \eqref{eq:nearhor} are still unsolved once the parameters characterizing the model \eqref{deformedstu} have been imposed. In particular the method studying near-horizon geometries developed in \cite{Halmagyi:2013qoa} for the case of symmetric special K\"ahler manifolds doesn't work in this case, since the model is neither symmetric nor homogeneous.

\subsection{\label{Solution}The Black Hole Solution}

The present section is devoted to the presentation of an exact black hole
solution for the non-homogeneous $STU$ model \eqref{deformedstu}. In order to
simplify the BPS equations \eqref{eq:to_solve}, we introduce the functions\footnote{A common choice
for the functions $H_{i}$ is to make them coincide with the
components of the symplectic sections \cite{Cacciatori:2009iz}. For the present situation, we
preferred to choose $H_{3}$ in a different way, in order to simplify the
structure of the equations.}
\begin{equation}
\begin{split}
H^{0}& =\frac{e^{-U}}{\sqrt{2}}\left( \lambda ^{1}\lambda ^{2}\lambda ^{3}-%
\frac{A}{3}(\lambda ^{3})\mbox{}^{3}\right) ^{\!-\frac{1}{2}}, \\
H_{1}& =\lambda ^{2}\lambda ^{3}H^{0}\,,\qquad H_{2}=\lambda ^{1}\lambda
^{3}H^{0}\,,\qquad H_{3}=(\lambda ^{3})\mbox{}^{2}H^{0}\,.  \label{eq:def_H}
\end{split}
\end{equation}
In terms of the latter, the equations \eqref{eq:to_solve} become
\begin{equation}
\begin{split}
& (H^{0})^{\prime }+2g_{0}(H^{0})^{2}=-e^{-2\psi }p^{0}\,, \\
& H_{1}^{\prime 2}H_{1}^{2}+\frac{2}{3}Ag^{2}H_{3}^{2}-\frac{4}{3}%
Ag^{3}H_{1}H_{3}=e^{-2\psi }q_{1}\,, \\
& H_{2}^{\prime 2}H_{2}^{2}+\frac{2}{3}Ag^{1}H_{3}^{2}-\frac{4}{3}%
Ag^{3}H_{2}H_{3}=e^{-2\psi }q_{2}\,, \\
& H_{3}^{\prime }+2H_{3}(g^{1}H_{1}+g^{2}H_{2})-2g^{3}\left( H_{1}H_{2}+%
\frac{A}{3}H_{3}^{2}\right) = \\
& \hspace*{0.5cm}=e^{-2\psi }\frac{H_{3}}{H_{1}H_{2}+AH_{3}^{2}}%
(q_{1}H_{2}+q_{2}H_{1}-q_{3}H_{3})\,, \\
& \psi ^{\prime }=g_{0}H^{0}+g^{1}H_{1}+g^{2}H_{2}+g^{3}\left( \frac{%
H_{1}H_{2}}{H_{3}}-AH_{3}\right) \,.  \label{eq:eq_H}
\end{split}
\end{equation}
A remarkable feature of the this model is that, contrary to e.g.~the case considered
in \cite{Cacciatori:2009iz}, the equations \eqref{eq:eq_H} cannot be decoupled, due to the nondiagonal
terms in the metric (\ref{eq:Kaehler_metric}).
Following the strategy of \cite{Cacciatori:2009iz}, we introduce the following Ansatz
\begin{equation}
\begin{split}
& \psi =\log \left( a\,r^{2}+c\right) \,, \\
& H^{0}=e^{-\psi }\left( \alpha ^{0}r+\beta ^{0}\right) \,, \\
& H_{i}=e^{-\psi }\left( \alpha _{i}r+\beta _{i}\right) ,\qquad i=1,2,3\,.
\label{eq:Ansatz_H}
\end{split}
\end{equation}
The solution for the fields is then expressed in terms of the functions $H^0,H_i$ by inverting the
relations (\ref{eq:def_H}). This yields
\begin{equation}
e^{2U}=\frac{1}{2}\left( \frac{H_{3}}{H^{0}}\right)^{\frac12}\left(
H_{1}H_{2}-\frac{A}{3}H_{3}^{2}\right)^{-1}\,,
\end{equation}
and
\begin{equation}
\lambda^1 = H_{2}\left( H_{3}H^{0}\right) ^{-\frac{1}{2}}\,, \qquad
\lambda^2 = H_{1}\left( H_{3}H^{0}\right) ^{-\frac{1}{2}}\,, \qquad
\lambda^3 = \left(\frac{H_{3}}{H^{0}}\right) ^{\frac{1}{2}}\,,
\label{eq:lambda_H}
\end{equation}
for the warp factor $U$ and the scalars $\lambda^i$ respectively.
By means of the Ansatz (\ref{eq:Ansatz_H}), the differential equations (\ref{eq:to_solve}) boil down
to a system of algebraic conditions on the
parameters and the charges characterizing the solution,
i.e. $\{\alpha^0,\alpha_i,\beta^0,\beta_i,a,c,p^0,q_i\}$. The set
of equations obtained in this way reduces, after some algebraic
manipulations, to
\begin{equation}
\begin{split}
\alpha ^{0}=& \,\frac{a}{2g_{0}}\,,\qquad \alpha _{1}=\,\frac{g^{2}}{g^{3}}%
\,\alpha _{3}\,,\qquad \alpha _{2}=\,\frac{g^{1}}{g^{3}}\,\alpha _{3}\,, \qquad
\alpha _{3} = \frac{a\,g^{3}}{2\left( g^{1}g^{2}-\frac{A}{3}\left(
g^{3}\right) ^{2}\right) }\,, \\
\beta _{1}=& \,\frac{g^{2}}{g^{3}}\,\beta _{3}\,,\qquad \beta _{2}=\,-\frac{1%
}{2}\,\beta _{3}\left( \frac{g^{1}}{g^{3}}-A\frac{g^{3}}{g^{2}}\right) -%
\frac{1}{2}\,\beta ^{0}\,\frac{g_{0}}{g^{2}}\,, \\
q_{1}=& \,2\,\beta _{3}^{2}\,\frac{g^{2}}{\left( g^{3}\right) ^{2}}\left(
g^{1}g^{2}-\frac{A}{3}\left( g^{3}\right) ^{2}\right) +g^{2}\,\frac{ac}{%
2\left( g^{1}g^{2}-\frac{A}{3}\left( g^{3}\right) ^{2}\right) }\,, \\
q_{2}=& \,\frac{1}{2g^{2}}\left( \beta ^{0}g_{0}+\beta _{3}\,\frac{g^{1}g^{2}%
}{g^{3}}\right) ^{2}+g^{1}\,\frac{ac}{2\left( g^{1}g^{2}-\frac{A}{3}\left(
g^{3}\right) ^{2}\right) } \\
& +\frac{A}{3}\,\beta _{3}\,\frac{g^{3}}{g^{2}}\left( \beta _{3}\,\frac{%
g^{1}g^{2}}{g^{3}}-\beta ^{0}g_{0}-\frac{A}2\beta_3 g^3\right)\,, \\
q_{3}=& \,\frac{g^{2}}{g^{3}}\,q_{2}-A\,\frac{g^{3}}{g^{2}}\,q_{1}\,, \qquad
p^0 = -\frac{ac}{2g_{0}}-2g_{0}\left( \beta ^{0}\right) ^{2}\,. \\
\label{eq:solution}
\end{split}
\end{equation}
The solution for the scalars is obtained by plugging the parameters written in (\ref{eq:solution}) into
the expressions (\ref{eq:lambda_H}). In this way, the scalars assume the explicit form
\begin{equation} \label{expl-form-scalars}
\begin{split}
& \lambda ^{1}=\frac{a\,\frac{g^{1}}{g^{3}}\left( \lambda _{\infty
}^{3}\right) ^{2}r-g_{0}\,\beta _{3}\left( \frac{g^{1}}{g^{3}}-A\,\frac{g^{3}%
}{g^{2}}\right) -\beta ^{0}\,\frac{g_{0}^{2}}{g^{2}}}{\sqrt{\left(
2g_{0}\,\beta ^{0}+a\,r\right) \left( 2g_{0}\,\beta _{3}+a\,r\left( \lambda
_{\infty }^{3}\right) ^{2}\right) }}\,, \\
& \lambda ^{2}=\frac{g^{2}}{g^{3}}\lambda ^{3}\,, \qquad
\lambda ^{3}=\lambda _{\infty }^{3}\sqrt{\frac{ar+\,\frac{2\,g_{0}}{%
(\lambda _{\infty }^{3})\mbox{}^{2}}\,\beta _{3}}{ar+2g_0\beta^0}}\,,
\end{split}
\end{equation}
where $\lambda _{\infty }^{3}$ is the asymptotic value of $\lambda ^{3}$,
\begin{equation}
\lambda_{\infty }^3 = \sqrt{\frac{g_0 g^3}{g^1 g^2 - \frac A3 (g^3)^2}}\,.
\end{equation}
The warp factor in the metric reads
\begin{equation}
\begin{split}
& \mathrm{e}^{2U}=\frac{2g_0 g^{3}(ar^2+c)\mbox{}^{2}}{\lambda
_{\infty }^{3}\left( ar-g_0\beta^0 - \frac{\,g_{0}}{(\lambda_{\infty}^3)^2}\beta_3\right)
\sqrt{\left( ar+2g_0\beta^0
\right) \left( ar+\frac{2 g_0}{(\lambda _{\infty }^{3})^2}\beta_3\right)}}\,.
\end{split}
\label{eq:e2U}
\end{equation}
This solution represents a black hole, with a horizon at the largest zero of $e^{2U}$, i.e. at
$r_{H}=\sqrt{-c/a}$, where we assumed $a>0$ and $c<0$. The curvature invariants diverge
where the angular component of the metric $e^{2\psi-2U}$ vanishes.
Note that all the scalar fields $\lambda _{i}$ should be well-defined and
positive outside the horizon. Moreover, we still have to impose the condition (\ref{eq:kappa}), i.e.
\begin{equation}
g_0 p^0 - g^i q_i = -\kappa \label{cond-Dirac}
\end{equation}
on the solution (\ref{eq:solution}).
Moreover these requirements are compatible with any of the three
possible choices for $\kappa =0\,,\pm 1$, i.e. the horizon
topology can be either spherical, flat or hyperbolic.

The Dirac quantization condition \eqref{cond-Dirac} fixes one of the four parameters
$\{a,c,\beta^0,\beta_3\}$ that determine the solution, for example $c$ . Furthermore, one easily sees
that the solution enjoys the scaling symmetry
\begin{equation}
(t,r,\theta,\phi,a,c,\beta^0,\beta_3,\kappa) \mapsto (t/s,sr,\theta,\phi,a/s,sc,\beta^0,\beta_3,\kappa)\,,
\qquad s\in\mathbb{R}\,,
\end{equation}
that can be used to set $a=1$ without loss of generality. Consequently, there are only two
physical parameters left, on which the solution depends.\\
Notice that the solution (\ref{eq:solution}) is characterized by the proportionality
between the scalars $\lambda_2$ and $\lambda_3$, as is evident from \eqref{expl-form-scalars}.
However, it is worth stressing that this fact does not trivialize our results,
since the locus $\lambda^2=\frac{g^2}{g^3}\lambda^3$ in the scalar manifold does not yield a
consistent two-moduli truncation for the $STU$ model, namely $ST^2$. In other
words, the K\"{a}hler geometry that can be derived from the truncated model
$F\left( X^{1},X^{2},X^{3})\right\vert _{\lambda ^{2}\propto \lambda^3}$
is not equivalent to the two-dimensional one characterized by the prepotential
\begin{equation}
F=\frac{\tilde{X}^{1}\left( X^{3}\right)^2}{X^0}\,, \qquad \mbox{with}
\qquad \tilde{X}^{1}=X^{1}-\frac{A}{3}\,X^{3}\,,  \label{eq:troncato}
\end{equation}
which is homogeneous and symmetric (see for example \cite{Bellucci:2007zi}). This difference is evident, for example, in terms of the K\"{a}hler metric.
In fact one has
\begin{equation}
g_{ij}^{(3)}d\lambda^i d\lambda^j\rvert_{\lambda_2\propto\lambda_3} \neq g_{MN}^{(2)}
d\lambda^M d\lambda^N\,, \qquad i,j=1,2,3\,, \qquad M,N=1,2\,,
\end{equation}
where the left-hand side is the line element obtained with the metric (\ref{eq:Kaehler_metric}) when the condition $\lambda_2\propto\lambda_3$ is imposed, while the right-hand side describes the geometry associated to the prepotential (\ref{eq:troncato}).

We conclude this section with a comment on the behaviour of the
solution for $A=0$. Due to the particular definition of $H_3$ we have chosen (with respect to
the more common one used for example in \cite{Dall'Agata:2010gj,Cacciatori:2009iz,Gnecchi:2013mta}), setting $A=0$ and $\lambda^2=\frac{g^2}{g^3}\lambda^3$ is not sufficient to match
exactly the $STU$ black hole solution with two independent parameters that can be derived from \cite{Cacciatori:2009iz}. However, the parameters in (\ref{eq:Ansatz_H}) can
be redefined as
\begin{equation}
\alpha_3^{\prime} = \frac{\alpha_1\alpha_2}{\alpha_3} - \frac A3\alpha_3\,, \qquad
\beta_3^{\prime} = \frac{\beta_1\beta_2}{\beta_3} - \frac A3\beta_3\,,
\end{equation}
in terms of which the solution (\ref{eq:solution}) matches explicitly the
known one when $A=0$. This redefinition of the parameters is a way to
recover the choice for the functions that is usually made when solving the
BPS equations (\ref{eq:BPS}), whose analogue for the present case is
\begin{equation}
H_3^{\prime} = \left(\lambda^1\lambda^2 - A(\lambda^3)^2\right)H^0\,, \quad \text{or}
\quad H_3^{\prime} = e^{-\psi}(\alpha_3^{\prime}r + \beta_3^{\prime})\,.
\end{equation}

\subsection{\label{Phys-Discussion}The IR and UV Analysis}

In this section, we discuss some properties of our solution, like near-horizon limit, entropy or
area-product formula.

In the asymptotic limit $r\rightarrow\infty$, the metric (\ref{eq:e2U}) becomes $\mathrm{AdS}_4$, i.e.,
at leading order one has
\begin{equation}
\mathrm{d}s^2 = -\frac{r^2}{ R_{\scriptsize{\mrm{AdS}_4}}^2}\mathrm{d}t^2 +  R_{\scriptsize{\mrm{AdS}_4}}^2\frac{\mathrm{d}r^2}{r^2} +
r^2\mathrm{d}\Omega_{\kappa}^2\,,
\end{equation}
where the asymptotic $\mrm{AdS}_4$ curvature radius $ R_{\scriptsize{\mrm{AdS}_4}}$ is given by
\begin{equation}
R_{\scriptsize{\mrm{AdS}_4}}^2 = \frac{\lambda_\infty^3}{2g_0g^3}\,, \label{ell}
\end{equation}
and the coordinates have been rescaled according to $t\to R_{\scriptsize{\mrm{AdS}_4}} t$, $r\to r/ R_{\scriptsize{\mrm{AdS}_4}}$. Notice that the asymptotic value
of the cosmological constant is
\begin{equation}
\Lambda = -\frac3{ R_{\scriptsize{\mrm{AdS}_4}}^2} = -\frac{6g_0g^3}{\lambda_\infty^3}\,.
\end{equation}
On the other hand, when $r$ approaches the horizon, the functions $U$ and $\psi$ assume,
after shifting $r\to r+r_{ H}$, the form \eqref{eq:att_metric_limits}, with $R_{\scriptsize{\mrm{AdS}_2}}$ and $R_H$ given by
\begin{equation}
R_{\scriptsize{\mrm{AdS}_2}}^2 = -\frac{\lambda_\infty^3 f(r_H)}{8g_0g^3c}\,, \qquad R_H^2 =
\frac{\lambda_\infty^3 f(r_H)}{2g_0g^3}\,,
\end{equation}
where
\begin{displaymath}
f(r_H) = \left(r_H - g_0\beta^0 - \frac{g_0}{(\lambda_\infty^3)^2}\beta_3\right)
\sqrt{(r_H + 2g_0\beta^0)\left(r_H + \frac{2g_0}{(\lambda_\infty^3)^2}\beta_3\right)}\,.
\end{displaymath}
In this limit, the spacetime becomes $\mathrm{AdS}_2\times\Sigma_\kappa$, with metric
\begin{equation}
\mathrm{d}s^2 = -\frac{r^2}{R_{\scriptsize{\mrm{AdS}_2}}^2}\mathrm{d}t^2 + \frac{R_{\scriptsize{\mrm{AdS}_2}}^2}{r^2}\mathrm{d}r^2
+ R_H^2\mathrm{d}\Omega_{\kappa}^2\,.
\end{equation}
The Bekenstein-Hawking entropy is given by
\begin{equation}
S_{BH} = \frac{A}{4} = \frac{R_H^2 \,\text{vol}(\Sigma_\kappa)}4\,.
\end{equation}
This expression can be written in terms of the charges $p^0,q_i$ and the
gaugings $g_0,g^i$ only. To this aim, the equation (\ref{eq:solution})
need to be inverted, in order to use the charges $p^0,q_1,q_2$ as
parameters. This sustains the fact that this black hole is an attractor, which is a non-trivial statement, due to
the non-homogeneity of the model we have been discussing.
Finally, the product of the areas of all the horizons $r=r_I$, $I=1,\ldots,4$ (i.e., all the roots of
$e^{2U}$) assumes the remarkably simple form
\begin{equation} \label{area-prod}
\prod_{I=1}^4 A(r_I) = -\frac{36}{\Lambda^2}\frac{\text{vol}(\Sigma_\kappa)^4g^2}{g^3}p^0q_1\tilde q_2^2\,,
\end{equation}
where we defined
\begin{equation}
\tilde q_2 = q_2 - \frac A3\left(\frac{g^3}{g^2}\right)^2 q_1\,.
\label{eq:q_modf}
\end{equation}
Note that \eqref{area-prod} depends only on the charges and the gauging parameters.
Similar formulas have been proven to be true in a number of examples (see
for instance \cite{Bellucci:2009pg,Toldo:2012ec,Cvetic:2010mn,Galli:2011fq,Castro:2012av,Klemm:2012vm,Gnecchi:2013mja}), a fact that calls for an underlying microscopic interpretation.

\section{A Black Hole with the Universal Hypermultiplet}

In this section we will present an exact black hole solution in $\ma N=2$, $d=4$ gauged supergravity coupled to a vector multiplet with special geometry defined by a prepotential of the type
\begin{equation}
 F=-iX^0X^1\,,
\label{eq:hypersol_prepot}
 \end{equation}
and to the universal hypermultiplet \eqref{uhm}. The moduli space for this model is given by
\begin{equation}
\frac{\mrm{SU(1,1)}}{\mrm{U(1)}}\times \frac{\mrm{SU(2,1)}}{\mrm{SU(2)}\times\mrm{U(1)}}\,,
\label{modelhyper}
\end{equation}
where the terms in the product are in order the special K\"ahler manifold and the quaternionic moduli space of the universal hypermultiplet.

In spite of the form of its coset, the special K\"ahler structure of the moduli space identified by the prepotential \eqref{eq:hypersol_prepot} doesn't represent the reduction of five-dimensional $\ma N=2$ supergravity leading to the completely truncated $STU$ model, i.e. the $T^3$ model, and it has not an interpretation as truncation on a Calabi-Yau threefold of type $\mrm{II}$ string theory or $\mrm{M}$-theory as for the other cases studied in this thesis. However it can be related by a rigid symplectic transformation to the so-called $\mathbb{CP}_{1,1}$ model \cite{Sabra:1996xg}. This belongs to a general class of $\ma N=2$ models called $\mathbb{CP}_{n-1,1}$ that appear in orbifold compactifications of heterotic string theory \cite{LopesCardoso:1994is}.

In this section we will consider an electric abelian gauging on the universal hypermultiplet of the non-compact group $\mathbb{R}\times \mrm{U(1)}$ and we will solve explicitly the supersymmetric first-order equations \eqref{eq:flow-equ-electr}. This will produce an exact black hole solution characterized by a radial flow of the dilaton included in the universal hypermultiplet and an hyperscaling-violating asymptotic geometry.

\subsection{The Model and the $\mathbb{R}\times \mrm{U(1)}$ Gauging}
\label{X0X1UHM}

Let's consider a general setup of four-dimensional $\ma N=2$ gauged supergravity \eqref{eq:mainaction} coupled to $n_V=1$ vector multiplet, parametrized by a complex scalar $z$, and to the universal hypermultiplet presented in section \ref{quaternionic}, spanned by the four real scalars $q^u=(\phi, \xi^0, \tilde{\xi}_0, a)$.
Let's analyze separately the two sectors of the scalar geometry.

The complex modulus $z$ can be defined in terms of two homogeneous coordinates $X^\Lambda$ such that
\begin{equation}
\frac{X^{1}}{X^{0}}=z \,,
\label{X0X1par}
\end{equation}
while the symplectic sections (\ref{specialembedding}) are given by
\begin{equation}
\begin{split}
& L^{\Lambda }=e^{\mathcal{K}/2}\left(1, z\right) ^{t}\,, \\
& M_{\Lambda }=e^{\mathcal{K}/2}\left( -i z, -i\right) ^{t}\,,
\end{split}
\end{equation}
This define the following forms of the K\"{a}hler potential and the metric,
\begin{equation}
e^{-\mathcal{K}}=4\,\mrm{Re}\,z \qquad \text{and} \qquad g_{z\bar{z}}=\frac{1}{4\,(\mrm{Re}\,z)^2 }\,,
\end{equation}
Finally the period matrix $\ma{N}_{\Lambda\Sigma}$ has the form
\begin{equation}
 \mathcal{N}_{\Lambda \Sigma}=-i\left(\begin{array}{cc}
                    z&0\\
                    0&\frac{1}{z}
                   \end{array}\right)\,.
\end{equation}
Regarding the universal hypermultiplet we already introduced the metric \eqref{eq:hypersol_hyper_metric} and the $\mrm{SU(2)}$ connection $\omega^x$ in \eqref{su2connectionuhm}, so we can concentrate on the gauging.
Since the theory includes two vector fields, we can choose to gauge up two isometries of the metric \eqref{eq:hypersol_hyper_metric}. We choose to gauge the (commuting) electric isometries generated by the following Killing vectors on the quaternionic manifold \cite{Ceresole:2001wi}
\begin{equation}
\label{eq:hypersol_killing_vecs}
 k_\Lambda=g_\Lambda\partial_a+\delta^0{}_\Lambda c \left( \tilde{\xi}_0\partial_{\xi^0}-\xi^0\partial_{\tilde{\xi}_0} \right)\,,
\end{equation}
where $g_\Lambda$ and $c$ are constants. This means that we are gauging the $\mathbb{R}$ group of the translations along $a$ with
the combination $A^\Lambda g_\Lambda$, and the $\mrm{U(1)}$ group of rotations in the $\xi^0$--$\xi_0$ plane with the field $A^0$.
The triholomorphic moment maps associated with the Killing vectors \eqref{eq:hypersol_killing_vecs} can be obtained from \eqref{eq:mommaps}, and are
\begin{align}
 P_\Lambda^1&=-\delta^0_\Lambda c\, \xi^0 e^\phi\,,\qquad\qquad P_\Lambda^2=\delta^0_\Lambda c\, \tilde{\xi}_0 e^\phi\,,\nonumber\\
 &\label{eq:hypersol_tri_mommaps}\\
 P_\Lambda^3&=\delta^0_\Lambda c\left[1-\frac14 e^{2\phi}\left((\xi^0)^2+(\tilde{\xi}_0)^2\right)\right]+\frac12 g_\Lambda e^{2\phi}
 \nonumber\,.
\end{align}
With these choices the scalar potential \eqref{eq:scal_pot} reads
\begin{equation}
\label{eq:hypersol_scalpot}
 V_g=\frac{1}{z+\bar{z}}\left[\frac{e^{4\phi}}{4}\left[g_0-\frac{c}{2}\left((\xi^0)^2+(\tilde{\xi}_0)^2 \right) \right]^2
    - c^2 - g_0 c\, e^{2\phi}  \right] +\frac{z\bar z}{z+\bar z}\frac{e^{4\phi}}{4} g_1^2 -\, g_1 c\,e^{2\phi}\,.
\end{equation}
For simplicity we will look for solutions with axion vanishing. We point out that for this model this condition consists\footnote{This is due to the parametrization of the special geometry defined in \eqref{X0X1par}.} in requiring that $z$ is real. Moreover we further impose that $z> 0$ for the positivity of the kinetic terms.

We are interested in static solutions with spherical/hyperbolic symmetry thus we impose the Ansatz on the fields introduced in \eqref{eq:Ansatzmet}, \eqref{eq:Ansatzscalars} and \eqref{eq:Ansatzgaugefields}.
The first and main consequence of this request consists in the truncation of the hypermultiplet to the dilaton $\ma \phi$. Let's consider the first-order analysis of section \ref{effectivebh} and remind the two relations \eqref{eq:maxAnsatz-phi} and \eqref{eq:maxAnsatz-r}, coming from the Maxwell equations for static gauge fields with spherical/hyperbolic symmetry, given by
\begin{equation}
p^\Lambda k_\Lambda^u=0 \qquad \text{and} \qquad  k_{\Lambda u} q'^u=0\,,
\label{truncationhyper}
\end{equation}
where $p^\Lambda=(p^0, p^1)^t$ are the magnetic charges. Plugging the Killing vectors \eqref{eq:hypersol_killing_vecs} into \eqref{truncationhyper}, it follows that
\begin{equation}
\xi^0=0\,,\qquad \tilde{\xi}_0=0\, \qquad \text{and} \qquad a^\prime=0\,.
\label{truncationuhm}
\end{equation}
From \eqref{truncationhyper} we obtain the following relation
\begin{equation}
p^\Lambda g_\Lambda=0\,.
\label{quatizationhyper1}
\end{equation}
Hence, after this truncation, the electric moment maps can be organized as a symplectic vector such that $\ma P^1=\ma P^2=0$ and with
\begin{equation}
\mathcal P^3 = \left(\begin{array}{c} 0 \\ P^3_\Lambda \end{array}\right)\,,
\label{eq:gaugingsamu}
\end{equation}
where the moment map $P^3_\Lambda$ is given by
\begin{equation}
P^3_0=c+\frac{g_0}{2}\,e^{2\phi} \qquad \text{and}\qquad P^3_0=\frac{g_1}{2}\,e^{2\phi}\,.
\end{equation}
The moment maps and the charges obey the quantization condition \eqref{eq:quantization} that, in the case of a generic vector of charges $\ma Q$ and electric moment maps, takes the form
\begin{equation}
\langle \ma P^3, \ma Q\rangle=p^\Lambda P^3_\Lambda=-\kappa\,.
\end{equation}
If we plug into the quantization condition the explicit form of the moment maps of our model \eqref{eq:gaugingsamu} and use the relation \eqref{quatizationhyper1}, we obtain the quantization condition of the model
\begin{equation}
p^0=-\frac{\kappa}{c}\,.
\label{quanthorizonhyper}
\end{equation}
With the truncation \eqref{quatizationhyper1} and axions vanishing, the scalar potential takes the following form
\begin{equation}
\begin{split}
\label{eq:hypersol_scalpot1}
 V_g=\frac{1}{2z}\left[\frac{e^{4\phi}}{4}g_0^2
    - c^2 - g_0 c\, e^{2\phi}  \right] +\frac{g_1^2\,z}{8}\,e^{4\phi}  - g_1 c\,e^{2\phi}\,.
    \end{split}
\end{equation}
This potential admits an $\mrm{AdS}_4$ vacuum given by
\begin{equation}
z=-\frac{g_0}{g_1}\qquad \text{and} \qquad e^{-2\phi}=-\frac{g_0}{c}\,,
\end{equation}
where, for the positivity of the moduli, we have $\frac{g_0}{g_1}<0$ and $\frac{g_0}{c}<0$.

The scalar potential \eqref{eq:hypersol_scalpot1} can be rewritten in the covariant form \eqref{eq:superV1}. In fact one can derive the expressions of $\ma L$ from \eqref{Lsuperpotential} and of the central charge $\ma Z=\langle \ma Q, \ma V \rangle$ obtaining
\begin{equation}
\begin{split}
&\ma L=-\kappa\,e^{\frac{\scriptsize{\ma K}}{2}}(c+\frac{g_0}{2}\,e^{2\phi} +\frac{g_1}{2}\,e^{2\phi}\,z )\,,\\
&\ma Z= i\,e^{\frac{\scriptsize{\ma K}}{2}}(p^1 +p^0\,z )\,.
\label{ZLhyper}
\end{split}
\end{equation}
Thus one can derive the superpotential $W$ given in \eqref{eq:superpotential} and the phase $\alpha$ \eqref{eq:defal}. As in the case of non-homogeneous $STU$, from the axions vanishing condition it follows that $\alpha$ is constant, i.e. $\alpha=\pm \pi/2$, with this particular gauging. Moreover, with these prescriptions the first-order equations \eqref{eq:flowequations} take the form
\begin{equation}
\begin{split}  
& 2 e^{2\psi}\left(e^{-U}\mathrm{Re}\mathcal{V}\right)^{\prime}+e^{2(\psi-U)}\Omega\mathcal{M}\mathcal{P}^3 + \mathcal{Q} = 0\,, \\
& (e^\psi)^{\prime}=2e^{\psi-U } \mathrm{Re}\,(\ma P^{3\,t} \Omega \ma V)\,, \\
&q^{\prime\,u} = - e^{-U} h^{uv}\partial_v\mathrm{Re}\,(\ma P^{3\,t} \Omega \ma V)\,, \\
&e^\prime_\Lambda = 4 e^{2\psi-3U}\ma H_{\Lambda\Sigma}\mathrm{Im} L^\Sigma \,.
\label{flowhyper}
\end{split}
\end{equation}
From the last equation it follows that $e_\Lambda^\prime=0$ since $L^\Lambda$ is real in this particular model. Furthermore, if we recall the relation \eqref{eq:ReL-A}, for the same reason also the electric components of the gauge fields vanish, i.e. $A_t^\Lambda=0$. Thus it follows that the electric charges $e_\Lambda$ can be consistently set to zero, thus the configuration of charges $\ma Q=(p^\Lambda, 0)^t$ is completely magnetic.

\subsection{The Black Hole Solution}

In this section we explicitly solve the equations \eqref{flowhyper} for the model \eqref{modelhyper}. First of all we are going to write down the first-order equations in a similar way of \eqref{eq:eq_H}. Let's introduce the new variables 
\begin{equation}
H^0=e^{-U}L^0 \qquad \text{and} \qquad H^0=e^{-U}L^1\,.
\label{Hhyper}
\end{equation}
Thus the first equation of \eqref{flowhyper} becomes 
\begin{equation}
\begin{split}
&2 (H^{0})^\prime+4 c (H^0)^2+ 2g_0 (H^0)^2\,e^{2\phi}+p^0e^{-2\psi}=0\,,\\
&2 (H^{1})^\prime+ 2g_1 (H^1)^2 \,e^{2\phi}+p^1e^{-2\psi}=0\,
\label{floweqH}
\end{split}
\end{equation}
that can be rewritten in the more compact fashion as follows
\begin{equation}
2 \,(H^{\Lambda})^\prime+4 c (H^0)^2\delta_0^\Lambda+ 2g_\Lambda (H^\Lambda)^2\,e^{2\phi}+p^\Lambda e^{-2\psi}=0\,,
\end{equation}
with no summation over the index $\Lambda$. Regarding the equation for $\psi$ in \eqref{flowhyper}, it takes the form
\begin{equation}
\psi^\prime=2c\,H^0+ g_\Lambda H^\Lambda e^{2\phi} \,,
\label{floweqpsi}
\end{equation}
where we used \eqref{ZLhyper} and \eqref{Hhyper}. Finally, because of the truncation \eqref{truncationuhm}, the equations for the hyperscalars boil down to the equation for the dilaton that takes the form
\begin{equation}
\phi^\prime=- g_\Lambda H^\Lambda e^{2\phi}\,,
\label{floweqphi}
\end{equation}
where we used \eqref{ZLhyper} and the metric of the universal hypermultiplet \eqref{eq:hypersol_hyper_metric}.

We point out that the relation 
\begin{equation}
\psi^\prime=2c\,H^0-\phi^\prime\,,
\end{equation}
obtained by merging the two flow equations \eqref{floweqpsi} and \eqref{floweqphi} is characteristic of the abelian $\mathbb{R}\times \mrm{U(1)}$ gauging in the sense that it is possible to verify that the same relation between $\psi^\prime$, $\phi^\prime$ and $H^0$ (except for different numerical factors) holds also for the $\mrm{M}$-theory truncations of section \ref{Mtheorytruncation}.

Let' consider now the equations \eqref{floweqH} and multiply respectively the first and the second by $g_0$ and $g_1$. By summing the two resulting equations we obtain
\begin{equation}
(g_\Lambda H^{\Lambda})^\prime+ (g_\Lambda H^\Lambda)^2\,e^{2\phi}+2\, g_0H^0e^{2\phi}\left(c\,H^0 e^{-2\phi}-g_1H^1 \right)=0\,,
\label{floweqH2}
\end{equation}
where the term in $g_\Lambda p^\Lambda$ vanishes thanks to the relation \eqref{quatizationhyper1}. If we impose the following relation between the scalar $z$ and the dilaton $\phi$,
\begin{equation}
z=\frac{H^1}{H^0}=\frac{c}{g_1}\,e^{-2\phi}\,,
\end{equation}
the equation \eqref{floweqH2} takes the form
\begin{equation}
\partial_\phi \left[(g_\Lambda H^{\Lambda})^2  \right]-2\,(g_\Lambda H^{\Lambda})^2=0\,,
\label{floweqH3}
\end{equation}
where we used the chain rule and the equation \eqref{floweqphi} for trading the coordinate $r$ for $\phi$. The equation \eqref{floweqH3} is solved by
\begin{equation}
H^0=\frac{e^\phi}{g_0+c\,e^{-2\phi}}\,,\qquad H^1=\frac{c}{g_1}\frac{ e^{-\phi}}{g_0+c\,e^{-2\phi}}\,.
\label{solutionH}
\end{equation}
Plugging back \eqref{solutionH} into the equations \eqref{floweqH}, one can show that the the warp factor $\psi$ is real only if $\kappa=-1$, namely the solution we are going to construct must have hyperbolic symmetry.
Finally one can solve the equation of the dilaton \eqref{floweqphi} obtaining\footnote{This expression for the dilaton can be obtained by performing the redifinition $r \rightarrow \frac13 r^3$.}
\begin{equation}
\phi=-\log \, r\,
\end{equation}
and derive the expression for the warp factor $U$ in terms of \eqref{Hhyper},
\begin{equation}
e^{2U}=\frac{1}{4\,H^0H^1}\,.
\end{equation}
Thus the final result is given by\footnote{Provided that one rescales the time coordinate as $t \rightarrow \frac{g_0}{4\, p^1}\,t$.}
\begin{equation}
\D s^2 = \frac{-4 p^1}{g_0} r^2\left[-\left(1 + \frac{g_0}{c\, r^2}\right)^2 r^2\D t^2 + \left(1 +
\frac{g_0}{c\, r^2}\right)^{-2}\frac{\D r^2}{r^2} + \frac12\D\Omega^2_{-1}\right]\,, \label{eq:solution-metr-hyper}
\end{equation}
\begin{equation}
z = \frac c{g_1} r^2\,, \qquad e^{-2\phi} = r^2\,, \qquad A^\Lambda =  p^\Lambda\cosh\theta\,\D\phi\,,
\label{eq:solution-scal-A-hyper}
\end{equation}
where the magnetic charges respect the algebraic conditions \eqref{quanthorizonhyper} and \eqref{quatizationhyper1} with $\kappa=-1$, and are defined in the following intervals
\begin{equation}
\frac{g^0}{c}<0\,,\qquad \frac{c}{g_1}>0\,,\qquad \frac{p^1}{g_0}<0\,.
\end{equation}
With these values for the parameters the solution describe a black hole since the metric \eqref{eq:solution-metr-hyper} has a background singularity in $r=0$ and a well-defined event-horizon in $r_H=\sqrt{-g_0/c}$ defining an hyperbolic geometry of the type $\mrm{AdS}_2\times H^2$.

\subsection{\label{Phys-DiscussionHyper}The IR and UV Analysis}

Let's consider the asymptotic regime, $r\rightarrow +\infty$, of the metric \eqref{eq:solution-metr-hyper}. It is immediate to see that it reduces to
\begin{equation}
\label{eq:hypersol_asympt_metric}
 \D s^2=\frac{-4 p^1}{g_0} r^2\left[-r^2 \D t^2+\frac{\D r^2}{r^2}
 +\frac{1}{2}\left( \D \vartheta^2 +\sinh^2\vartheta \, \D \varphi^2 \right)\right]\,,
\end{equation}
which is manifestly conformally equivalent to $\mrm{AdS}_2\times H^2$. Note that
\eqref{eq:hypersol_asympt_metric} is very similar to hyperscaling violating geometries, which in $d$
dimensions have the form
\begin{equation}
\D s^2 = r^{-\frac{2\theta}{d-2}}\left(-r^{2z} \D t^2 + \frac{ \D r^2}{r^2} + r^2( \D x^i)^2\right)\,, \label{hyp-viol}
\end{equation}
where $i=1,\ldots,d-2$. Here, $z$ is the dynamical critical exponent and $\theta$ the so-called
hyperscaling violation exponent. Under the scaling $r\to r/\lambda$, $x^i\to\lambda x^i$,
$t\to\lambda^z t$, \eqref{hyp-viol} is not invariant, but transforms covariantly,
$ \D s\to\lambda^{\theta/(d-2)} \D s$. The metric \eqref{eq:hypersol_asympt_metric} exhibits actually a scaling behaviour similar to that of
\eqref{hyp-viol}. To see this, introduce new coordinates $x$, $y$ on $H^2$ according to
\begin{equation}
x + iy = \frac{e^{i\varphi}\tanh\frac{\vartheta}2 + 1}{e^{i\varphi}\tanh\frac{\vartheta}2 - 1}\,,
\end{equation}
which casts \eqref{eq:hypersol_asympt_metric} into the form
\begin{equation}
\D s^2 =\frac{-4 p^1}{g_0} \left[-r^2  \D t^2+\frac{ \D r^2}{r^2} + \frac{ \D x^2 +  \D y^2}{2x^2}\right]\,.
\label{hyp-viol-prime}
\end{equation}
Under the scaling
\begin{equation}
r \to \frac r\lambda\,, \qquad t \to \lambda t\,, \qquad x \to \lambda x\,, \qquad y \to \lambda y\,,
\end{equation}
\eqref{hyp-viol-prime} transforms as $\D s\to \D s/\lambda$.

In the near-horizon limit, $r\rightarrow r_H$, the metric takes the form\footnote{We rescaled $t\rightarrow \frac{t}{4}$.}
\begin{equation}
 \D s^2=\frac{p^1}{c}\,\left[-r^2  \D t^2+\frac{  \D r^2}{r^2}\right]+\frac{2p^1}{c}\left[  \D \vartheta^2 +
\sinh^2\vartheta\,  \D \varphi^2 \right]\,,
\end{equation}
which is $\mrm{AdS}_2\times H^2$, while the scalar fields take the near-horizon values
\begin{equation}
  e^{-2\phi_H}= -\frac{g_0}{c}\,,\qquad\qquad z_H=-\frac{g_0}{g_1}\,.
  \label{nhhypersol}
\end{equation}
It is easy to show that this black hole is an attractor. Using \eqref{Vbh} and \eqref{eq:hypersol_scalpot1}, we can obtain the values of $V_{\text{BH}}$ and $V_g$ at the horizon values \eqref{nhhypersol},
\begin{equation}
V_{\text{BH}}=p^0p^1\qquad \text{and} \qquad V_g=-\frac{3\,c^2\,p^0}{4\,p^1}\,.
\end{equation}
Thus one can derive the near-horizon value of $V_{eff}$ given in \eqref{effective} and obtain the entropy density of the black hole
\begin{equation}
 s_{BH}=\frac{S_{BH}}{\mrm{vol}(H^2)}=\frac14\,V_{eff}(z_H, \phi_H)=\frac{p^0p^1}{2}\,,
\end{equation}
where $\mrm{vol}(H^2)$ is obtained by compacting $H^2$ on a Riemann surface of genus $\mathfrak{g}$.
The entropy is thus strictly charge-dependent and, thus, the scalars $z$ and $\phi$ flow to the horizon driven by the attractor mechanism.

%% file: chap6.tex

\chapter{BPS Objects in $\ma N=1$, $d=7$ Supergravity}
\label{7Dobjects}
\thispagestyle{plain}


In this chapter we move on with the analysis on $\mrm{BPS}$ objects in gauged supergravities by walking up towards higher-dimensional supergravities. The general framework will be now given by $\ma N=1$ minimal gauged supergravity in seven dimensions \cite{Townsend:1983kk}. 

Recalling the classification of dimensional reductions on spheres presented in section \ref{geometry}, we will consider two particular consistent truncations leading to this seven-dimensional gauged supergravity.
The first is the truncation of eleven-dimensional supergravity on a 4-sphere $S^{4}$ \cite{Freund:1980xh,Nastase:1999cb,Chamseddine:1999uy,Lu:1999bc} and the second is realized by massive type $\mrm{IIA}$ supergravity on a squashed 3-sphere $\tilde{S}^{3}$ \cite{Passias:2015gya}.

Together with the seven-dimensional graviton, a triplet of $\mrm{SU(2)}$ gauge vector fields and the dilaton, the degrees of freedom describing these truncations include also a tensor field in the supergravity multiplet given by a 3-form gauge potential. The presence of the 3-form gives rise to many possible physical configurations like black brane solutions and, generally, a rich class of flows interpolating between various types of closed string vacua (see for example \cite{Pernici:1984nw,Maldacena:2000mw,Gauntlett:2000ng,Acharya:2000mu}). Moreover the gauge group $\mrm{SU(2)}$ together with a topological mass deformation for the 3-form implies a scalar potential for the dilaton, thus without the necessity of including the coupling to matter multiplets. This potential possesses two $\mrm{AdS}_7$ extrema \cite{Mezincescu:1984ta}, one has spontaneously broken supersymmetry, while the other is supersymmetric and with uplifts respectively in the $D=11$ Freund-Rubin solution $\mrm{AdS}_7\times S^4$ \eqref{freundrubin}, and in the massive $\mrm{IIA}$ warped solution $\mrm{AdS}_7\times_w \tilde{S}^3$ presented in \eqref{massiveIIAtruncations}. 

The first aim of this chapter is to present a class of $\mrm{BPS}$ flows in $\mathcal{N}=1$, $d=7$ gauged supergravity including a non-trivial profile for the three-form gauge potential \cite{Dibitetto:2017tve}. These solutions will
be derived thanks to the formulation of suitable Ansatz{\"e} on the fields and by solving $\mrm{BPS}$ equations obtained by the Killing spinor analysis\footnote{As far as we know, the Hamilton-Jacobi analysis involving supergravities coupled to tensor fields has not been formulated yet.}.
Most of the flow equations, that we will obtain, will be integrated analytically, while for some of them we will have to employ numerical integration methods. Moreover we will discuss the eleven-dimensional origin of the various aforementioned solutions especially in relation to a particular example of $\mrm{M}2$-$\mrm{M}5$ bound state preserving 16 supercharges \cite{Izquierdo:1995ms,Dibitetto:2017tve}.
 
Among these flows we will encounter $\mrm{AdS}_{3}$ solutions slicing an asymptotically $\mrm{AdS}_7$ background. These are of particular interest from the point of view of $\mrm{AdS}/\mrm{CFT}$ correspondence since they represent an ideal framework to study {\itshape conformal defects} \cite{Karch:2000gx,Lunin:2007ab} within a non-lagrangian $d=6$ superconformal field theory dual to the $\mrm{AdS}_7$ vacuum. In section \ref{flowacrossdimensions} we introduced the conformal defects as a way to interpret a lower-dimensional $\mrm{SCFT}_p$ arising in the IR of wrapped branes' constructions as a theory living on a hypersurface within the manifold on which a ``mother" $\mrm{SCFT}_d$, dual to the UV asymptotics, is defined. The internal $\mrm{SCFT}_p$ manifests its presence through the breaking of the conformal symmetry of the $\mrm{SCFT}_d$. This induces an $\mrm{RG}$ flow across dimensions between a $\mrm{UV}$ fixed point represented by the $\mrm{SCFT}_d$ and a $\mrm{IR}$, described by the lower-dimensional $\mrm{SCFT}_p$. The existence of the defect can be immediately verified since it determines the breaking of the conformal invariance implying non-vanishing values of the one-point functions and the non-conservation of the stress-energy tensor of the $\mrm{SCFT}_d$.

In our case we will deal with a two-dimensional $\ma N=(4, 0)$ $\mrm{SCFT}_2$ dual to the $\mrm{AdS}_{3}$ and identifying a {\itshape surface defect} within the six-dimensional $\ma N=(1,0)$ $\mrm{SCFT}_6$ associated to the $\mrm{AdS}_7$ vacuum \cite{Dibitetto:2017klx}. This particular setup will be realized starting by considering a particular flow with an $\mrm{AdS}_{3}$-slicing choosen between the seven-dimensional solutions presented. We will construct the brane picture describing the uplift of this flow that it turns out to be defined by the wrapping of the bound state $\mrm{D}2$-$\mrm{D}4$-$\mrm{NS}5$-$\mrm{D}6$-$\mrm{D}8$ in massive type $\mrm{IIA}$ supergravity \cite{Dibitetto:2017klx}. This bound state, when considered in two particular limits of its coordinates, reproduces both the $\mrm{AdS}_7$ vacua and the $\mrm{AdS}_3$ slicing, while outside of these limits it defines an $\mrm{RG}$ flow across dimensions from the $\mrm{UV}$ given by the $\ma N=(1,0)$ $\mrm{SCFT}_6$ towards the $\mrm{IR}$ limit described by the defect $\mathcal{N}=(4,0)$ $\mrm{SCFT}_2$ \cite{Dibitetto:2017klx}.

\section{The Minimal Gauged Supergravity in $d=7$}
\label{sec:gauged_sugra}

Gauged supergravities in seven dimensions may be divided into maximal theories \cite{Pernici:1984xx,Pernici:1984zw}, i.e. with $32$ real supercharges, and half-maximal ones \cite{Townsend:1983kk} with only $16$. Since Majorana
spinors do not exist in $1+6$ dimensions, it is impossible to further go down to $8$ supersymmetries. 
While the complete embedding tensor formulation of the maximal gauged theories has been worked out in all details in \cite{Samtleben:2005bp}, such a complete formulation is lacking in the context of 
theories with $16$ supercharges \cite{Dibitetto:2012rk,Dibitetto:2015bia}.

The theory we want to focus on can be obtained as a particular truncation of half-maximal supergravity obtained by restricting oneself to the $\mathcal{N}=1$ supergravity multiplet. The ungauged theory is characterized by a bosonic field content including the metric, a three-form gauge potential, three vector fields and one scalar, and is usually referred to as \emph{minimal}, (see the classification of section \ref{survey}).
\begin{table}[h!]
\renewcommand{\arraystretch}{1}
\begin{center}
\scalebox{1}[1]{
\begin{tabular}{c|c|c|c|c}
fields and $\textrm{SO}(5)$ irrep's & $\mathbb{R}^{+} \, \times \, \mrm{SO(3)}$ irrep's & $\mrm{SU(2)}_{R}$ irrep's & \# dof's  \\
\hline \hline
${e_{\mu}}^{m}$ & $\textbf{14}$ & $\textbf{1}_{(0)}$ & $\textbf{1}$ & $14$ \\
\hline
${A_{\mu}}^{i}$ & $\textbf{5}$ & $\textbf{3}_{(+1)}$ & $\textbf{1}$ & $15$ \\
\hline
$B_{\m\n\r}$ & $\textbf{10}$ & $\textbf{1}_{(-2)}$ & $\textbf{1}$ & $10$ \\
\hline
$X$ & $\textbf{1}$ & $\textbf{1}_{(+1)}$ & $\textbf{1}$ & $1$ \\
\hline
\hline
$\psi_{\m\a}$ & $\textbf{16}$ & $\textbf{1}_{(0)}$ & $\textbf{2}$ & $32$ \\
\hline
${\chi}_{\a}$ & $\textbf{4}$ & $\textbf{1}_{(0)}$ & $\textbf{2}$ & $8$ \\
\end{tabular}
}
\end{center}
\caption[Field content of ungauged minimal $\mathcal{N}=1$ supergravity in $d=7$. Table from \cite{Dibitetto:2017tve}.]{{\it The on-shell field content of ungauged minimal $\mathcal{N}=1$ supergravity in $d=7$. Each field is massless and hence transforms in some irrep of the corresponding little group $\textrm{SO}(5)$ w.r.t.
spacetime diffeomorphisms and local Lorentz transformations. Table from \cite{Dibitetto:2017tve}.}} \label{Table:dofs}
\end{table}
The most general consistent deformation turns out to be a combination of a gauging of the $\textrm{SU}(2)$ R-symmetry group and a St\"uckelberg-like massive deformation for the three-form potential.
The purely gauged minimal theory was found to stem from a reduction of type I supergravity on $S^{3}$ \cite{Chamseddine:1999uy}, while the purely massive theory may be obtained as a reduction of 
eleven-dimensional supergravity on a $T^4$ with non-vanishing four-form flux. 
However, none of the above limiting cases allows for moduli stabilization, since the induced scalar potential always exhibits a run-away direction.

In this case, the full Lagrangian enjoys a global symmetry given by
\be
G_{0} \ = \ \mathbb{R}^{+}_{X} \, \times \, \mrm{SO(3)} \ .
\notag
\ee

As for $\ma N=2$ supergravities, the $(40_{B} \ + \ 40_{F})$ bosonic and fermionic propagating degrees of freedom of the theory are rearranged into irreducible representations of $G_{0}$ as described in table~\ref{Table:dofs}. We refer to 
the appendix \ref{app:SM_spinors} for a summary of our notations concerning symplectic-Majorana (SM) spinors.
In such a minimal setup, the possible consistent deformations of the theory associated with a generalized embedding tensor are of the following two different types:
\begin{itemize}
\item An $\mrm{SU(2)}$ gauging realized by the three vector fields in table~\ref{Table:dofs} and controlled by the gauge coupling constant $g$.
\item A St\"uckelberg-like coupling $h$ giving a mass to the 3-form gauge potential $B_{(3)}$ in the gravity multiplet. 
\end{itemize}

The bosonic Lagrangian for the deformed theory then reads \cite{Townsend:1983kk}
\be
\label{Lagrangian_seven-dimensional}
\begin{split}
&\mathcal{L}  =  R\,\star_{7}1\,-\,5\,X^{-2}\,\star_{7}dX\,\wedge\,dX \,-\,\frac{1}{2}\,X^{4}\,\star_{7}\mathcal{F}_{(4)}\,\wedge\,\mathcal{F}_{(4)} \,-\, V_g\,\star_{7}1  \\[1mm]
&  - \,\frac{1}{2}\,X^{-2}\,\star_{7}\mathcal{F}_{(2)}^{i}\,\wedge\,\mathcal{F}_{(2)}^{i} \,-\, h\,\mathcal{F}_{(4)}\,\wedge\,B_{(3)} \,+\, \frac{1}{2}\,\mathcal{F}_{(2)}^{i}\,\wedge\,\mathcal{F}_{(2)}^{i} \,\wedge\,B_{(3)} \, ,
\end{split}
\ee
where $R$ denotes the seven-dimensional Ricci scalar and $V_g(X)$ is the scalar potential. The quantities $\mathcal{F}_{(2)}$ and $\mathcal{F}_{(4)}$ are the (modified) field strengths of the 1- and 3-form gauge potentials, respectively. 
Their explicit form is given by
\be
\mathcal{F}_{(2)}^{i}  =  \D A^{i} \, - \, \frac{g}{2} \, \epsilon^{ijk}\,A^{j}\wedge A^{k} \qquad \text{ and } \qquad\mathcal{F}_{(4)}  = \D B_{(3)} \, .
\ee
The explicit form of the scalar potential induced by the two aforementioned deformations reads
\be
V_g(X) \ = \ 2h^{2} \, X^{-8} \, - \, 4\sqrt{2}\,gh\, X^{-3} \, - \, 2g^{2} \,X^{2} \ ,
\label{potential}
\ee
which may be, in turn, rewritten in terms of a real superpotential
\be
\label{superpotential}
f(X) \ = \ \frac{1}{2}\,\left(h\,X^{-4}\,+\,\sqrt{2}\,g\,X\right) \ ,
\ee
through the relation
\be
V_g(X) \ = \ \frac{4}{5}\,\left(-6f(X)^{2}\,+\,X^{2}\,\left(D_{X}f\right)^{2}\right) \ .
\ee

Finally, due to the presence of the topological term in \eqref{Lagrangian_seven-dimensional} induced by $h$ and $B_{(3)}$, one has to impose an {\itshape odd-dimensional self-duality condition} \cite{Townsend:1983xs} of the form
\be
\label{SD_cond}
X^{4} \, \star_{7}\mathcal{F}_{(4)}\ \overset{!}{=} \ -2h \, B_{(3)} \, + \, \frac{1}{2} \, A^{i} \, \wedge \, \mathcal{F}_{(2)}^{i} \,+ \, 
\frac{g}{12} \epsilon_{ijk} \, A^{i} \, \wedge \, A^{j} \, \wedge \, A^{k} \ .
\ee

This supergravity theory enjoys $\mathcal{N}=1$ supersymmetry, which can be made manifest by checking the invariance of its full Lagrancian with respect to the following supersymmetry transformations
\be
\label{SUSY_eqns_7D}
\begin{split}
&\delta_{\zeta}{e_{\m}}^{m}  =  \bar{\zeta}_{a}\,\g^{m}\,{\psi_{\m}}^{a}  , \\
&\delta_{\zeta}X  =  \frac{X}{2\sqrt{10}}\,\bar{\zeta}_{a}\,\chi^{a}  , \\[2mm]
&\delta_{\zeta}{A_{\m}}^{i} =  i\,\frac{X}{\sqrt{2}} \,\left[\left(\bar{\psi}_{\m b}\,\zeta^{a}\,-\,\frac{1}{2}\,d^{a}_{b}\,\bar{\psi}_{\m c}\,\zeta^{c}\right)
\,-\,\frac{1}{\sqrt{5}}\,\left(\bar{\chi}_{b}\,\g_{\m}\,\zeta^{a}\,-\,\frac{1}{2}\,d^{a}_{b}\,\bar{\chi}_{c}\,\g_{\m}\,\zeta^{c}\right)\right]  , \\[2mm]
&\delta_{\zeta}{B_{\m\n\r}}  =  \frac{X^{-2}}{\sqrt{2}} \, \left(\frac{3}{2}\,\bar{\psi}_{\m a}\,\g_{\n\r}\,\zeta^{a}
\,+\,\frac{1}{\sqrt{5}}\,\bar{\chi}_{a}\,\g_{\m\n\r}\,\zeta^{a}\right)  , \\[2mm]
&\delta_{\zeta}{\psi_{\m}}^{a} =  \nabla_{\m}\zeta^{a} \, + \, ig\,{\left(A_{\m}\right)^{a}}_{b}\,\zeta^{b}  \, + \, 
i\,\frac{X^{-1}}{10\sqrt{2}}\,\left({\g_{\m}}^{mn}\,-\,8\,{e_{\m}}^{m}\,\g^{n}\right) \, {\left(\mathcal{F}_{(2)\, mn}\right)^{a}}_{b}\,\zeta^{b}    \\[1mm]
& \qquad + \,\frac{X^{2}}{160}\,\left({\g_{\m}}^{mnpq}\,-\,\frac{8}{3}\,{e_{\m}}^{m}\,\g^{npq}\right) \, \mathcal{F}_{(4)\, mnpq}\,\zeta^{a} \, - \, \frac{1}{5}\,f(X)\,\g_{\m}\,\zeta^{a}  , \\[2mm]
&\delta_{\zeta}{\chi^{a}} =   \frac{\sqrt{5}}{2}\,X^{-1}\slashed{\partial}X \, \zeta^{a} \, - \, i\, \frac{X^{-1}}{\sqrt{10}}\,  {\left(\slashed{\ma F}_{(2)}\right)^{a}}_{b}\,\zeta^{b} 
\, + \, \frac{X^{2}}{2\sqrt{5}}\,  \slashed{\ma F}_{(4)}\,\zeta^{a} \, - \, \frac{X}{5} \, D_{X}f \, \zeta^{a}  ,
\end{split}
\ee
where we introduced the following notation $\slashed{\omega}_{(p)}\,=\,\frac{1}{p!}\,\g^{m_{1}\cdots m_{p}}\,\omega_{(p)\,m_{1}\cdots m_{p}}$, $\omega_{(p)}$ being a $p$-form, and the 
$\mrm{SU(2)}$-valued vector fields read
\be
A \ = \ \frac{1}{2} \, A^{i} \, \s^{i} \ ,
\ee
where $\left\{\sigma^{i}\right\}$ being the Pauli matrices given in \eqref{Pauli}.

The bosonic field equations obtained by varying the action \eqref{Lagrangian_seven-dimensional} are given by  
\be
\label{EOM_7D}
\begin{split}
&R_{\m\n} \, - \, 5X^{-2}\,\partial_{\m}X\,\partial_{\n}X \, - \, \frac{1}{5} \, V_g(X) \, g_{\m\n} \, - \, \frac{1}{2}\,X^{-2} \, \left(\mathcal{F}_{(2)}\right)^{2}_{\m\n}
 \, - \, \frac{1}{2}\,X^{4} \, \left(\mathcal{F}_{(4)}\right)^{2}_{\m\n} = 0 , \\[2mm]
&\nabla_{\m}\left(X^{-1}\,\nabla^{\m}X\right) \, - \, \frac{X^{4}}{120}\,\left|\mathcal{F}_{(4)}\right|^{2} \, + \, \frac{X^{-2}}{20}\,\left|\mathcal{F}_{(2)}\right|^{2} \, - \, \frac{X}{10}\,D_{X}V_g  =  0\, , \\[2mm]
&\D \left(X^{4}\,\star_{7}\mathcal{F}_{(4)}\right) \, + \, \frac{g}{\sqrt{2}}\,\mathcal{F}_{(4)} \, - \, \frac{1}{2} \, \mathcal{F}_{(2)}^{i} \, \wedge \, \mathcal{F}_{(2)}^{i}  = 0 \, , \\[2mm]
&\D_{A}\left(X^{-2}\,\star_{7}\mathcal{F}_{(2)}^{i}\right) \, - \, \mathcal{F}_{(2)}^{i} \, \wedge \, \mathcal{F}_{(4)}= 0 \, , \\[2mm]
\end{split}
\ee
where 
\be
\begin{split}
&\left(\mathcal{F}_{(2)}\right)^{2}_{\m\n} = {\mathcal{F}_{(2)}^{i}}_{\m\r}\,{{\mathcal{F}_{(2)}^{i}}_{\n}}^{\r}  -  \frac{1}{10}\,\left|\mathcal{F}_{(2)}\right|^{2} \, g_{\m\n} \,,\qquad \qquad \, \, \, \, \, 
\left|\mathcal{F}_{(2)}\right|^{2}  = {\mathcal{F}_{(2)}^{i}}_{\m\n}\,{{\mathcal{F}_{(2)}^{i}}}^{\m\n} \, , \\[2mm]
&\left(\mathcal{F}_{(4)}\right)^{2}_{\m\n}   =  {\mathcal{F}_{(4)}}_{\m\r\s\kappa}\,{{\mathcal{F}_{(4)}}_{\n}}^{\r\s\kappa}  -  \frac{1}{40}\,\left|\mathcal{F}_{(4)}\right|^{2} \, g_{\m\n} \, ,\quad \qquad
\left|\mathcal{F}_{(4)}\right|^{4}  =  {\mathcal{F}_{(4)}}_{\m\n\r\s}\,{{\mathcal{F}_{(4)}}}^{\m\n\r\s} \, ,
\end{split}
\ee
 and $\D _{A}$ denotes the gauge-covariant differential.
Note that the equations of motion in \eqref{EOM_7D} are implied by the $\mathrm{SUSY}$ conditions \eqref{SUSY_eqns_7D} written down for a purely bosonic background.

\section{BPS Objects and Dyonic 3-Form}
\label{sec:flows}

In this section we will derive a class of $\mrm{BPS}$ flows within minimal gauged supergravity in $d=7$ by solving the first-order equations. These will be obtained through the formulation of a suitable Ansatz on the bosonic fields and on the Killing spinor associated to the flow.
In particular these flows will involve a non-trivial profile for the 3-form potential $B_{(3)}$ and with 8 real preserved supercharges. Moreover, in this section we will keep the vectors inactivated, leaving to the next section their inclusion. 

There are two crucially different possibilities when considering a flow with a running 3-form potential $B_{(3)}$ and vanishing vectors:

\begin{itemize}
\item  {\itshape Vanishing topological mass}: For these models the self-duality condition \eqref{SD_cond} is trivially satisfied by an electric profile for the 3-form potential and these cases are well described by the
already known solutions of ungauged supergravity (see e.g. \cite{Townsend:1995de}).

\item  {\itshape Non-vanishing topological mass}: For these models the condition \eqref{SD_cond} requires a more complicated {\itshape dyonic} Ansatz for the 3-form potential. This is the situation 
that we are going to consider in this section and, in this context, we will derive a new class of $\mrm{BPS}$ solutions with 8 real supercharges ($\mrm{BPS}$/2) \footnote{This situation was originally considered in \cite{Liu:1999ai}, where some insights were
given concerning the search for dyonic membrane solutions.}. 
\end{itemize} 

\subsection{The Domain Wall Solution}

We first start constructing the domain wall (DW) solution as the most simple of supersymmetric solution of this theory \cite{Bergshoeff:2004nq}.
Supersymmetric DW's are $\mrm{BPS}$ flows where the only excited degrees of freedom are the metric and the scalar fields. 
In this case we consider the following Ansatz for the $d=7$ fields
\be
\begin{split}
& \D s^2   =  e^{2U(r)}\,\D s_{\scriptsize{\mathbb{R}}^{1,5}}^{2} \, + \, e^{2V(r)} \,  \D r^{2}  \, , \\[2mm]
&X  =  X(r) \, ,
\label{DWansatz}
\end{split}
\ee
where $\D s_{\scriptsize{\mathbb{R}}^{1,5}}^{2}$ denotes the flat six-dimensional Minkowski metric,
while both vector and 3-form fields are kept vanishing. Note that the arbitrary function $V(r)$ is in fact non-dynamical and could be set to zero by means of a suitable gauge choice. 
However, when solving this type of problems, it is often convenient to keep such a gauge freedom in order to simplify the resulting flow equations such in way that they may be integrated analytically.

By choosing a Killing spinor of the form
\be
\zeta(r) \, = \, Y(r)\,\zeta_{0} \ ,
\ee
where $\zeta_{0}$ is a constant SM spinor (i.e. obeying \eqref{SM_cond}) and further satisfying the following projection condition\footnote{In what follows we shall adopt the following notation: 
$\Pi(\mathcal{O}) \, \equiv \, \frac{1}{2} \, (\mathbb{I}+\mathcal{O})$, where $\mathcal{O}$ denotes an idempotent spinorial operator.}
\be
\Pi(\g^{6}) \, \zeta_{0} \, = \, \zeta_{0} \ , 
\ee
the $\mathrm{SUSY}$ equations are fully implied by the following first-order flow equations
\be
\begin{array}{lclclc}
U' \, = \, \frac{2}{5} \, e^{V} \, f & , & Y' \, = \, \frac{Y}{5} \, e^{V} \, f & , & X' \, = \, -\frac{2}{5} \, e^{V} \,X^{2} \,D_{X}f & .
\label{eqDW}
\end{array}
\ee
If we make the gauge choice $e^{V}=-\frac{5\, X^{-2}}{2\, D_{X}f}\,$, the general solution of \eqref{eqDW} is given by
\be 
e^{2U}= \left(\frac{r}{\sqrt{2}\, g \, r^5-4 \,h}\right)^{1/2}\ , \qquad X=r\ ,
\label{DWflow}
\ee
where one can further consider special cases where $g\,=\,0$, $h\,\neq\,0$, $g\,\neq\,0$, $h\,=\,0$, and $g\,\neq\,0$ and $h\,\neq\,0$. 

\subsection{Charged Flow on the Background $\mathbb{R}^{1,2}\times \mathbb{R}^3$}

Let us now consider a non-trivial dyonic profile for the 3-form potential $B_{(3)}$ for a seven-dimensional background including the flat manifold $\mathbb{R}^{3}$.
We make the following Ansatz for the fields
\be
\begin{split}
& \D s^2      =  e^{2U(r)}\, \D s_{\mathbb{R}^{1,2}}^{2} \, + \, e^{2V(r)} \,  \D r^{2}  \, + \, e^{2W(r)}\,\D s_{\mathbb{R}^3}^{2} \, , \\[2mm]
&X  =  X(r) \, , \\[2mm]
&B_{(3)}  =  k(r)\, \mrm{vol}_{\mathbb{R}^{1,2}} \, + \, l(r)\, \mrm{vol}_{\mathbb{R}^3} \,,
\end{split}
\ee
where $ \D s_{\mathbb{R}^{1,2}}^{2}$ and $ \D s_{\mathbb{R}^{3}}^{2}$ respectively denote the flat $\mathbb{R}^{1,2}$ and the flat $\mathbb{R}^{3}$ metric,
while the vector fields are still kept vanishing. Note that $V(r)$ is an arbitrary non-dynamical function 
and can be set to zero with a suitable gauge choice. 

By choosing a Killing spinor of the form
\be
\zeta(r) \, = \, Y(r)\,\left(\cos\theta(r)\,\mathbb{I}_{8}\,+\,\sin\theta(r)\,\g^{012}\right)\,\zeta_{0} \ ,
\label{Kspinor}
\ee
where $\zeta_{0}$ is a constant SM spinor (i.e. obeying \eqref{SM_cond}) and further satisfying the following projection condition
\be
\Pi(\g^{3}) \, \zeta_{0} \, = \, \zeta_{0} \ , 
\label{gamma3proj}
\ee
the Killing spinor equations are fully implied by the following first-order flow equations 
\be
\begin{split}
&U' =  \frac{1}{5} \, e^{V} \, f \, \frac{(3 \cos(4 \theta)\,-\,1)}{\cos(2 \theta)}  \, , \\[2mm]
&W'  =  -\frac{2}{5} \, e^{V} \, f \, \frac{(\cos (4 \theta)\,-\,2)}{\cos(2 \theta)}  \, , \\[2mm]
&Y' =  \frac{1}{10} \, e^{V} \, Y \, f \, \frac{(3 \cos (4 \theta)\,-\,1)}{\cos(2 \theta)}  \, , \\[2mm]
\end{split}
\nonumber
\ee
\be
\begin{split}
&\theta' =  -e^{V} \, f \, \sin (2 \theta) \, , \\[2mm] 
&k'  =  -\frac{4 f\, e^{3 U+V}}{X^2} \, \tan (2 \theta) \, , \\[2mm] 
&l'  =  \frac{4 f\, e^{V+3 W}}{X^2} \, \sin (2 \theta) \, , \\[2mm] 
&X'  = -\frac{2}{5} \, e^{V} \, X \, \left(X\, D_{X}f\,-\,8 f\, \frac{\sin^4\theta}{\cos (2 \theta)}\right) \, ,\label{floweqDW}
\end{split}
\ee
provided that the following extra differential constraint
\be
\label{constraintDW}
X \, D_{X}f \, + \, 4\,f \, \overset{!}{=} \, 0 \ ,
\ee
holds along the flow.
It can be shown that \eqref{constraintDW} is solved by a superpotential 
of the original form given in \eqref{superpotential} by setting $g\,=\,0$. This situation corresponds to having a pure St\"uckelberg deformation associated with the 
parameter $h$, without any gauging.

After performing the following gauge choice for the function $V$
\be
e^{V} \, \overset{\textrm{gauge fix.}}{=} \, f^{-1} \ ,
\ee
the above flow equations may be integrated analytically and the solution reads
\be
\label{ch_DW_noAdS}
\begin{split}
&e^{2U}  =  \sinh(4r)^{1/5}\,\coth(2r)\, , \\[2mm]
 &e^{2V} \, = \frac{4}{h^2}\, \sinh(4r)^{16/5} \, , \\[2mm]
&e^{2W} \, = \, \sinh(4r)^{1/5}\,\tanh(2r) \, , \\[2mm]
&X \, = \, \sinh(4r)^{2/5} \, , \\[2mm]
&k \, = \, \frac{1}{\sqrt{2}\,\sinh^{2}(2r)}\, , \\ 
&l \, = \, -\frac{1}{\sqrt{2}\,\cosh^{2}(2r)} \, , \\[2mm]
&Y \, = \, \sinh(4r)^{1/20}\,\coth(2r)^{1/4} \, , \\[2mm]
& \theta \, = \, \arctan \left(e^{-2r}\right) \, .
\end{split}
\ee

One may check that \eqref{ch_DW_noAdS} correctly satisfies the bosonic field equations in \eqref{EOM_7D} as well as the odd-dimensional self-duality condition \eqref{SD_cond}. Note that this solution is
not asymptotically $\textrm{AdS}_{7}$, consistently with the fact that the monomial scalar potential induced by the only contribution of the topological mass has a run-away behavior in $X$. 

\subsection{Charged Flow on the Background $\mathbb{R}^{1,2}\times S^3$}

It is now natural to wonder if flows driven by the complete profile of the potential \eqref{potential} exist and, then, if asymptotically $\mathrm{AdS}_7$ solutions with a running profile for the 3-form exist in the considered theory.

It is well known that one of the main features of the first-order formulation of supergravity is the gauge-dependence: the profile
of the Killing spinor directly determines the background through the first-order flow equations, which turn out to explicitly depend on the spin connection of the background itself. Adapting this story to 
our case, this implies that searching for 
 Killing spinors associated to asymptotically $\mathrm{AdS}_7$ flows is equivalent to look for a background parametrization such that the correspondent flow equations are
driven by the complete superpotential \eqref{superpotential}. 

We claim that, in our case, this happens only if the locally Euclidean part of the background admits an $\mathrm{SO(3)}$-covariant {\itshape parallelized basis}, i.e. we need a
field configuration parametrized in such a way the spin connection of the Euclidean part of the metric takes non-zero constant values once expressed in flat coordinates\footnote{This basis parametrizes to the Hopf-fibration of the 3-sphere.}.
From these considerations it follows that the presence of $\mathrm{AdS}_7$ is excluded for a metric containing $\mathbb{R}^3$ since it is flat and also for $H^3$ since its parallelized basis is $\mathrm{SO(2,1)}$-covariant.

Thus we consider an Ansatz of the form,
\be
\begin{split}
& \D s^2     =  e^{2U(r)}\,\D s_{\mathbb{R}^{1,2}}^{2} \, + \, e^{2V(r)} \,  \D r^{2}  \, + \, e^{2W(r)}\,  \D s_{S^{3}}^{2} \,, \\[2mm]
&X  =  X(r) \, , \\[2mm]
&B_{(3)}  =  k(r)\,  \mrm{vol}_{\mathbb{R}^{1,2}} \, + \, l(r)\, \mrm{vol}_{S^{3}} \, ,
\label{ansatz:mkw21}
\end{split}
\ee
where $  \D s_{S^{3}}^{2}$ is the metric of a unit $S^{3}$ and $\text{vol}_{S^{3}}$ its volume.
We choose the set of Hopf coordinates $(\theta_1,\theta_2, \theta_3)$ on $S^3$, such that
\be
\D s_3^2=\frac{1}{\kappa^2}\left[ d\theta_2^2+\cos^2\theta_2 d\theta_3^2+\left(d\theta_1+\sin\theta_2\theta_3\right)^2   \right] \, .
\label{S3}
\ee
The dreibein corresponding to this parametrization of the $S^3$ is non-diagonal\footnote{The dreibein \eqref{S3vielbein} corrects the expression (4.10) of \cite{Dibitetto:2017tve}.},
\be
 \begin{split}
&e^{1}  =  \frac{1}{\kappa} \,(\D \theta_{1}+\sin \theta_2 \D \theta_3)  ,\\
&e^{2}  = \frac{1}{\kappa} \, \left(\cos\theta_{1}\D \theta_{2}\,-\,\sin\theta_{1}\cos\theta_2\D \theta_{3}\right)  ,\\
&e^{3}  = \frac{1}{\kappa} \,(\cos\theta_1\cos\theta_2\D\theta_3+\sin\theta_1\D \theta_2) ,
\end{split}
\label{S3vielbein}
\ee
and the corresponding spin connection is constant if expressed in the flat basis \eqref{S3vielbein} and given by
\be
\omega_{i\,jk}=\frac{\kappa}{2}\,\epsilon_{ijk} \quad \text{ with } \quad i,j,k=1,2,3\,.
\label{spinconnS3}
\ee
In what follows the $\mathrm{\mathrm{SO(3)}}$ indices $(i,j,k)=1,2,3$ must be identified with the $(4,5,6)$ components of the flat basis of the whole seven-dimensional metric.

By choosing a Killing spinor with the same profile of \eqref{Kspinor} and satisfying 
the projection condition \eqref{gamma3proj}, the Killing spinor equations are satisfied if the following 
system of first-order flow equations hold,
\be
\begin{split}
\label{floweqAdSDW}
&U'  =  \frac{1}{25} \, e^{V} \, \frac{(3 \cos(4 \theta)\,+\,7)\, f+6\sin^2(2 \theta)\, X\, D_{X}f}{\cos(2 \theta)}  , \\[2mm]
&W'  =   -\frac{2}{25} \, e^{V} \, \frac{(\cos(4 \theta)\,-\,6)\, f+2\sin^2(2 \theta)\, X\, D_{X}f}{\cos(2 \theta)}   , \\[2mm]
&Y'  =  \frac{1}{50} \, e^{V} \,Y\, \frac{(3 \cos(4 \theta)\,+\,7)\, f+6\sin^2(2 \theta)\, X\, D_{X}f}{\cos(2 \theta)}   , \\[2mm]
&\theta'  =  -\frac15 \,e^{V} \,  \sin (2 \theta)\,\left(f-\, X\, D_{X}f\right)  , \\[2mm] 
&k'  =  \frac25 \frac{e^{V+3 U}\tan(2\theta)\,\left(2\,f+3\, X\, D_{X}f\right)}{X^2}  , \\[2mm] 
&l'  =  \frac45 \frac{e^{V+3 W}\sin(2\theta)\,\left(f-\, X\, D_{X}f\right)}{X^2}  , \\[2mm] 
&X'  =  -\frac{2}{25} \, e^{V} \, X \,\frac{(4+\cos(4\theta))\,X\, D_{X}f\,+\,2 \sin^2(2 \theta)\,f}{\cos (2 \theta)}  ,
\end{split}
\ee
where the constraint
\be
\label{constraintAdSDW}
\kappa+\frac25 e^{W}\tan(2\theta)\left(X \, D_{X}f \, + \, 4\,f \right)\overset{!}{=} \, 0 \ ,
\ee
has been imposed along the whole flow. Imposing the constraint \eqref{constraintAdSDW} and the flow
\eqref{floweqAdSDW} on the equations of motion \eqref{EOM_7D}, it follows that they are satisfied if the superpotential is given by the 
\eqref{superpotential} with arbitrary values of $g$ and $h$.

As in the case of the previous section, we perform the following gauge choice for the function $V$
\be
e^{-V} \, \overset{\textrm{gauge fix.}}{=} \, \frac15 \left (f\,-\,X\,D_Xf\right) \ ,
\label{gauge}
\ee
the above flow equations may be integrated analytically and the solution reads
\be
\label{ch_DW_AdS}
\begin{split}
&e^{2U} \, = \,\left(\frac{\left(\rho ^4+1\right)^3 \left(\rho ^{16}+4 \rho ^{12}+4 \rho ^4+1\right)}{\rho ^{10}
   \left(\rho ^4-1\right)^2}\right)^{2/5} \, ,\\
  & e^{2V}\,=\, \frac{ 2^{2/5} \left(\rho ^8-1\right)^{16/5}}{h^{2/5} g^{8/5} \left(\rho
   ^{16}+4 \rho ^{12}+4 \rho ^4+1\right)^{8/5}} \, ,\\
&e^{2W} \, = \, \left(\frac{\left(\rho ^4-1\right)^3 \left(\rho ^{16}+4 \rho ^{12}+4 \rho ^4+1\right)}{\rho ^{10}
   \left(\rho ^4+1\right)^2}\right)^{2/5}  \, ,\\
&X \, = \, \frac{2^{3/10} h^{1/5} \left(\rho ^8-1\right)^{2/5}}{
   g^{1/5} \left(\rho ^{16}+4 \rho
   ^{12}+4 \rho ^4+1\right)^{1/5}} \, ,\\
        \end{split}\nonumber
   \ee
   \be
   \begin{split}
&k \, = \, \frac{2^{2/5} g^{2/5} \left(\rho ^{16}+4 \rho ^{12}+4 \rho ^4+1\right)}{h^{2/5} \rho ^4
   \left(\rho ^4-1\right)^2}   \, ,\\
&    l \, = \,\frac{2^{2/5} g^{2/5} \left(\rho ^{16}-4 \rho ^{12}-4 \rho ^8-4 \rho ^4+1\right)}{h^{2/5}
   \rho ^4 \left(\rho ^4+1\right)^2}  \, ,\\
&Y \, = \, \left(\frac{\left(\rho ^4+1\right)^3 \left(\rho ^{16}+4 \rho ^{12}+4 \rho ^4+1\right)}{\rho ^{10}
   \left(\rho ^4-1\right)^2}\right)^{1/10}   \, ,\\
  &  \theta \, = \, \arctan \left(\rho^{-2}\right)  \, ,
\end{split}
\ee
where $r=\log\,\rho$ and from \eqref{constraintAdSDW} one obtains $\kappa=- 2^{9/5} g^{4/5} h^{1/5}$.

In the asymptotic region, the flow \eqref{ch_DW_AdS} turns out to locally reproduce $\mathrm{AdS_7}$, in fact the 
contribution of $\mathcal{F}_{(4)}$ turns out to be sub-leading when $r \rightarrow +\infty$. 
In this limit one has
\be
\theta=0\, ,\quad X= 1\,, \quad \mathcal{F}_{0123}=0\,, \quad \mathcal{F}_{3456}=0\, ,
\ee
where we made the choice for the parameters\footnote{The explicit dependence on the parameters $h\text{ and }g$
of the flow is related to the gauge choice \eqref{gauge}.
Given this particular gauge choice, one can always choose $h=\frac{g}{2\sqrt{2}}$ in order to obtain $X=1$ as an asymptotic of
value for the $\mathbb{R}^{+}$ dilaton.} $h$ and $g$ such that $h=\frac{g}{2\sqrt{2}}$.
In the limit $r \rightarrow 0$ the flow is singular. Finally it is easy to verify that \eqref{ch_DW_AdS} correctly satisfies the equations of motion
in \eqref{EOM_7D} and the odd-dimensional self-duality condition \eqref{SD_cond}.

\subsection{$\mathrm{AdS}_7$ Charged Domain Wall}
\label{AdS3defect}

We want now to consider a slightly more complicated system such that the whole background is curved. This is achieved by considering an $\mathrm{AdS}_3$ slicing of the seven-dimensional background. In this section we will consider
for simplicity a background depending only on a independent warp factor $U(r)$, thus the configuration of the fields has the form,
\be
\begin{split}
&\D s^2    =  e^{2U(r)}\, \D s_{\mathrm{AdS}_3}^{2} \, + \, e^{2V(r)} \,  \ \D r^{2}   \, + \, e^{2U(r)}\,  \D s_{S^{3}}^{2} \, , \\[2mm]
&X  =  X(r) \, , \\[2mm]
&B_{(3)}  =  k(r)\, \mrm{vol}_{\mathrm{AdS}_3} \, + \, l(r)\, \mrm{vol}_{S^{3}} \, ,
\label{defectAnsatz}
\end{split}
\ee
where $  \D s_{S^{3}}^{2}$ is again the metric of the $S^3$ parametrized as in \eqref{S3}, while $ \ \D s_{\mathrm{AdS}_3}^{2}$ is the metric of $\mathrm{AdS}_3$ in the parallelized basis $(t,x^1,x^2)$ such that
\be
 \D s_{\mathrm{AdS}_3}^{2}=\frac{1}{L^2}\left[ (dx^1)^2+\cosh^2x^1 (dx^2)^2-\left(d t-\sinh x^1 d x^2\right)^2   \right] \, .
\label{AdS3}
\ee
The non-symmetric dreibein associated to this parametrization is given by\footnote{The dreibein \eqref{AdS3vielbein} corrects the expression (4.19) of \cite{Dibitetto:2017tve}.}
\be
 \begin{split}
&e^{0}  =  \frac{1}{L} \,\left(\D t -\sinh x^1 \D x^2\right)\, ,\\
&e^{1}  =  \frac{1}{L} \, \left(\cos t \,\D x^{1}\,-\,\sin t \cosh x^1 \,\D x^{2}\right) \, ,\\
&e^{2}  = \frac{1}{L}\left(\cos t \cosh x^1\D x^2 + \sin t \, \D x^1 \right) \, ,
\end{split}
\label{AdS3vielbein}
\ee
and defines a constant spin connection as in the case of $S^3$.

Keeping the same Killing spinor given in \eqref{Kspinor} with 
the projection condition \eqref{gamma3proj}, the Killing spinor equations determine a
system of first-order flow equations for the superpotential \eqref{superpotential} if 
\be
\theta(r)=0\ ,\quad k(r)=l(r)\ ,\quad \kappa=L\ .
\ee
In this case the $\mrm{BPS}$ equations take the simple form
\be
\begin{split}
&U'  =  \frac{2}{5} \, e^{V} \, f \,,\\
&Y'   =  \frac{Y}{5} \, e^{V} \, f  \,,\\
&k'  = - \frac{e^{2U+V}\,L}{ X^2}  \,,\\
&X'  =  -\frac{2}{5} \, e^{V} \,X^{2} \,D_{X}f\,.
\end{split}
\label{floweqDW_U=W}
\ee
Choosing the gauge
\be
e^{-V}\,\overset{\textrm{gauge fix.}}{=}\,-\frac25\,X^2\, D_{X}f\ ,
\ee
 and taking the parameters as $h=\frac{g}{2\sqrt{2}}$, the equations \eqref{floweqDW_U=W} are easly integrated in the interval $r\in (0,1)$, yielding
 \be
\begin{split}
&e^{2U} =  \frac{2^{-1/4}}{\sqrt{g}}\,\left(\frac{r}{1\,-\,r^{5}}\right)^{1/2}\, ,\\
&e^{2V}  =  \frac{25}{2g^{2}}\,\frac{r^{6}}{\left(1\,-\,r^{5}\right)^{2}}\, ,\\
&Y  =    \frac{2^{-1/16}}{g^{1/8}}\,\left(\frac{r}{1\,-\,r^{5}}\right)^{1/8}\, ,\\
&k =  -\frac{2^{1/4}\,L}{g^{3/2}}\left(\frac{r^{5}}{1\,-\,r^{5}}\right)^{1/2}\, ,\\
&X =  r\,. 
\label{AdS3_DW_AdS_U=V}
\end{split}
\ee
This solution turns out to be asymptotically locally $\mathrm{AdS}_7$. In particular, in the limit $r \rightarrow 1$ one has
\be
 X= 1 \ , \quad \mathcal{F}_{0123}=0\ , \quad \mathcal{F}_{3456}=0\ ,
\ee
while for $r \rightarrow 0$ the solution is singular.

\subsection{The General Flow $\mathrm{AdS}_7 \rightarrow \mathrm{AdS}_3\times T^4$}

Let us now consider a slightly more complicated background where the warping is determined by two independent functions $U$ and $W$,
\be
\begin{split}
\label{backgroundAdS3}
&\D s^2    =  e^{2U(r)}\, \D s_{\mathrm{AdS}_3}^{2} \, + \, e^{2V(r)} \,  \ \D r^{2}   \, + \, e^{2W(r)}\,  \D s_{S^{3}}^{2} \, , \\[2mm]
&X  =  X(r) \, , \\[2mm]
&B_{(3)}  =  k(r)\, \mrm{vol}_{\mathrm{AdS}_3} \, + \, l(r)\, \mrm{vol}_{S^{3}} \, ,
\end{split}
\ee
where $  \D s_{S^{3}}^{2}$ is again the metric of the $S^3$ parametrized as in \eqref{S3}, while $ \D s_{\mathrm{AdS}_3}^{2}$ is the metric of $\mathrm{AdS}_3$ parametrized as in \eqref{AdS3}.

Given the usual Killing spinor \eqref{Kspinor} with 
the projection condition \eqref{gamma3proj}, the first-order flow equations are given by
\be
\begin{split}
\label{flowWarpedAdS}
&U'  =  \frac{1}{25} \, e^{V} \, \frac{(3 \cos(4 \theta)\,+\,7)\, f+6\sin^2(2 \theta)\, X\, D_{X}f-\,5L\,e^{-U} \sin(2\theta)}{\cos(2 \theta)}  \, ,\\
&W' =  -\frac{1}{25} \, e^{V} \, \frac{2(\cos(4 \theta)\,-\,6)\, f+4\sin^2(2 \theta)\, X\, D_{X}f+\,5L\,e^{-U}\sin(2\theta)}{\cos(2 \theta)}  \, ,\\
&Y' = \frac{1}{50} \, e^{V} \,Y\, \frac{(3 \cos(4 \theta)\,+\,7)\, f+6\sin^2(2 \theta)\, X\, D_{X}f-\,5L\,e^{-U} \sin(2\theta)}{\cos(2 \theta)}  \, ,\\
&\theta' = -\frac15 \,e^{V} \,  \sin (2 \theta)\,\left(f-\, X\, D_{X}f\right) \, ,\\
&k' =  \frac{e^{3U+V}}{5\, X^2}\left[ 2 \tan(2\theta)\, \left(2f+\,3\, X\, D_{X}f\right) -\frac{5\,L\,e^{-U}}{\cos(2\theta)} \right] \, ,\\
&l'  =  \frac{e^{3W+V}}{5\, X^2}\left[ 4 \sin(2\theta)\, \left(f\,-\, X\, D_{X}f\right) -5\,L\,e^{-U} \right]\, ,\\
&X'  =  -\frac{1}{25} \, e^{V} \, X \,\frac{2\,(4+\cos(4\theta))\,X\, D_{X}f\,+\,4 \sin^2(2 \theta)\,f-\,5\,L\,e^{-U}\sin(2\theta)}{\cos (2 \theta)} \, ,
\end{split}
\ee
where the constraint
\be
\label{constraintWarpedAdS}
\kappa-L\,\frac{e^{W-U}}{\cos(2\theta)}+\frac25 e^{W}\tan(2\theta)\left(X \, D_{X}f \, + \, 4\,f \right)\overset{!}{=} \, 0 \ ,
\ee
has been imposed along the whole flow.
Imposing the constraint \eqref{constraintWarpedAdS} the equations of motion \eqref{EOM_7D} are fully satisfied imposing \eqref{flowWarpedAdS} if the superpotential is given by the 
\eqref{superpotential} with arbitrary values of $g$ and $h$.

Performing the usual gauge choice for the function $V$
\be
e^{-V} \, \overset{\textrm{gauge fix.}}{=} \, \frac15 \left (f\,-\,X\,D_Xf\right) \ ,
\label{gauge}
\ee
the above flow equations are solved by
\be
\label{AdS3_DW_AdS}
\begin{split}
&e^{2U}= \frac{\left(\rho ^4+1\right)^2 \left(\sqrt{2} \,g\, \left(\rho ^{16}+4\,
   \rho ^{12}+4\, \rho ^4+1\right)-8 \,L \,\rho ^4 \left(\rho
   ^8+1\right)\right)^{2/5}}{ 2^{14/5} \,h^{2/5}\, \rho ^4  \left(\rho
   ^8-1\right)^{4/5}}  \,,\\
&e^{2V} = \frac{ 2^{26/5}\, \left(\rho ^8-1\right)^{16/5}}{h^{2/5}
   \left(\sqrt{2}\,g\, \left(\rho ^{16}+4 \rho ^{12}+4 \rho
   ^4+1\right)-8 \,L \,\rho ^4\, \left(\rho ^8+1\right)\right)^{8/5}}   \,,\\
&e^{2W}  =\frac{\left(\rho ^4-1\right)^2 \left(\sqrt{2}\, g\, \left(\rho
   ^{16}+4\, \rho ^{12}+4\, \rho ^4+1\right)-8\, L\, \rho ^4 \left(\rho
   ^8+1\right)\right)^{2/5}}{ 2^{14/5}\, h^{2/5} \rho ^{4}
   \left(\rho ^8-1\right)^{4/5}}   \,,\\ 
&X  = \frac{2^{2/5} \,h^{1/5}\,\left(\rho ^8-1\right)^{2/5}}{\left(\sqrt{2}\,
   g\, \left(\rho ^{16}+4\, \rho ^{12}+4\, \rho ^4+1\right)-8\, L\, \rho ^4
   \left(\rho ^8+1\right)\right)^{1/5}}  \,,\\
      \end{split}\nonumber
   \ee
   \be
   \begin{split}
&k  =\frac{\sqrt{2} \,g\, \left(\rho ^{16}+4\, \rho ^{12}+4 \rho ^4+1\right)-2\, L\, \left(\rho
   ^{16}+4 \rho ^{12}-2 \,\rho ^8+4\, \rho ^4+1\right)}{16\, h\, \rho ^4 \left(\rho ^4-1\right)^2}   \,,\\
&  l  =\frac{\sqrt{2}\,g\, \left(-\rho ^{16}+4\, \rho ^{12}+4\, \rho ^8+4\, \rho ^4-1\right)+2\, L\,
   \left(\rho ^{16}-4\, \rho ^{12}-2\, \rho ^8-4\, \rho ^4+1\right)}{16 \,h\, \rho ^4 \left(\rho
   ^4+1\right)^2}  \,,\\
&Y  =\left(\frac{\left(\rho ^4+1\right)^2 \left(\sqrt{2} \,g\, \left(\rho ^{16}+4\,
   \rho ^{12}+4\, \rho ^4+1\right)-8 \,L \,\rho ^4 \left(\rho
   ^8+1\right)\right)^{2/5}}{ 2^{14/5} \,h^{2/5}\, \rho ^4  \left(\rho
   ^8-1\right)^{4/5}}\right)^{1/4}  \,,\\
&\theta   =  \arctan \left(\rho^{-2}\right) \,,
\end{split}
\ee
where $r=\log\,\rho$ and, from \eqref{constraintWarpedAdS}, one obtains 
\label{k&Lrelation}
\be
\kappa+L\,=\,\sqrt{2}\,g\ .
\ee
This flow is asymptotically locally $\mathrm{AdS_7}$: for any values of $\kappa$ and $L$ respecting \eqref{k&Lrelation} and 
for $h=\frac{g}{2\,\sqrt{2}}$, one has 
\be
\theta=0\ ,\quad X= 1\ , \quad \mathcal{F}_{0123}=0\ , \quad \mathcal{F}_{3456}=0\  ,
\ee
in the limit $r \rightarrow +\infty$.

The study of the limit $r\rightarrow 0$ crucially depends on the relation between $\kappa$ and $L$. The general leading-order behavior of the scalar potential \eqref{potential} is given by
\be
V_g \,=\, \frac{ h^{2/5}\, \left(5 \,\sqrt{2}\, g-8\, L\right)^{8/5}  }{ 2^{3/10}\, r^{16/5}}\,+\,\cdots\ .
\ee
From this expression we conclude that the behavior of the flow in the limit $r \rightarrow 0$ is singular except for the special value
\be
L\,=\,\frac{5\,g}{4\,\sqrt{2}}\ ,
\label{magicrelation}
\ee
where the scalar potential takes a constant value and the flow turns out to be described locally by $\mathrm{AdS}_3\times T^4$, where the main difference with respect to the asymptotics is the fact that this geometry is
not a solution per se, as $\mathrm{AdS}_7$, but only the infrared (leading) profile of the flow \eqref{AdS3_DW_AdS} when the radii of $\mathrm{AdS}_3$ and $S^3$ are related by \eqref{magicrelation}.

In this limit, we have
\be
\theta=\frac{\pi}{4}\  ,\quad X= \frac{2^{2/5} }{3^{1/5}}\ , \quad \mathcal{F}_{0123}=0\ , \quad \mathcal{F}_{3456}=-\frac{3^{1/5}}{2^{19/10} }\,g\  .
\ee
Finally one can verify that \eqref{AdS3_DW_AdS} solves the equations of motion \eqref{EOM_7D} and the odd-dimensional self-duality condition \eqref{SD_cond}.

\section{The Coupling to the SU(2) Vectors}
\label{sec:vectors}

In this section we will extend our analysis including the coupling to the $\mathrm{\mathrm{SU(2)}}$ vectors $A^i$. In particular, the aim is finding solutions described by the
backgrounds \eqref{ansatz:mkw21} and \eqref{backgroundAdS3}, with running 3-form field,
including three non-Abelian vectors associated to the Hopf fibration of the 3-sphere $S^3$. Extending the set of excited fields produces a partial supersymmetry breaking. 
On one hand this is due to the 
presence of new terms in the Killing spinor equations \eqref{SUSY_eqns_7D}, on the other hand, the stucture of \eqref{SUSY_eqns_7D} tells how the profile of vectors should be in order to preserve some amount of supersymmetry. 

\subsection{Killing Spinors and Twisting Condition}

Let us consider the backgrounds \eqref{ansatz:mkw21} or \eqref{backgroundAdS3} with the $S^3$ metric parametrized as in \eqref{S3}, together with an Ansatz for the vectors given by
\be
A_j^i=\frac{A(r)}{2\,g}\,\epsilon^{i\,k\,l}\,\omega_{j\,kl}\ ,
\label{ansatzvector}
\ee
where $\omega_i$ are the components of the spin connection of the $S^3$ and the last three values of the curved index $\mu=4,5,6$ have been identified with the $\mathrm{SO(3)}$ indices $i,j\cdots=1,2,3$.

Given the Ansatz \eqref{ansatzvector}, we notice that the SM structure of the spinors turns out to be crucial in order to avoid a complete SUSY breaking. This may be seen explicitly by looking at the
 gravitinos supersymmetry variations $\delta_{\zeta}{\psi_{\m}}^{a}$, which acquire now the following new terms depending on $A^i$
\be
\cdots +\frac14\,\omega_{i\,\,jk}\,\gamma^{j\,k}\zeta^a + \, ig\,{\left(A_{i}\right)^{a}}_{b}\,\zeta^{b}+ \, 
i\,\frac{X^{-1}}{10\sqrt{2}}\,\left({\g_{i}}^{mn}\,-\,8\,{e_{i}}^{m}\,\g^{n}\right) \, {\left(\mathcal{F}_{mn}\right)^{a}}_{b}\,\zeta^{b}+\cdots\  ,
\label{susyvariationsvector}
\ee
which are characterized by a non-trivial action of the vectors on the $\mathrm{\mathrm{SU(2)}}$ structure of the spinor $\zeta^a$. If one looks at first contribution in \eqref{susyvariationsvector} 
coming from the spin connection of the $S^3$ in relation to the 
second term, we see that the only way of preserving some supersymmetry is to take the Killing spinor \emph{oriented along} the direction identified by the vectors. This happens only if one imposes three new projection conditions on the spinor. 
In terms of the SM spinor $\zeta^a$ defined in \eqref{Kspinor} and satisfying \eqref{gamma3proj}, these new conditions are given by
\be
\gamma^{5\,6}\, \zeta_0^a=-i\,\left(\sigma^1 \right )^a_{\,\,\,b}\zeta_0^b\ ,\quad \gamma^{4\,6}\, \zeta_0^a=i\,\left(\sigma^2 \right )^a_{\,\,\,b}\zeta_0^b\ , \quad
 \gamma^{4\,5}\, \zeta_0^a=-i\,\left(\sigma^3 \right )^a_{\,\,\,b}\zeta_0^b\ ,
 \label{vectorproj}
\ee
which may be reexpressed as
\be
\Pi\left(i\,\g^{ij}\,\otimes\,\s^{k}\right)\,\zeta_0 \overset{!}{=} \, \zeta_0 \ ,
\ee
with $i,\,j,\,k$ chosen to be all different and in all possible permutations.
It easy to show that the SM condition \eqref{SM_cond} is given exactly by the second projection condition in \eqref{vectorproj} if one represents the spinor $\zeta_0^a$ as a $\mathrm{\mathrm{SU(2)}}$ doublet. 
Thus \eqref{vectorproj} reduce the total amount of supersymmetry to two real supercharges ($\mrm{BPS}/8$).

It has been shown \cite{Acharya:2000mu} that the projection conditions \eqref{vectorproj} are naturally realized from those configurations with $A(r)=1$ and, then, with gauge fields independent of the radial coordinate.
In this case, the effect of the vectors \eqref{ansatzvector} is to exactly compensate
the contribution in \eqref{susyvariationsvector} due to the spin connection of $S^3$. This can be understood by recalling the expression of the spin connection of
$S^3$ given in \eqref{spinconnS3} and comparing the first two terms of \eqref{susyvariationsvector}. It is easy to show that \eqref{vectorproj} 
are implied by a \emph{twisting condition} \cite{Maldacena:2000mw} given by
\be 
-\frac12 \,\omega_{i\,\,jk}\,\gamma^{j\,k}\zeta^a=i\,g\,  A^j_i \,\left(\sigma^j \right )^a_{\,\,\,b}\zeta^b\ .
\label{twisting}
\ee
Thus, in this case, the effect of the coupling to the vector fields is literarly to twist the Killing spinor in order to compensate the contribution coming from the curvature of the background and preserving
a certain amount of supersymmetry. 

It is worth mentioning that including of the 3-form implies a non-trivial radial dependence for the gauge fields. In the next sections we will provide some examples of this fact. Generally the special form 
of the Killing spinor \eqref{Kspinor}, which is needed in order to include the 3-form, implies a non-trivial profile for $A(r)$ and from this it follows that all the solutions of the type $\mathrm{AdS}_{p+2}\times \Sigma_{7-p-2}$ are either
characterized by a constant value for $A(r)$ and a vanishing 4-form field strength, or by a non-constant profile for the gauge fields and a non-trivial 3-form.

\subsection{Vectors Coupled to the Background $\mathbb{R}^{1,2}\times S^3$}
\label{mkw21vector}

Let us consider the background \eqref{ansatz:mkw21}, and furthermore include vectors given by the Ansatz \eqref{ansatzvector}. Thus one has
\be
\begin{array}{lcll}
\D s^2   & = & e^{2U(r)}\,\D s_{\mathbb{R}^{1,2}}^{2} \, + \, e^{2V(r)} \, \D  r^{2} \, + \, e^{2W(r)}\,\ \D s_{S^{3}}^{2}  , \\[2mm]
X & = & X(r)  , \\[2mm]
B_{(3)} & = & k(r)\,  \mrm{vol}_{\mathbb{R}^{1,2}} \, + \, l(r)\, \mrm{vol}_{S^{3}}  , \\[2mm]
A^i\,&=\,&\frac{A(r)}{2\,g}\,\epsilon^{i\,k\,l}\,\omega_{j\,kl}\,\D \,\theta^j ,
\label{ansatz:mkw21vector}
\end{array}
\ee
where $S^3$ is parametrized by the parallelized basis $\{\theta_i\}$ introduced in \eqref{S3}.

As we mentioned in the previous section, we consider a Killing spinor $\zeta^a$ of the form \eqref{Kspinor} and satisfying \eqref{gamma3proj} and \eqref{vectorproj}. 
Thus $\zeta^a$ has two real independent components.
Plugging this Ansatz into the Killing spinor equations \eqref{SUSY_eqns_7D}, we obtain the set of consistent first-order flow equations given in \eqref{flow:mkw21vector}. 
Remarkably, the coupling to 
the vector fields produces a set of consistent flow equations without any additional constraint as opposed to what happened in section \ref{sec:flows} for flows without vectors.

By solving the flow equations on 
a background of the form $\mathbb{R}^{1,2}\times H^3$ with a vanishing 4-form field strength and $A(r)=1$ we know that $\mathrm{AdS}_{4}\times H^3$ solutions exist \cite{Acharya:2000mu}.
Thus it resoneable to wonder whether a solution of the same type with an $\mathrm{AdS}_{4}\times S^3$ background exists as a 
particular solution for \eqref{ansatz:AdS3vector}. However, one gets easily
convinced that such a solution cannot exist within the $\mathcal{N}=1$ truncation of the theory\footnote{Imposing $A(r)=1$, we found that the flow equations \eqref{flow:mkw21vector} and the equations
of motion \eqref{EOM_7D} are satisfied by a constant 3-form, by a linear dependence on $r$ of $U$ and by an imaginary constant value of $W$. }.
Moreover we observed that imposing $A(r)=1$ in \eqref{flow:mkw21vector} without any other specifications on the fields, the equations of motion do not admit any solutions. 

Thus we are forced to keep a non-trivial radial dependence for the gauge fields. In this case the flow equations \eqref{flow:mkw21vector} can be intregrated numerically. We are interested in those solutions that are
asymptotically locally $\mathrm{AdS}_7$, which means that we  first have to verify if there is a particular limit of the background in \eqref{ansatz:mkw21vector} reproducing $\mathrm{AdS}_7$ at the leading order in its asymptotic expansion.

In order to be able to perform numerical integration, we also need to make a choice of the value of the free parameter in the system. In particular, we impose for simplicity $g=1$, $h=\frac{1}{2\sqrt{2}}$ and $\kappa=1$ and we make the gauge choice $V(r)=0$.
Then, it is possible to verify that the following configuration
\be
\begin{split}
&U=\frac{r}{2\sqrt{2}}\ ,\quad W=\frac{r}{2\sqrt{2}}\ ,\quad X=1\ ,\quad \theta=0\  ,\quad Y=e^{U/2}\ ,\\
&k=0\ ,\quad \quad \quad l=0\ ,\quad \quad \quad A=1\ ,
\end{split}
\label{AdS7asymptotics}
\ee
solves \eqref{flow:mkw21vector} at the leading order when $r\rightarrow+\infty$. One can intregrate numerically \eqref{flow:mkw21vector}, by using the asymptotic behavior of the fields given in 
\eqref{AdS7asymptotics} as initial data. 

\begin{figure}[htbp]
\centering
\vspace{0.1cm}
\includegraphics[width=55mm]{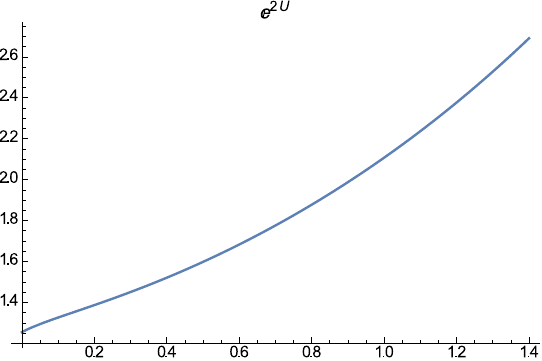}
\vspace{0.7cm}
\hspace{0.5cm}
\includegraphics[width=55mm]{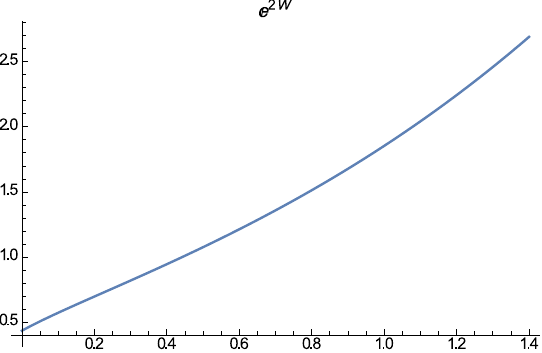}
\vspace{0.7cm}
\hspace{0.5cm}
\includegraphics[width=55mm]{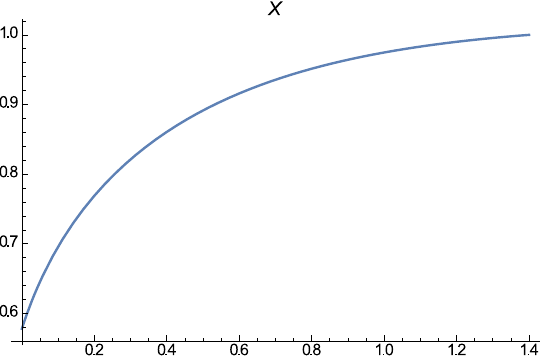}
\vspace{0.7cm}
\hspace{0.5cm}
\includegraphics[width=55mm]{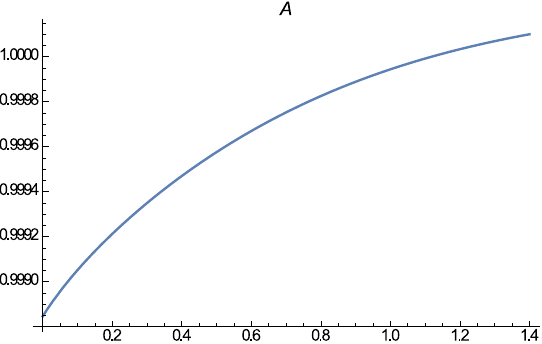}
\vspace{0.7cm}
\hspace{0.5cm}
\includegraphics[width=55mm]{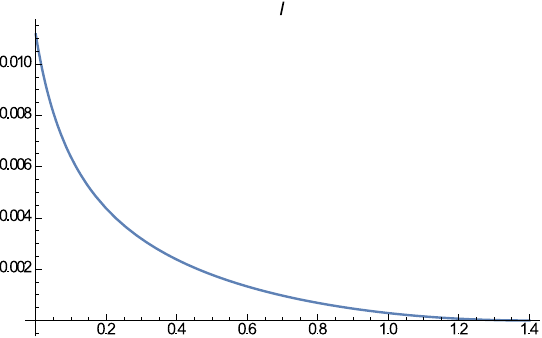}
\hspace{0.5cm}
\includegraphics[width=55mm]{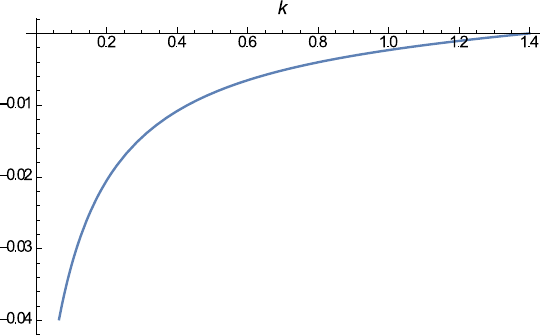}
\caption[Numerical flows for the background $\mathbb{R}^{1,2}\times S^3$. Plots from \cite{Dibitetto:2017tve}.]{\it Flows of $e^{2U(r)},\,e^{2W(r)},\,X(r),\,A(r),\,l(r),\,k(r)$ plotted in the interval $r\in (0,1.4]$ with $g=1$, $h=\frac{1}{2\sqrt{2}}$, $\kappa=1$ and $V(r)=0$. Plots from \cite{Dibitetto:2017tve}.}
\label{mk21numerical}
\end{figure}

By doing so, one obtains a profile for the fields that is singular in $r\rightarrow 0$ and locally $\mathrm{AdS}_7$ for $r\rightarrow +\infty$. The explicit radial profile of the fields for this solution
is plotted in figure~\ref{mk21numerical}.

\clearpage
\subsection{Vectors Coupled to the Background $\mathrm{AdS}_3\times S^3$ }
\label{AdS3vector}

Let us now consider the background \eqref{backgroundAdS3} coupled to the vectors as given in \eqref{ansatzvector}. The complete Ansatz is now given by
\be
\begin{array}{lcll}
\D s^2   & = & e^{2U(r)}\, \D s_{\mathrm{AdS}_3}^{2} \, + \, e^{2V(r)} \,  \D r^{2}  \, + \, e^{2W(r)}\,  \D s_{S^{3}}^{2} , \\[2mm]
X & = & X(r)  , \\[2mm]
B_{(3)} & = & k(r)\, \mrm{vol}_{\mathrm{AdS}_3} \, + \, l(r)\, \mrm{vol}_{S^{3}}  , \\[2mm]
A^i\,&=\,&\frac{A(r)}{2\,g}\,\epsilon^{i\,k\,l}\,\omega_{j\,kl}\,\D \,\theta^j\,,
\label{ansatz:AdS3vector}
\end{array}
\ee
where $\mathrm{AdS}_3$ and $S^3$ are respectively parametrized as in \eqref{AdS3} and \eqref{S3}. 

Given the Killing spinor $\zeta^a$ of the form \eqref{Kspinor} and satisfying \eqref{gamma3proj}, the set of the first-order flow equations describing the background \eqref{ansatz:AdS3vector} is
given in \eqref{flow:AdS3vector}.

The background \eqref{ansatz:AdS3vector} admits, among others, an $\mathrm{AdS}_3\times H^4$ solution. This is an example of $\mathrm{AdS}_{d<7}$ solution with a non-constant
profile for both the gauge fields and the 3-form \cite{Gauntlett:2000ng}. In particular the following expressions for the fields,

\be
\begin{split}
&e^{2U}=\frac{L^2}{7}\, ,\quad \quad \quad e^{2W}=\sinh\,(r)\, ,\quad \quad \quad e^{2V}=1\, ,\quad \quad \quad X=\left(\frac{7}{12}\right)^{1/5}\,,\\[1.5mm]
&k=-\frac{2^{4/5}\, 3^{2/5}\, L^3}{ 7^{19/10}}\, ,\quad \quad \quad l=\frac{3^{2/5}}{2^{1/5}\, 7^{9/10}}\, \left(9 \cosh\, (r)-\cosh \, (3r)+8\right)\,,\\[2mm]
& A=1+\cosh\,(r)\,, \quad \quad \quad \theta=\frac{\pi}{4}\,,\quad \quad \quad Y=e^{U/2}\,,
\end{split}
\label{AdS3H4}
\ee
provide a solution for both \eqref{flow:AdS3vector} and \eqref{EOM_7D} with $h=\frac{g}{2\sqrt{2}}$, $\kappa=2$ and $g=\frac{3^{1/5}\,7^{3/10}}{2^{1/10}}$.

\begin{figure}[htbp]
\centering
\vspace{0.1cm}
\includegraphics[width=55mm]{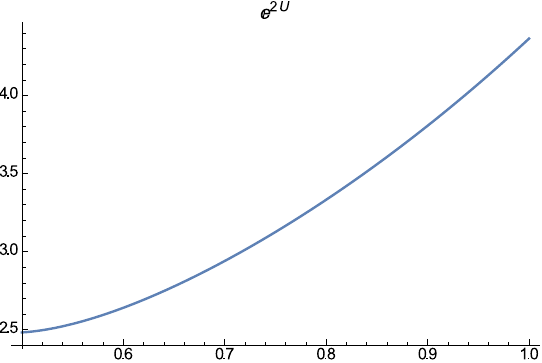}
\vspace{0.7cm}
\hspace{0.5cm}
\includegraphics[width=55mm]{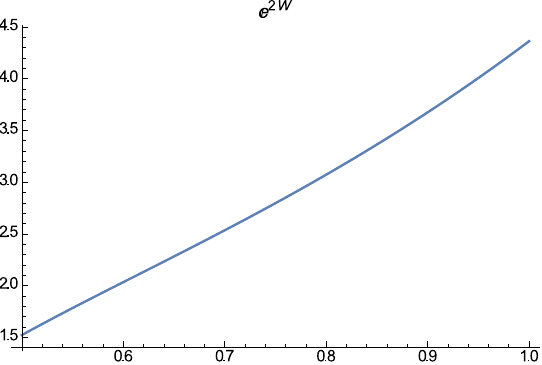}
\vspace{0.7cm}
\hspace{0.5cm}
\includegraphics[width=55mm]{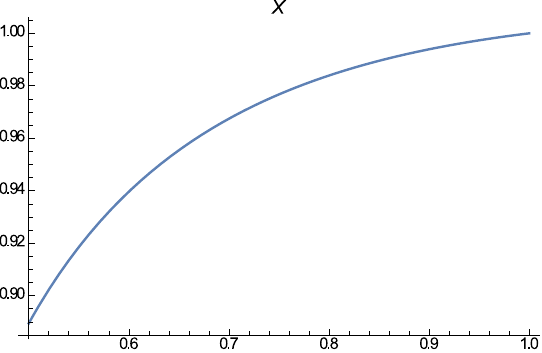}
\vspace{0.5cm}
\hspace{0.5cm}
\includegraphics[width=55mm]{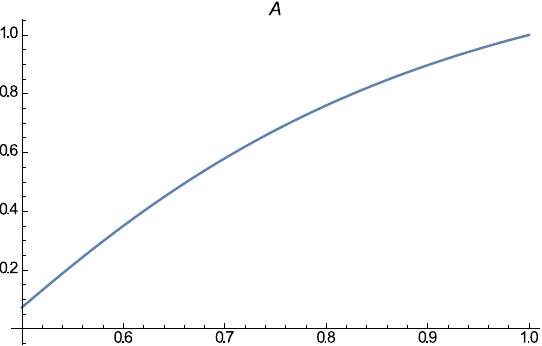}
\vspace{0.5cm}
\hspace{0.5cm}
\includegraphics[width=55mm]{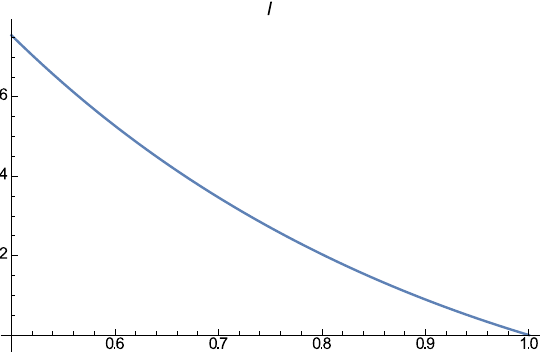}
\hspace{0.5cm}
\includegraphics[width=55mm]{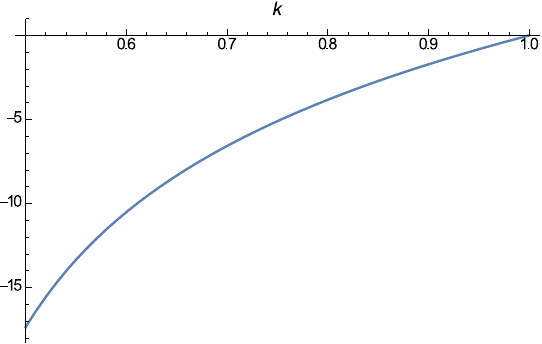}
\caption[Numerical flows for the background $\mathrm{AdS}_3\times S^3$. Plots from \cite{Dibitetto:2017tve}.]{\it Flows of $e^{2U(r)},\,e^{2W(r)},\,X(r),\,A(r),\,l(r),\,k(r)$ plotted in the interval $r\in (0.5,1]$ with $g=\frac{3^{1/5}\,7^{3/10}}{2^{1/10}}$, $h=\frac{g}{2\sqrt{2}}$, $\kappa=2$, $L=0.3$ and $V(r)=0$. Plots from \cite{Dibitetto:2017tve}.}
\label{AdS3numerical}
\end{figure}

As in the previous section, we integrate numerically the flow equations \eqref{flow:AdS3vector} by starting from the locally $\mathrm{AdS}_7$ asymptotics. Choosing the same values for $g,\,h$ and $\kappa$ characterizing the 
solution \eqref{AdS3H4} and $V(r)=0$, it is possibile to show that the locally $\mathrm{AdS}_7$ configuration
\be
\begin{split}
&U=\frac{3^{1/5}\,7^{3/10}}{2^{8/5}}\,r\ ,\quad W=\frac{3^{1/5}\,7^{3/10}}{2^{8/5}}\,r\ ,\quad X=1\ ,\quad \theta=0\, ,\quad Y=e^{U/2}\ ,\\
&k=0\ ,\quad l=0\ ,\quad A=1\ ,
\end{split}
\label{AdS7asymptoticsAdS3}
\ee
solves \eqref{flow:AdS3vector} in the limit $r \rightarrow +\infty$.
We can intregrate numerically the flow equations in \eqref{flow:AdS3vector} starting from \eqref{AdS7asymptoticsAdS3}. In this way we obtained
a flow that shows a singular behavior as $r\rightarrow 0$, while clearly keeping its locally $\mathrm{AdS}_7$ structure in its asymptotic region.
The explicit profile of the seven-dimensional fields is shown in figure~\ref{AdS3numerical}.

 It may be worth noticing that the above flow does not describe $\mathrm{AdS}_3\times H^4$ in the limit where $r\rightarrow 0$, but this should not be a surprise since this solution describes
an $\mathrm{AdS}_3$ slicing of the seven-dimensional background, where the radial coordinate of the seven-dimensional background does not coincide with the radial coordinate of $\mathrm{AdS}_3$.
It is in fact this latter one which is expected to parametrize the flow where $\mathrm{AdS}_3$ emerges in the IR limit.
As for the complete flow realizing the full interpolation
between $\mathrm{AdS}_7$ and $\mathrm{AdS}_3\times H^4$, it should be represented by a more general $\mrm{BPS}$ background describing a $\mathbb{R}^{1,1}$ slicing of the seven-dimensional background dependent on both coordinates.

\clearpage

\section{BPS Objects in $d=7$ and their M-Theory Origin}
\label{sec:Mtheory_lifts}

Given the solutions derived in section~\ref{sec:flows}, we will now try to give them an intepretation in terms of bound states in $\mathrm{M}$-theory. 
It is well known that the equations of motion of 
the minimal gauged supergravity in $d=7$ written in \eqref{EOM_7D} are obtained by reducing eleven-dimensional supergravity on $S^4$ \cite{Lu:1999bc}. 
This consistent truncation produces the scalar potential \eqref{potential} depending on the parameters $g$ and $h$, where $h$ is related to the eleven-dimensional $F_4$ flux and $g$ is the gauge parameters of the $\mathrm{\mathrm{SU(2)}}$ vectors describing 
the squashing of the 3-sphere with respect to which the $S^4$ is written as an $S^3$-fibration over a segment. 

This can be explicitly checked by means of a simple group-theoretical argument. To this end, we decompose the embedding tensor piece of the maximal theory with eleven-dimensional origin from $S^{4}$, i.e. the
$\textbf{15}$ of $\textrm{SL}(5,\mathbb{R})$, and identify the $\mrm{SO(3)}$-singlets corresponding to $h$ and $g$. This procedure yields
\be
\begin{array}{ccccccc}
\textrm{SL}(5,\mathbb{R}) & \supset & \mathbb{R}^{+}_{1}\times\textrm{SL}(4,\mathbb{R}) & \supset & \mathbb{R}^{+}_{1}\times\mathbb{R}^{+}_{2}\times\textrm{SL}(3,\mathbb{R}) & \supset & \mathbb{R}^{+}_{1}\times\mathbb{R}^{+}_{2}\times\mrm{SO(3)} \\[3mm]
\textbf{15} & \rightarrow & \textbf{1}_{(+4)}\,\oplus\,\textbf{10}_{(-1)} & \rightarrow & \textbf{1}_{(+4;0)}\,\oplus\,\textbf{6}_{(-1;-2)} & \rightarrow & 
\underbrace{\textbf{1}_{(+4;0)}}_{h}\,\oplus\,\underbrace{\textbf{1}_{(-1;-2)}}_{g}
\end{array}\notag
\ee
Now one gets easily convinced that the eleven-dimensional $F_4$ flux is to be identified with the embedding tensor piece which was already a singlet of $\textrm{SL}(4,\mathbb{R})$, i.e. $h$, while the $\mrm{SU(2)}$
curvature is only expected to be a singlet of $\mrm{SO(3)}$, and hence is naturally identified with $g$.
 
Since the above gauging parameters are related to the 
flux configurations in the higher-dimensional theory, other reductions can be in principle considered and the simplest of those is certainly the one on the torus $T^4$, yielding the potential \eqref{potential} with $g=0$.

 If one considers the flows obtained as solutions in $\mathcal{N}=1$ gauged supergravity in $d=7$, the existence of consistent truncations implies that
 the physics of some solitonic objects in $\mathrm{M}$-theory is captured by the solutions in seven-dimensional supergravity in the low-energy limit. 

The simplest example of this is given by the DW solutions in \eqref{DWflow} that describe three possible configurations in $\mathrm{M}$-theory, depending on how the gauging in the seven-dimensional supergravity is further specified. 
All of them consist of M5-branes reduced in different ways on their transverse space. In particular, one may easily see that \cite{Bergshoeff:2004nq}:
\begin{itemize}
 \item  $h\neq0$ and $g=0$: The DW \eqref{DWflow} describes an $\mathrm{M5}$-brane with four of its transverse coordinates reduced on a $T^4$.
  \item $h=0$ and $g\neq0$: The DW \eqref{DWflow} describes an $\mrm{NS}5$-brane in $\mathrm{\mrm{IIA}}$ string theory reduced on an $S^3$
or an $\mathrm{M5}$-brane with four of its transverse coordinates reduced on $S^1\times S^3$.
   \item  $h\neq0$ and $g\neq 0$: The DW \eqref{DWflow} describes an $\mathrm{M5}$-brane with four of its transverse coordinates reduced on an $S^4$.
 \end{itemize}

As a general fact, not all of the truncations of higher-dimensional theories admit solutions with an $\mathrm{AdS}_7$ asymptotic behavior. 
In fact, since only the complete form of the potential \eqref{potential} admits $\mathrm{AdS}_7$ critical points,
 only the last DW solution (with $h\neq0$ and $g\neq 0$) will asymptote to the $\mathrm{AdS}_7$ that is associated with the $\mathrm{AdS}_7\times S^4$ Freund-Rubin vacuum. 
Such a vacuum can be indeed obtained by taking the near-horizon limit of the $\mathrm{M5}$-brane geometry.

\subsection{Dyonic Solutions and the M2-M5 Bound State}

Moving to the flows involving a dyonic profile for the 3-form potential, let us first consider the solution presented in \eqref{ch_DW_noAdS}. In this case the potential driving the solution is given by
\be
V_g(X) \ = \ 2h^{2} \, X^{-8}\ ,
\label{runawaypotential}
\ee
which, due to its run-away behavior, has no critical points. As we said, the truncation producing a potential with $g= 0$ is obtained by considering the low-energy limit of $\mathrm{M}$-theory on a 4-torus 
$T^4$ with non-vanishing 4-form flux. 
In this section we want to show that the flow in \eqref{ch_DW_noAdS} is the low-energy description of a supersymmetric 
$\mrm{M2}$-$\mrm{M5}$ bound state discovered in \cite{Izquierdo:1995ms} by uplifting to eleven dimensions a dyonic membrane solution obtained in 
$\mathcal{N}=2$, $d=8$ supergravity.

The corresponding eleven-dimensional background reads \cite{Izquierdo:1995ms}
\be
\begin{split}
 \D s^2_{11}&\,=\,H^{-2/3}\,\left(\sin^2 \xi +H \, \cos^2\xi \right)^{1/3}\,\D s^2_{\mathbb{R}^{1,2}}+
 H^{1/3}\,\left(\sin^2 \xi +H \, \cos^2\xi \right)^{1/3}\,ds^2_{\mathbb{R}^5}\\
 &\ +\,H^{1/3}\,\left(\sin^2 \xi +H \, \cos^2\xi \right)^{-2/3}\,ds^2_{\mathbb{R}^3}\ ,
\end{split}
\label{lambert:metric}
\ee
where $H$ is a harmonic function on $\mathbb{R}^5$ and $\xi$ is a constant angle. The 4-form field strength is given by
\be
F_4\,=\,\frac12 \,\cos \xi \, \star_5 \D\,H+\frac12\,\sin \xi \, \D \,H^{-1} \wedge \, \mathrm{vol}_{\mathbb{R}^{1,2}}-\frac{3\,\sin(2\,\xi)}{2\,\left(\sin^2 \xi +H \, \cos^2\xi \right)^{2}}\,\mathrm{vol}_{\,\mathbb{R}^3}\,\wedge\, \D H\ ,
\label{lambert:F11}
\ee
where $\mathrm{vol}_{\mathbb{R}^{1,2}}$ and $\mathrm{vol}_{\mathbb{R}^3}$ are respectively the volume of the three-dimensional Minkowski space and the volume of $\mathbb{R}^3$.

Since $H$ is defined on $\mathbb{R}^5$, the solution may be interpreted as the effective 
description of an $\mathrm{M2}$-brane completely smeared over the worldvolume of an $\mathrm{M5}$ or, equivalently, of an $\mathrm{M5}$-brane
carrying a dissolved $\mathrm{M2}$ charge. This configuration preserves $16$ supercharges. Note that it is not the mere
superposition between the $\mathrm{M2}$ and the $\mathrm{M5}$-brane and this is due to the presence
of the third term of \eqref{lambert:F11} accounting for M2-M5 interactions.
There are two particular values for the parameter $\xi$:
\begin{itemize}
 \item $\cos \xi=0$: Purely electric case corresponding to a pure (smeared) $\mathrm{M2}$-brane.
  \item $\sin \xi=0$: Purely magnetic case corresponding to a pure $\mathrm{M5}$-brane.
\end{itemize}
Because of the intrinsic structure of bound state of the solution \eqref{lambert:metric}, its brane
interpretation for general values of $\xi$ remains somewhat obscure\footnote{This issue was originally discussed in \cite{Niarchos:2012pn,Niarchos:2013ia}, where this eleven-dimensional solution at generic angles $\xi$ 
was given an interpretation in terms of an M2-M5 funnel geometry.}, 
but it can be shown that it has a smooth horizon for any $\xi\neq\frac{\pi}{2}$, the corresponding near-horizon geometry
being $\mathrm{AdS}_7\times S^4$. 

From \cite{Izquierdo:1995ms} we know that, by compactifying\footnote{Giving a periodic identification on the coordinates of $\mathbb{R}^3$.} \eqref{lambert:metric} on a $T^3$, one obtains a flow in $\mathcal{N}=2$, $d=8$ supergravity featured by a dyonic
3-form and an axio-dilaton. The eight-dimensional flow trasforms under $\textrm{SL}(2,\mathbb{R})$ and this means
that one can always find a transformation such that the eight-dimensional 3-form is completely electric.

Let us now reduce the \eqref{lambert:metric} and \eqref{lambert:F11} on a $T^4$.
In order for this procedure to be consistent, a smearing of the charge distrubution is required along all the $T^4$ coordinates. To implement this, choose the $\mathbb{R}^5$ coordinates such that
\be
\D s^2_{\mathbb{R}^5}\,=\,\D z^2+\D s^2_{T^4}\,,\qquad \text{with} \qquad H=1+\alpha \, z\ ,
\ee
with $\alpha$ real parameter. The form of \eqref{lambert:F11} suggests a dyonic profile for the corresponding seven-dimensional 3-form, but in this case the odd-dimensional self-duality conditions \eqref{SD_cond} spoil the possibility of rotating the dyonic 3-form
into a completely electric one as it was done in the eight-dimensional case.

The reduction on $T^4$ of eleven-dimensional supergravity can be performed directly at the level of the 
eleven-dimensional action
 with the following reduction Ansatz on the metric,
\be
\D s_{11}^2\,=\,X^{-4/3}\, \D s^2_{7}\,+\, X^{5/3} \,\D s^2_{T^4}\ ,
\label{reductedmetric}
\ee
and including a 4-form field strength wrapping the $T^4$,
\be
F_4=q\,\mathrm{vol}_{\,T^4}\ ,
\label{G4}
\ee
where $q$ is the flux associated to the eleven-dimensional 3-form and $X$ is
the scalar field belonging to the supergravity multiplet of the seven-dimensional minimal supergravity associated to the volume modulus of $T^4$. Imposing\footnote{We imposed the relation
 $\frac{\mathrm{vol}_{T^4}}{2\,\kappa_{11}^2}=\frac{1}{2\,\kappa_7^2}$ between the gravitational couplings.}  this reduction Ansatz we obtain the action \eqref{Lagrangian_seven-dimensional} with $g=0$, $A^i=0$ and
a scalar potential given by \eqref{runawaypotential}.

Using the reduction Ansatz \eqref{reductedmetric} and \eqref{G4}, we want to compare \eqref{lambert:metric} and \eqref{lambert:F11} with \eqref{ch_DW_noAdS}. We start by extracting the seven-dimensional flow from 
\eqref{lambert:metric} and \eqref{lambert:F11}.

Let us begin with
the first term of \eqref{lambert:F11} placed on $T^4$, i.e. $\star_5 \D\,H\,=\,\alpha\,\mathrm{vol}_{T^4}$.
Comparing it with \eqref{G4}, we immediately obtain $q=\frac{\cos \xi}{2}\,\alpha$.
 By a comparison with \eqref{reductedmetric}, it is possibile to extract a seven-dimensional metric and the
expression for $X$ from \eqref{lambert:metric}, one obtains
\be
\begin{split}
\D s_7^2 &\,=\,  H^{-2/5}\,\left(\sin^2 \xi +H \, \cos^2\xi \right)^{3/5}\,\D s^2_{\mathbb{R}^{1,2}}+
 H^{3/5}\,\left(\sin^2 \xi +H \, \cos^2\xi \right)^{3/5}\,\D z^2\\
 &\ +\,H^{3/5}\,\left(\sin^2 \xi +H \, \cos^2\xi \right)^{-2/5}\,\D s^2_{\mathbb{R}^3}\ ,\\
 X&\,=\, H^{1/5}\,\left(\sin^2 \xi +H \, \cos^2\xi \right)^{1/5}\ .
 \end{split}
 \label{lambertseven-dimensional:metric}
\ee
The seven-dimensional 4-form field strength is simply given by the second and the third terms of \eqref{lambert:F11},
in particular one has
\be
\mathcal{F}_{(4)}\,=\,\frac{\sin\xi}{2}\,\D \left( H^{-1}\,\mathrm{vol}_{\mathbb{R}^{1,2}} \right)+\frac{3\,\sin\xi}{\cos \xi}
\,\D \left(\left(\sin^2 \xi +H \, \cos^2\xi \right)^{-1}\,\mathrm{vol}_{\mathbb{R}^3} \right)\ .
 \label{lambertseven-dimensional:4form}
\ee
We can now consider the flow \eqref{ch_DW_noAdS} and compare it with \eqref{lambertseven-dimensional:metric} and \eqref{lambertseven-dimensional:4form}.
We firstly rewrite \eqref{ch_DW_noAdS} with a more general dependence on the integration constants that will be
fixed by the matching,
\be
\label{ch_DW_noAdS_constants}
\begin{array}{lclclc}
e^{2U} \, = 2^{1/5}\,e^{2\alpha_1} \sinh(2hr)^{1/5}\,\coth(hr)\, , & & & & e^{V} \, = 2^{8/5}\,\alpha_3^4 \sinh(2hr)^{8/5}  , \\[2mm]
e^{2W} \, = 2^{1/5}\,e^{2\alpha_2}\, \sinh(2hr)^{1/5}\,\tanh(hr)\,  ,& &  & & X \, = 2^{2/5}\,\alpha_3 \sinh(2hr)^{2/5}  , \\[2mm]
k \, = \,\frac{e^{3\alpha_1}}{2\,\alpha_3^2} \,\sinh(hr)^{-2}\,  ,& &  & & l \, = -\frac{e^{3\alpha_2}}{2\,\alpha_3^2}\, \cosh(hr)^{-2}  . \\[2mm]
\end{array}
\ee
One finds that for the following values of the constants $\alpha_1,\,\alpha_2,\,\alpha_3$ and of $H$,
\be
\begin{split}
 &e^{\alpha_1}\,=\,2^{-1/5}\cos(\xi)^{2/5}\sin(\xi)^{1/5}\ ,\\
 &e^{\alpha_2}\,=\,-2^{-1/5}\cos(\xi)^{-3/5}\sin(\xi)^{1/5}\ ,\\
 &\alpha_3^2\,=\,2^{-8/5}\cos(\xi)^{-4/5}\sin(\xi)^{8/5}\ ,\\
 &H\,=\,2\,\sin\xi \, \alpha_3^2 \,e^{-3\alpha_1}\,\sinh(hr)^{2}\,,
 \label{constants}
\end{split}
\ee
the functions $U,\,W,\,X,\,l,\,k$ describing the flow \eqref{ch_DW_noAdS_constants} match exactly with \eqref{lambertseven-dimensional:metric} and \eqref{lambertseven-dimensional:4form}.
We can finally derive the relation between the seven-dimensonal radial coordinate $r$ and
the radial coordinate of $\mathrm{M}$-theory $z$ by comparing the radial parts of the seven-dimensional metrics,
\be
e^V\,\D r\overset{!}{=}H^{3/10}\,\left(\sin^2 \xi +H \, \cos^2\xi \right)^{3/10}\,\D z\ .
\label{eqzr}
\ee
Using \eqref{constants} and integrating \eqref{eqzr} we obtain
\be
z\,=\,\frac{\sin^2\xi}{4\,h\,\cos\xi}\,\cosh(2hr)\,+\,z_0\ .
\label{Mradius}
\ee
The constant $z_0$ can be determined by comparing $H=1+\alpha\,z$ with the expression of $H$ written in \eqref{constants} obtaining
\be
\alpha=\frac{2\,h}{\cos \xi}\,, \qquad z_0=-\frac{1+\cos^2\xi}{4h\cos\xi}\ .
\ee
Recalling that $q=\frac{\cos \xi}{2}\,\alpha$, one finds
\be
q\,=\,h\ .
\ee
We conclude that the flow \eqref{ch_DW_noAdS_constants} obtained in minimal supergravity in $d=7$ and described
by a dyonic 3-form and by the potential \eqref{runawaypotential} describes the low-energy limit of 
the bound state $\mathrm{M2}$-$\mathrm{M5}$ reduced on $T^4$. In particular the St{\"u}ckelberg mass $h$ is identified with the flux associated to 
the eleven-dimensional 4-form field strength wrapping the 4-torus.

\subsection{Asymptotically $\mathrm{AdS}_7$ Flows and $S^4$ Reductions}

Let us now move to considering the asymptotically $\mathrm{AdS}_7$ flows derived in section \ref{sec:flows}
and their $\mathrm{M}$-theory picture. The main difference with respect to the case
of \eqref{ch_DW_noAdS} is the $\mathrm{AdS}_7$ asymptotic behavior that extremizes the potential
\be
V_g(X) \ = \ 2h^{2} \, X^{-8} \, - \, 4\sqrt{2}\,gh\, X^{-3} \, - \, 2g^{2} \,X^{2} \  .
\label{pot}
\ee
The truncation of eleven-dimensional supergravity describing \eqref{pot} is the one on a squashed $S^4$ \cite{Lu:1999bc} and 
it is defined by the complete $\mathcal{N}=1$, $d=7$ supergravity multiplet $(g_{\mu\nu},\,X,\,B_{(3)},\,A^i)$ whose
equations of motion, supplemented with the odd-dimensional self-duality conditions are written in \eqref{EOM_7D} and \eqref{SD_cond}. 

The metric of the internal $S^4$ is given by a foliation
of 3-spheres and its deformations are parametrized by the seven-dimensional scalar $X$. The squashing
leaves the 3-sphere foliations preserved. Thus,
introducing the basis of left-invariant forms $\eta^i$ on the 3-sphere, the seven-dimensional gauge fields $A^i$
describe the $\mathrm{\mathrm{\mathrm{SU(2)}}}$ bundle over the $S^3$ and the metric of the internal space is given by
\be
   \D s_4^2 \,=\,X^3\,\Delta \,  \D \psi ^2\,+\,\frac{X^{-1}}{4}\,\cos^2\psi \sum_{i=1}^3\,(\eta^i-g\,A^i)^{2} \ , \quad \text{with}\quad 
 \Delta\,= \,X^{-4}\sin^2\psi+X\,\cos^2\psi\ .
 \label{S^4squashed}
\ee
The truncation holds at the level of the equations of
motion and of the odd-dimensional self-duality conditions \eqref{EOM_7D} and \eqref{SD_cond}, and it is specified by the following eleven-dimensional Ansatz,
\be
\begin{split}
 \D s_{11}^2&\,=\,\Delta^{1/3}\, \D s^2_7+2\,g^{-2}\, \Delta^{-2/3}\,  \D s_4^2\ ,\\
 A_{(3)}&\,=\,\sin\psi \, B_{(3)}+\frac{g^{-3}}{2\sqrt{2}}\,\left(2\,\sin\psi+\sin\psi\,\cos^2\psi\,\Delta^{-1}\,X^{-4} \right)\,\epsilon_{(3)}\\
 &-\frac{g^{-2}}{\sqrt{2}}\,\sin\psi \,F^i_{(2)}\wedge h^i-\frac{g^{-1}}{\sqrt{2}}\,\sin\psi \, \omega_{(3)}\ ,
 \label{S^4reduction}
\end{split}
\ee
where $h^i\,=\,\eta^i-g\,A^i$, $\epsilon_{(3)}\,=\,h^1\wedge h^2 \wedge h^3$ and 
$\omega_{(3)}\,=\,A^i\wedge \mathcal{F}_{(2)}^i-\frac16\,g\,\epsilon_{ijk}\,A^i\wedge A^j\wedge A^k$, and the fields
$X,\,B_{(3)}$ and $A^i$ are functions of the seven-dimensional background.

The flows with an $\mathrm{AdS}_7$ asymptotic behavior obtained in section \ref{sec:flows} can be organized in the following two groups:
\begin{itemize}
 \item $\mathbb{R}^{1,2}\times S^3$ backgrounds\ .
 \item $\mathrm{AdS}_3\times S^3$ backgrounds\ .
\end{itemize}
Furthermore in both cases we presented flows with and without the coupling to $\mathrm{\mathrm{\mathrm{SU(2)}}}$ vectors. By means of the uplift formula in \eqref{S^4reduction}, it is possibile to lift the seven-dimensional flows
given in \eqref{ch_DW_AdS}, \eqref{AdS3_DW_AdS_U=V}, \eqref{AdS3_DW_AdS} to eleven dimensions, while the existence of numerical flows obtained by solving \eqref{flow:mkw21vector} and \eqref{flow:AdS3vector} ensures the existence of 
corresponding eleven-dimensional configurations.

We know that all the $\mathrm{AdS}_7$
flows of section \ref{sec:flows} are described by a dyonic profile for the 3-form that cannot be recast into a purely electric form because of the odd-dimensional self-duality conditions in \eqref{SD_cond}. 
Due to this argument, we are then again forced into considering
 $\mathrm{M2}$-$\mathrm{M5}$ bound states described in eleven-dimensional supergravity by the solution \eqref{lambert:metric} and \eqref{lambert:F11}. 
This solution has an $\mathrm{AdS}_7\times S^4$ near-horizon geometry compatible
with the asymptotics of our seven-dimensional flows and a dyonic profile of the 3-form once compactified on $S^4$, but the issue here is to find a suitable coordinate system for the uplifted solutions such that a clean brane picture arises.
This is particularly manifest for the flow \eqref{ch_DW_AdS} 
coming from the $\mathbb{R}^{1,2}\times S^3$ where such diffeomorphisms on the uplifted flow should relate the coordinates $(r,\,\psi)$ with the radial coordinate of $\mathrm{M}$-theory. 

Giving an interpretation of the warped solutions
\eqref{AdS3_DW_AdS_U=V} and \eqref{AdS3_DW_AdS} is more difficult since the presence of the $\mathrm{AdS}_3$ slicing implies a modification of the brane picture. For example, the semi-localized intersection of a pp-wave with an $\mathrm{M5}$-brane 
would modify the geometry of the worldvolume of the $\mathrm{M5}$ producing $\mathrm{AdS}_3$ in the near-horizon limit \cite{Cvetic:2000cj}. 
This may in principle hold true even when constructing an intersection of the $\mathrm{M2}$-$\mathrm{M5}$ bound state with a pp-wave, but it is in general difficult to specify the concrete momentum charge
distribution realizing it.

Finally the flows involving vectors should describe the wrapping of the worldvolume of the $\mathrm{M2}$-$\mathrm{M5}$ bound state on an $S^3$. However, since in this case we are even lacking the analytic form of the flows, 
it becomes technically impossible to search for the correct coordinate system which could verify our expectations. 
On the hand of course, the presence of the twisting condition \eqref{twisting} guaranteeing some residual supersymmetry suggests some spontaneous brane wrapping mechanism. 

\section{Charged Domain Wall and Holographic Conformal Defects }

In this section we are going to construct the holographic interpretation of the charged domain wall given in \eqref{AdS3_DW_AdS_U=V} as the supergravity configuration defining a conformal surface defect within the $\ma N=(1, 0)$ $\mrm{SCFT}_6$ dual to the $\mrm{AdS}_7$ asymptotics. The conformal defect will be described in terms of a $\ma N=(4, 0)$ $\mrm{SCFT}_2$ dual to the $\mrm{AdS}_3$ included in the seven-dimensional background and arising when the radial flow of the domain wall moves from the $\mrm{AdS}_7$ asymptotics to the $\mrm{IR}$ singularity of the solution.

As we mentioned at the beginning of this chapter, this is a complementary interpretation of the $\mrm{RG}$ flow across dimensions between different vauca and consists in viewing the lower-dimensional $\mrm{SCFT}_2$ as a theory inducing a position-dependent coupling in the ``mother" $\mrm{SCFT}_6$ and living on a surface within the six-dimensional Minkowski space.
The typical signature of the breaking of higher-dimensional conformal symmetry induced by the defect is the presence of non-vanishing one-point correlators as well as a displacement operator associated
with a non-conserved energy-momentum tensor. Starting from the seminal work of \cite{Karch:2000gx}, many stringy realizations of defect $\mrm{CFT}$'s have been given in the literature. All of them rely on 
the study of boundary conditions of branes ending on other branes (see for example the case of wrapped $\mrm{D}5$ branes ending on $\mrm{D}3$ branes, describing codimension 1 defects inside $\mathcal{N}=4$ $\mrm{SYM}$ \cite{Gutperle:2012hy},
or  $\mrm{M}2$- ending on $\mrm{M}5$ branes describing surface defects in the $\ma N=(2,0)$ $\mrm{SCFT}_2$ \cite{Lunin:2007ab}, or the more recent constructions by \cite{Maruyoshi:2016caf} of surface defects in class-$\mathcal{S}$ theories).
From the viewpoint of the supergravity dual description, such physical situations are described by asymptotically $\mrm{AdS}_{d+1}$ flows involving a special slicing where the otherwise-flat slices are replaced by a lower-dimensional $\mrm{AdS}_{p+2}$ space. 

The case of our charged domain wall is a perfect example of this particular supergravity configuration.
Hence, in this section, we are going to propose a construction of conformal surface defects in the $\ma N=(1,0)$ theory arising from $\mrm{NS}5$-$\mrm{D}6$-$\mrm{D}8$ systems in massive type $\mrm{IIA}$ string theory \cite{Hanany:1997gh}. In particular our description will be given in terms of $\mrm{D}2$- and wrapped $\mrm{D}4$ branes ending on the above brane intersection. 
In other words, we will give the ten-dimensional lift of the domain wall \eqref{AdS3_DW_AdS_U=V} by explicitly constructing the brane intersection $\mrm{D}2$-$\mrm{D}4$-$\mrm{NS}5$-$\mrm{D}6$-$\mrm{D}8$ in massive type $\mrm{IIA}$ supergravity containing both $\mrm{AdS}_{7}$ and $\mrm{AdS}_{3}$ in different limits.
Then we will conclude by showing how a $\mathcal{N}=(4, 0)$ $\mrm{SCFT}_2$ defect theory comes out in the low-energy description of this bound state by sketching a computation of its one-point functions.

\subsection{The $\ma N=(1,0)$ $\mrm{SCFT}_6$ and $\mrm{AdS}_7/\mrm{CFT}_6$ Correspondence}
\label{ads6/cft7}

Within the plethora of $\mrm{SCFT}$ constructions obtained by taking the decoupling limit of string and $\mrm{M}$-theory the most known example of $\mrm{SCFT}_6$ is the $\ma N=(2,0)$ theory living on the worldvolume of a stack of
coincident M5-branes \cite{Strominger:1995ac}. In this context many examples of $\mrm{RG}$ flows across dimensions to lower-dimensional superconformal field theories have been studied. 
As an example we cite the spontaneous wrapping of $\mrm{M}5$ branes describing flows from the $\ma N=(2,0)$ $\mrm{UV}$ theory to $d=4, 3, 2$ $\mrm{IR}$ conformal fixed points \cite{Gauntlett:2006ux,Bah:2012dg,Benini:2013cda}.
Some more recent and exotic examples of such flows involving brane wrapping in massive type $\mrm{IIA}$ on punctured Riemann surfaces may be found in \cite{Bah:2017wxp}.

More in general, the range of all possible maximally supersymmetric conformal theories is exhausted by the so-called ADE classification, which can be 
accessed through geometric engineering in type $\mrm{IIB}$ string theory \cite{Witten:1995zh}. The recent interesting proposals of \cite{Heckman:2013pva,DelZotto:2014hpa,Heckman:2015bfa} represent substancial progress towards a complete classification. 

When moving to the understanding of the dynamical process through which fixed points are reached, the issue that becomes of utmost importance is that of relating $\mrm{RG}$ flows to brane movements in the underlying
brane picture of a given field theory. At the level of the gravity dual, such a process is tipically captured by domain walls. This fact is sometimes referred to as
the $\mrm{DW/QFT}$ correspondence \cite{Boonstra:1998mp}. In this holographic description of $\mrm{RG}$ flows, the scalar fields which assume a non-trivial profile are identified with relevant operators realizing a 
marginal deformation of the original $\mrm{CFT}$, thus triggering the flow. The space of all exactly marginal deformations is called the {\itshape conformal manifold} and in the case of minimal supersymmetry in $d=6$, the conformal manifold has been recently proven to be empty \cite{Cordova:2015fha,Louis:2015mka}.

Moreover, when decreasing the amount of supersymmetry down to $\ma N=(1,0)$, there appears to be a much richer structure.
The constructions yielding $\ma N=(1,0)$ theories which are relevant for our purposes were carried out in the context of massive type $\mrm{IIA}$ by employing $\mrm{NS}5$-$\mrm{D}6$-$\mrm{D}8$ brane systems 
\cite{Hanany:1997gh,Brunner:1997gf,Brunner:1997gk}. According to the analysis performed by the works \cite{Apruzzi:2013yva,Gaiotto:2014lca,Apruzzi:2014qva,Apruzzi:2015zna}, 
all of the field theory models obtained from the above brane intersection admit a holographic description in terms of supersymmetric $\mrm{AdS}_{7}\times \tilde{S}^{3}$ solutions of massive type $\mrm{IIA}$ 
supergravity. In particular, in the above references, firstly an exhaustive classification of all supersymmetric $\mrm{AdS}_{7}$ solutions was developed and secondly their holographic interpretation was proposed. 
 
Six-dimensional $\mathcal{N}=(1,0)$ $\mrm{QFT}$'s enjoy eight real chiral supercharges transforming as a doublet of the R-symmetry group $\textrm{SU}(2)_{R}$. The standard vector multiplets can be coupled to extra
matter such as hypermultiplets and the more exotic {\itshape tensor multiplets}. The bosonic field content of these latter ones comprizes a real scalar $\phi$ and a self-dual 2-form field $b_{(2)}$.
The special branch of moduli space called the {\itshape tensor branch} is precisely parametrized by the vacuum expectation value of $\phi$.
The bosonic Lagrangian describing the coupling between the vector multiplet's and one abelian tensor multiplet is sketchily given by 
\be
\label{Lagrangian_$d=6$}
\mathcal{L}_{6d} \, = \, \phi\,\textrm{Tr}\left(\left|F_{(2)}\right|^{2}\right)\,+\,\left(\partial\phi\right)^{2}\,+\,\left|db_{(2)}\right|^{2}\,+\,
\star_6\left(b_{(2)}\,\wedge\,F_{(2)}\,\wedge\,F_{(2)}\right)\ ,
\ee
where the vacuum expectation value $\langle\phi\rangle$ parametrizing the tensor branch may be seen as the effective gauge coupling $g_{\textrm{YM}}^{-2}$.
From this perspective, it appears clear that the singular point $\langle\phi\rangle\,=\,0$ is related to physics at strong coupling, even though taking such a limit cannot be done naively since it involves
some subtleties. As in other examples of stringy constructions of supersymmetric $\mrm{QFT}$'s, this strongly coupled regime corresponds to a fixed point of an $\mrm{RG}$ flow described by an 
interacting\footnote{Though several constructions in string and M-theory yielding $\ma N=(1,0)$ $\mrm{SCFT}_6$'s are available in the literature, there is still no exhaustive classification. However, a significant step in this direction has
been recently taken in \cite{Chang:2017xmr} by applying conformal bootstrap techniques.} $\mrm{SCFT}$ \cite{Seiberg:1996vs}.

\begin{table}[h!]
\renewcommand{\arraystretch}{1}
\begin{center}
\scalebox{1}[1]{
\begin{tabular}{c||c c c c c c|c||c c c}
branes & $t$ & $y^{1}$ & $y^{2}$ & $y^{3}$ & $y^{4}$ & $y^{5}$ & $z$ & $r$ & $\theta^{1}$ & $\theta^{2}$ \\
\hline \hline
$\mrm{NS}5$ & $\times$ & $\times$ & $\times$ & $\times$ & $\times$ & $\times$ & $-$ & $-$ & $-$ & $-$ \\
$\mrm{D}6$ & $\times$ & $\times$ & $\times$ & $\times$ & $\times$ & $\times$ & $\times$ & $-$ & $-$ & $-$ \\
$\mrm{D}8$ & $\times$ & $\times$ & $\times$ & $\times$ & $\times$ & $\times$ & $-$ & $\times$ & $\times$ & $\times$ \\
\end{tabular}
}
\end{center}
\caption[The brane picture of the system NS5-D6-D8 underlying the $\ma N=(1,0)$ $\mrm{SCFT}_6$. Table from \cite{Dibitetto:2017klx}.]{{\it The brane picture underlying the $\ma N=(1,0)$ $\mrm{SCFT}_6$ described by a $\mrm{NS}5$-$\mrm{D}6$-$\mrm{D}8$ system. The above system is $\mrm{BPS}/4$.
Note that the radial coordinate realizing the dual $\mrm{AdS}_{7}$ geometry turns out to be a combination of $z$ and $r$. Table from \cite{Dibitetto:2017klx}.}} \label{Table:GT}
\end{table}

Let us move further to the details of the stringy construction of $\ma N=(1,0)$, $d=6$ $\mrm{SCFT}$. The original construction proposed in \cite{Hanany:1997gh} in the context of massive type \mrm{IIA} string theory realizes a class of linear quivers which may be regarded as the six-dimensional analog of the constructions of \cite{Hanany:1996ie} obtained in a three-dimensional case. The brane system underlying these $d=6$ field theories is made of $\mrm{NS}5$-, $\mrm{D}6$- and $\mrm{D}8$-branes, placed such in a way as
to preserve eight real supercharges, as shown in table~\ref{Table:GT}.

The most general linear quiver arising as the low-energy description of the brane system in table~\ref{Table:GT} contains $N$ vector multiplet's which stem from $\mrm{D}6$-brane worldvolume dynamics (and giving rise to the 
$\mrm{SU(N)}$ gauge symmetry), hypermultiplet's in the fundamental representation due to the presence of $\mrm{D}8$-branes, and finally a bi-fundamental hypermultiplet's for each $\mrm{NS}5$ and a tensor multiplet for each pair of $\mrm{NS}5$'s. 
In the stringy picture, the real scalar in the tensor multiplet represents the relative distance between the two $\mrm{NS}5$-branes, which happens to be finite in the tensor branch. Subsequently, the fixed point is
reached in the limit where the $\mrm{NS}5$-branes collide and tensionless string states appear in the spectrum of the effective field theory, which is therefore conformal.
This situation is depicted in figure~\ref{fig:(1,0)quiver}.
\begin{figure}[h]
\begin{center}
\scalebox{1}[1]{
\includegraphics[width=120mm]{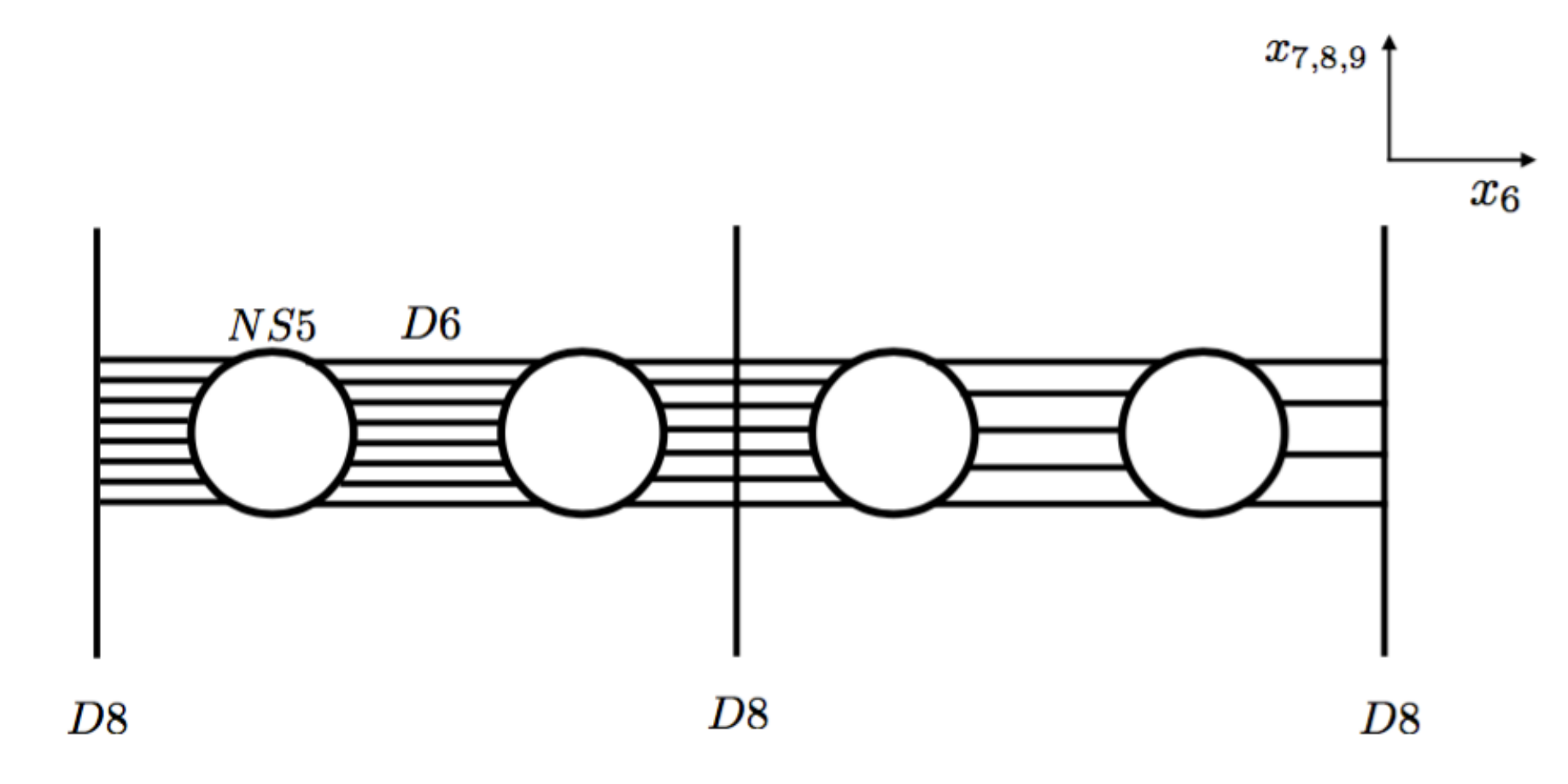}} 
\caption[Intuitive representation of the linear quiver realizing the $\ma N=(1,0)$ $\mrm{SCFT}_6$. Image from \cite{Dibitetto:2017klx}.]{\it The brane picture of the linear quiver realizing the above $\ma N=(1,0)$ $\mrm{SCFT}$'s. In this diagram the $x^{6}$ coordinate represents $z$, while the directions $x^{7,8,9}$ are to be identified with 
$(r,\theta^{1},\theta^{2})$. The $\mrm{D}8$-branes are vertical lines, $\mrm{D}6$-branes are horizontal lines, and finally the ``fat'' bubbles represent point-like $\mrm{NS}5$-branes which collide at the fixed point. Image from \cite{Dibitetto:2017klx}.}
\label{fig:(1,0)quiver}
\end{center}
\end{figure}
As originally conjectured and motivated in \cite{Gaiotto:2014lca}, by making use of the AdS/CFT correspondence, the above linear quivers admit $\textrm{AdS}_{7}$ duals which were originally found in 
\cite{Apruzzi:2013yva} as \mrm{BPS} solutions of massive type \mrm{IIA} supergravity by using the pure spinor formalism (see also \cite{Apruzzi:2014qva,Apruzzi:2015zna,Passias:2016fkm} for further details).
These solutions stem from a warped compactification on a squashed $S^{3}$ obtained from a fibration of a round $S^{2}$ over a segment. In this gravity dual description $\mrm{D}6$-branes fill $\textrm{AdS}_{7}$
and are completely localized at the poles of $S^{3}$, while $\mrm{D}8$-branes wrap $S^{2}$ equators with finite volume. The $\textrm{SU}(2)_{R}$ R-symmetry group emerges here as the unbroken isometry group of 
the $S^{2}$.
In \cite{Gaiotto:2014lca} the relation of these solutions to a near-horizon limit of the above brane system was hinted at and later further clarified in \cite{Bobev:2016phc}.

\subsection{$\mrm{AdS}_3$ Slicing and Uplift to Massive IIA}

Our present goal will be that of giving a holographic description of conformal defects within the $d=6$ $\ma N=(1,0)$ theories which are dual to the aforementioned $\mrm{AdS}_7$ vacua. 
Following the logic in \cite{Clark:2004sb}, this in turn implies that we need to look for Janus-type solutions to the effective supergravity model in question. These solutions describe a foliation of
spacetime in terms of lower-dimensional $\mrm{AdS}_3$ space slices. 

In our setup, we consider the flow obtained in section \ref{AdS3defect} and given in \eqref{AdS3_DW_AdS_U=V}. This solution is asymptotically locally $\textrm{AdS}_{7}$ and furthermore characterized by an $\mrm{AdS}_{3}\times S^{3}$ foliation.
Let us then recall the Ansatz for the seven-dimensional fields
\be
\begin{split}
&\D s^2    =  e^{2U(r)}\,\left( \D s_{\mathrm{AdS}_3}^{2}  +  \D s_{S^{3}}^{2}\right)+ \, e^{2V(r)} \,  \D r^{2}\, , \\[2mm]
&X  =  X(r) \, , \\[2mm]
&B_{(3)}  =  k(r)\, \left(\mrm{vol}_{\mathrm{AdS}_3}  + \, \mrm{vol}_{S^{3}}\right) \, ,
\label{defectAnsatz}
\end{split}
\ee
where $\D s_{\scriptsize{\mrm{AdS}}_{3}}^{2}$ and $\D s_{S^{3}}^{2}$ respectively denote the unit $\mrm{AdS}_{3}$ and the unit $S^{3}$ metric while $\mrm{vol}_{\mathrm{AdS}_3}$ and $ \mrm{vol}_{S^{3}}$ represent the corresponding volume forms. 
If we fix $h=\,\frac{g}{2\sqrt{2}}$, so as to have the $\mrm{SUSY}$ $\mrm{AdS}_7$ vacuum at $X=1$, and we set the radius of $\mathrm{AdS}_3$ as $L=1$, the flow has the following form
\be
\label{seven-dimensionalSOL}
\begin{split}
&e^{2U}  =  \frac{2^{-1/4}}{\sqrt{g}}\,\sqrt{\frac{r}{1\,-\,r^{5}}} \,,\\
& e^{2V}  =  \frac{25}{2g^{2}}\,\frac{r^{6}}{\left(1\,-\,r^{5}\right)^{2}} \,, \\
&k  =  -\frac{2^{1/4}}{g^{3/2}}\sqrt{\frac{r^{5}}{1\,-\,r^{5}}} \,,\\
& X  =  r \,,
\end{split}
\ee
where $r$ ranges from $0$ to $1$.
As we showed in section \ref{AdS3defect}, this solution is asymptotically locally $\textrm{AdS}_{7}$ since, as $r\,\rightarrow\,1^{-}$,
\be
\begin{split}
&X  =  1 \, + \, \mathcal{O}(1-r) \,, \\[2mm]
&R  =  -\frac{21}{4}\,g^{2} \, + \, \mathcal{O}\left((1-r)^{2}\right)\,,
\end{split}
\ee
while as $r\,\rightarrow\,0$ it has an $\mrm{IR}$ singularity of the form
\be
\label{IR_sing}
\begin{split}
&e^{2U}  = 6 \frac{r^{1/2}}{2^{1/4}\sqrt{g}}\, + \, \mathcal{O}(r^{5/2}) \,,\\
&e^{2V}  =  \frac{25r^{6}}{2g^{2}}\, + \, \mathcal{O}(r^{7}) \,,\\
&k  =  -\frac{2^{1/4}}{g^{3/2}}\,r^{5/2} \, + \, \mathcal{O}(r^{7/2}) \,,\\
&  X  =  r \, .
\end{split}
\ee
This solution can be uplifted consistently to massive $\mrm{IIA}$ supergravity \eqref{massiveIIAaction} since minimal $\textrm{SU}(2)$ gauged seven-dimensional $\mathcal{N}=1$ supergravity is known to arise from a consistent truncation of massive type IIA on a squashed $S^{3}$ \cite{Passias:2015gya}.
By employing the uplift formulae to our flow \eqref{seven-dimensionalSOL}, we find a ten-dimensional solution given by
\be
\label{brane_sol}
\begin{split}
&\D s_{10}^{2}  =  \frac{\sqrt{2}}{g}  X^{-1/2}  e^{2A}  \left(e^{2U}\left(\D s_{\scriptsize{\mrm{AdS}}_{3}}^{2} +\D s_{S^{3}}^{2} \right) +  e^{2V} \D r^{2}\right) +  X^{5/2}  \left(\D y^{2}+\frac{1-x^{2}}{16 w}e^{2A}\D s_{S^{2}}^{2} \right),\\
&e^{2\Phi}  =  \frac{8\sqrt{2}}{g^{3}}\,\frac{X^{5/2}}{w}\,e^{2\Phi_0} \,,\\
&B =  \frac{1}{\sqrt{2}g^{3}}\,e^{2A}\,\frac{x\sqrt{1-x^{2}}}{w}\,\mrm{vol}_{S^{2}} 
\,-\, \frac{4\sqrt{2}}{g^{3}}\,e^A\,\D y\,\wedge\,\psi \,,\\
&F_0 =  m  \,,\\
&F_2  =  e^{A-\Phi_0}\,\sqrt{1-x^{2}}\,\left(-\frac{1}{4}\,+\,\frac{m}{\sqrt{2}g^{3}\,w}\,e^{A+\Phi_0}\,x\right)\,\mrm{vol}_{S^{2}}  \,,\\
&F_4 =  -e^{2A-\Phi_0}\,\left(\frac{4\sqrt{2}}{g}X^{4}\,\sqrt{1-x^{2}}\,\D y\,\wedge\,\star_7\mathcal{F}_{(4)}\,+\,\frac{x}{2}\,e^{A}\,\mathcal{F}_{(4)}\right) \,,\\
\end{split}
\ee
with $w\,=\,X^{5}\,\left(1-x^{2}\right)\,+\,x^{2}$, and the 1-form $\psi$ is such that $-2\,\D \psi\,=\,\mrm{vol}_{S^{2}}$.
In the above ten-dimensional solution $U$, $V$, $X$ and $\mathcal{F}_{(4)}$ are the radial functions given in \eqref{seven-dimensionalSOL}, while $A$, $x$ and $\Phi_{0}$ are functions of the $y$ coordinate satisfying the following first-order flow equations
\be\label{eq:oder}
\begin{split} 
&\frac{\D \Phi_{0}}{\D y}  =  \frac{1}{4} \frac{e^{-A}}{\sqrt{1-x^2}} \, \left(12 \, x \, + \, \left(2x^2-5\right)\,m \, e^{A+\Phi_{0}}\right)\,,\\
&\frac{\d x}{\D y}  =  -\frac{1}{2} e^{-A}\sqrt{1-x^2} \, \left(4 \, + \, x \, m\, e^{A+\Phi_{0}}\right) \,,\\
&\frac{\D A}{\D y}  =  \frac{1}{4} \frac{e^{-A}}{\sqrt{1-x^2}} \, \left(4\,x \, - \, m \, e^{A+\Phi_{0}}\right) \,,
\end{split}
\ee
and thus completely specifying the warping. 

\subsection{Brane Picture and Surface Defect}

In this section we will consider a particular bound state of branes in massive $\mrm{IIA}$ realizing the brane picture of the seven-dimensional flow \eqref{seven-dimensionalSOL}. 
In particular we will see how the $\mrm{IR}$ singularity \eqref{IR_sing} can be interpreted as a brane singularity in ten dimensions and how the dyonic $\mathcal{F}_{(4)}$ singularity is related to $\mrm{D}2$- and $\mrm{D}4$-branes filling $\mrm{AdS}_{3}$.
Generically, away from the two above limits, this bound state solution describes a seven-dimensional supersymmetric background obtained as warped product of $\mrm{AdS}_{3}$ times a four-dimensional hyperbolic space constructed as a fibration of $S^{3}$ over a segment.

The brane construction underlying the flow \eqref{seven-dimensionalSOL} can be understood in two steps. The first part of the construction is the one setting up the boundary SCFT, i.e. the 
$\ma N=(1,0)$ theory and it is made of the $\mrm{D}6$-$\mrm{NS}5$-$\mrm{D}8$ bound state as explained in section \ref{ads6/cft7}. 
The corresponding massive type \mrm{IIA} supergravity background can be described in terms of a special class of $ \mrm{BPS}/4$ flows as those introduced in \cite{Imamura:2001cr}.

The second part of our construction is here given by $\mrm{D}2$-branes and wrapped $\mrm{D}4$-branes ending on the above massive brane intersection. The physics of the boundary conditions of these branes in the context of massive type \mrm{IIA} string theory turns out to be described by a $\mrm{SCFT}_{2}$ living on a codimension 4 defect inside the original $d=6$ spacetime.

The complete brane system realizing this mechanism is sketched in table~\ref{Table:branes}.
\begin{table}[h!]
\renewcommand{\arraystretch}{1}
\begin{center}
\scalebox{1}[1]{
\begin{tabular}{c||c c|c c c c|c||c c c}
branes & $t$ & $y$ & $\rho$ & $\varphi^{1}$ & $\varphi^{2}$ & $\varphi^{3}$ & $z$ & $r$ & $\theta^{1}$ & $\theta^{2}$ \\
\hline \hline
$\mrm{NS}5$ & $\times$ & $\times$ & $\times$ & $\times$ & $\times$ & $\times$ & $-$ & $-$ & $-$ & $-$ \\
$\mrm{D}6$ & $\times$ & $\times$ & $\times$ & $\times$ & $\times$ & $\times$ & $\times$ & $-$ & $-$ & $-$ \\
$\mrm{D}8$ & $\times$ & $\times$ & $\times$ & $\times$ & $\times$ & $\times$ & $-$ & $\times$ & $\times$ & $\times$ \\
\hline
$\mrm{D}2$ & $\times$ & $\times$ & $-$ & $-$ & $-$ & $-$ & $\times$ & $-$ & $-$ & $-$ \\
$\mrm{D}4$ & $\times$ & $\times$ & $-$ & $-$ & $-$ & $-$ & $-$ & $\times$ & $\times$ & $\times$ \\
\end{tabular}
}
\end{center}
\caption[The brane picture of the system D2-D4-NS5-D6-D8 underlying the $\mrm{SCFT}_2$ defect theory. Table from \cite{Dibitetto:2017klx}.]{{\it The brane picture underlying the $\mrm{SCFT}_2$ described by $\mrm{D}2$- and $\mrm{D}4$-branes ending on an $\mrm{NS}5$-$\mrm{D}6$-$\mrm{D}8$ intersection. The above system is $\mrm{BPS}/8$. Table from \cite{Dibitetto:2017klx}.}} \label{Table:branes}
\end{table}

The ten-dimensional supergravity background corresponding to the brane system sketched in table~\ref{Table:branes} can be constructed as a non-harmonic superposition of a solution from \cite{Imamura:2001cr} with $\mrm{D}2$ and $\mrm{D}4$-branes, yielding
\be
\label{brane_sol}
\begin{split}
&\D s_{10}^{2}  =  S^{-1/2}H_{\scriptsize{\mrm{D}2}}^{-1/2}H_{\scriptsize{\mrm{D}4}}^{-1/2}\,\D s_{\scriptsize{\mathbb{R}^{1,1}}}^{2}\,+\, 
S^{-1/2}H_{\scriptsize{\mrm{D}2}}^{1/2}H_{\scriptsize{\mrm{D}4}}^{1/2}\,\left(\D \rho^{2}+\rho^{2}\,\D s_{\scriptsize{\mrm{S}^3}}^{2}\right) \,\\
& +  K\,S^{-1/2}H_{\scriptsize{\mrm{D}2}}^{-1/2}H_{\scriptsize{\mrm{D}4}}^{1/2}\,\D z^{2}\,+\,K\,S^{1/2}H_{\scriptsize{\mrm{D}2}}^{1/2}H_{\scriptsize{\mrm{D}4}}^{-1/2}\,\left(\D r^{2}+r^{2}\,\D s_{\scriptsize{\mrm{S}^2}}^{2}\right) \,,\\
&e^{\Phi} =  g_{s}\,K^{1/2}\,S^{-3/4}H_{\scriptsize{\mrm{D}2}}^{1/4}H_{\scriptsize{\mrm{D}4}}^{-1/4}  \,,\\
&H_{3}  =  \frac{\partial}{\partial z}\left(KS\right)\mrm{vol}_{(3)} \,-\,dz\,\wedge\,\star_{3}\, \D K  \,,\\
&F_{0}  =  m  \,,\\
&F_{2}  =  -g_{s}^{-1}\,\star_{3}\, \D S \,,\\
&F_{4}  =  g_{s}^{-1}\,\mrm{vol}_{(1,1)}\,\wedge\,\D z\,\wedge\, \D H_{\scriptsize{\mrm{D}2}}^{-1} \, + \, 
\star_{10}\left(\textrm{vol}_{(1,1)}\,\wedge\,\mrm{vol}_{(3)}\wedge\,\D H_{\scriptsize{\mrm{D}4}}^{-1}\right)  \,,
\end{split}
\ee
where the functions $K(z,r)$ and $S(z,r)$ satsify \cite{Imamura:2001cr}
\be
\begin{split}
&mg_{s}\,K \,-\, \frac{\partial S}{\partial z}  =  0 \, , \\
&\nabla_{(3)}S \, + \, \frac{1}{2}\frac{\partial^{2}}{\partial z^{2}} S^{2}  =  0 \,,
\end{split}
\ee
while 
\be
\begin{split}
&H_{\scriptsize{\mrm{D}2}}(\rho,r) =  \left(1+\frac{Q_{\scriptsize{\mrm{D}4}}}{\rho^{2}}\right)\left(1+\frac{Q_{\scriptsize{\mrm{D}6}}}{r}\right)\,,\\
&H_{\scriptsize{\mrm{D}4}}(\rho) = \left(1+\frac{Q_{\scriptsize{\mrm{D}4}}}{\rho^{2}}\right) \, .
\end{split}
\ee
The above background may be regarded as a massive generalization of the special ``non-standard" $\mrm{D}2$-$\mrm{D}4$-$\mrm{NS}5$-$\mrm{D}6$ intersection found in \cite{Boonstra:1998yu}.

Now we show that in some particular regimes of the coordinates $r$, $z$ and $\rho$, the solution \eqref{brane_sol} is described respectively by an $\mrm{AdS}_7$ and $\mrm{AdS}_3$ vacua. In particular these regimes are obtained in the following way:
\begin{itemize}
\item  {\itshape The $\mrm{AdS}_7$ regime}: Take $K\, \sim \, \frac{2}{z^{3}} \, G\left(\frac{r}{z^{2}}\right)$, and $S\, \sim \, \frac{1}{4r} \, H\left(\frac{r}{z^{2}}\right)$, for some 
suitable functions $G$ and $H$. Now take $\rho\,\rightarrow\,\infty$ (which effectively gets rid of the $\mrm{D}4$-brane charge) and perform the following coordinate redefiniton (see e.g. \cite{Cvetic:2000cj})
\be
\begin{split}
&r^{1/2}  = 6 \frac{\sin\alpha}{\zeta} \, , \\
&z  =  \frac{\cos\alpha}{\zeta} \,,
\end{split}
\ee
after which, upon choosing $H\,=\,1$ and $G\,=\,\frac{1}{2}\,\cos^3\alpha$, the metric of \eqref{brane_sol} in this limit reads
\be
\D s_{10}^{2}\ \sim\ 2\,\cos\alpha \,\left(\tan\alpha\,\D s_{\scriptsize{\mrm{AdS}_7}}^{2} \,+\,\tan\alpha\,\D\alpha^{2}\,+\,\frac{1}{4}\,\sin^{2}\alpha\,\D s_{\scriptsize{S}^2}^{2}\right) \ ,
\ee
which is nothing but $\mrm{AdS}_7\times_w \tilde{S}^{3}$ introduced in \eqref{massiveIIAtruncations}, where $\tilde{S}^{3}$ is a 3-manifold topologically spherical and obtained as a fibration of a round $S^{2}$ over a segment.
\begin{figure}
\begin{center}
\scalebox{1}[1]{\xymatrix{*+[F-,]{\textrm{AdS}_{7}\times_{w}\tilde{S}^{3}} & *+[F-,]{\begin{array}{c} \textrm{pure $\mrm{D}8$}\\ \textrm{limit}\end{array}} \\ 
*+[F-,]{\begin{array}{c} \textrm{$\mrm{D}2$ -- $\mrm{D}4$ -- $\mrm{D}6$ -- $\mrm{NS}5$ -- $\mrm{D}8$}\\ \textrm{bound state}\end{array}} \ar[r]_{\hspace{13mm}\rho}\ar[u]^{\zeta\,\equiv\,\left(r\,+\,z^{2}\right)^{-1/2}}\ar[ru]^{z} 
 &  *+[F-,]{\mrm{AdS}_{3}\times_{w}\mathcal{M}_{7}}
 }}
\end{center}
\caption[Schematic representation of the different limits of the brane system realizing the conformal defect $\mrm{SCFT}_2$ in massive IIA. Image from \cite{Dibitetto:2017klx}.]{{\it The three different limits of the brane system represented in \eqref{brane_sol} depending on the three coordinates $(\rho,z,r)$, respectively yielding $\textrm{AdS}_{7}$, 
the asymptotic domain-wall behavior typical of massive type \mrm{IIA} solutions, and $\mrm{AdS}_{3}$. Each limit is controlled by a different combination of the above coordinates. Image from \cite{Dibitetto:2017klx}.}}\label{fig:three_limits} 
\end{figure}
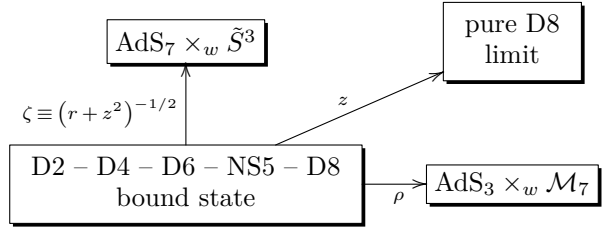
\item { \itshape The $\mrm{AdS}_3$ regime}: Take $z, r \rightarrow \infty$ (while still keeping $\frac{r}{z^{2}}$ finite) and send $\rho\,\rightarrow\,0$. In this limit, the metric \eqref{brane_sol} looks like a warped 
product of an effective seven-dimensional metric and the above $\tilde{S}^{3}$ with the warping parametrized by the $\alpha$ coordinate.
If we focus on the seven-dimensional block of the metric, we find
\be
ds_{7}^{2}\ \sim \ \zeta^{-1/4}\,\underbrace{\left(\frac{\rho^{2}}{Q_{\scriptsize{\mrm{D}4}}}\,\D s_{\scriptsize{\mathbb{R}^{1,1}}}^{2}\,+\,\frac{Q_{\scriptsize{\mrm{D}4}}}{\rho^{2}}\,d\rho^{2}\right)}_{\D s_{\scriptsize{\mrm{AdS}_{3}}}^{2}}
\,+\,\frac{d\zeta^{2}}{\zeta^{2}}\,+\,Q_{\scriptsize{\mrm{D}4}}\,\zeta^{-1/4}\, \D s_{\scriptsize{S^3}}^{2}\ ,
\ee
which is the warped product of $\mrm{AdS}_{3}$ with a four-manifold $\mathcal{M}_{4}$ constructed as a fibration of $S^{3}$ over a segment.
\end{itemize}
To summarize, the brane bound state given in \eqref{brane_sol} comprizes three different limits that can be taken, two of which are holographically relevant. These respectively correspond to 
taking a near-horizon limit of the $\mrm{NS}5$-$\mrm{D}6$-$\mrm{D}8$ bound state to access the $\ma N=(1,0)$ $\mrm{SCFT}_{6}$, and approaching the $\mrm{D}2$-$\mrm{D}4$ bound state to probe the defect $\ma N=(4,0)$ $\mrm{SCFT}_{2}$. 
The third limit just describes how to move far away from all sources in the system to recover the correct domain-wall asymptotic behavior which is expected from the presence of $\mrm{D}8$ branes.
This situation is sketched in figure~\ref{fig:three_limits}.

\subsection{The One-Point Correlation Function}

In this section we will try to test the above holographic picture by sketching the derivation of the one-point correlation functions of the $\ma N=(1,0)$ $\mrm{SCFT}_6$ in presence of the $\ma N=(4,0)$ $\mrm{SCFT_2}$ surface
defect. Even in the case of a non-Lagrangian theory like ours, one can still think of the defect as some operator insertion realizing a deformation in the original $\mrm{SCFT}_6$.
Non-vanishing one-point correlators can be seen as a typical consequence of the fact that $\mrm{SO(2,6)}$ symmetry is broken by the aforementioned deformation.
Following the philosophy in \cite{Clark:2004sb}, we will sketch the computation by using two different methods: a holographic one which uses the standard holographic dictionary to extract the desired information 
from the gravity dual, and a field-theoretical one, which relies on the conformal symmetry preserved by the defect.
Of course, as opposed to \cite{Clark:2004sb}, due to a lack of a Lagrangian description, we will only be able to derive a matching constraint between unknown parameters on the two sides.

Starting from the solution in equation~\eqref{seven-dimensionalSOL}, we first extract the boundary metric to get
\be
\D s^{2}\ = \ F^{-2} \,\bigg(\underbrace{\D s_{\scriptsize{\mathbb{R}^{1,1}}}^{2}+\D \rho^{2}\,+\,\rho^{2}\D s_{\scriptsize{S^3}}^{2}}_{ds_{\scriptsize{\mathbb{R}^{1,5}}}^{2}}\,+\,\rho^{2} \D R^{2}\bigg) \ ,
\ee
where $F\,=\,e^{-U}\,\rho$ and the new coordinate $R$ has been introduced such that $\D R\,\overset{!}{=}\,e^{V}\,\D r$. One gets easily convinced that, by performing the different gauge choice 
$e^{V} \, \overset{\textrm{gauge fix.}}{=} \,1$, the metric in the $(\rho,R)$-plane has a conical defect where the ``angular" coordinate $r$ ranges from $0$ to $1$, thus identifying an angular wedge.

Now pick the scalar $X$ as the responsible for the deformation driving the flow under consideration. Near the boundary, its normalized mass reads \cite{Apruzzi:2016rny}
\be
m_{X}^{2}\,\ell^{2}  =  -8  \overset{!}{=}  \Delta_{X}\,(\Delta_{X}-6) \, ,
\ee
whence $\Delta_{X}\,=\,4$. In terms of the $R$ coordinate its asymptotic bahavior is then given by
\be
X(R) \ \sim \ 1 \, - \, c\,e^{\frac{R}{\sqrt{2}g}} \ , \ \ \textrm{as } \ R\,\rightarrow\,-\infty \ ,
\ee
where $c$ is an arbitrary constant. Now using the holographic prescription we can relate the vacuum expectation value of $X$ to the one-point function of its dual operator $\mathcal{O}_{X}$ through
\be
X(R) \ = \ 1 \, - \, b\,\langle\mathcal{O}_{X}\rangle \, F^{\Delta_{X}} \, + \, \dots \ ,
\ee
whence $\langle\mathcal{O}_{X}\rangle\,=\,\frac{1}{5\sqrt{2}g\,b}\,\rho^{-4}$.

An alternative way of computing correlators for a $\mrm{CFT}$ deformed by some operator insertion is that of using {\itshape conformal perturbation theory}. 
Inspired by cases admitting a Lagrangian description, the aforementioned perturbation may be viewed as an extra term in the Lagrangian of the form $\gamma\,\phi\,\mathcal{O}_{X}$, where $\phi$ is our 
position-dependent coupling (mass dimension $2$), and $\gamma$ is a dimensionless quantity, which is usually related to the anomalous dimension of $\mathcal{O}_{X}$. The only difference here with respect to the aforementioned cases is that the overall 
constant appearing in front of the one-point correlation functions cannot be fixed by using this method since no coupling should be assumed small. 
Based on typical field theory intuition, one can treat the above marginal deformation as a perturbation which appears an operator insertion inside corrections of $n$-point correlators as
\be
\hspace{-4mm}
\begin{split}
\langle\mathcal{O}_{1}(x_{1})\cdots&\mathcal{O}_{n}(x_{n})\rangle_{\textrm{def.}}  =  \langle\mathcal{O}_{1}(x_{1})\cdots\mathcal{O}_{n}(x_{n})\rangle_{0} \\
& +  \gamma\,\int \D^{6}z \,\phi(z)\,\langle\mathcal{O}_{1}(x_{1})\cdots\mathcal{O}_{n}(x_{n})\,\mathcal{O}_{X}(z)\rangle_{0}  \, \\
 &+ \frac{\gamma^{2}}{2!}\,\int \D^{6}z\int d^{6}w \,\phi(z)\phi(w)\,\langle\mathcal{O}_{1}(x_{1})\cdots\mathcal{O}_{n}(x_{n})\,\mathcal{O}_{X}(z)\,\mathcal{O}_{X}(w)\rangle_{0} \, + \, \dots 
\end{split}
\ee
By applying the above formula for a one-point function of $\mathcal{O}_{1}\,=\,\mathcal{O}_{X}$, we find
\be
\langle\mathcal{O}_{X}(\rho)\rangle_{\textrm{def.}}\,=\,\underbrace{\langle\mathcal{O}_{X}(\rho)\rangle_{0}}_{0}
\,+\,\gamma\,\int \D^{6}z\,\phi(z)\,\underbrace{\langle\mathcal{O}_{X}(\rho)\mathcal{O}_{X}(z)\rangle_{0}}_{\frac{a}{|\rho-z|^{8}}}\,+\,\dots
\ee
Now we pick coordinates such that $\D^{6}z\,\rightarrow\,\rho'^{3}\,\D \rho'\,\D s_{\scriptsize{S^3}}\,\D^{2}z$, and we find that we need to have a position-dependent coupling behaving as $\phi(\rho)\,\sim\,\rho^{-2}$ in
order to match the gravity result. By making this assumption, we find
\be
\langle\mathcal{O}_{X}(\rho)\rangle_{\textrm{def.}}\,=\,\frac{\pi^{3}}{30}\,a\gamma\,\rho^{-4}\ ,
\ee
which exactly matches the holographic prediction, provided that $(5\sqrt{2}g\,b)^{-1}\,=\,\frac{\pi^{3}}{30}\,a\gamma$.

%% file: App.tex
\part*{Appendices}

%% file: App1.tex
\chapter{Symplectic-Majorana Spinors in $d=7$}
\label{app:SM_spinors}
\thispagestyle{plain}

In this appendix we summarize the set of relevant conventions concerning irreducible spinors in $1+6$ dimensions and the corresponding representation of the Dirac matrices which we adopt throughout this
work. In $d=7$ with Lorentzian signature, the irreducible spinors are of Dirac type and carry $2^{\textrm{\textlbrackdbl} 7/2 \textrm{\textrbrackdbl}}\,=\,8$ complex components. 
The same degrees of freedom may be then rearranged into a symplectic-Majorana (SM) spinor, i.e. an $\mrm{SU}(2)_{R}$ doublet of spinors $\zeta^{a}$ satisfying a \emph{pseudo-reality condition}
of the form
\be
\label{SM_cond}
\zeta_{a}\ \equiv \ \left(\zeta^{a}\right)^{*} \ \overset{!}{=} \ \epsilon_{ab}\,B\,\zeta^{b} \ ,
\ee
where $\epsilon_{ab}$ denotes the $\mrm{SU}(2)$-invariant Levi-Civita symbol, and $B$ is the matrix that controls complex conjugation of Dirac spinors \cite{VanProeyen:1999ni}. 
Note that the condition \eqref{SM_cond} makes sure that the amount of on-shell real degrees of freedom described by $\zeta$ be $16$.
The Dirac matrices $\left\{\gamma^{m}\right\}_{m\,=\,0,\,\cdots\,6}$ satisfy
\be
\left\{\gamma^{m},\,\gamma^{n}\right\} \ = \ 2\,\eta^{mn} \, \mathbb{I}_{8} \ ,
\ee
where $\eta\,=\,\textrm{diag}(-1,+1,+1,+1,+1,+1,+1)$.

We adopted the following explicit representation for the Clifford algebra \cite{VanProeyen:1999ni}
\be
\label{rep_gamma}
\begin{split}
&\gamma^{0} =  i \, \sigma^{2} \, \otimes \, \mathbb{I}_{2} \, \otimes \, \mathbb{I}_{2} \,,\\
&\gamma^{1}  =  \,\, \sigma^{1} \, \otimes \, \mathbb{I}_{2} \, \otimes \, \mathbb{I}_{2}  \,,\\
&\gamma^{2}  =  \,\, \sigma^{3} \, \otimes \, \sigma^{1} \, \otimes \, \mathbb{I}_{2} \,,\\
&\gamma^{3}  =  \,\, \sigma^{3} \, \otimes \, \sigma^{3} \, \otimes \, \mathbb{I}_{2}  \,,\\
&\gamma^{4}  =  \,\, \sigma^{3} \, \otimes \, \sigma^{2} \, \otimes \, \sigma^{1}  \,,\\
&\gamma^{5}  =  \,\, \sigma^{3} \, \otimes \, \sigma^{2} \, \otimes \, \sigma^{2}  \,,\\
&\gamma^{6}  =  \,\, \sigma^{3} \, \otimes \, \sigma^{2} \, \otimes \, \sigma^{3}  \,,
\end{split}
\ee
where $\left\{\sigma^{i}\right\}_{i\,=\,1,\,2,\,3}$ are the Pauli matrices
\be
\label{Pauli}
\s^{1} =
\left(
\begin{array}{cc}
0 & 1  \\
1 & 0
\end{array}
\right)
\,, \hspace{10mm}
\s^{2} =
\left(
\begin{array}{cc}
0 & -i  \\
i & 0
\end{array}
\right)
\,, \hspace{10mm}
\s^{3} =
\left(
\begin{array}{cc}
1 & 0  \\
0 & -1
\end{array}
\right) \ .
\ee
One can check that the representation given in \eqref{rep_gamma} satisfies the following identity
\be
\g_{*} \ \equiv \ \g^{0}\,\g^{1}\,\g^{2}\,\g^{3}\,\g^{4}\,\g^{5}\,\g^{6} \ = \ \mathbb{I}_{8} \ .
\ee
In this spacetime signature the matrices $A$, $B$ and $C$ which respectively realize Dirac, complex and charge conjugation of spinors, have the following defining properties \cite{VanProeyen:1999ni}
\be
\begin{array}{lclclclclclc}
\left(\g^{m}\right)^{\dagger} & = & -A \, \g^{m} \, A^{-1} \, , & \left(\g^{m}\right)^{*} & = & B \, \g^{m} \, B^{-1} \, , & \left(\g^{m}\right)^{T} & = & -C \, \g^{m} \, C^{-1} \, . 
\end{array}
\ee
One can check that a consistent choice for the above operators with respect to our representation given in \eqref{rep_gamma} is given by
\be
\begin{array}{lclclc}
A \ = \ \g^{0} \, , & &B \ = \ -i \, \g^{46}  \, , & &C \ = \ i \, \g^{046} \, , 
\end{array}
\ee
which satisfy
\be
\begin{array}{lclclc}
B^{T} \ = \ C \, A^{-1} \, ,& & B^{*}\,B \ = \ -\mathbb{I}_{8}  \, ,& & C^{T} \ = \ -C^{-1} \ = \ -C^{\dagger} \ = \ C \, .
\end{array}
\ee

%% file: App2.tex
\chapter{Flow Equations in $d=7$ and SU(2) Vectors}
\label{floweqvector7d}
\thispagestyle{plain}

In this appendix the systems of the first-order equations that are numerically integrated in \ref{mkw21vector} and \ref{AdS3vector} are presented.

\section{Flow Equations and SU(2) Vectors: $\mathbb{R}^{1,2}\times S^3$}
\label{floweq:mk21_S3_vector}

In what follows the first-order equations that are integrated numerically in section \ref{mkw21vector} are given.
Let's consider the Ansatz \eqref{ansatz:mkw21vector} with Killing spinor $\zeta^a$ of the form \eqref{Kspinor} and satisfying \eqref{gamma3proj}, \eqref{vectorproj}.
From \eqref{SUSY_eqns_7D} we obtain the following system of BPS equations, 
\begin{equation}
\begin{split}
\label{flow:mkw21vector}
U'\,=\,&\frac{ e^{V-2 W}}{80\, g\, X\,\cos (2 \theta )}\,\biggl[ 6\, \kappa \, A \,\left(4\, g\, e^{W}\, X\, \sin (4 \theta)+\sqrt{2}\, \kappa\,  (\cos (4 \theta )-3)\right)\\
&-3\, \sqrt{2}\, \kappa ^2\, A^2\, (\cos (4 \theta)-3)+ 8\, g\, e^{W}\,   X \,\left(2 \,e^{W}\, f\, (3\, \cos (4 \theta )-1)-3\, \kappa \, \sin (4 \theta )\right)]\biggr]\,,\\
W'\,=\,& \frac{ e^{V-2 W}}{40\, g\, X\, \cos (2 \theta)}\, \biggl[-2\, \kappa \, A\, \left(4 \,g\, e^{W} X\, \sin (4 \theta )+\sqrt{2}\, \kappa \, (\cos (4 \theta)-8)\right)\\
&+\sqrt{2}\, \kappa ^2 \,A^2 \,(\cos (4 \theta)-8)-8\, g\, e^{W} \,X \, \left(2 e^{W} \,f\, (\cos (4 \theta)-2)-\kappa \,  \sin (4 \theta)\right)\biggr]\,,\\
Y'\,=\,&\frac{ e^{V-2 W}\,Y}{160\, g\, X\,\cos (2 \theta )}\,\biggl[ 6\, \kappa \, A \,\left(4\, g\, e^{W}\, X\, \sin (4 \theta)+\sqrt{2}\, \kappa\,  (\cos (4 \theta )-3)\right)\\
&-3\, \sqrt{2}\, \kappa ^2\, A^2\, (\cos (4 \theta)-3)+ 8\, g\, e^{W}\,   X \,\left(2 \,e^{W}\, f\, (3\, \cos (4 \theta )-1)-3\, \kappa \, \sin (4 \theta )\right)]\biggr]\,,\\
\theta'\,=\,&\frac{ e^{V-2 W}}{80\, g\, X}\, \biggl[-30\, \kappa \, A \left(\sqrt{2}\, \kappa \, \sin (2 \theta )-4\, g \,e^{W} \,X \,\cos (2 \theta )\right)+15\, \sqrt{2}\, \kappa ^2\, A^2 \,\sin (2 \theta)\\
&-8\, g\, e^{W}\, X\,   \left(4 \,e^{W} \,X\, \sin (2 \theta )\, D_X\,f+26\, e^{W} \,f\, \sin (2 \theta)+15\, \kappa \, \cos (2 \theta)\right)\biggr]\,,\\
k'\,=\,& -\frac{3 \,e^{3 U+V-2 W}}{2\, g\, X^3}\,\biggl[6\, \kappa \, A \,\left(\sqrt{2} \,\kappa \, \tan (2 \theta)-2 \, g\, e^{W}\, X\right)\\
&-3 \, \sqrt{2}\, \kappa ^2\, A^2\, \tan (2 \theta )+4\, g\, e^{W}\, X\, \left(4\, e^{W} \,f\,
   \tan (2 \theta)+3 \,\kappa \,\right)\biggr]\,,\\
l'\,=\,&\frac{24 \, \sin (2 \theta ) \,e^{V+3 W} }{5\, X^2}\,\left(f\,-\,X\, D_X\,f\right)\,,\\ \nonumber
\end{split}
\end{equation}
\begin{equation}
\begin{split}
X'\,=\,&\biggl[- 6\, \kappa \, A \, \left(4 \, g \, e^{W} \,X \, \sin (4 \theta )+\sqrt{2} \, \kappa \, (\cos (4 \theta )-3)\right)
+3 \, \sqrt{2} \,\kappa ^2 \,A^2\, (\cos (4 \theta )-3)\\
&-8\, g\, e^{W}\,  X\, \left(4 \,e^{W} \,X\, \cos ^2(2 \theta) \,D_X\,f-8\, e^{W}\, f\, \sin ^2(2 \theta )-3\, \kappa \, \sin (4 \theta)\right)\biggr]\,\frac{ e^{V-2 W}}{80\, g\,\cos (2 \theta )} \,,\\
A'\,=\,&-\frac{ e^{V- W}}{10\,\sqrt{2}\,\kappa}\, \biggl[10\, \kappa \, A\, \left(\sqrt{2}\, \kappa \, \sin (2 \theta )-4\, g\, e^{W}\, X\, \cos (2 \theta)\right)-5\, \sqrt{2} \,\kappa ^2 \,A^2\, \sin (2 \theta)\\
&+8 \,g\, e^{W}\, X\,  \left(2\, e^{W} \,X\, \sin (2 \theta) \,D_X\,f+8\, e^{W}\, f\, \sin (2 \theta )+5 \,\kappa \, \cos (2 \theta )\right) \biggr]\,. 
\end{split}
\end{equation}

\section{Flow Equations and SU(2) Vectors: $\mathrm{AdS}_{3}\times S^3$}
\label{floweq:AdS3_S3_vector}

In this appendix we present the first-order equations that are integrated numerically in section \ref{AdS3vector}.
Given the Ansatz \eqref{ansatz:AdS3vector} with the usual Killing spinor of the form \eqref{Kspinor} and satisfying \eqref{gamma3proj}, \eqref{vectorproj}, the system of BPS equations is given by,
\begin{equation}
\begin{split}
\label{flow:AdS3vector}
 U'\,=\,&\frac{ e^{V-2 W}}{80\, g\, X\,\cos (2 \theta )}\,\biggl[ 6\, \kappa \, A \,\left(4\, g\, e^{W}\, X\, \sin (4 \theta)+\sqrt{2}\, \kappa\,  (\cos (4 \theta )-3)\right)-3\, \sqrt{2}\, \kappa ^2\, A^2\, (\cos (4 \theta)-3)\\
&+ 8\, g\, e^{W}\,   X \,\left(2 \,e^{W}\, f\, (3\, \cos (4 \theta )-1)-3\, \kappa \, \sin (4 \theta )+4\,e^{W-U}\,L\,\sin(2\theta)\right)\biggr]\,,\\
W'\,=\,& \frac{ e^{V-2 W}}{40\, g\, X\, \cos (2 \theta)}\, \biggl[-2\, \kappa \, A\, \left(4 \,g\, e^{W} X\, \sin (4 \theta )+\sqrt{2}\, \kappa \, (\cos (4 \theta)-8)\right)+\sqrt{2}\, \kappa ^2 \,A^2 \,(\cos (4 \theta)-8)\\
&-8\, g\, e^{W} \,X \, \left(2 e^{W} \,f\, (\cos (4 \theta)-2)-\kappa \,  \sin (4 \theta)+3\,e^{W-U}\,L\,\sin(2\theta)\right)\biggr]\,,\\
Y'\,=\,&\frac{ e^{V-2 W}\,Y}{160\, g\, X\,\cos (2 \theta )}\,\biggl[ 6\, \kappa \, A \,\left(4\, g\, e^{W}\, X\, \sin (4 \theta)+\sqrt{2}\, \kappa\,  (\cos (4 \theta )-3)\right)-3\, \sqrt{2}\, \kappa ^2\, A^2\, (\cos (4 \theta)-3)\\
&+ 8\, g\, e^{W}\,   X \,\left(2 \,e^{W}\, f\, (3\, \cos (4 \theta )-1)-3\, \kappa \, \sin (4 \theta )+4\,e^{W-U}\,L\,\sin(2\theta)\right)\biggr]\,,\\
\theta'\,=\,&\frac{ e^{V-2 W}}{80\, g\, X}\, \biggl[-30\, \kappa \, A \left(\sqrt{2}\, \kappa \, \sin (2 \theta )-4\, g \,e^{W} \,X \,\cos (2 \theta )\right)+15\, \sqrt{2}\, \kappa ^2\, A^2 \,\sin (2 \theta)\\
&-8\, g\, e^{W}\, X\,   \left(4 \,e^{W} \,X\, \sin (2 \theta )\, D_X\,f+26\, e^{W} \,f\, \sin (2 \theta)+15\, \kappa \, \cos (2 \theta)-15\, L\, e^{W-U}\right)\biggr]\,,\\
k'\,=\,& -\frac{3 \,e^{3 U+V-2 W}}{2\, g\, X^3}\,\biggl[6\, \kappa \, A \,\left(\sqrt{2} \,\kappa \, \tan (2 \theta)-2 \, g\, e^{W}\, X\right)\\ 
&-3 \, \sqrt{2}\, \kappa ^2\, A^2\, \tan (2 \theta )+4\, g\, e^{W}\, X\, \left(4\, e^{W} \,f\,
   \tan (2 \theta)+3 \,\kappa \,-2\, L\, e^{W-U} \,\sec (2 \theta )\right)\biggr]\,,\nonumber
\end{split}
\end{equation}
\begin{equation}
\begin{split}
l'\,=\,&-\frac{6  \,e^{V+3 W} }{5\, X^2}\,\biggl[5\,L\,e^{-U}-4\, \sin (2 \theta )\,\left(f\,-\,X\, D_X\,f\right)\biggr]\,,\\
X'\,=\,&\frac{ e^{V-2 W}}{80\, g\,\cos (2 \theta )}\,\biggl[- 6\, \kappa \, A \, \left(4 \, g \, e^{W} \,X \, \sin (4 \theta )+\sqrt{2} \, \kappa \, (\cos (4 \theta )-3)\right)+3 \, \sqrt{2} \,\kappa ^2 \,A^2\, (\cos (4 \theta )-3)\\
&-8\, g\, e^{W}\,  X\, \left(4 \,e^{W} \,X\, \cos ^2(2 \theta) \,D_X\,f-8\, e^{W}\, f\, \sin ^2(2 \theta )-3\, \kappa \, \sin (4 \theta)+4\, L\, e^{W-U}\, \sin (2 \theta)\right)\biggr] \,,\\
A'\,=\,&-\frac{ e^{V- W}}{10\,\sqrt{2}\,\kappa}\, \biggl[10\, \kappa \, A\, \left(\sqrt{2}\, \kappa \, \sin (2 \theta )-4\, g\, e^{W}\, X\, \cos (2 \theta)\right)-5\, \sqrt{2} \,\kappa ^2 \,A^2\, \sin (2 \theta)\\
&+8 \,g\, e^{W}\, X\,  \left(2\, e^{W} \,X\, \sin (2 \theta) \,D_X\,f+8\, e^{W}\, f\, \sin (2 \theta )+5 \,\kappa \, \cos (2 \theta )-5 L e^{W-U}\right) \biggr]\,.\\  
\end{split}
\end{equation}

%% file: pubs.tex
\chapter*{Publications}
\thispagestyle{plain}
\vspace{0.2cm}

\begin{enumerate}

 \item G.~Dibitetto and N.~Petri, {\itshape 6d surface defects from massive type IIA}, JHEP {\bf 1801} (2018) 039, hep-th/1707.06154. 
 \vspace{2mm}

 \item G.~Dibitetto and N.~Petri, {\itshape BPS objects in D = 7 supergravity and their M-theory origin}, JHEP {\bf 1712} (2017) 041, hep-th/1707.06152.
 \vspace{2mm}

  \item  D.~Klemm, A.~Marrani, N.~Petri and M.~Rabbiosi, {\itshape Nonlinear symmetries of black hole entropy in gauged supergravity}, JHEP {\bf 1704} (2017) 013, hep-th/1701.08536.
   \vspace{2mm}
 
 \item D.~Klemm, N.~Petri and M.~Rabbiosi, { \itshape Black string first order flow in $N = 2, d = 5$ abelian gauged supergravity}, JHEP {\bf 1701} (2017) 106, hep-th/1610.07367.
  \vspace{2mm}
\item D.~Klemm, N.~Petri and M.~Rabbiosi, {\itshape Symplectically invariant flow equations for $N = 2$, $D = 4$ gauged supergravity with hypermultiplets}, JHEP {\bf 1604} (2016) 008, hep-th/1602.01334.
 \vspace{2mm}
 \item D.~Klemm, A.~Marrani, N.~Petri and C.~Santoli, {\itshape BPS black holes in a non-homogeneous deformation of the stu model of $N=2$, $D=4$ gauged supergravity}, JHEP {\bf 1509} (2015) 205, hep-th/1507.05553.
 \vspace{2mm}
  \item S.~Chimento, D.~Klemm and N.~Petri, {\itshape Supersymmetric black holes and attractors in gauged supergravity with hypermultiplets}, JHEP {\bf 1506} (2015) 150, hep-th/1503.09055.
 
\end{enumerate}